\documentclass[12pt]{article}  
\usepackage{cite}
\usepackage{epsfig}
\usepackage{graphicx}
\usepackage{amsmath}
\usepackage{amssymb}
\usepackage{color}  

\usepackage{a41}
\usepackage{color}
\usepackage[rflt]{floatflt}              
\usepackage{float}
\usepackage{slashed}

\usepackage{breqn}
  
\def\asr{\left( \frac{\alpha_s}{4 \pi} \right)}
\def\b0{\beta_0}

\def\q{\slashed{q}}

\usepackage{array}

\usepackage[english]{babel}
\usepackage[latin1]{inputenc}
\usepackage[T1]{fontenc}
\usepackage{ae}

\usepackage{url}

\usepackage{amsmath, amsthm, amssymb}
\newtheorem{thm}{Theorem}[section]

\newtheorem{definition}[thm]{Definition}

 \newcommand{\GeV}{\mathrm{GeV}}

\newcommand{\abs}[1]{\left|{#1}\right|}

\newcommand{\F}{{\mathcal F}}

\newcommand{\ep}{\varepsilon}

\usepackage{rotating}

\usepackage{graphicx}

\newcounter{mmacnt}
\def\restartmma{\setcounter{mmacnt}{0}}
\restartmma \catcode`|=\active
\def|#1|{\mathrm{#1}}
\catcode`|=12
\newenvironment{mma}{
 \par\smallskip
 \catcode`|=\active
 \parskip=0pt\parindent=0pt 
 \small
 \def\In##1\\{%
\def\linebreak{\hfill\break\null\qquad}%
\refstepcounter{mmacnt}
\hangindent=2.5em\hangafter=0
\leavevmode
\llap{\tiny\sffamily n[\arabic{mmacnt}]:=\kern.5em}%
\mathversion{bold}\footnotesize$\displaystyle##1$\normalsize
\mathversion{normal}\par
 }%
 \def\Print##1\\{%
\def\linebreak{\hfill\break}%
\hangindent=2.5em\hangafter=0
\leavevmode ##1\par}%
 \def\Out##1\\{%
\def\linebreak{$\hfill\break\null\hfill$}%
\kern\abovedisplayskip\par
\hangindent=2.5em\hangafter=0
\leavevmode
\llap{\tiny\sffamily Out[\arabic{mmacnt}]=\kern.5em}
\footnotesize$\displaystyle##1$\normalsize\hfill\null\par
\kern\belowdisplayskip
 }%
 \def\Warning##1##2\\{%
\def\linebreak{\hfill\break}%
\hangindent=2.5em\hangafter=0
\leavevmode
{\scriptsize##1 : ##2}\par}%
}{%
 \par\smallskip
}

\usepackage{color}

\newenvironment{fshaded}{%
\MakeFramed {\FrameRestore}
}%
{\endMakeFramed}

\allowdisplaybreaks[4]

\begin{document}
\setlength{\baselineskip}{0.515cm}
\sloppy
\thispagestyle{empty}
\begin{flushleft}
DESY 17-102\\DO-TH 17/16\\Nikhef 2017-063\\TTK-17-44
\end{flushleft}

\mbox{}
\vspace*{\fill}
\begin{center}

{\LARGE\bf The Heavy Quark Form Factors at Two Loops}

\vspace*{3mm} 

\vspace{3cm}
\large
{\large 
J.~Ablinger$^a$,
A.~Behring$^{b,c}$,
J.~Bl\"umlein$^b$, 
G.~Falcioni$^{b,d}$,\\
A.~De Freitas$^b$, 
P.~Marquard$^b$,
N.~Rana$^b$
and 
C.~Schneider$^a$
}

\vspace{1.cm}
\normalsize
{\it $^a$~Research Institute for Symbolic Computation (RISC),\\
  Johannes Kepler University, Altenbergerstra{\ss}e 69,
  A--4040, Linz, Austria}

\vspace*{3mm}
{\it  $^b$ Deutsches Elektronen--Synchrotron, DESY,}\\
{\it  Platanenallee 6, D-15738 Zeuthen, Germany}

\vspace*{3mm}
{\it $^c$~Institut f\"ur Theoretische Teilchenphysik und Kosmologie,\\ 
RWTH Aachen University, D-52056 Aachen, Germany}

\vspace*{3mm}
{\it $^d$~Nikhef Theory Group, Science Park 105, 1098 XG Amsterdam, The Netherlands}


\end{center}
\normalsize
\vspace{\fill}
\begin{abstract}
\noindent
 We compute the two-loop QCD corrections to the heavy quark form factors in case of the vector, 
 axial-vector, scalar and pseudo-scalar currents up to second order in the dimensional 
 parameter $\ep = (4-D)/2$. These terms are required in the renormalization of the higher order 
 corrections to these form factors.
\end{abstract}

\vspace*{\fill}
\noindent
\numberwithin{equation}{section}
\newpage 

\section{Introduction}
\label{sec:1}

\vspace*{1mm}
\noindent
Since its discovery \cite{Abe:1995hr,D0:1995jca} in 1995, the top quark has been 
studied extensively both in theoretical and experimental premises. These studies of the heaviest 
particle of the Standard Model (SM) provide as well a detailed probe onto some aspects of 
electro-weak symmetry breaking (EWSB). Due to its very short lifetime, the top quark decays 
before hadronizing, and thus provides a window to study its production dynamics widely
without accounting for the hadronization effects and therefore as a quark being directly accessible 
as such.\footnote{The reconstruction of the $t$-quark
itself, and in particular its mass, however, requires to study the hadronization effects.}
The experiments carried out at the Tevatron and later at the LHC, have already measured many 
observables which allow to extract the properties of the top quark with remarkable 
accuracy. Compared to the Tevatron, the LHC offers an abundant rate of top quark pair 
and single production, hence providing a perfect ground for precision tests. Due to the combined effort 
from 
both the theoretical and experimental sides, a striking accuracy has been achieved in many 
observables, e.g.~the uncertainties on the predictions of the inclusive production cross-section of a 
top quark pair, now are around 5 $\%$ at a fixed top quark mass of $m_t = 172.5~\GeV$.
While these precise measurements provide a strong ground for testing the predictions within the 
SM, beyond the Standard Model (BSM) physics scenarios can as well hide under those small 
uncertainties.
To find a hint of BSM physics or to rule some hypotheses out, we need more precision 
and certainly a future linear or circular $e^+e^-$ collider can achieve that. In order 
to match the experimental accuracy, precise predictions are required on the theoretical side.

In this paper, we focus on perturbative Quantum Chromodynamics (QCD) corrections to 
the form factors involving heavy quarks which are basic building blocks 
of various physical quantities concerning top quark pair production.
The massive vector and axial vector form factors play an important role in the forward-backward asymmetry 
of bottom or top quark production at electron-positron colliders. Likewise
the decay of a scalar or pseudo-scalar particle to a pair of heavy quarks
could also play a very important role in shedding light on the quantum nature of the Higgs boson.
There are also static quantities like the anomalous magnetic moment, which receive
contributions from such massive form factors. For these reasons, the phenomenology and 
the perturbative QCD corrections to these form factors
have gained much attention during the last decade.

In Refs.~\cite{Arbuzov:1991pr, Djouadi:1994wt}, the first order QCD corrections were obtained for 
the vector and axial-vector form factors. 
A massless approximation was considered to obtain the next-to-next-to-leading order (NNLO)
QCD corrections in \cite{Altarelli:1992fs} numerically, later followed by an analytic computation in \cite{Ravindran:1998jw}.
Another numerical computation was performed in \cite{Catani:1999nf} at NNLO using a different 
formalism.
On the other hand, the next-to-leading order (NLO) contributions to the scalar and pseudo-scalar form factors
were known \cite{Braaten:1980yq, Sakai:1980fa, Inami:1980qp, Drees:1990dq} for long and NNLO corrections 
by employing quark mass expansion to various orders, 
c.f.~\cite{Gorishnii:1983cu,Gorishnii:1991zr,Surguladze:1994em,Surguladze:1994gc,Larin:1995sq,Chetyrkin:1995pd,Harlander:1997xa,Harlander:2003ai}. 
A series of papers followed obtaining the two-loop QCD corrections for the vector form factor \cite{Bernreuther:2004ih}, 
the axial-vector form factor \cite{Bernreuther:2004th}, the anomaly contributions \cite{Bernreuther:2005rw}
and the scalar and pseudo-scalar form factors \cite{Bernreuther:2005gw}. 
An independent cross-check of the vector form factor has been performed in \cite{Gluza:2009yy}
with the addition of the ${\mathcal O}(\ep)$ contribution, where $\ep = (4-D)/2$ and $D$ is the 
space-time dimension.
Recently, the calculation of a subset of the three-loop master integrals \cite{Henn:2016kjz} has made it possible 
to obtain the vector form factor 
at three loops \cite{Henn:2016tyf} in the color-planar limit. The large $\beta_0$ limit has been considered in \cite{Grozin:2017aty}.
While the main goal is to obtain the complete three-loop corrections for the form factors, the ${\cal O}(\ep)$ pieces 
at two-loop order are important ingredients.
In addition, we compute the master integrals to the required order in $\ep$ with a different technique. 

In the present paper, we compute the contributions to the massive form factors
up to ${\cal O}(\ep^2)$ for different currents, namely, vector, axial-vector, scalar 
and pseudo-scalar currents, which serve as input for ongoing and future 3- and 4-loop calculations.
We also perform the expansion of the exact results in different kinematic regions.
In Section~\ref{sec:theory}, we briefly describe the theoretical formalism for all the 
currents and corresponding
form factors, followed by their renormalization procedure and a description of the 
universal infrared (IR) structure. 
Section~\ref{sec:comp} contains the details about the computational technique. Here we 
describe especially how we 
have computed the
master integrals. In Section~\ref{sec:result} we present the results, also expanding the complete 
expressions in regions, which are kinematically relevant. Finally we conclude in 
Section~\ref{sec:conclu}. Various of the expressions are rather voluminous. A part of it is presented 
in
the appendices and the $O(\varepsilon^2)$ terms are only given in computer readable form in a file 
attached to this paper.

\section{The heavy quark form factors} \label{sec:theory}

\vspace*{1mm}
\noindent
We consider the decay of a virtual massive boson of momentum $q$ into
a pair of heavy quarks of mass $m$, momenta $q_1$ and $q_2$ and color
$c$ and $d$, through a vertex $X_{cd}$, where
$X_{cd} = \Gamma^{\mu}_{V,cd}, \Gamma^{\mu}_{A,cd}, \Gamma_{S,cd}$ and
$\Gamma_{P,cd}$ indicates the coupling to a vector, an axial-vector, a
scalar and a pseudo-scalar boson, respectively.  Here 
$q^2 = 
(q_1+q_2)^2$ is the center of mass
energy squared and the dimensionless variable $s$ is defined by
\begin{equation}
 s = \frac{q^2}{m^2}\,.
\end{equation}
The amplitude takes the following general form 
\begin{equation}
 \bar{u}_c (q_1) X_{cd} v_d (q_2) \,,
\end{equation}
where $\bar{u}_c (q_1)$ and $v_d (q_2)$ are the bi-spinors of the quark and the anti-quark, 
respectively.
We denote the corresponding UV renormalized form factors by $F_{I}$, with $I = V, A, S, P$.
They are expanded in the strong coupling constant $\alpha_s = g_s^2/(4\pi)$ as follows
\begin{equation}
 F_{I} = \sum_{n=0}^{\infty} \asr^n F_{I}^{(n)} \,.
\end{equation}
The unrenormalized form factors are denoted by $\hat{F}_{I}$. 
In the following sub-sections, we discuss the  properties of each current, 
the corresponding renormalization procedure and their universal infrared structure.

\subsection{The vector and axial-vector current}

\vspace*{1mm}
\noindent
In this section, we consider the spin 1 case, i.e. the vertex $\Gamma^{\mu}$ of a $Z$-boson or photon
coupling to a pair of heavy quarks. 
The general structure of $\Gamma^{\mu}$ consists of six form factors, 
two of which are CP odd. We consider only higher order QCD effects 
and SM neutral current interactions to lowest order. Since CP invariance holds, we only take into 
account the four 
CP even form factors, $F_{V,i} (s), F_{A,i} (s) ~ i = 1,2$ in the following.
They can be cast in the following general form
\begin{align}
 \Gamma_{cd}^{\mu} &= \Gamma_{V,cd}^{\mu} + \Gamma_{A,cd}^{\mu}
 \nonumber\\
 &= -i \delta_{cd}\Big[ v_Q \Big( \gamma^{\mu}~F_{V,1} + \frac{i}{2 m} \sigma^{\mu\nu} q_{\nu} ~F_{V,2} \Big)
+ a_Q \Big( \gamma^{\mu} \gamma_5~F_{A,1} 
         + \frac{1}{2 m} q^{\mu} \gamma_5 ~ F_{A,2}  \Big) \Big], 
\end{align}
where $\sigma^{\mu\nu} = \frac{i}{2} [\gamma^{\mu},\gamma^{\nu}]$, $q=q_1+q_2$, and $v_Q$ and $a_Q$ are the 
SM vector and axial-vector coupling constants as defined by 
\begin{equation}
 v_Q = \frac{e}{\sin \theta_w \cos \theta_w} \Big( \frac{T_3^Q}{2} - \sin^2 \theta_w Q_Q \Big) \,,
 \qquad
 a_Q = - \frac{e}{\sin \theta_w \cos \theta_w} \frac{T_3^Q}{2} \,.
\end{equation}
$e$ denotes the elementary charge, $\theta_w$ the weak mixing angle, $T_3^Q$ the third component 
of the weak isospin, and $Q_Q$ the charge of the heavy quark.

To extract the form factors $F_{I,i}, ~ I=V,A$, we multiply
$\Gamma^{\mu}$ by the following projectors and perform a trace over
the spinor and color indices
\begin{align}
 P_{V,i} &= \frac{i}{v_Q} \frac{\delta_{cd}}{N_c} \frac{\q_2 - m}{m} \Big( \gamma_{\mu} g_{V,i}^{1} + \frac{1}{2 m} (q_{2 \mu} - q_{1 \mu}) g_{V,i}^{2} \Big) \frac{\q_1 + m}{m} \,,
 \nonumber\\
 P_{A,i} &= \frac{i}{a_Q} \frac{\delta_{cd}}{N_c} \frac{\q_2 - m}{m} \Big( \gamma_{\mu} \gamma_5 g_{A,i}^{1} 
          + \frac{1}{2 m} (q_{1 \mu} + q_{2 \mu}) \gamma_5 g_{A,i}^{2} \Big) \frac{\q_1 + m}{m} \,,
\end{align}
where, {\footnote{In \cite{Bernreuther:2004ih}, the expression for $g_{V,2}^{2}$ contains a 
typographical error.}}
\begin{align}
 &g_{V,1}^{1} = - \frac{1}{4 (1-\ep)} \frac{1}{(s-4)} \,,
& &g_{V,1}^{2} = \frac{(3-2\ep)}{(1-\ep)} \frac{1}{(s-4)^2} \,,
 \nonumber\\ 
 &g_{V,2}^{1} = \frac{1}{(1-\ep)} \frac{1}{s(s-4)} \,,
& &g_{V,2}^{2} = - \frac{1}{(1-\ep)} \frac{1}{(s-4)^2} \Bigg( \frac{4}{s} + 2 - 2 \ep \Bigg) \,,
 \nonumber\\ 
 &g_{A,1}^{1} = - \frac{1}{4 (1-\ep)} \frac{1}{(s-4)} \,,
& &g_{A,1}^{2} = - \frac{1}{(1-\ep)} \frac{1}{s(s-4)} \,,
 \nonumber\\ 
 &g_{A,2}^{1} = \frac{1}{(1-\ep)} \frac{1}{s(s-4)} \,,
& &g_{A,2}^{2} = \frac{1}{(1-\ep)} \frac{1}{s^2 (s-4)} \Bigg( 4 (3-2 \ep) - 2 s (1-\ep) \Bigg) \,,
\end{align}
and $N_c$ denotes the number of colors. Later on we will also use the Casimir operators
$C_A = N_c, C_F = (N_c^2-1)/(2 N_c), T_F = 1/2$ for $SU(N_c)$, with $N_c = 3$ in the case of QCD.

\subsection{The scalar and pseudo-scalar current}

\vspace*{1mm}
\noindent
We consider the current implied by a general neutral spin 0 particle $h$
that couples to heavy quarks through the Yukawa interaction
\begin{equation}
 {\cal L}_{int} = - \frac{m}{v} \Big[ s_Q \bar{Q}Q + i p_Q \bar{Q} \gamma_5 Q  \Big] h,
\end{equation}
where $v = (\sqrt{2} G_F)^{-1/2}$ is
the SM Higgs vacuum expectation value, with $G_F$ being the Fermi
constant, $s_Q$ and $p_Q$ are the scalar and pseudo-scalar coupling,
respectively, and $Q$ and $h$ are the heavy quark and scalar and
pseudo-scalar field, respectively.
The vertex for $h \rightarrow \bar{Q} + Q$,
$X_{cd} \equiv \Gamma_{cd}$ consists of two form factors with the
following general structure
\begin{align}
 \Gamma_{cd} &= \Gamma_{S,cd} +  \Gamma_{P,cd}
 \nonumber\\
 &= - \frac{m}{v} \delta_{cd} ~ \Big[ s_Q \, F_{S} + i p_Q \gamma_5 \, F_{P} \Big] \,,
\end{align}
where $F_S$ and $F_P$ denote the renormalized scalar and pseudo-scalar
form factors, respectively. As before, the form factors can be
obtained from $\Gamma_{cd}$ through suitable projectors as given below
and performing the trace over the spinor and color indices 
\begin{align}
 P_{S} &= \frac{v}{2 m s_Q} \frac{\delta_{cd}}{N_c} \frac{\q_2 - m}{m} \Bigg( - \frac{1}{(s-4)} \Bigg) \frac{\q_1 + m}{m} \,,
 \nonumber\\
 P_{P} &= \frac{v}{2 m p_Q} \frac{\delta_{cd}}{N_c} \frac{\q_2 - m}{m} \Bigg( - \frac{i}{s} \gamma_5 \Bigg) \frac{\q_1 + m}{m} \,.
\end{align}

\subsection{Anomaly and Ward identities}
\newcommand{\ns}{\mathrm{ns}}
\newcommand{\sing}{\mathrm{s}}

\vspace*{1mm}
\noindent
Since we use dimensional regularization \cite{tHooft:1972tcz} in $D=4-2\ep$ space-time dimensions, 
one important point is to define a proper description for the treatment of $\gamma_5$. 
In the case of the axial-vector and the pseudo-scalar form factors, two types of Feynman diagrams
contribute: the non-singlet diagrams containing only open fermion lines,
and the singlet diagrams where a fermion loop is attached to the axial-vector or pseudo-scalar vertex.
It is convenient to  separate the two contributions and write,
\begin{equation}
 \Gamma_{A,cd}^{\mu} = \Gamma_{A,cd}^{\mu, \ns} + \Gamma_{A,cd}^{\mu, \sing} \,, \quad
 \Gamma_{P,cd} = \Gamma_{P,cd}^{\ns} + \Gamma_{P,cd}^{\sing} \,,
\end{equation}
where $\ns$ and $\sing$ denote the non-singlet and the singlet contributions, respectively.

In the non-singlet case, we use an anticommuting $\gamma_5$ in $D$ space-time dimensions, with $\gamma_5^2 = 1$,
as it does not lead to any spurious singularities. This approach 
respects chiral invariance and leaves us with the Ward identity 
\begin{equation} \label{eq:cwi}
 q^{\mu} \Gamma_{A,cd}^{\mu, \ns} = 2 m \Gamma_{P,cd}^{\ns} \,,
\end{equation}
which in terms of the form factors, takes the  form
\begin{equation} \label{eq:cwiFF}
 2 F_{A,1}^{\ns} + \frac{s}{2} F_{A,2}^{\ns} = 2 m F_{P}^{\ns} \,.
\end{equation}

On the other hand, the singlet pieces for the axial-vector and the pseudo-scalar vertex are related 
to each other through the Adler-Bell-Jackiw (ABJ) anomaly \cite{Adler:1969gk,Bell:1969ts}. 
With this constraint, we use the following prescription as presented in 
\cite{Akyeampong:1973xi,Larin:1993tq}, which mostly followed \cite{tHooft:1972tcz}:
For a single $\gamma_5$ in a fermion loop, we use
\begin{equation}
 \gamma_5 = \frac{i}{4!} \epsilon_{\mu\nu\rho\sigma} \gamma^{\mu} \gamma^{\nu} \gamma^{\rho} 
\gamma^{\sigma},
\end{equation}
where $\varepsilon_{\mu\nu\rho\sigma}$ is the completely antisymmetric Levi-Civita tensor and all  
Lorentz indices are taken $D$-dimensional. Finally, the contraction of two $\epsilon$-tensors is expressed 
in terms of products of  $D$-dimensional metric tensors.
This prescription of $\gamma_5$ needs a special treatment during renormalization, as discussed later.

The ABJ anomaly involves the truncated matrix element of the gluonic operator between the vacuum and a pair of heavy quark states.
The gluonic operator is given by
\begin{equation}
 G(x) \tilde{G} (x) \equiv \varepsilon_{\mu\nu\rho\sigma} G^{a,\mu\nu} (x) G^{a,\rho\sigma}(x) \,,
\end{equation}
where $G^{a,\mu\nu}$ represents the gluonic field strength tensor.
Denoting its contribution by $F_{G,Q}$, we can immediately write down the anomalous Ward 
identity for the singlet case as follows 
\begin{equation} \label{eq:awi}
 q_{\mu} \Gamma_{A,cd}^{\mu, s} = 2 m \Gamma_{P,cd}^{s} - i \asr T_F \langle G\tilde{G} \rangle_Q \,,
\end{equation}
which implies 
\begin{equation}
 2 F_{A,1}^{s} + \frac{s}{2} F_{A,2}^{s} = 2 m F_{P}^{s} - i \asr T_F F_{G,Q} \,.
\label{FGQ}
\end{equation}

\subsection{Renormalization}
\label{sec:reno}

\vspace*{1mm}
\noindent
The UV renormalization of the form factors has been performed in a mixed scheme.
We renormalize the heavy quark mass and wave function in the on-shell (OS)
renormalization scheme, while the strong coupling constant is renormalized 
in the modified minimal subtraction ($\overline{\rm MS}$) scheme \cite{tHooft:1973mfk, Bardeen:1978yd}.
The corresponding renormalization constants are well known and are denoted by 
$Z_{m, {\rm OS}}$ \cite{Broadhurst:1991fy, Melnikov:2000zc,Marquard:2007uj,
Marquard:2015qpa,Marquard:2016dcn}, 
$Z_{2,{\rm OS}}$ \cite{Broadhurst:1991fy, Melnikov:2000zc,Marquard:2007uj,Marquard:2017XXX} and 
$Z_{a_s}$ \cite{Gross:1973id, Politzer:1973fx, Caswell:1974gg, Jones:1974mm, Egorian:1978zx}
for the heavy quark mass, wave function and strong coupling constant, respectively. 
All renormalization constants follow a perturbative expansion in $\alpha_s$ 
\begin{equation}
 Z_{I} = \sum_{n=0}^{\infty} \Big( \frac{\alpha_s}{4 \pi} \Big)^n Z_{I}^{(n)} \,.
\end{equation}
For reference, we  present $Z_{m, {\rm OS}}^{(n)}$ and $Z_{2, {\rm OS}}^{(n)}$ in appendix \ref{app:renz} up to $n=2$
and ${\cal O}(\ep^2)$. 
\begin{align}
\beta_0 &= \frac{11}{3} C_A - \frac{4}{3} T_F ( n_l + n_h ) \,.
\end{align}
Here $n_l$ and $n_h$ denote the number of light and heavy quarks, respectively. In the following we 
will set $n_h = 1$. 

While the renormalization of the heavy-quark wave function and the
strong coupling constant can be done  multiplicatively, the mass renormalization
requires the explicit calculation  of counterterm diagrams.  Hence, the bare and
renormalized vector form factors are at two loops related by
\begin{equation} \label{eq:vecren}
 F_{V,i} = Z_{2, {\rm OS}} \hat{F}_{V,i} + \asr^2 Z_{m, {\rm OS}}^{(1)} \hat{F}_{V,i}^{ct,(1)} + {\cal O} (\alpha_s^3) \,,
\end{equation}
where the unrenormalized form factors $\hat{F}_{V,i}$ are expanded in the unrenormalized strong 
coupling constant $\hat{\alpha}_s = \alpha_s Z_{a_s}$, and
$\hat{F}_{V,i}^{\mathrm{ct},(1)}$ denotes the bare contribution from counterterm diagrams at one loop. 

\newcommand{\Zfin}{Z_5^{\mathrm{fin}}} The non-singlet contributions
to the axial-vector form factor can be renormalized in the same way.
The singlet part requires extra care due to the prescription employed
for $\gamma_5$.  It is infrared finite and the UV pole is renormalized
by the multiplicative renormalization constant $Z_J$ as
\begin{equation}
 F_{A,i}^{s} = Z_J \Zfin \hat{F}_{A,i}^{s} \,,
\end{equation}
where $\Zfin$ is a finite renormalization constant which restores the
anomalous Ward identity Eq.~(\ref{eq:awi}).  We would like to remark
that the Ward identities are valid for physical quantities.
Therefore, it is not reasonable to study them at higher orders in
$\ep$, and neither it is   to consider $\ep$-dependent pieces of
$\Zfin$.  In the $\overline{\rm MS}$ scheme for the form factors, 
\begin{align}
 Z_J = 1 + \asr^2 \frac{6 C_F T_F}{\ep} + {\cal O}(\alpha_s^3)
\end{align}
and 
\begin{equation}
 \Zfin = 1 + \asr^2 \Big( 3 C_F T_F \Big) + {\cal O}(\alpha_s^3) \,.
\end{equation}
The remaining finite renormalization has to be carried out later for the corresponding observables of 
which the form factors form a part.

The renormalization of the quantity $F_{G,Q}$ appearing in Eq.~(\ref{FGQ}) involves the mixing of the gluonic operator $G\tilde{G}$ 
with another operator, namely, $\partial_{\mu} \bar{\psi} \gamma^{\mu} \gamma_5 \psi$, as discussed in 
\cite{Larin:1993tq, Zoller:2013ixa, Ahmed:2015qpa}, where $\psi$ indicates all quark flavors including 
the massive one. We get,
\begin{equation}
 F_{G,Q} = Z_{GG} \hat{F}_{G,Q} + Z_{GJ} \hat{F}_{J,Q} \,,
\end{equation}
where $\hat{F}_{J,Q}$ indicates the bare contribution from the second operator, while $Z_{GG}$ and $Z_{GJ}$
are the corresponding renormalization constants. 

The renormalization of the scalar and pseudo-scalar (non-singlet) vertices also follows a similar 
procedure, except for the presence of the heavy quark mass in the Yukawa coupling. Thus the renormalized
form factors are given by 
\begin{align} \label{eq:scalarren}
 F_{S,i} &= Z_{m, {\rm OS}} Z_{2, {\rm OS}} \hat{F}_{S,i} + \asr^2 Z_{m, {\rm OS}}^{(1)} \hat{F}_{S,i}^{ct,(1)} + {\cal O} (\alpha_s^3) \,,
 \nonumber\\
 F_{P,i}^{\ns} &= Z_{m, {\rm OS}} Z_{2, {\rm OS}} \hat{F}_{P,i}^{\ns} + \asr^2 Z_{m, {\rm OS}}^{(1)} \hat{F}_{P,i}^{ct,\ns,(1)} + {\cal O} (\alpha_s^3) \,,
\end{align}
On the other hand, the singlet piece of the pseudo-scalar vertex is both IR and UV finite, hence 
no additional renormalization is necessary.

\subsection{The infrared structure}

\vspace*{1mm}
\noindent
The study of the IR behavior of the form factors has attracted a lot of
attention in the past few decades. A plethora of works on massless
scattering amplitudes
\cite{Catani:1998bh,Sterman:2002qn,Becher:2009cu,Gardi:2009qi,Ravindran:2004mb}
has already provided a remarkable understanding on the universal IR
pattern characterized by soft and collinear dynamics.  In
\cite{Mitov:2006xs}, the first step was taken to generalize this for
the two-loop scattering amplitudes with massive partons.  Later, in
\cite{Becher:2009kw}, following a soft-collinear effective theory
(SCET) approach, the general IR structures have been
presented.

The IR singularities of the massive form factors can be factorized as
a multiplicative renormalization factor. Its structure is
constrained by the renormalization group equation (RGE), as follows,
\begin{equation}
 F_{I} = Z (\mu) F_{I}^{\mathrm{fin}} (\mu)\, ,
\end{equation}
where $F_{I}^{\mathrm{fin}}$ is finite as $\ep \rightarrow 0$. The RGE for $Z$ reads
\begin{equation} \label{eq:rgeZ}
 \frac{d}{d \ln \mu} \ln Z(\ep, x, m, \mu)  = - \Gamma (x,m,\mu) \,,
\end{equation}
where $\Gamma$ is the corresponding anomalous dimension.
Notice that $Z$ does not carry any information regarding the vertex. 
$\Gamma$ can be identified as the massive cusp anomalous dimension,
which is by now available up to the three-loop level
\cite{Korchemsky:1987wg,Korchemsky:1991zp,Grozin:2014hna,Grozin:2015kna}.
Both $Z$ and $\Gamma$ can be expanded in a perturbative series in
$\alpha_s$ as follows
\begin{equation}
 Z = \sum_{n=0}^{\infty} \asr^n Z^{(n)} \,, \qquad
 \Gamma = \sum_{n=0}^{\infty} \asr^{n+1} \Gamma_{n}
\end{equation}
and we find the following solution for Eq.~(\ref{eq:rgeZ})
\begin{equation} \label{eq:solnZ}
 Z = 1 + \asr \Bigg[ \frac{\Gamma_0}{2 \ep} \Bigg] 
   + \asr^2 \Bigg[ \frac{1}{\ep^2} \Big( \frac{\Gamma_0^2}{8} - \frac{\beta_0 \Gamma_0}{4} \Big) + \frac{\Gamma_1}{4 \ep} \Bigg] 
   + {\cal O} (\alpha_s^3) \,.
\end{equation}
Eq.~(\ref{eq:solnZ}) correctly predicts the infrared singularities for all massive form factors at the
two-loop level. 

\section{Details of the calculation}
\label{sec:comp}

\vspace*{1mm}
\noindent
The Feynman diagrams were generated using {\tt QGRAF}
\cite{Nogueira:1991ex}, the output of which was then processed using
{\tt Q2e/Exp} \cite{Harlander:1997zb,Seidensticker:1999bb} and {\tt
  FORM} \cite{Vermaseren:2000nd, Tentyukov:2007mu} in order to express
the diagrams in terms of a linear combination of a large set of scalar
integrals.  These integrals were then reduced to a much smaller set of
master integrals (MIs) using integration by parts identities (IBPs)
\cite{Lagrange:IBP, Gauss:IBP, Green:IBP, Ostrogradski:IBP,
  Chetyrkin:1980pr} with the help of the program {\tt Crusher}
\cite{Crusher}. Since all this is common practice, we refrain from going
into any detail.

After performing the reductions, all that remains to be done is to
calculate the master integrals. In the following sections, we present
the methods we used to achieve this.

\subsection{The conventional differential equations method}
\label{standardDEQsection}

We computed the two-loop master integrals contributing to the massive
fermion form factors as Laurent expansions in the dimensional parameter $\varepsilon$
by means of the differential equation method
\cite{Kotikov:1990kg,Kotikov:1991hm,Kotikov:1991pm,Remiddi:1997ny,Kotikov:2010gf,Henn:2013pwa,Ablinger:2015tua}. 
This
technique has already been applied to massive form factor integrals at
two and three loops in
\cite{Bonciani:2003te,Bonciani:2003hc,Henn:2016kjz}. In this work, we
calculate the two-loop master integrals up to a sufficiently high order in
$\varepsilon$ to obtain ${\cal O}(\varepsilon^2)$ accuracy in the form
factors.

\begin{figure}[ht]
\begin{center}
\begin{minipage}[c]{0.22\linewidth}
     \includegraphics[width=1\textwidth]{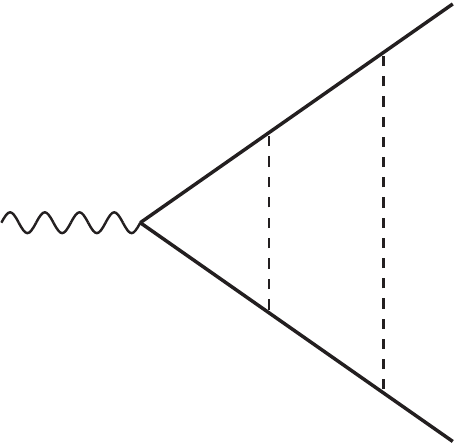}
\vspace*{-11mm}
\begin{center}
{\footnotesize (a)}
\end{center}
\end{minipage}
\hspace*{2mm}
\begin{minipage}[c]{0.22\linewidth}
     \includegraphics[width=1\textwidth]{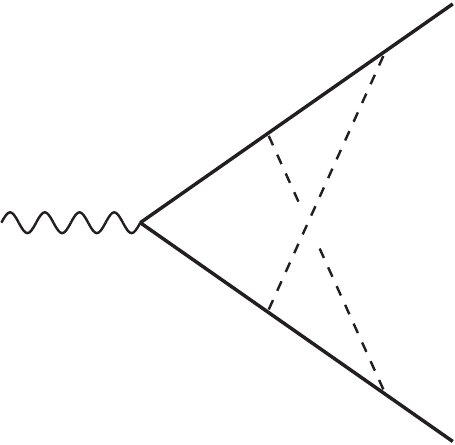}
\vspace*{-11mm}
\begin{center}
{\footnotesize (b)}
\end{center}
\end{minipage}
\hspace*{2mm}
\begin{minipage}[c]{0.22\linewidth}
     \includegraphics[width=1\textwidth]{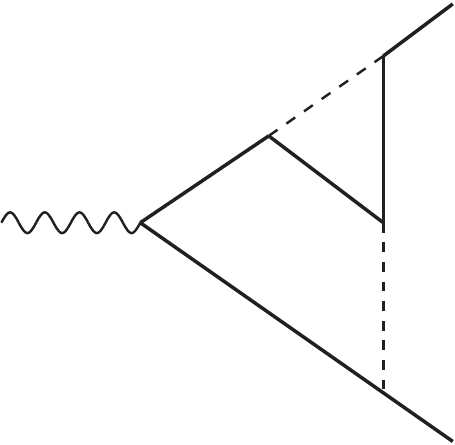}
\vspace*{-11mm}
\begin{center}
{\footnotesize (c)}
\end{center}
\end{minipage}
\hspace*{2mm}
\begin{minipage}[c]{0.22\linewidth}
     \includegraphics[width=1\textwidth]{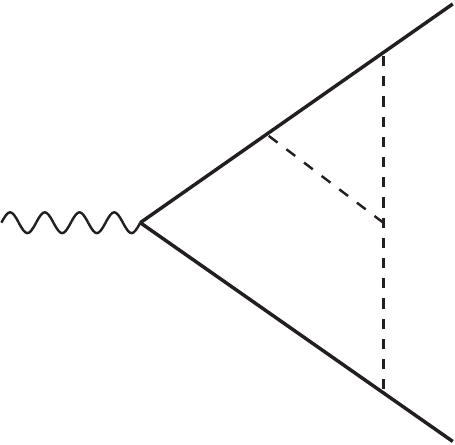}
\vspace*{-11mm}
\begin{center}
{\footnotesize (d)}
\end{center}
\end{minipage}
\caption{\sf \small The non-singlet topologies required for the calculation of two-loop form factors. 
Solid 
lines represent
massive particles in external or internal lines, while dashed lines correspond to massless propagators. 
The external vector current 
can also be replaced by an axial-vector, scalar or pseudo-scalar. The master integrals associated to 
topologies (c) 
and (d) are a subset
of the master integrals required for topologies (a) and (b).
}
\label{NStopologies}
\end{center}
\end{figure}

\begin{figure}[ht]
\begin{center}
     \includegraphics[width=0.3\textwidth]{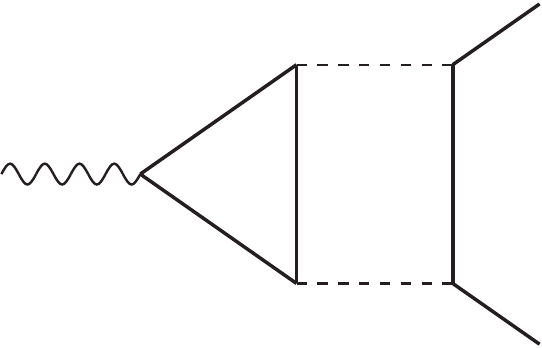}
\end{center}
\caption{\sf \small The singlet topology.}
\label{Stopology}
\end{figure}

In this section, we briefly review the main steps of this
calculation. The master integrals are classified according to their
underlying topology. In particular, we distinguish the non-singlet
topologies (Figure \ref{NStopologies}) from the singlet topology
(Figure \ref{Stopology}) according to whether the external current
does or does not couple  to the external massive quark.

In the case of the non-singlet topologies depicted in Figure \ref{NStopologies}, it turns out that the master integrals associated to the two topologies
on the right of that figure represent a subset of the ones required to calculate the topologies on the left. 
We therefore concentrate on the topologies in Figures \ref{NStopologies}(a) and \ref{NStopologies}(b).
The master integrals for both topologies can be expressed in terms of a single integral family with seven propagators given by
\begin{equation}
J(\nu_1,\dots,\nu_7) = \Big((4\pi)^{2-\varepsilon}e^{\varepsilon\gamma_E)}\Big)^2\,\int\frac{d^Dl_1d^Dl_2}{(2\pi)^{2D}}\,\frac{1}{D_1^{\nu_1} \dots D_7^{\nu_7}}, 
\end{equation}
where 
\begin{eqnarray}
&& D_1 = (l_1+q_1)^2-m^2, \quad D_2 = (l_2+q_1)^2-m^2, \quad D_3 = (l_1-q_2)^2-m^2, \nonumber \\
&& D_4 = (l_2-q_2)^2-m^2, \quad D_5 =l_1^2, \quad D_6 = (l_1-l_2)^2, \quad D_7 = (l_1-l_2+q_2)^2-m^2.
\end{eqnarray}
Here the $q_i$'s with $i=1,2$ are the external momenta, which are taken on-shell ($q_1^2=q_2^2=m^2$).
The MIs are therefore labeled by the exponents $\nu_1, \ldots, \nu_7$ of the denominators $D_1, \ldots, D_7$. 
The integrals corresponding to the topology in Figure \ref{NStopologies}(a) will have $\nu_7=0$, while the ones
corresponding to the topology in Figure \ref{NStopologies}(b) will have $\nu_3=0$. There are several master integrals where $\nu_3$ and $\nu_7$ are both
equal to zero, which are therefore common to both topologies\footnote{This is the reason we chose to group here both topologies within a single integral family, 
although the actual reductions performed with {\tt Crusher} were done using two separate families.}. 
The list of master integrals required to reduce these topologies is given in Table \ref{NonSingletMIlist}. 
Notice that there are several sets of master integrals that have the same set of non-vanishing positive
powers of propagators. Such a set of integrals is called a {\it sector}. For example, integrals $J_6$, $J_7$ and $J_8$ belong to the same sector, since for all of them
$\nu_1,\nu_4,\nu_6 \geq 1$ and $\nu_2=\nu_3=\nu_5=\nu_7=0$. A given sector is said to be a subsector of another sector if the propagators in the first sector 
are a subset of the propagators in the second.

\begin{table}[htb]
\begin{center}
\begin{tabular}{|c|ccccccc|c|}
\hline
MI        & $\nu_1$ & $\nu_2$ & $\nu_3$ & $\nu_4$ & $\nu_5$ & $\nu_6$ & $\nu_7$ & $k$  \\
\hline
$J_1$     &    1    &    1    &    0    &    0    &    0    &    0    &    0    &        6          \\ 
$J_2$     &    0    &    1    &    0    &    0    &    1    &    1    &    0    &        6          \\
$J_{3}$  &    1    &    0    &    0    &    1    &    0    &    0    &    1    &        4          \\
$J_4$     &    1    &    1    &    1    &    0    &    0    &    0    &    0    &        4          \\
$J_5$     &    1    &    1    &    1    &    1    &    0    &    0    &    0    &        4          \\
$J_6$     &    1    &    0    &    0    &    1    &    0    &    1    &    0    &        4          \\
$J_7$     &    1    &    0    &    0    &    2    &    0    &    1    &    0    &        3          \\
$J_8$     &    1    &    0    &    0    &    1    &    0    &    2    &    0    &        4          \\
$J_9$     &    0    &    1    &    1    &    0    &    1    &    1    &    0    &        3          \\
$J_{10}$  &    0    &    1    &    0    &    1    &    1    &    1    &    0    &        4          \\
$J_{11}$  &    1    &    1    &    0    &    1    &    0    &    0    &    0    &        3          \\
$J_{12}$  &    1    &    0    &    0    &    1    &    0    &    1    &    1    &        3          \\
$J_{13}$  &    1    &    0    &    0    &    2    &    0    &    1    &    1    &        3          \\
$J_{14}$  &    1    &    0    &    0    &    1    &    0    &    2    &    1    &        3          \\
$J_{15}$  &    1    &    1    &    0    &    0    &    1    &    0    &    1    &        3          \\
$J_{16}$  &    1    &    1    &    0    &    0    &    2    &    0    &    1    &        3          \\
$J_{17}$  &    1    &    1    &    0    &    0    &    1    &    0    &    2    &        3          \\
$J_{18}$  &    1    &    1    &    0    &    1    &    0    &    0    &    1    &        3          \\
$J_{19}$  &    1    &    1    &    0    &    2    &    0    &    0    &    1    &        3          \\
$J_{20}$  &    1    &    1    &    0    &    1    &    1    &    1    &    0    &        2          \\
$J_{21}$  &    1    &    1    &    0    &    1    &    1    &    2    &    0    &        4          \\
$J_{22}$  &    1    &    1    &    0    &    1    &    1    &    1    &    1    &        3          \\
$J_{23}$  &    1    &    1    &    0    &    1    &    1    &    2    &    1    &        3          \\
\hline
\end{tabular}
\end{center}
\caption{\sf \small The list of the non-singlet master integrals identified by the indices $\nu_1$ to 
$\nu_7$.
In the last column, we indicate the order $k$ in $\varepsilon$ to which each integral needs to be 
expanded
in order to calculate the form factors to ${\cal O}(\varepsilon^2)$.}
\label{NonSingletMIlist}
\end{table}

In the case of the singlet topology, shown in Figure \ref{Stopology}, the master integrals are given by
\begin{equation}
K(\nu_1,\dots,\nu_6) = \Big((4\pi)^{2-\varepsilon}e^{\varepsilon\gamma_E)}\Big)^2\,\int\frac{d^Dl_1d^Dl_2}{(2\pi)^{2D}}\,\frac{1}{D_1^{'\,\nu_1} \dots D_6^{'\,\nu_6}}, 
\end{equation}
where
\begin{eqnarray}
&& D_1' = (l_1+q_1)^2, \quad D_2' = (l_2+q_1)^2-m^2, \quad D_3' = (l_1-q_2)^2, \nonumber \\
&& D_4' = (l_2-q_2)^2-m^2, \quad D_5' =l_1^2-m^2, \quad D_6' = (l_1-l_2)^2-m^2.
\end{eqnarray}

The list of master integrals required in this case is given in Table \ref{SingletMIlist}. From now on, instead of using the exponents $\nu_i$ to
identify the master integrals, we will use the single subindex we have assigned to each integral according to the leftmost columns 
of Tables \ref{NonSingletMIlist} and \ref{SingletMIlist}.

\begin{table}[htb]
\begin{center}
\begin{tabular}{|c|cccccc|c|}
\hline
MI        & $\nu_1$ & $\nu_2$ & $\nu_3$ & $\nu_4$ & $\nu_5$ & $\nu_6$  & $k$  \\
\hline
$K_1$     &    0    &    1    &    0    &    0    &    1    &    0    &         6          \\ 
$K_2$     &    0    &    1    &    0    &    0    &    1    &    1    &         4          \\ 
$K_3$     &    1    &    1    &    1    &    0    &    0    &    0    &         3          \\ 
$K_4$     &    1    &    1    &    1    &    1    &    0    &    0    &         3          \\ 
$K_5$     &    0    &    1    &    1    &    0    &    0    &    1    &         3          \\
$K_6$     &    0    &    1    &    1    &    0    &    0    &    2    &         3          \\
$K_7$     &    0    &    1    &    0    &    1    &    1    &    0    &         3          \\
$K_8$     &    1    &    1    &    1    &    0    &    1    &    0    &         3          \\
$K_9$     &    0    &    1    &    1    &    0    &    1    &    1    &         3          \\
$K_{10}$  &    0    &    1    &    1    &    0    &    1    &    2    &         3          \\
$K_{11}$  &    0    &    1    &    0    &    1    &    1    &    1    &         3          \\
$K_{12}$  &    0    &    1    &    0    &    1    &    1    &    2    &         3          \\
$K_{13}$  &    1    &    1    &    1    &    1    &    1    &    0    &         2          \\
$K_{14}$  &    1    &    1    &    1    &    1    &    0    &    1    &         2          \\
$K_{15}$  &    1    &    1    &    1    &    1    &    1    &    1    &         2          \\
\hline
\end{tabular}
\end{center}
\caption{\sf \small The list of singlet master integrals identified by the indices $\nu_1$ to $\nu_6$.
In the last column, we indicate the order $k$ in $\varepsilon$ to which each integral needs to be 
expanded
in order to calculate the form factors to ${\cal O}(\varepsilon^2)$.}
\label{SingletMIlist}
\end{table}

All of the master integrals can be expressed in terms of harmonic polylogarithms (HPLs) 
\cite{Remiddi:1999ew} in the kinematic variable \cite{Barbieri:1972as,Barbieri:1972hn}
\begin{equation} \label{eq:varx}
 x=\frac{\sqrt{q^2-4m^2}-\sqrt{q^2}}{\sqrt{q^2-4m^2}+\sqrt{q^2}}\quad \leftrightarrow 
\quad\frac{q^2}{m^2}=-\frac{(1-x)^2}{x}.
\end{equation}
In particular, we focus on the Euclidean
region, $q^2<0$, corresponding to $x$ ranging in $(0,1)$. A large
center of mass energy $\abs{q^2}\gg m^2$ is equivalent to the boundary
$x\rightarrow 0$, while the large-mass limit $m^2\gg \abs{q^2}$ is
mapped to the endpoint $x\rightarrow 1$.

We derived a system of coupled linear differential equations for each
topology by reducing the derivative with respect to $x$ of each MI to
a linear combination of the master integrals themselves, with the help
of \verb+Crusher+.  The derivative of an integral cannot produce
integrals in this linear combination with more propagators than the
original one, so they are all either in the same sector as the
integral to which one takes the derivative or they belong to a
subsector.  This means that the integrals in the subsectors need to be
solved first, and the differential equations within a given sector
will constitute a coupled subsystem.  In the case of the singlet
master integrals, there are three such coupled subsystems, namely, the
subsystems formed by the set of integrals $\{K_5, \, K_6\}$,
$\{K_9, \, K_{10}\}$ and $\{K_{11}, \, K_{12}\}$.  In all other
sectors, only one integral is present.  The differential equations in
each coupled subsystem will therefore need to be decoupled in order to
solve them.  The strategy for solving the whole system consists then
in solving first the simplest sectors (with fewer propagators), and
move up in the chain of subsystems, decoupling and solving each one of
them until all integrals are obtained.  The starting point are the
integrals for which the derivative with respect to $x$ equals
zero\footnote{These are $J_1$, $J_2$ and $J_3$ in the case of the
  non-singlet topologies, and $K_1$ and $K_2$ in the singlet case.},
which must therefore be obtained not by the differential equations
method but by other means.

In the case of the non-singlet topologies, by expanding the MIs in a
Laurent series around $\varepsilon=0$, the systems greatly simplify,
assuming an almost complete block-triangular form. Only one $2\times2$
coupled subsystem remains after the expansion in $\varepsilon$. We
therefore solve the system order-by-order in $\varepsilon$ by
integrating each equation by quadratures. These steps are automated
and results are efficiently simplified using a minimal set of
independent HPLs by means of the \verb+Mathematica+ packages
\verb+Sigma+ \cite{Schneider:sigma1,Schneider:sigma2} and
\verb+HarmonicSums+ \cite{Ablinger:2014rba, Ablinger:2010kw,
  Ablinger:2013hcp, Ablinger:2011te, Ablinger:2013cf,
  Ablinger:2014bra}. This procedure is slightly modified for the
aforementioned $2\times2$ coupled system, which does not assume a
triangular form after expanding in $\varepsilon$

\begin{align}
  \begin{split}
    \frac{d}{dx}\left(
    \begin{array}{l}
      J_{22}\\
      J_{23}
    \end{array}
    \right)&=
    \left[
      \begin{array}{cc}
        \frac{1+x^2}{x(1-x^2)}&\frac{1-x^2}{x^2}\\
        -\frac{1}{1-x^2}&\frac{4(1+x^2)}{x(1-x^2)}\\
      \end{array}
      \right]
    \left(
    \begin{array}{l}
      J_{22}\\
      J_{23}
    \end{array}
    \right)
    +\left(
    \begin{array}{l}
      R_{1}(\varepsilon,x)\\
      R_{2}(\varepsilon,x)
    \end{array}
    \right),
    \label{coupsyst}
  \end{split}
\end{align}
where the inhomogeneities $R_1(\varepsilon,x)$, $R_2(\varepsilon,x)$
are determined at each order in $\varepsilon$ by subsector MIs. The
most general solution of the homogeneous system involves only
logarithms and rational functions
\begin{align}
  \begin{split}
    J_{22}&=c_1\frac{x^2}{(1-x^2)^2}+c_2\frac{x^2\,\ln(x)}{(1-x^2)^2},\\
    J_{23}&=c_1\frac{x^2(1+x^2)}{(1-x^2)^4}+c_2\frac{x^3(x^2\,\ln(x)+\ln(x)+2)}{(1-x^2)^4},
    \label{hom26}
  \end{split}
\end{align}
therefore we integrate the inhomogeneous system order-by-order in
$\varepsilon$ using the method of variation of constants. For example,
at leading order in $\varepsilon$, the inhomogeneous parts of the
system (\ref{coupsyst}) are
\begin{eqnarray}
  R_1(\varepsilon,x)&=&\frac{1}{\varepsilon}\bigg\{\frac{x^4+x^3+6 x^2+x+1}{(1-x)^2 (x+1)^4} \ln(x)
\nonumber \\ &&
-\frac{x^6+2 x^5-25 x^4-4 x^3-25 x^2+2 x+1}{16 x^2(1-x) (x+1)^3}\bigg\}+{\cal O}(\varepsilon^0),\\
  R_2(\varepsilon,x)&=&\frac{1}{\varepsilon}\bigg\{\frac{x  \ln(x)}{(1-x)^4 (x+1)^6}\left(2 x^6+x^5-4 x^4+6 x^3-4 x^2+x+2\right)\nonumber\\
  &&-\frac{x^8+2 x^7-8 x^6-10 x^5+22 x^4-10 x^3-8 x^2+2 x+1}{4 (1-x)^3 x (x+1)^5}\bigg\}+{\cal O}(\varepsilon^0).
\end{eqnarray}

By introducing Eq.~(\ref{hom26}) with $c_i\rightarrow c_i(x)$ into the
system (\ref{coupsyst}), we get first order differential equations for
$c_1(x)$ and $c_2(x)$ that can be solved by quadratures using the same
automated tools introduced above. The integration constants are fixed
by imposing the regularity of the functions $J_{22}$ and $J_{23}$ in
the limit of vanishing space-like momentum $q^2\rightarrow 0$,
corresponding to $x\rightarrow 1$, giving
\begin{eqnarray}
J_{22} &=& \frac{1}{\varepsilon}\biggl\{
-\frac{x^2}{3 (1-x^2)^2} \big[ H_0^3(x)-3 H_0(x) \big(2 H_{0,-1}(x)-2 H_{0,1}(x)-\zeta_2\big) 
\nonumber \\ && 
+12 H_{0,0,-1}(x)-12 H_{0,0,1}(x)+3 \zeta_3\big]\biggr\}+{\cal O}(\varepsilon^0), 
\nonumber \\
J_{23} &=& \frac{1}{\varepsilon}\biggl\{
\frac{x^3 (1+x^2)}{(1-x^2)^4} \biggl[
2 H_0(x) H_{0,-1}(x)-2 H_0(x) H_{0,1}(x)-4 H_{0,0,-1}(x)+4 H_{0,0,1}(x)
\nonumber \\ &&
-\zeta_2 H_0(x)-\frac{1}{3} H_0^3(x)-\zeta_3
\biggr]
+\frac{x^3}{(1-x^2)^3} \big[2 H_{0,1}(x)-2 H_{0,-1}(x)+2 H_{-1}(x) H_0(x)
\nonumber \\ &&
-2 H_0(x) H_1(x)-H_0^2(x)-\zeta_2\big]
+\frac{x^6+2 x^5-25 x^4-4 x^3-25 x^2+2 x+1}{16(1-x)^2 (x+1)^4}
\nonumber \\ &&
-\frac{x^2 \big(x^4+x^3+6 x^2+x+1\big)}{(1-x)^3 (x+1)^5}H_0(x)
\biggr\}+{\cal O}(\varepsilon^0) \,,
\end{eqnarray}
where the harmonic polylogarithms, $H_{a_1, a_2, \ldots, a_n}(x)$, are defined by
\begin{eqnarray}
H_{a_1, a_2, \ldots, a_n}(x) &=& \int_0^x dy f_{a_1}(y) H_{a_2,
                                 \ldots, a_n}(y),~~~H_\emptyset =
                                 1,~a_i~\in~\{0,1,-1\}~,
\label{HPLdef}
\end{eqnarray}
with
\begin{eqnarray}
\label{eq:hpol}
f_0(x) = \frac{1}{x},~~~f_1(x) = \frac{1}{1-x},~~~f_{-1}(x) = \frac{1}{1+x}~,
\end{eqnarray}
and
\begin{equation}
H_{\underbrace{\text{\scriptsize 0},\ldots,\text{\scriptsize 0}}_{\text{\scriptsize n times}}}(x)
=
\frac{1}{n!} \ln^n(x)~.
\end{equation}
$\zeta_n$ denotes the Riemann $\zeta$-function
\[
\zeta_n = \sum_{k=1}^\infty \frac{1}{k^n}, n \geq 2, n \in \mathbb{N}.
\]

We proceed similarly at higher orders, getting expansions of the MIs
$J_{22}$ and $J_{23}$ up to ${\cal O}(\varepsilon^3)$. It should be
noticed that solving a system of coupled differential equations is in
general far from trivial. When the solution is written in terms of
multiple polylogarithms or related iterated integrals, as in the case of the two-loop massive form
factor \cite{Bonciani:2003te,Bonciani:2003hc} and of the planar
three-loop massive form factor \cite{Henn:2016kjz}, it is possible to
properly choose the set of MIs such that the system doesn't have any
coupled equation, assuming for example the \textit{canonical} form of
\cite{Kotikov:2010gf,Henn:2013pwa}. In this work we are interested in
applying the methods developed in
\cite{Bluemlein:2014qka,Ablinger:2015tua,Schneider:2016szq}, to solve
the systems of coupled differential equations algorithmically, as
discussed in the next sections.

\begin{figure}[h!]
\centering
\includegraphics{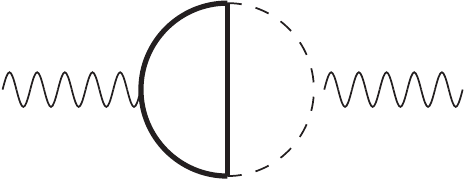}
\caption{\sf \small The master integral $K_{14}$. Massive (massless) propagators are represented by 
solid 
(dashed) lines. The presence of a massless cut determines the asymptotic behaviour $K_{14}\sim \log(1-x)$.}
\label{SingletBC}
\end{figure}

In order to solve the differential equations, boundary conditions
have to be determined. As it was observed in \cite{Bonciani:2003te,
  Bonciani:2003hc}, the analytic structure of the master integrals
strongly constrains the choice of the integration constants. In
particular, boundary conditions of the master integrals of the
non-singlet topologies are completely determined by requiring the
regularity of the functions in $x=1$, as will be discussed in the example
above.  However, we cannot use the same argument for some of the
master integrals of the singlet topology, which are characterized by a
branch cut at $x=1$, as occurs, for example, in the case of the
integral $K_{14}=K(1,1,1,1,0,1)$ depicted in Figure \ref{SingletBC}.
We fixed the boundary conditions of such integrals by matching the
general solutions of the differential equations with the asymptotic
expansions of the corresponding integrals around $x\rightarrow 1$. The
latter were computed by means of the large-mass expansion $m^2\gg q^2$
\cite{Smirnov:2002pj}, or with a Mellin Barnes representation, as
described e.g. in \cite{Smirnov:2012gma}.


\subsection{Calculation of the master integrals using difference equations}
\label{differenceSection}

\vspace*{1mm}
\noindent
In the following we describe an alternative method for calculating the master integrals. 
The idea of the method is to write all integrals in terms of series expansions
and then use the differential equations obeyed by the MIs to derive difference equations satisfied by the coefficients of these series.
In the non-singlet case, this can be done using the fact that the MIs are regular around $x=1$. In terms of the variable $y=1-x$, we can 
therefore write
\begin{equation}
J_i(y)=\sum_{n=0}^{\infty} \sum_{j=-2}^r \varepsilon^j C_{i,j}(n) y^n,
\label{NSMIyexp}
\end{equation}
where we have included the expansion in $\varepsilon$ up to ${\cal O}(\varepsilon^r)$. We can introduce Eq.~(\ref{NSMIyexp}) in the
differential equations after rewriting them in terms of the variable $y$, leading to a system of coupled difference equations for 
the different $C_{i,j}$'s.

The method of solving differential equations by introducing series expansions and then finding the 
solutions to the resulting recursion relations is, of course well-known. However, to the best of our knowledge, this method has 
not been used before to calculate master integrals in perturbative quantum field theory. We propose
this method here since it can be useful also for higher loop calculations, and we can take advantage of the powerful mathematical tools implemented in {\tt Sigma} and
{\tt HarmonicSums} to solve the difference equations.

Whenever we are given subsystems of differential equations, it is usually more convenient to uncouple them and then insert Eq.~(\ref{NSMIyexp}) in 
the corresponding uncoupled equations, as supposed to inserting Eq.~(\ref{NSMIyexp}) first and then uncouple the resulting coupled difference equations. 
This is so because usually the later approach will lead to difference equations of higher order, although occasionally, this might not be the case and this
later approach might turn out to be preferable.

In the singlet case, some of the master integrals will not be regular at $x=1$, due to the presence of the logarithm $\ln(1-x)=\ln(y)$ discussed in the previous section. 
We therefore include a formal expansion around powers of this 
logarithm\footnote{In two particular cases (integrals $K_8$ and $K_{13}$), the sum in $n$ actually 
starts from $n=-1$.},
\begin{equation}
K_i(y)=\sum_{n=0}^{\infty} \sum_{k=0}^3 \sum_{j=-2}^r \varepsilon^j C_{i,j,k}(n) \ln^k(y) y^n.
\label{SMIyexp}
\end{equation}

The integrals belonging to the three multi-integral sectors in the singlet case have no logarithmic singularities at $y=0$, 
and can therefore all be written as in Eq.~(\ref{NSMIyexp}).
Only some of the integrals that are the sole representative of their sector in Table \ref{SingletMIlist} need to be written according to Eq.~(\ref{SMIyexp}). Since Eq.~(\ref{NSMIyexp})
is a particular case of Eq.~(\ref{SMIyexp})\footnote{That is, the case where $C_{i,j,k}(n)=0$ for $k>0$.}, we will consider for illustration purposes one of the integrals that require
an expansion of the form (\ref{SMIyexp}). The integral $K_4=K(1,1,1,1,0,0)$ is simple enough (it consists of a product of two one-loop integrals) to be calculated just through Feynman parameters,
but precisely because of this simplicity, it allows us to describe the main features of the method without unnecessary complications or long formulae.

The differential equation associated to $K_4$ is given by
\begin{equation}
\frac{d K_4}{dy}
+\left(\frac{2 \varepsilon y}{(2-y) (1-y)}+\frac{2 (1+2 \varepsilon)}{y (2-y)}\right) K_4 =
\frac{2-2 \varepsilon}{y (2-y)} K_3.
\label{diffeqK4}
\end{equation}

As we discussed above, in order to solve the differential equations, we must first obtain all
of the integrals associated to the subtopologies of the system under consideration. In the case of Eq.~(\ref{diffeqK4}), this means obtaining $K_3$. 
In terms of the variable $y$, it is given by 
\begin{eqnarray}
K_3(y) &=&-\frac{1}{\varepsilon^2}+\frac{1}{\varepsilon} \big[-3-\ln(1-y)+2 \ln(y)\big]
-7-3 \ln(1-y)-\frac{1}{2} \ln^2(1-y)
\nonumber \\ &&
+\big[6+2 \ln(1-y)\big] \ln(y)-2 \ln^2(y)
+\varepsilon \biggl[-15+\frac{8}{3} \zeta_3-7 \ln(1-y)
\nonumber \\ &&
-\frac{3}{2} \ln^2(1-y)-\frac{1}{6} \ln^3(1-y)
+\big[14+6 \ln(1-y)+\ln^2(1-y)\big] \ln(y)
\nonumber \\ &&
-\big[6+2 \ln(1-y)\big] \ln^2(y)+\frac{4}{3} \ln^2(y) \biggl]
+{\cal O}(\varepsilon^2),
\end{eqnarray}
which, in expanded form, as in Eq.~(\ref{SMIyexp}), can be written as
\begin{eqnarray}
K_3(y) &=& -\frac{1}{\varepsilon^2}+\frac{1}{\varepsilon} \biggl(-3+2 \ln(y)+\sum_{n=1}^{\infty} \frac{y^n}{n}\biggr)
-7+6 \ln(y)-2 \ln^2(y)
\nonumber \\ &&
-\sum_{n=1}^{\infty} \frac{2}{n} y^n \ln(y)+\sum_{n=1}^{\infty} \biggl(\frac{1+3 n}{n^2}-\frac{1}{n} S_1(n)\biggr) y^n
\nonumber \\ &&
+\varepsilon \Biggl[
\frac{4}{3} \ln^3(y)-6 \ln^2(y)+14 \ln(y)-15+\frac{8}{3} \zeta_3
\nonumber \\ &&
+\sum_{n=1}^{\infty} \frac{2}{n} y^n \ln^2(y)-2 \sum_{n=1}^{\infty} \biggl(\frac{1+3 n}{n^2}-\frac{1}{n} S_1(n)\biggr) y^n \ln(y)
\nonumber \\ &&
+\sum_{n=1}^{\infty} \biggl(\frac{1+3 n+7 n^2}{n^3}-\frac{1+3 n}{n^2} S_1(n)+\frac{1}{2 n} S_1^2(n)-\frac{1}{2 n} S_2(n)\biggr) y^n\Biggr]
\nonumber \\ &&
+{\cal O}(\varepsilon^2),
\label{K3yexp}
\end{eqnarray}
where
\begin{equation}
S_k(n) = \sum_{i=1}^n \frac{1}{i^k}.
\end{equation}

We proceed as follows: First, we clear the denominators in Eq.~(\ref{diffeqK4}), which in this case means multiplying by $(2-y) (1-y) y$. 
After that, we can then insert the expanded version of $K_3$ given in Eq.~(\ref{K3yexp}) together with the generic expanded version of $K_4$ according to Eq.~(\ref{SMIyexp}). 
The resulting equation will be satisfied if the coefficients in the expansion in $y$, for each power in $\ln(y)$, are equal on both sides of the equation.
This leads to the following system of difference equations
\begin{eqnarray}
0 &=& (2-n) C_{4,-2,3}(n-2)+(3 n-1) C_{4,-2,3}(n-1)-2 (n+1) C_{4,-2,3}(n), \label{diffeq1} \\
0 &=& (2-n) C_{4,-2,k}(n-2)+(3 n-1) C_{4,-2,k}(n-1)-2 (n+1) C_{4,-2,k}(n) \nonumber \\ &&
- (k+1) C_{4,-2,k+1}(n-2)+3 (k+1) C_{4,-2,k+1}(n-1)-2 (k+1) C_{4,-2,k+1}(n) \nonumber \\ &&
{\rm for} \quad k=0,1,2, \label{diffeq2} \\
0 &=& -2 C_{4,j,3}(n-2)+4 C_{4,j,3}(n-1)-4 C_{4,j,3}(n)+(2-n) C_{4,j+1,3}(n-2) \nonumber \\ &&
+(3 n-1) C_{4,j+1,3}(n-1)-2 (n+1) C_{4,j+1,3}(n) \nonumber \\ &&
{\rm for} \quad j=-2,-1,0, \label{diffeq3} \\
0 &=& -2 C_{4,j,k}(n-2)+4 C_{4,j,k}(n-1)-4 C_{4,j,k}(n)+(2-n) C_{4,j+1,k}(n-2) \nonumber \\ &&
+(3 n-1) C_{4,j+1,k}(n-1)-2 (n+1) C_{4,j+1,k}(n)-(k+1) C_{4,j+1,k+1}(n-2) \nonumber \\ &&
+3 (k+1) C_{4,j+1,k+1}(n-1)-2 (k+1) C_{4,j+1,k+1}(n) \nonumber \\ &&
{\rm for} \quad j=-2,\, k=1,2 \quad {\rm and} \quad j=-1, \, k=2, \label{diffeq4} \\
0 &=& -2 C_{4,j,k}(n-2)+4 C_{4,j,k}(n-1)-4 C_{4,j,k}(n)+(2-n) C_{4,j+1,k}(n-2) \nonumber \\ &&
+(3 n-1) C_{4,j+1,k}(n-1)-2 (n+1) C_{4,j+1,k}(n)- (k+1) C_{4,j+1,k+1}(n-2) \nonumber \\ &&
+3 (k+1) C_{4,j+1,k+1}(n-1)-2 (k+1) C_{4,j+1,k+1}(n)+\frac{j-2 k-7 j k}{(n-1) n} \nonumber \\ &&
{\rm for} \quad j=-2,\, k=0; \,\, j=-1, \, k=1 \quad {\rm and} \quad j=0, \, k=2, \label{diffeq5} \\
0 &=& -2 C_{4,j,k}(n-2)+4 C_{4,j,k}(n-1)-4 C_{4,j,k}(n)+(2-n) C_{4,j+1,k}(n-2) \nonumber \\ &&
+(3 n-1) C_{4,j+1,k}(n-1)-2 (n+1) C_{4,j+1,k}(n)-(k+1) C_{4,j+1,k+1}(n-2) \nonumber \\ &&
+3 (k+1) C_{4,j+1,k+1}(n-1)-2 (k+1) C_{4,j+1,k+1}(n) \nonumber \\ &&
+\frac{2 (-1)^k (k+1)}{(n-1) n} S_1(n)-2 (-1)^k (k+1) \frac{3 n^2-n-1}{(n-1)^2 n^2} \nonumber \\ &&
{\rm for} \quad j=-1, \, k=0 \quad {\rm and} \quad j=0, \, k=1, \label{diffeq6} \\
0 &=& -2 C_{4,0,0}(n-2)+4 C_{4,0,0}(n-1)-4 C_{4,0,0}(n)+(2-n) C_{4,1,0}(n-2) \nonumber \\ &&
+(3 n-1) C_{4,1,0}(n-1)-2 (n+1) C_{4,1,0}(n)-C_{4,1,1}(n-2) \nonumber \\ &&
+3 C_{4,1,1}(n-1)-2 C_{4,1,1}(n)+\frac{2 \left(3 n^2-n-1\right)}{(n-1)^2 n^2} S_1(n)-\frac{S_1^2(n)}{(n-1) n} \nonumber \\ &&
+\frac{S_2(n)}{(n-1) n}-\frac{2 \left(6 n^4-6 n^3+1\right)}{(n-1)^3 n^3}. \label{diffeq7} 
\end{eqnarray}

This system is triangular. So, the coefficients $C_{4,j,k}(n)$ can be obtained successively by solving 
Eqs. (\ref{diffeq1}) to (\ref{diffeq7}) 
one after the other, inserting the results of the $C_{4,j,k}$'s obtained at each step in subsequent equations. We start with the coefficient for which 
the value of $j$ is the lowest and the value of $k$ is the largest (in this case, $C_{4,-2,3}(n)$), and proceed to obtain the coefficients for lower values
of $k$, keeping $j$ fixed, until all coefficients for $k=3$ to $k=0$ are determined. After that, we increase the value of $j$ by one and repeat the procedure until
all values of $j$ are exhausted and all coefficients are determined. All of this can be done automatically using {\tt Sigma} and {\tt HarmonicSums}. 
The results will be given in terms of harmonic sums \cite{HSUM},
 \begin{equation}
  S_{b,\vec{a}}(n) = \sum_{k=1}^n 
\frac{({\rm sign}(b))^k} {k^{ \abs{b} }} 
S_{\vec{a}}(k),~~~S_\emptyset = 1,~b, a_i~\in \mathbb{Z} \backslash \{0\}~,
 \end{equation}
and generalized harmonic sums \cite{Ablinger:2013cf,Moch:2001zr},
\begin{equation}
S_{b,\vec{a}}\big(\big\{c, \vec{d} \hspace*{0.7mm}\big\}, n\big) = \sum_{k=1}^n \frac{c^k}{k^b} S_{\vec{a}}\big(\big\{\vec{d} \hspace*{0.7mm}\big\}, k\big),~~
b, a_i \in \mathbb{N} \backslash \{0\},~~c, d_i \in \mathbb{Z}  \backslash \{0\},~S_\emptyset = 1~.
\end{equation}

Since all difference equations are of second order, we need at least two initial values in order to solve them. The first few expansion coefficients of $K_4(y)$
are given by
\begin{eqnarray}
K_4(y) &=& -\frac{1}{\varepsilon^2}
+\frac{1}{\varepsilon} \left(-2+y+\frac{2}{3} y^2+\frac{y^3}{2}+2 \ln(y)\right)
-4+2 y+\frac{5}{6} y^2+\frac{y^3}{3}
\nonumber \\ &&
+\left(4-2 y-\frac{4}{3} y^2-y^3\right) \ln(y)-2 \ln^2(y)
+\varepsilon \biggl[-8+\frac{8}{3} \zeta_3+4 y+\frac{5}{3} y^2
\nonumber \\ &&
+\frac{5}{6} y^3
+\left(-4+2 y+\frac{4}{3} y^2+y^3\right) \ln^2(y)
+\left(8-4 y-\frac{5}{3} y^2-\frac{2}{3} y^3\right) \ln(y)
\nonumber \\ &&
+\frac{4}{3} \ln^3(y)
\biggr]+{\cal O}(\varepsilon^2,y^4).
\end{eqnarray}
These were obtained as described in the previous section. In general, the solutions to the difference 
equations (\ref{diffeq1}--\ref{diffeq7}) for general values
of $n$ will be valid starting from a certain value $n=n_0$, and therefore the initial values used to solve the difference equations must also be taken starting from $n \geq n_0$. 
There are a few cases above where we can start from $n_0=0$, but in most cases we must take the initial values starting from $n_0=1$. 
Later, when we formally perform the sum (\ref{SMIyexp}), the expansion terms for $n<n_0$ will have to be added separately.

We obtain the following results for the expansion coefficients,
\begin{eqnarray}
C_{4,-2,k}(n) &=& 0 \quad {\rm for} \quad k \geq 1, \, n \geq 0 \quad {\rm and} \quad k=0, \, n \geq 1, \label{res1} \\ 
C_{4,-1,k}(n) &=& 0 \quad {\rm for} \quad k \geq 2, \, n \geq 0, \quad {\rm and} \quad k=1, \, n \geq 1, \\ 
C_{4,-1,0}(n) &=& \frac{2}{n+1} \quad {\rm for} \quad n \geq 1, \\ 
C_{4,0,k}(n) &=& 0 \quad {\rm for} \quad k=3, \, n \geq 0, \quad {\rm and} \quad k=2, \, n \geq 1, \\ 
C_{4,0,1}(n) &=& -\frac{4}{n+1} \quad {\rm for} \quad n \geq 1, \\ 
C_{4,0,0}(n) &=& \frac{8}{n+1}+\frac{2 (1-2 n)}{n (n+1)} S_1(n)-\frac{2^{1-n}}{n (n+1)} S_1(\{2\},n) \quad {\rm for} \quad n \geq 1, \\ 
C_{4,1,3}(n) &=& 0 \quad {\rm for} \quad n \geq 1, \\ 
C_{4,1,2}(n) &=& \frac{4}{n+1} \quad {\rm for} \quad n \geq 1, \\ 
C_{4,1,1}(n) &=& -2 C_{4,0,0}(n) \quad {\rm for} \quad n \geq 1, \\ 
C_{4,1,0}(n) &=& \frac{24}{n+1}+\frac{4 n-3}{n (n+1)} S_1^2(n)-\frac{S_2(n)}{n (n+1)}-\frac{8}{n (n+1) 2^n} S_1(\{2\},n)
\nonumber \\ && 
-\frac{4 S_1(n)}{n (n+1)} \left[4 n-2+\frac{1}{2^n} S_1(\{2\},n)\right]-\frac{6}{n (n+1) 2^n} S_2(\{2\},n)
\nonumber \\ && 
-\frac{4 (n-1)}{n (n+1)} S_{1,1}(\{{\textstyle \frac{1}{2}},2\},n)+\frac{10}{n (n+1) 2^n} S_{1,1}(\{2,1\},n) \quad {\rm for} \quad n \geq 1,
\end{eqnarray}
with the separate values,
\begin{eqnarray}
&& C_{4,-2,0}(0) = -1, \quad C_{4,-1,1}(0) = 2, \quad C_{4,-1,0}(0) = -2, \quad C_{4,0,2}(0) = -2, \nonumber \\
&& C_{4,0,1}(0) = 4, \quad C_{4,0,0}(0) = -4, \quad C_{4,1,3}(0) = \frac{4}{3}, \quad C_{4,1,2}(0) = -4, \nonumber \\
&& C_{4,1,1}(0) = 8, \quad C_{4,1,0}(0) = \frac{8}{3} \zeta_3-8, 
\label{firstexpcoeff}
\end{eqnarray}

We can insert the results from Eq.~(\ref{res1}) to Eq.~(\ref{firstexpcoeff}) into the expansion (\ref{SMIyexp}) and perform the sums
using {\tt Sigma}, {\tt HarmonicSums}, {\tt EvaluateMultiSums} and {\tt SumProduction} \cite{Ablinger:2010pb, Blumlein:2012hg, Schneider:2013zna}. We obtain
\begin{eqnarray}
K_4(y) &=& -\frac{1}{\varepsilon^2}
+\frac{1}{\varepsilon} \left[\frac{2}{y} H_1(y)+2 \ln(y)-4\right]
-12-\frac{2}{y} (y-2) H_{2,1}(y)+\frac{(y-3)}{y} H_1^2(y)
\nonumber \\ &&
+\frac{8}{y} H_1(y)+\left(8-\frac{4}{y} H_1(y)\right) \ln (y)-2 \ln^2(y)
+\varepsilon \biggl[
\frac{y-2}{y} \big(
-8 H_{2,1}(y)
\nonumber \\ &&
+4 H_{1,2,1}(y)+6 H_{2,1,1}(y)-4 H_{2,2,1}(y)\big)
-\frac{3 y-7}{3 y} H_1^3(y)+\frac{4}{y} (y-3) H_1^2(y)
\nonumber \\ &&
+\frac{24}{y} H_1(y)
+\frac{8}{3} \zeta_3-32
+\biggl(\frac{4}{y} (y-2) H_{2,1}(y)-\frac{2}{y} (y-3) H_1^2(y)-\frac{16}{y} H_1(y)
\nonumber \\ &&
+24\biggr) \ln(y)
+\left(\frac{4}{y} H_1(y)-8\right) \ln^2(y)+\frac{4}{3} \ln^3(y)
\biggr]+{\cal O}(\varepsilon^2).
\end{eqnarray}
Notice the presence of the letters 
\begin{equation}
\frac{1}{2-y},~~~\frac{1}{1-y}~\text{and}~~~\frac{1}{y}.
\end{equation}
After we go back to the original variable $x=1-y$, we 
obtain a representation in terms of the standard harmonic polylogarithms with the letters 
(\ref{eq:hpol}),
\begin{eqnarray}
K_4(x) &=& -\frac{1}{\varepsilon^2}
-\frac{1}{\varepsilon} \left[\frac{2}{1-x} H_0(x)+2 H_1(x)+4\right]
-\frac{2 (1+x)}{1-x} H_{0,-1}(x)-\frac{2+x}{1-x} H_0^2(x)
 \nonumber \\ &&
+\frac{1}{1-x} \big(2 (1+x) H_{-1}(x)-4 H_1(x)-8\big) H_0(x)
-2 H_1^2(x)-8 H_1(x)-12
 \nonumber \\ &&
+\frac{1+x}{1-x} \zeta_2
+\varepsilon \biggl[
\left(\frac{1+x}{1-x} \big(2 \zeta_2-4 H_{0,-1}(x)\big)-24\right) H_1(x)
-\frac{4}{3} H_1^3(x)
 \nonumber \\ &&
-\frac{1+x}{1-x} \big[H_{-1}(x) \left(-4 H_{0,-1}(x)-3 H_0^2(x)-\left(4 H_1(x)+8\right) H_0(x)+2 \zeta_2\right)
 \nonumber \\ &&
+2 H_0(x) \left(H_{0,-1}(x)-\zeta_2\right)
+8 H_{0,-1}(x)+4 H_{0,-1,-1}(x)+2 H_{0,0,-1}(x)
\nonumber \\ &&
+2 H_0(x) H_{-1}^2(x)-4 \zeta_2\big]
-\frac{4+3 x}{3 (1-x)} H_0^3(x)
-\frac{2+x}{1-x} \left(2 H_1(x)+4\right) H_0^2(x)
\nonumber \\ &&
-\frac{2}{1-x} \left(2 H_1^2(x)+8 H_1(x)+12\right) H_0(x)-8 H_1^2(x)+\frac{2 (7-x)}{3 (1-x)} \zeta_3-32
\biggr]
\nonumber \\ &&
+{\cal O}(\varepsilon^2).
\end{eqnarray}

\section{Results} \label{sec:result}

\vspace*{1mm}
\noindent
We calculated the heavy-quark form factors $F_I$,
$I=V,A,S,P$, up to two loops and ${\cal O}(\ep^2)$. Due to the length
of the expressions we list here only the expansion corresponding to
the low-energy ($0 < q^2 \ll m^2$), high-energy ($ \abs{q^2} \gg m^2$)
and threshold ($ \abs{q^2} \approx 4 m^2$) region up to
$\mathcal{O}(\ep)$. In appendices \ref{app:vFF}, \ref{app:avFF},
\ref{app:sFF}, and \ref{app:pFF} we present the complete analytic
expressions up to ${\cal O}(\ep)$.  The full expressions up to
${\cal O}(\ep^2)$ are provided as supplemental material together with
this publication. We present renormalized results for all form factors but
the singlet contributions to the axial-vector and pseudo-scalar currents
$\hat{F}_{A,i}^{\sing,(2)}$ and $\hat{F}_{P}^{\sing,(2)}$ for which
we present the bare results as discussed in Section \ref{sec:reno}.
The expansions have been obtained with the help of the Mathematica
packages \verb|Sigma| and \verb|HarmonicSums|.

For convenience, we collect here the notation used in the
presentation of the results. We use the dimensionless variable $x$, Eq.~(\ref{eq:varx}).  
The kinematic regions of interest correspond to
$x\to 1$ ($q^2 = 0$), $x\to \pm 0$ ($q^2 = \mp \infty$) and  $x\to -1$
($q^2 = 4 m^2$). 

Since the region $0 < q^2 < 4 m^2$ corresponds to the
upper  half of the unit circle in the complex plane it
is convenient to define the variable $\phi$ by
  \begin{equation}
x = e^{i\phi}
\label{eq:xphi}
      \end{equation}
and to expand around $\phi = 0 $ instead.

In the threshold region we use the velocity of the heavy quarks as basic variable
\begin{equation}
 \beta = \sqrt{1 - \frac{4m^2}{q^2}} \quad \leftrightarrow \quad x =
 \frac{\beta -1 }{ \beta + 1}
\end{equation}
and expand around $\beta  = 0$. This avoids the appearance of square roots.

For the presentation of the complete analytic results in appendices
\ref{app:vFF}, \ref{app:avFF}, \ref{app:sFF} and \ref{app:pFF} we
introduce the following subsidiary variables
\begin{equation}
 x_{+} = \frac{1}{1+x} \,, \quad \eta = \frac{1}{(1-x)(1+x)}\,, \quad \xi = \frac{(1+x^2)}{(1-x)(1+x)} \,.
\end{equation}
Furthermore,  we use the abbreviations 
\begin{align}
 {c}_1 &= 12 \zeta_2 \ln^2(2) + \ln^4(2) + 24 {\rm Li}_4 \Big( \frac{1}{2} \Big)
\nonumber\\
 c_2 &= 26 \zeta_2^2 \ln (2) - 20 \zeta_2 \ln^3(2) - \ln^5(2) + 120 {\rm Li}_5 \Big( \frac{1}{2} \Big) \,
\end{align}
and
\begin{equation}
H_{a_1,\ldots,a_n} \equiv H_{a_1,\ldots,a_n}(x)\,, \quad   L \equiv H_0(x) = \ln(x) \,, \quad \bar{H}_0 
(\phi) \equiv H_{0} (\phi) - \frac{i \pi}{2} \,,
\end{equation}
with the harmonic polylogarithms $H_{a_1,\ldots,a_n}(x)$ as defined in Eq.~(\ref{HPLdef}).
%

To validate our results we compare them to the existing literature.
Up to ${\cal O}(\ep^0)$ we agree with all available unrenormalized
results for the various form factors. Note that in
Refs.~\cite{Bernreuther:2004ih, Bernreuther:2004th,
  Bernreuther:2005gw} a different normalization for the master
integrals has been used, resulting in a difference proportional to
$(\Gamma(1 + \ep)/\exp(\gamma_E \ep))^2$, where $\gamma_E$ denotes the Euler-Mascheroni constant.

At ${\cal O}(\ep)$ we can
compare our results for the vector form factors with the results given
in Ref.~\cite{Gluza:2009yy} and find a difference 
\begin{equation}
\label{diffGluza}
 - C_F C_A \Bigg\{ \ep \left[ \frac{1037 x^3}{(1+x)^6} \right] \Bigg\} \,,
\end{equation}
which has been reported in \cite{Henn:2016tyf} already. In addition we
compared our analytic results as well as the corresponding
expansions with the results for the color-planar limit given in
\cite{Henn:2016tyf} and find agreement.

Comparing the renormalized results, we found that the wave function
renormalization in Refs.~\cite{Bernreuther:2004ih, Bernreuther:2004th,
  Bernreuther:2005gw} has been performed incorrectly, resulting in a
difference proportional to $\zeta_2$ at ${\cal O}(\ep^0)$. For the
vector form factor we agree with the renormalized results in the
color-planar limit given in Ref.~\cite{Henn:2016tyf} and up to the term
mentioned in Eq.~(\ref{diffGluza}) above with the results given in Ref.~\cite{Gluza:2009yy}.

Although we do not present results for the renormalized singlet
contributions we like to point out, that we cannot reproduce the
result for the singlet axial-vector contribution presented in
Ref.~\cite{Bernreuther:2005rw}.

\subsection{Low energy region \boldmath $0< q^2 \ll m^2$}

\vspace*{1mm}
\noindent
The low energy limit of the space-like form factors is given by
$x\rightarrow1$.  To facilitate the expansion of the HPLs in the
region, we use the variable $x$ as defined in Eq.~(\ref{eq:xphi}) and expand around 
$\phi=0$.
In the following we present the series expansion of the one and two-loop form factors, denoted by
$\bar{F}$, for all the currents up to 4th order in $\phi$.
\subsubsection{Vector form factor}

\vspace*{1mm}
\noindent
For the vector form factors we find 
\begin{align}
 \bar{F}_{V,1}^{(1)} &=
C_F \Bigg\{
\frac{1}{\ep} \Bigg[
   -\frac{2}{3} \phi^2 - \frac{2}{45} \phi^4
\Bigg]
+  \Bigg[
  -\frac{1}{2} \phi^2 - \frac{17}{120} \phi^4
\Bigg]
+ \ep  \Bigg[
\phi ^2 \Big(
        -2
        -\frac{\zeta_2}{3}
\Big)
\nonumber\\
&
+\phi ^4 \Big(
        -\frac{1}{4}
        -\frac{\zeta_2}{45}
\Big)
\Bigg]
+ \ep^2  \Bigg[
\phi ^2 \Big(
        -4
        -\frac{\zeta_2}{4}
        +\frac{2 \zeta_3}{9}
\Big)
+ \phi ^4 \Big(
        -\frac{2}{3}
        -\frac{17 \zeta_2}{240}
        +\frac{2 \zeta_3}{135}
\Big)
\Bigg]
\Bigg\} \,.
%
\\
\bar{F}_{V,1}^{(2)} &=
C_F^2 \Bigg\{
\frac{2}{9 \ep^2} \phi^4  
+ \frac{1}{3 \ep} \phi^4
+  \Bigg[
\phi ^2 \Big(
        -\frac{47}{36}
        -\frac{175}{9} \zeta_2
        +48 \ln (2) \zeta_2
        -12 \zeta_3
\Big)
+\phi ^4 \Big(
        \frac{14473}{6480}
\nonumber\\&
        -\frac{34243 \zeta_2}{5400}
        +\frac{68}{5} \ln (2) \zeta_2
        -\frac{17}{5} \zeta_3
\Big)
\Bigg]
+ \ep  \Bigg[
\phi ^2 \Big(
        \frac{11713}{216}
        -8 c_1
        -\frac{6763}{54} \zeta_2
        +\frac{578}{3} \ln (2) \zeta_2
        +\frac{504}{5} \zeta_2^2
\nonumber\\&
        -\frac{1567}{18} \zeta_3
\Big)
+\phi ^4 \Big(
        \frac{3508637}{194400}
        -\frac{34 c_1}{15}
        -\frac{758317 \zeta_2}{20250}
        +\frac{10139}{180} \ln (2) \zeta_2
        +\frac{714}{25} \zeta_2^2
        -\frac{295457 \zeta_3}{10800}
\Big)
\Bigg]
\Bigg\}
\nonumber\\&
+ C_F C_A \Bigg\{
\frac{1}{\ep^2}  \Bigg[
  \frac{11}{9} \phi ^2 + \frac{11}{135} \phi^4
\Bigg]
+ \frac{1}{\ep} \Bigg[
\phi ^2 \Big(
        -\frac{94}{27}
        +\frac{4 \zeta_2}{3}
\Big)
+ \phi ^4 \Big(
        -\frac{91}{405}
        +\frac{4 \zeta_2}{45}
\Big)
\Bigg]
\nonumber\\&
+  \Bigg[
\phi ^2 \Big(
        -\frac{2579}{324}
        +\frac{155}{18} \zeta_2
        -24 \ln (2) \zeta_2
        +\frac{26}{3} \zeta_3
\Big)
+
\phi ^4 \Big(
        -\frac{36239}{19440}
        +\frac{7447 \zeta_2}{2160}
        -\frac{34}{5} \ln (2) \zeta_2
\nonumber\\&
        +\frac{169}{90} \zeta_3
\Big)
\Bigg]
+ \ep  \Bigg[
\phi ^2 \Big(
        -\frac{134327}{1944}
        +4 c_1
        +\frac{1297}{27} \zeta_2
        -\frac{289}{3} \ln (2) \zeta_2
        -\frac{608}{15} \zeta_2^2
        +\frac{1487}{36} \zeta_3
\Big)
\nonumber\\&
+
\phi ^4 \Big(
        -\frac{278341}{23328}
        +\frac{17 c_1}{15}
        +\frac{110029 \zeta_2}{6480}
        -\frac{10139}{360} \ln (2) \zeta_2
        -\frac{613}{45} \zeta_2^2
        +\frac{12041}{864} \zeta_3
\Big)
\Bigg]
\Bigg\}
\nonumber\\&
+
C_F n_l T_F \Bigg\{
\frac{1}{\ep^2}  \Bigg[
   -\frac{4}{9} \phi ^2
   -\frac{4}{135} \phi ^4
\Bigg]
+ \frac{1}{\ep} \Bigg[
   \frac{20}{27} \phi ^2
+  \frac{4}{81} \phi ^4
\Bigg]
+  \Bigg[
\phi ^2 \Big(
        \frac{283}{81}
        +\frac{16 \zeta_2}{9}
\Big)
\nonumber\\&
+
\phi ^4 \Big(
        \frac{3139}{4860}
        +\frac{16 \zeta_2}{135}
\Big)
\Bigg]
+ \ep  \Bigg[
\phi ^2 \Big(
        \frac{8827}{486}
        +\frac{181 \zeta_2}{27}
        +\frac{32 \zeta_3}{9}
\Big)
+\phi ^4 \Big(
        \frac{95527}{29160}
        +\frac{1777 \zeta_2}{1620}
        +\frac{32 \zeta_3}{135}
\Big)
\Bigg]
\Bigg\}
\nonumber\\& 
+
C_F T_F \Bigg\{
  \Bigg[
\phi ^2 \Big(
        -\frac{1099}{81}
        +9 \zeta_2
\Big)
+
\phi ^4 \Big(
        -\frac{21019}{4860}
        +\frac{53 \zeta_2}{20}
\Big)
\Bigg]
+ \ep  \Bigg[
\phi ^2 \Big(
        -\frac{635}{18}
        +\frac{937}{27} \zeta_2
\nonumber\\&
        -\frac{154}{3} \ln (2) \zeta_2
        +\frac{1601}{54} \zeta_3
\Big)
+
\phi ^4 \Big(
        -\frac{8293}{648}
        +\frac{9821}{810} \zeta_2
        -\frac{283}{18} \ln (2) \zeta_2
        +\frac{29651 \zeta_3}{3240}
\Big)
\Bigg]
\\
\bar{F}_{V,2}^{(1)} &=
C_F \Bigg\{
 \Bigg[
2 +  \frac{1}{3} \phi ^2 + \frac{7}{180} \phi ^4
\Bigg]
+ \ep  \Bigg[
8 + \frac{5}{3} \phi ^2 + \frac{41}{180} \phi ^4
\Bigg]
+ \ep^2  \Bigg[
16
+\zeta_2
+\phi ^2 \Big(
        4
        +\frac{\zeta_2}{6}
\Big)
\nonumber\\&
+\phi ^4 \Big(
        \frac{19}{30}
        +\frac{7 \zeta_2}{360}
\Big)
\Bigg]
\Bigg\} \,.
%
\\
\bar{F}_{V,2}^{(2)} &=
C_F^2 \Bigg\{
 \frac{1}{\ep} \Bigg[
 -\frac{4}{3} \phi ^2  -\frac{14}{45} \phi ^4 
\Bigg]
+  \Bigg[
-31
+40 \zeta_2
-48 \ln (2) \zeta_2
+12 \zeta_3
+\phi ^2 \Big(
        -\frac{77}{5}
        +\frac{122}{5} \zeta_2
\nonumber\\&
        -\frac{184}{5} \ln (2) \zeta_2
        +\frac{46}{5} \zeta_3
\Big)
+\phi ^4 \Big(
        -\frac{4931}{1260}
        +\frac{2963}{350} \zeta_2
        -\frac{1478}{105} \ln (2) \zeta_2
        +\frac{739}{210} \zeta_3
\Big)
\Bigg]
+ \ep  \Bigg[
-\frac{1243}{6}
\nonumber\\&
+8 c_1
+\frac{944}{3} \zeta_2
-384 \ln (2) \zeta_2
-\frac{504}{5} \zeta_2^2
+176 \zeta_3
+\phi ^2 \Big(
        -\frac{19666}{225}
        +\frac{92 c_1}{15}
        +\frac{3704}{25} \zeta_2
\nonumber\\&
        -\frac{14164}{75} \ln (2) \zeta_2
        -\frac{1932}{25} \zeta_2^2
        +\frac{7201}{75} \zeta_3
\Big)
+\phi ^4 \Big(
        -\frac{9903863}{396900}
        +\frac{739 c_1}{315}
        +\frac{10057561 \zeta_2}{220500}
\nonumber\\&
        -\frac{125887 \ln (2) \zeta_2}{2205}
        -\frac{739}{25} \zeta_2^2
        +\frac{1376111 \zeta_3}{44100}
\Big)
\Bigg]
\Bigg\}
+
C_F C_A \Bigg\{
  \Bigg[
\frac{317}{9}
-12 \zeta_2
+24 \ln (2) \zeta_2
-6 \zeta_3
\nonumber\\&
+\phi ^2 \Big(
        \frac{1699}{270}
        -\frac{137}{15} \zeta_2
        +\frac{92}{5} \ln (2) \zeta_2
        -\frac{23}{5} \zeta_3
\Big)
+\phi ^4 \Big(
        \frac{11927}{22680}
        -\frac{21269 \zeta_2}{6300}
        +\frac{739}{105} \ln (2) \zeta_2
\nonumber\\&
        -\frac{739}{420} \zeta_3
\Big)
\Bigg]
+ \ep  \Bigg[
\frac{12881}{54}
-4 c_1
-\frac{313}{3} \zeta_2
+192 \ln (2) \zeta_2
+\frac{252}{5} \zeta_2^2
-72 \zeta_3
\nonumber\\&
+\phi ^2 \Big(
        \frac{485453}{8100}
        -\frac{46 c_1}{15}
        -\frac{8983}{150} \zeta_2
        +\frac{7082}{75} \ln (2) \zeta_2
        +\frac{966}{25} \zeta_2^2
        -\frac{6281}{150} \zeta_3
\Big)
\nonumber\\&
+\phi ^4 \Big(
        \frac{50620531}{4762800}
        -\frac{739 c_1}{630}
        -\frac{4335431 \zeta_2}{220500}
        +\frac{125887 \ln (2) \zeta_2}{4410}
        +\frac{739}{50} \zeta_2^2
        -\frac{1224967 \zeta_3}{88200}
\Big)
\Bigg]
\Bigg\}
\nonumber\\&
+ C_F n_l T_F \Bigg\{
  \Bigg[
-\frac{100}{9} 
-\frac{62}{27} \phi ^2
- \frac{253}{810} \phi ^4
\Bigg]
+ \ep  \Bigg[
-\frac{1922}{27}
-12 \zeta_2
\nonumber\\&
+\phi ^2 \Big(
        -\frac{1405}{81}
        -2 \zeta_2
\Big)
+\phi ^4 \Big(
        -\frac{13147}{4860}
        -\frac{7 \zeta_2}{30}
\Big)
\Bigg]
\Bigg\}
+ C_F T_F \Bigg\{
  \Bigg[
\frac{476}{9}
-32 \zeta_2
+\phi ^2 \Big(
        \frac{622}{27}
        -14 \zeta_2
\Big)
\nonumber\\&
+\phi ^4 \Big(
        \frac{4841}{810}
        -\frac{109 \zeta_2}{30}
\Big)
\Bigg]
+ \ep  \Bigg[
\frac{2254}{27}
-\frac{308}{3} \zeta_2
+192 \ln (2) \zeta_2
-112 \zeta_3
+\phi ^2 \Big(
        \frac{4247}{81}
        -\frac{490}{9} \zeta_2
\nonumber\\&
        +84 \ln (2) \zeta_2
        -49 \zeta_3
\Big)
+\phi ^4 \Big(
        \frac{16753}{972}
        -\frac{2203}{135} \zeta_2
        +\frac{109}{5} \ln (2) \zeta_2
        -\frac{763}{60} \zeta_3
\Big)
\Bigg]
\Bigg\} \,.
\end{align}
Note, that $\bar{F}_{V,2}$ is UV and IR finite in this limit and the leading
term agrees with the computation of 
the anomalous magnetic moment in \cite{Grozin:2007fh}.

\subsubsection{Axial-vector form factor}
For the axial-vector form factor we present the renormalized results
for the non-singlet contributions  and the unrenormalized one for the
singlet parts.
\begin{align}
\bar{F}_{A,1}^{(1),\ns} &=
C_F \Bigg\{
\frac{1}{\ep} \Bigg[
 -\frac{2}{3} \phi ^2 -\frac{2}{45} \phi ^4
\Bigg]
+  \Bigg[
-2 -\frac{5}{6} \phi ^2 -\frac{13}{72} \phi ^4
\Bigg]
+ \ep  \Bigg[
\phi ^2 \Big(
        -\frac{7}{3}
        -\frac{\zeta_2}{3}
\Big)
+\phi ^4 \Big(
        -\frac{29}{90}
\nonumber\\&
        -\frac{\zeta_2}{45}
\Big)
\Bigg]
+ \ep^2  \Bigg[
-\zeta_2
+\phi ^2 \Big(
        -4
        -\frac{5 \zeta_2}{12}
        +\frac{2 \zeta_3}{9}
\Big)
+\phi ^4 \Big(
        -\frac{7}{10}
        -\frac{13 \zeta_2}{144}
        +\frac{2 \zeta_3}{135}
\Big)
\Bigg]
\Bigg\} \,.
%
\\
\bar{F}_{A,1}^{(2),\ns} &=
C_F^2 \Bigg\{
\frac{1}{\ep^2}  \Bigg[
  \frac{2}{9} \phi ^4
\Bigg]
+ \frac{1}{\ep} \Bigg[
  \frac{4}{3} \phi ^2 + \frac{29}{45} \phi ^4
\Bigg]
+  \Bigg[
-\frac{29}{3}
+32 \zeta_2
-32 \ln (2) \zeta_2
+8 \zeta_3
+\phi ^2 \Big(
        -\frac{1121}{180}
\nonumber\\&
        -\frac{217}{45} \zeta_2
        +\frac{128}{5} \ln (2) \zeta_2
        -\frac{32}{5} \zeta_3
\Big)
+\phi ^4 \Big(
        \frac{95341}{45360}
        -\frac{5423 \zeta_2}{5400}
        +\frac{376}{105} \ln (2) \zeta_2
        -\frac{94}{105} \zeta_3
\Big) 
\Bigg]
\nonumber\\&
+ \ep  \Bigg[
-\frac{1889}{18}
+\frac{16 c_1}{3}
+56 \zeta_2
-\frac{160}{3} \ln (2) \zeta_2
-\frac{336}{5} \zeta_2^2
+\frac{232}{3} \zeta_3
+\phi ^2 \Big(
        -\frac{18997}{5400}
        -\frac{64 c_1}{15}
\nonumber\\&
        -\frac{49259 \zeta_2}{1350}
        +\frac{6958}{75} \ln (2) \zeta_2
        +\frac{1344}{25} \zeta_2^2
        -\frac{14777}{450} \zeta_3
\Big)
+\phi ^4 \Big(
        -\frac{5022433}{9525600}
        -\frac{188 c_1}{315}
        -\frac{44327 \zeta_2}{20250}
\nonumber\\&
        +\frac{721523 \ln (2) \zeta_2}{44100}
        +\frac{188}{25} \zeta_2^2
        -\frac{131151 \zeta_3}{19600}
\Big)
\Bigg]
\Bigg\}
+ C_F C_A \Bigg\{
\frac{1}{\ep^2}  \Bigg[
  \frac{11}{9} \phi ^2 + \frac{11}{135} \phi ^4
\Bigg]
+ \frac{1}{\ep} \Bigg[
\phi ^2 \Big(
        -\frac{94}{27}
\nonumber\\&
        +\frac{4 \zeta_2}{3}
\Big)
+
\phi ^4 \Big(
        -\frac{91}{405}
        +\frac{4 \zeta_2}{45}
\Big)
\Bigg]
+  \Bigg[
-\frac{143}{9}
-8 \zeta_2
+16 \ln (2) \zeta_2
-4 \zeta_3
+\phi ^2 \Big(
        -\frac{19813}{1620}
\nonumber\\&
        +\frac{317}{90} \zeta_2
        -\frac{64}{5} \ln (2) \zeta_2
        +\frac{88}{15} \zeta_3
\Big)
+\phi ^4 \Big(
        -\frac{413831}{136080}
        +\frac{14479 \zeta_2}{10800}
        -\frac{188}{105} \ln (2) \zeta_2
        +\frac{197}{315} \zeta_3
\Big)
\Bigg]
\nonumber\\&
+ \ep  \Bigg[
-\frac{887}{54}
-\frac{8 c_1}{3}
-\frac{59}{3} \zeta_2
+\frac{80}{3} \ln (2) \zeta_2
+\frac{168}{5} \zeta_2^2
-\frac{68}{3} \zeta_3
+\phi ^2 \Big(
        -\frac{3362201}{48600}
        +\frac{32 c_1}{15}
\nonumber\\&
        +\frac{1753}{150} \zeta_2
        -\frac{3479}{75} \ln (2) \zeta_2
        -\frac{1276}{75} \zeta_2^2
        +\frac{16777}{900} \zeta_3
\Big)
+\phi ^4 \Big(
        -\frac{281936227}{28576800}
        +\frac{94 c_1}{315}
        +\frac{583651 \zeta_2}{378000}
\nonumber\\&
        -\frac{721523 \ln (2) \zeta_2}{88200}
        -\frac{698}{225} \zeta_2^2
        +\frac{5002453 \zeta_3}{1058400}
\Big)
\Bigg]
\Bigg\}
+ C_F n_l T_F \Bigg\{
\frac{1}{\ep^2}  \Bigg[
  -\frac{4}{9} \phi ^2 -\frac{4}{135} \phi ^4
\Bigg]
\nonumber\\&
+ \frac{1}{\ep} \Bigg[
 \frac{20}{27} \phi ^2 + \frac{4}{81} \phi ^4
\Bigg]
+  \Bigg[
\frac{28}{9}
+\phi ^2 \Big(
        \frac{361}{81}
        +\frac{16 \zeta_2}{9}
\Big)
+\phi ^4 \Big(
        \frac{3901}{4860}
        +\frac{16 \zeta_2}{135}
\Big)
\Bigg]
\nonumber\\&
+ \ep  \Bigg[
-\frac{34}{27}
+12 \zeta_2
+\phi ^2 \Big(
        \frac{9229}{486}
        +\frac{235 \zeta_2}{27}
        +\frac{32 \zeta_3}{9}
\Big)
+\phi ^4 \Big(
        \frac{105253}{29160}
        +\frac{431 \zeta_2}{324}
        +\frac{32 \zeta_3}{135}
\Big)
\Bigg]
\Bigg\}
\nonumber\\&
+ C_F T_F \Bigg\{
  \Bigg[
\frac{460}{9}
-32 \zeta_2
+\phi ^2 \Big(
        \frac{491}{81}
        -3 \zeta_2
\Big)
+\phi ^4 \Big(
        \frac{1343}{4860}
        -\frac{3 \zeta_2}{20}
\Big)
\Bigg]
+ \ep  \Bigg[
\frac{3998}{27}
-\frac{412}{3} \zeta_2
\nonumber\\&
+192 \ln (2) \zeta_2
-112 \zeta_3
+\phi ^2 \Big(
        \frac{4931}{162}
        -\frac{653}{27} \zeta_2
        +\frac{62}{3} \ln (2) \zeta_2
        -\frac{667}{54} \zeta_3
\Big)
\nonumber\\&
+\phi ^4 \Big(
        \frac{10963}{1944}
        -\frac{1436}{405} \zeta_2
        +\frac{97}{90} \ln (2) \zeta_2
        -\frac{2101 \zeta_3}{3240}
\Big)
\Bigg]
\Bigg\} \,.
\end{align}
\begin{align} 
%
{\bar{\hat{F}}}_{A,1}^{(2),\sing} &=
C_F T_F \Bigg\{
%
 -\frac{6}{\ep}
+  \Bigg[
-\frac{37}{3}
+\frac{64}{3} \zeta_2
+\phi ^2 \Big(
        -\frac{85}{18}
        +\frac{136 \zeta_2}{45}
\Big)
+\phi ^4 \Big(
        -\frac{341}{540}
        +\frac{2554 \zeta_2}{4725}
\Big)
\Bigg]
\nonumber\\&
+ \ep  \Bigg[
-\frac{817}{18}
+\frac{266}{9} \zeta_2
-128 \ln (2) \zeta_2
+\frac{224}{3} \zeta_3
+\phi ^2 \Big(
        -\frac{3659}{540}
        +\frac{4876}{675} \zeta_2
        -\frac{272}{15} \ln (2) \zeta_2
\nonumber\\&
        +\frac{476}{45} \zeta_3
\Big) 
-\frac{1}{2} i \phi ^3 \zeta_2
+\phi ^4 \Big(
        -\frac{14281}{16200}
        +\frac{985837 \zeta_2}{1984500}
        -\frac{5108 \ln (2) \zeta_2}{1575}
        +\frac{1277}{675} \zeta_3
\Big)
\Bigg]
\Bigg\} \,.
\end{align}
\begin{align}
 \bar{F}_{A,2}^{(1),\ns} &=
C_F \Bigg\{
  \Bigg[
\frac{14}{3} + \frac{11}{15} \phi ^2 + \frac{103}{1260} \phi ^4
\Bigg]
+ \ep  \Bigg[
8 + \frac{31}{15} \phi ^2 + \frac{389}{1260} \phi ^4 
\Bigg]
+ \ep^2  \Bigg[
16
+\frac{7}{3} \zeta_2
\nonumber\\&
+\phi ^2 \Big(
        4
        +\frac{11 \zeta_2}{30}
\Big)
+\phi ^4 \Big(
        \frac{47}{70}
        +\frac{103 \zeta_2}{2520}
\Big)
\Bigg]
\Bigg\} \,.
%
\\
\bar{F}_{A,2}^{(2),\ns} &=
C_F^2 \Bigg\{
 \frac{1}{\ep} \Bigg[
-\frac{28}{9} \phi ^2 - \frac{94}{135} \phi ^4
\Bigg]
+  \Bigg[
-\frac{23}{5}
+\frac{176}{5} \zeta_2
-\frac{176}{5} \ln (2) \zeta_2
+\frac{44}{5} \zeta_3
+\phi ^2 \Big(
        -\frac{11111}{945}
\nonumber\\&
        +\frac{592}{45} \zeta_2
        -\frac{88}{7} \ln (2) \zeta_2
        +\frac{22}{7} \zeta_3
\Big)
+\phi ^4 \Big(
        -\frac{251113}{56700}
        +\frac{208091 \zeta_2}{66150}
        -\frac{298}{105} \ln (2) \zeta_2
        +\frac{149}{210} \zeta_3
\Big)
\Bigg]
\nonumber\\&
+ \ep  \Bigg[
\frac{15527}{450}
+\frac{88 c_1}{15}
+\frac{28688}{225} \zeta_2
-\frac{6512}{25} \ln (2) \zeta_2
-\frac{1848}{25} \zeta_2^2
+\frac{3388}{25} \zeta_3
+\phi ^2 \Big(
        \frac{376}{11025}
        +\frac{44 c_1}{21}
\nonumber\\&
        +\frac{3776}{135} \zeta_2
        -\frac{235856 \ln (2) \zeta_2}{3675}
        -\frac{132}{5} \zeta_2^2
        +\frac{466972 \zeta_3}{11025}
\Big)
+\phi ^4 \Big(
        -\frac{47414267}{17860500}
        +\frac{149 c_1}{315}
\nonumber\\&
        +\frac{34009847 \zeta_2}{6945750}
        -\frac{25891 \ln (2) \zeta_2}{2205}
        -\frac{149}{25} \zeta_2^2
        +\frac{1220729 \zeta_3}{132300}
\Big)
\Bigg]
\Bigg\}
+ C_F C_A \Bigg\{
  \Bigg[
\frac{7663}{135}
-\frac{752}{45} \zeta_2
\nonumber\\&
+\frac{88}{5} \ln (2) \zeta_2
-\frac{22}{5} \zeta_3
+\phi ^2 \Big(
        \frac{5039}{378}
        -\frac{422}{75} \zeta_2
        +\frac{44}{7} \ln (2) \zeta_2
        -\frac{11}{7} \zeta_3
\Big)
+\phi ^4 \Big(
        \frac{27793}{12600}
        -\frac{56827 \zeta_2}{44100}
\nonumber\\&
        +\frac{149}{105} \ln (2) \zeta_2
        -\frac{149}{420} \zeta_3
\Big)
\Bigg]
+ \ep  \Bigg[
\frac{871991}{4050}
-\frac{44 c_1}{15}
-\frac{19517}{675} \zeta_2
+\frac{3256}{25} \ln (2) \zeta_2
+\frac{924}{25} \zeta_2^2
\nonumber\\&
-\frac{14846}{225} \zeta_3
+\phi ^2 \Big(
        \frac{20526143}{396900}
        -\frac{22 c_1}{21}
        -\frac{142673 \zeta_2}{15750}
        +\frac{117928 \ln (2) \zeta_2}{3675}
        +\frac{66}{5} \zeta_2^2
        -\frac{70838 \zeta_3}{3675}
\Big)
\nonumber\\&
+\phi ^4 \Big(
        \frac{548179231}{71442000}
        -\frac{149 c_1}{630}
        -\frac{5017571 \zeta_2}{3087000}
        +\frac{25891 \ln (2) \zeta_2}{4410}
        +\frac{149}{50} \zeta_2^2
        -\frac{118921 \zeta_3}{29400}
\Big)
\Bigg]
\Bigg\}
\nonumber\\&
+ C_F n_l T_F \Bigg\{
  \Bigg[
-\frac{412}{27}-\frac{466}{135} \phi ^2 - \frac{2761}{5670} \phi ^4
\Bigg]
+ \ep  \Bigg[
-\frac{5630}{81}
-28 \zeta_2
+\phi ^2 \Big(
        -\frac{7427}{405}
        -\frac{22 \zeta_2}{5}
\Big)
\nonumber\\&
+\phi ^4 \Big(
        -\frac{104863}{34020}
        -\frac{103 \zeta_2}{210}
\Big)
\Bigg]
\Bigg\}
%
+ C_F T_F \Bigg\{
  \Bigg[
-\frac{412}{27}
+\frac{32}{3} \zeta_2
+\phi ^2 \Big(
        -\frac{14}{27}
        +\frac{2 \zeta_2}{5}
\Big)
+\phi ^4 \Big(
        \frac{4601}{5670}
\nonumber\\&
        -\frac{103 \zeta_2}{210}
\Big)
\Bigg]
+ \ep  \Bigg[
-\frac{9230}{81}
+\frac{724}{9} \zeta_2
-64 \ln (2) \zeta_2
+\frac{112}{3} \zeta_3
+\phi ^2 \Big(
        -\frac{10703}{405}
        +\frac{146}{9} \zeta_2
\nonumber\\&
        -\frac{12}{5} \ln (2) \zeta_2
        +\frac{7}{5} \zeta_3
\Big)
+\phi ^4 \Big(
        -\frac{119839}{34020}
        +\frac{1238}{945} \zeta_2
        +\frac{103}{35} \ln (2) \zeta_2
        -\frac{103}{60} \zeta_3
\Big)
\Bigg]
\Bigg\} \,.
\end{align}
%
\begin{align}
\bar{\hat{F}}_{A,2}^{(2),\sing} &=
C_F T_F \Bigg\{
\Bigg[
\frac{24}{\phi ^2}
+\frac{20}{3}
+\frac{16}{15} \zeta_2
+4 i \phi  \zeta_2
+\phi ^2 \Big(
        -\frac{8}{15}
        -\frac{184}{315} \zeta_2
        +\frac{4 \bar{H}_0 (\phi )}{3}
\Big)
+\frac{13}{15} i \phi ^3 \zeta_2 
\nonumber\\&
+\phi ^4 \Big(
        -\frac{461}{6300}
        -\frac{1174 \zeta_2}{4725}
        +\frac{13 \bar{H}_0 (\phi )}{45}
\Big)
\Bigg]
+ \ep  \Bigg[
\frac{1}{\phi ^2} \Big(
124
-128 \zeta_2
\Big)
+\frac{3394}{45}
-\frac{11264}{225} \zeta_2
\nonumber\\&
-\frac{32}{5} \ln (2) \zeta_2
+\frac{56}{15} \zeta_3
+ i \phi  \Big(
        -10 \zeta_2
        +8 \ln (2) \zeta_2
        -8 \zeta_2 \bar{H}_0 (\phi )
\Big)
+\phi ^2 \Big(
        \frac{3173}{135}
        -\frac{221449 \zeta_2}{33075}
\nonumber\\&
        +\frac{368}{105} \ln (2) \zeta_2
        -\frac{92}{45} \zeta_3
        -4 \bar{H}_0 (\phi )
        -\frac{4}{3} \bar{H}_0 ^2(\phi )
\Big)
+ i \phi ^3 \Big(
        -\frac{9}{5} \zeta_2
        +\frac{26}{15} \ln (2) \zeta_2
        -\frac{26}{15} \zeta_2 \bar{H}_0 (\phi )
\Big)
\nonumber\\&
+\phi ^4 \Big(
        \frac{5566123}{1701000}
        -\frac{1063927 \zeta_2}{1984500}
        +\frac{2348 \ln (2) \zeta_2}{1575}
        -\frac{587}{675} \zeta_3
        -\frac{31 \bar{H}_0 (\phi )}{45}
        -\frac{13}{45} \bar{H}_0 ^2(\phi )
\Big)
\Bigg]
%
\Bigg\} \,.
\end{align}
%
\subsubsection{Scalar form factor}

\vspace*{1mm}
\noindent
For the scalar form factor in the low energy limit we obtain
\begin{align}
 \bar{F}_{S}^{(1)} &=
C_F \Bigg\{
\frac{1}{\ep} \Bigg[
-\frac{2}{3} \phi ^2 - \frac{2}{45} \phi ^4
\Bigg]
+  \Bigg[
-6-\frac{1}{3} \phi ^2 -\frac{31}{180} \phi ^4
\Bigg]
+ \ep  \Bigg[
-8
+\phi ^2 \Big(
        -\frac{7}{3}
        -\frac{\zeta_2}{3}
\Big)
\nonumber\\&
+\phi ^4 \Big(
        -\frac{49}{180}
        -\frac{\zeta_2}{45}
\Big)
\Bigg]
+ \ep^2  \Bigg[
-16
-3 \zeta_2
+\phi ^2 \Big(
        -4
        -\frac{\zeta_2}{6}
        +\frac{2 \zeta_3}{9}
\Big)
\nonumber\\&
+\phi ^4 \Big(
        -\frac{7}{10}
        -\frac{31 \zeta_2}{360}
        +\frac{2 \zeta_3}{135}
\Big)
\Bigg]
\Bigg\} \,.
%
\\
\bar{F}_{S}^{(2)} &=
C_F^2 \Bigg\{
\frac{2}{9 \ep^2} \phi^4
+ \frac{1}{\ep} \Bigg[
4 \phi ^2+\frac{22}{45} \phi ^4
\Bigg]
+  \Bigg[
33
+\phi ^2 \Big(
        \frac{62}{9}
        -\frac{71}{9} \zeta_2
        +28 \ln (2) \zeta_2
        -7 \zeta_3
\Big)
\nonumber\\&
+\phi ^4 \Big(
        \frac{5743}{1620}
        -\frac{7001 \zeta_2}{5400}
        +\frac{11}{3} \ln (2) \zeta_2
        -\frac{11}{12} \zeta_3
\Big)
\Bigg]
+ \ep  \Bigg[
\frac{135}{2}
-120 \zeta_2
+192 \ln (2) \zeta_2
-48 \zeta_3
\nonumber\\&
+\phi ^2 \Big(
        \frac{1297}{27}
        -\frac{14 c_1}{3}
        -\frac{2723}{54} \zeta_2
        +98 \ln (2) \zeta_2
        +\frac{294}{5} \zeta_2^2
        -\frac{725}{18} \zeta_3
\Big)
+\phi ^4 \Big(
        \frac{193531}{24300}
        -\frac{11 c_1}{18}
\nonumber\\&
        -\frac{141179 \zeta_2}{20250}
        +\frac{5971}{300} \ln (2) \zeta_2
        +\frac{77}{10} \zeta_2^2
        -\frac{29381 \zeta_3}{3600}
\Big)
\Bigg]
\Bigg\}
+ C_F C_A \Bigg\{
\frac{1}{\ep^2}  \Bigg[
\frac{11}{9} \phi ^2 + \frac{11}{135} \phi ^4
\Bigg]
\nonumber\\&
+ \frac{1}{\ep} \Bigg[
\phi ^2 \Big(
        -\frac{94}{27}
        +\frac{4 \zeta_2}{3}
\Big)
+
\phi ^4 \Big(
        -\frac{91}{405}
        +\frac{4 \zeta_2}{45}
\Big)
\Bigg]
+  \Bigg[
-\frac{185}{3}
+\phi ^2 \Big(
        -\frac{650}{81}
        +\frac{47}{18} \zeta_2
\nonumber\\&
        -14 \ln (2) \zeta_2
        +\frac{37}{6} \zeta_3
\Big)
+\phi ^4 \Big(
        -\frac{2389}{972}
        +\frac{11897 \zeta_2}{10800}
        -\frac{11}{6} \ln (2) \zeta_2
        +\frac{229}{360} \zeta_3
\Big)
\Bigg]
+ \ep  \Bigg[
-\frac{1463}{6}
\nonumber\\&
+21 \zeta_2
-96 \ln (2) \zeta_2
+24 \zeta_3
+\phi ^2 \Big(
        -\frac{32507}{486}
        +\frac{7 c_1}{3}
        +\frac{1873}{108} \zeta_2
        -49 \ln (2) \zeta_2
        -\frac{293}{15} \zeta_2^2
        +\frac{629}{36} \zeta_3
\Big)
\nonumber\\&
+\phi ^4 \Big(
        -\frac{2663879}{291600}
        +\frac{11 c_1}{36}
        +\frac{212591 \zeta_2}{81000}
        -\frac{5971}{600} \ln (2) \zeta_2
        -\frac{2873}{900} \zeta_2^2
        +\frac{101327 \zeta_3}{21600}
\Big)
\Bigg]
\Bigg\}
\nonumber\\&
+ C_F n_l T_F \Bigg\{
\frac{1}{\ep^2}  \Bigg[
-\frac{4}{9} \phi ^2 
- \frac{4}{135} \phi ^4
\Bigg]
+ \frac{1}{\ep} \Bigg[
\frac{20}{27} \phi ^2 + \frac{4}{81} \phi ^4 
\Bigg]
+  \Bigg[
\frac{52}{3}
+\phi ^2 \Big(
        \frac{316}{81}
        +\frac{16 \zeta_2}{9}
\Big)
\nonumber\\&
+\phi ^4 \Big(
        \frac{883}{1215}
        +\frac{16 \zeta_2}{135}
\Big)
\Bigg]
+ \ep  \Bigg[
\frac{206}{3}
+36 \zeta_2
+\phi ^2 \Big(
        \frac{4808}{243}
        +\frac{154 \zeta_2}{27}
        +\frac{32 \zeta_3}{9}
\Big)
+\phi ^4 \Big(
        \frac{2560}{729}
        +\frac{1037 \zeta_2}{810}
\nonumber\\&
        +\frac{32 \zeta_3}{135}
\Big)
\Bigg]
\Bigg\}
+ C_F T_F \Bigg\{
\Bigg[
\frac{52}{3}
+8 i \phi  \zeta_2
+\phi ^2 \Big(
        -\frac{1417}{81}
        +\frac{91}{9} \zeta_2
        +\frac{8}{3} \bar{H}_0(\phi )
\Big)
+\frac{17}{15} i \phi ^3 \zeta_2
\nonumber\\&
+\phi ^4 \Big(
        -\frac{25076}{6075}
        +\frac{6071 \zeta_2}{2700}
        +\frac{17}{45} \bar{H}_0(\phi )
\Big)
\Bigg]
+ \ep \Bigg[
\frac{350}{3}
-60 \zeta_2
+ i \phi  \Big(
        16 \ln (2) \zeta_2
        -16 \bar{H}_0(\phi ) \zeta_2
\Big)
\nonumber\\&
+\phi ^2 \Big(
        -\frac{787}{162}
        +23 \zeta_2
        -58 \ln (2) \zeta_2
        +\frac{1811}{54} \zeta_3
        +\frac{8}{3} \bar{H}_0(\phi )
        -\frac{8}{3} \bar{H}_0(\phi )^2
\Big)
+ i \phi ^3 \Big(
        \frac{17}{15} \zeta_2
\nonumber\\&
        +\frac{34}{15} \ln (2) \zeta_2
        -\frac{34}{15} \zeta_2 \bar{H}_0(\phi )
\Big)
+\phi ^4 \Big(
        -\frac{241487}{121500}
        +\frac{35681 \zeta_2}{6750}
        -\frac{1997}{150} \ln (2) \zeta_2
        +\frac{125491 \zeta_3}{16200}
\nonumber\\&
        +\frac{6}{5} \bar{H}_0(\phi )
        -\frac{17}{45} \bar{H}_0(\phi )^2
\Big)
\Bigg]
\Bigg\} \,.
\end{align}

\subsubsection{Pseudo-scalar form factor}
The non-singlet part of the pseudo-scalar form factor can be obtained by using Eq.~(\ref{eq:cwiFF})
\begin{align}
 \bar{F}_{P}^{(1),\ns} &= \bar{F}_{A,1}^{(1)} + \Big( \frac{\phi^2}{4} -
                     \frac{\phi^4}{48} \Big) \bar{F}_{A,2}^{(1)} + {\cal O}(\phi^6)\,,
 \nonumber\\
 \bar{F}_{P}^{(2),\ns} &= \bar{F}_{A,1}^{(2)} + \Big( \frac{\phi^2}{4}
                        - \frac{\phi^4}{48} \Big) \bar{F}_{A,2}^{(2)}
                        + {\cal O}(\phi^6) \,.
\end{align}
The unrernormalized singlet contribution to the pseudo-scalar form factor reads
\begin{align}
\bar{\hat{F}}_{P}^{(2),s} &=
C_F T_F \Bigg\{
\Bigg[
\frac{8}{3}
+\frac{64}{3} \zeta_2
+12 i \phi  \zeta_2
+\phi ^2 \Big(
        -\frac{68}{9}
        +4 \bar{H}_0(\phi )
        +\frac{148 \zeta_2}{45}
\Big)
+2 i \phi ^3 \zeta_2
+\phi ^4 \Big(
        -\frac{689}{540}
\nonumber\\&
        +\frac{1759 \zeta_2}{4725}
        +\frac{2}{3} \bar{H}_0(\phi )
\Big)
\Bigg]
+ \ep \Bigg[
\frac{88}{9}
-\frac{352}{9} \zeta_2
-128 \ln (2) \zeta_2
+\frac{224}{3} \zeta_3
+ i \phi  \Big(
        -36 \zeta_2
        +24 \ln (2) \zeta_2
\nonumber\\&
        -24 \zeta_2 \bar{H}_0(\phi )
\Big)
+\phi ^2 \Big(
        \frac{5227}{135}
        -\frac{3182}{675} \zeta_2
        -\frac{296}{15} \ln (2) \zeta_2
        +\frac{518}{45} \zeta_3
        -12 \bar{H}_0(\phi )
        -4 \bar{H}_0(\phi )^2
\Big)
\nonumber\\&
+ i \phi ^3 \Big(
        -7 \zeta_2
        +4 \ln (2) \zeta_2
        -4 \zeta_2 \bar{H}_0(\phi )
\Big)
+\phi ^4 \Big(
        \frac{101609}{16200}
        -\frac{255167 \zeta_2}{496125}
        -\frac{3518 \ln (2) \zeta_2}{1575}
\nonumber\\&
        +\frac{1759 \zeta_3}{1350}
        -2 \bar{H}_0(\phi )
        -\frac{2}{3} \bar{H}_0(\phi )^2
\Big)
\Bigg]
\Bigg\} \,.
\end{align}
%
\subsection{High energy region {\boldmath$\abs{q^2} \gg m^2$}}

\vspace*{1mm}
\noindent
We now present the expansion of all the form factors in the asymptotic limit
i.e. for $x \rightarrow 0^+$ up to ${\cal O}(x^2)$. The expanded form factors are denoted 
by $\F_I$. We use the abbreviation $L = \ln(x)$ in the following. The
correct analytic continuation to negative values of $x$ is given by $L \to L + i \pi$. 

\subsubsection{Vector form factor}
\begin{align}
\F_{V,1}^{(1)} &=
C_F  \Bigg\{
\frac{1}{\ep}  \Bigg[
-2
-2 L
-4 L x^2
\Bigg]
+
 \Bigg[
-4
+2 \zeta_2
-3 L
-L^2
+2 (-2+L) x
+x^2 \Big(
        1
        -8 L
        -2 L^2
\nonumber\\&
        +4 \zeta_2
\Big)
\Bigg]
+ \ep  \Bigg[
-8
+2 \zeta_2
+4 \zeta_3
+L \Big(
        -8
        +\zeta_2
\Big)
-\frac{3 L^2}{2}
-\frac{L^3}{3}
+x \Big(
        -10
        -2 \zeta_2
        +6 L
        +L^2
\Big)
\nonumber\\&
+x^2 \Big(
        4
        +8 \zeta_2
        +8 \zeta_3
        +L \Big(
                -19
                +2 \zeta_2
        \Big)
        -4 L^2
        -\frac{2 L^3}{3}
\Big)
\Bigg]
+ \ep^2  \Bigg[
-16
+6 \zeta_2
+\frac{14}{5} \zeta_2^2
+\frac{20}{3} \zeta_3
\nonumber\\&
+L \Big(
        -16
        +\frac{3 \zeta_2}{2}
        +\frac{14 \zeta_3}{3}
\Big)
+L^2 \Big(
        -4
        +\frac{\zeta_2}{2}
\Big)
-\frac{L^3}{2}
-\frac{L^4}{12}
+x \Big(
        -26
        -8 \zeta_2
        -4 \zeta_3
\nonumber\\&
        +L \Big(
                16
                -\zeta_2
        \Big)
        +3 L^2
        +\frac{L^3}{3}
\Big)
+x^2 \Big(
        17
        +\frac{39}{2} \zeta_2
        +\frac{28}{5} \zeta_2^2
        +16 \zeta_3
        +L \Big(
                -46
                +4 \zeta_2
                +\frac{28 \zeta_3}{3}
        \Big)
\nonumber\\&
        +L^2 \Big(
                -\frac{19}{2}
                +\zeta_2
        \Big)
        -\frac{4 L^3}{3}
        -\frac{L^4}{6}
\Big)
\Bigg]
\Bigg\}\,.
%
\\
\F_{V,1}^{(2)} &=
C_F^2  \Bigg\{
\frac{1}{\ep^2}  \Bigg[
2
+4 L
+2 L^2
+8 L (1+L) x^2
\Bigg]
+ \frac{1}{\ep}  \Bigg[
8
+14 L
+8 L^2
+2 L^3
-4 \zeta_2
-4 L \zeta_2
-4 (-2
\nonumber\\&
+L) (1+L) x
+x^2 \Big(
        -2
        +30 L
        +32 L^2
        +8 L^3
        -8 \zeta_2
        -16 L \zeta_2
\Big)
\Bigg]
+ \Bigg[
46
+\frac{85 L}{2}
+\frac{55 L^2}{2}
\nonumber\\&
+\frac{20 L^3}{3}
+\frac{7 L^4}{6}
+39 \zeta_2
-4 L^2 \zeta_2
-48 \ln (2) \zeta_2
-\frac{118}{5} \zeta_2^2
-44 \zeta_3
-32 L \zeta_3
+x \Big(
        -22
        +13 L
\nonumber\\&
        -37 L^2
        -\frac{28 L^3}{3}
        -\frac{L^4}{3}
        -30 \zeta_2
        +36 L \zeta_2
        +8 L^2 \zeta_2
        +288 \ln (2) \zeta_2
        +\frac{128}{5} \zeta_2^2
        -88 \zeta_3
        -48 L \zeta_3
\Big)
\nonumber\\&
+x^2 \Big(
        \frac{1307}{2}
        -365 L
        +214 L^2
        +32 L^3
        +\frac{26 L^4}{3}
        -980 \zeta_2
        -376 L \zeta_2
        -84 L^2 \zeta_2
        -576 \ln (2) \zeta_2
\nonumber\\&
        -\frac{1756}{5} \zeta_2^2
        +808 \zeta_3
        +496 L \zeta_3
\Big)
\Bigg]
+ \ep  \Bigg[
4
+8 c_1
+\frac{479 L}{4}
+\frac{153 L^2}{2}
+\frac{137 L^3}{6}
+\frac{11 L^4}{3}
+\frac{L^5}{2}
\nonumber\\&
+163 \zeta_2
+17 L \zeta_2
-\frac{8}{3} L^3 \zeta_2
-24 \ln (2) \zeta_2
-160 \zeta_2^2
-\frac{106}{5} L \zeta_2^2
-\frac{346}{3} \zeta_3
-\frac{284}{3} L \zeta_3
-\frac{112}{3} L^2 \zeta_3
\nonumber\\&
-12 \zeta_2 \zeta_3
-18 \zeta_5
+x \Big(
        210
        -48 c_1
        -\frac{335 L}{2}
        -96 L^2
        -\frac{131 L^3}{3}
        -6 L^4
        -\frac{L^5}{5}
        -580 \zeta_2
        +282 L \zeta_2
\nonumber\\&
        +34 L^2 \zeta_2
        +\frac{40}{3} L^3 \zeta_2
        +1728 \ln (2) \zeta_2
        +44 \zeta_2^2
        -\frac{668}{5} L \zeta_2^2
        -156 \zeta_3
        +300 L \zeta_3
        +4 L^2 \zeta_3
        -208 \zeta_2 \zeta_3
\nonumber\\&
        +480 \zeta_5
\Big)
+x^2 \Big(
        -\frac{1951}{4}
        +96 c_1
        +464 L
        +185 L^2
        +178 L^3
        +\frac{74 L^4}{3}
        +\frac{58 L^5}{15}
        -1164 \zeta_2
\nonumber\\&
        -2206 L \zeta_2
        -72 L^2 \zeta_2
        -\frac{476}{3} L^3 \zeta_2
        -4224 \ln (2) \zeta_2
        +3212 \zeta_2^2
        +\frac{5992}{5} L \zeta_2^2
        -2000 \zeta_3
        -\frac{7792}{3} L \zeta_3
\nonumber\\&
        -\frac{268}{3} L^2 \zeta_3
        +1656 \zeta_2 \zeta_3
        -6252 \zeta_5
\Big)
\Bigg]
\Bigg\}
+ C_F C_A  \Bigg\{
\frac{1}{\ep^2}  \Bigg[
\frac{11}{3}
+\frac{11 L}{3}
+\frac{22 L x^2}{3}
\Bigg]
+ \frac{1}{\ep}  \Bigg[
-\frac{49}{9}
-\frac{67 L}{9}
\nonumber\\&
+2 \zeta_2
+2 L \zeta_2
-2 \zeta_3
+x^2 \Big(
        4
        -\frac{188 L}{9}
        -4 L^2
        -\frac{4 L^3}{3}
        -4 \zeta_2
        -8 \zeta_3
\Big)
\Bigg]
+ \Bigg[
-\frac{1595}{27}
-\frac{2545 L}{54}
\nonumber\\&
-\frac{233 L^2}{18}
-\frac{11 L^3}{9}
-\frac{7}{9} \zeta_2
-\frac{22}{3} L \zeta_2
+2 L^2 \zeta_2
+24 \ln (2) \zeta_2
-\frac{3}{5} \zeta_2^2
+\frac{134}{3} \zeta_3
+26 L \zeta_3
+x \Big(
        -\frac{904}{9}
\nonumber\\&
        +\frac{341 L}{9}
        -\frac{25 L^2}{3}
        +\frac{8 L^3}{3}
        +\frac{L^4}{6}
        +\frac{494}{3} \zeta_2
        +44 L \zeta_2
        +8 L^2 \zeta_2
        -144 \ln (2) \zeta_2
        +28 \zeta_2^2
        -200 \zeta_3
\nonumber\\&
        -72 L \zeta_3
\Big)
+x^2 \Big(
        \frac{8723}{18}
        -\frac{8968 L}{27}
        +\frac{931 L^2}{9}
        -\frac{94 L^3}{9}
        -\frac{7 L^4}{3}
        -\frac{6848}{9} \zeta_2
        -\frac{1076}{3} L \zeta_2
        -72 L^2 \zeta_2
\nonumber\\&
        +288 \ln (2) \zeta_2
        -\frac{1408}{5} \zeta_2^2
        +\frac{5188}{3} \zeta_3
        +808 L \zeta_3
\Big)
\Bigg]
+ \ep  \Bigg[
-
\frac{28745}{162}
-4 c_1
-\frac{70165 L}{324}
-\frac{3337 L^2}{54}
\nonumber\\&
-\frac{565 L^3}{54}
-\frac{11 L^4}{12}
-\frac{71}{27} \zeta_2
-\frac{575}{18} L \zeta_2
-\frac{11}{2} L^2 \zeta_2
+\frac{4}{3} L^3 \zeta_2
+12 \ln (2) \zeta_2
+\frac{637}{5} \zeta_2^2
+\frac{88}
{5} L \zeta_2^2
\nonumber\\&
+\frac{1577}{9} \zeta_3
+\frac{260}{3} L \zeta_3
+26 L^2 \zeta_3
-2 \zeta_2 \zeta_3
-157 \zeta_5
+x \Big(
        -\frac{12683}{27}
        +24 c_1
        +\frac{5639 L}{54}
        +\frac{134 L^2}{9}
\nonumber\\&
        -\frac{17 L^3}{3}
        +\frac{11 L^4}{6}
        +\frac{L^5}{10}
        +\frac{4448}{9} \zeta_2
        +115 L \zeta_2
        +26 L^2 \zeta_2
        +\frac{16}{3} L^3 \zeta_2
        -864 \ln (2) \zeta_2
        -\frac{2944}{5} \zeta_2^2
\nonumber\\&
        -\frac{482}{5} L \zeta_2^2
        +272 \zeta_3
        +44 L \zeta_3
        -38 L^2 \zeta_3
        +8 \zeta_2 \zeta_3
        +768 \zeta_5
\Big)
+x^2 \Big(
        -\frac{25015}{27}
        -48 c_1
        +\frac{155567 L}{162}
\nonumber\\&
        -\frac{18413 L^2}{54}
        +\frac{3260 L^3}{27}
        -\frac{53 L^4}{6}
        -\frac{19 L^5}{15}
        -\frac{41933}{54} \zeta_2
        -\frac{6002}{9} L \zeta_2
        -125 L^2 \zeta_2
        -52 L^3 \zeta_2
\nonumber\\&
        +2112 \ln (2) \zeta_2
        +\frac{19996}{5} \zeta_2^2
        +1052 L \zeta_2^2
        +\frac{154}{9} \zeta_3
        -\frac{5168}{3} L \zeta_3
        +476 L^2 \zeta_3
        +148 \zeta_2 \zeta_3
\nonumber\\&
        -8178 \zeta_5
\Big)
\Bigg]
\Bigg\}
%
+ C_F n_l T_F  \Bigg\{
\frac{1}{\ep^2}  \Bigg[
-\frac{4}{3}
-\frac{4 L}{3}
-\frac{8 L x^2}{3}
\Bigg]
+ \frac{1}{\ep}  \Bigg[
\frac{20}{9}
+\frac{20 L}{9}
+\frac{40 L x^2}{9}
\Bigg]
\nonumber\\&
+ \Bigg[
\frac{424}{27}
+\frac{418 L}{27}
+\frac{38 L^2}{9}
+\frac{4 L^3}{9}
-\frac{28}{9} \zeta_2
+\frac{8}{3} L \zeta_2
-\frac{16}{3} \zeta_3
+x \Big(
        \frac{200}{9}
        -\frac{148 L}{9}
        -\frac{4 L^2}{3}
        +\frac{8}{3} \zeta_2
\Big)
\nonumber\\&
+x^2 \Big(
        -\frac{68}{9}
        +\frac{1064 L}{27}
        +\frac{88 L^2}{9}
        +\frac{8 L^3}{9}
        -\frac{176}{9} \zeta_2
        +\frac{16}{3} L \zeta_2
        -\frac{32}{3} \zeta_3
\Big)
\Bigg]
+ \ep  \Bigg[
\frac{5204}{81}
+\frac{5813 L}{81}
\nonumber\\&
+\frac{562 L^2}{27}
+\frac{94 L^3}{27}
+\frac{L^4}{3}
-\frac{176}{27} \zeta_2
+\frac{74}{9} L \zeta_2
+2 L^2 \zeta_2
-\frac{96}{5} \zeta_2^2
-\frac{280}{9} \zeta_3
-\frac{16}{3} L \zeta_3
+x \Big(
        \frac{3808}{27}
\nonumber\\&
        -\frac{2654 L}{27}
        -\frac{184 L^2}{9}
        -\frac{4 L^3}{3}
        +\frac{584}{9} \zeta_2
        -4 L \zeta_2
        +16 \zeta_3
\Big)
+x^2 \Big(
        -\frac{2698}{27}
        +\frac{18238 L}{81}
        +\frac{1406 L^2}{27}
\nonumber\\&
        +\frac{224 L^3}{27}
        +\frac{2 L^4}{3}
        -\frac{2974}{27} \zeta_2
        +\frac{184}{9} L \zeta_2
        +4 L^2 \zeta_2
        -\frac{192}{5} \zeta_2^2
        -\frac{896}{9} \zeta_3
        -\frac{32}{3} L \zeta_3
\Big)
\Bigg]
\Bigg\}
\nonumber\\&
+ C_F T_F  \Bigg\{
 \Bigg[
\frac{1532}{27}
+\frac{530 L}{27}
+\frac{38 L^2}{9}
+\frac{4 L^3}{9}
-\frac{8}{3} \zeta_2
+4 L \zeta_2
+x \Big(
        -\frac{784}{9}
        -\frac{436 L}{9}
        -\frac{52 L^2}{3}
\nonumber\\&
        -88 \zeta_2
\Big)
+x^2 \Big(
        \frac{1568}{9}
        +\frac{9604 L}{27}
        +\frac{808 L^2}{9}
        -\frac{16 L^3}{9}
        +\frac{2096}{3} \zeta_2
        -8 L \zeta_2
\Big)
\Bigg]
+ \ep  \Bigg[
\frac{4138}{27}
+\frac{191 L}{3}
\nonumber\\&
+\frac{562 L^2}{27}
+\frac{94 L^3}{27}
+\frac{L^4}{3}
-\frac{1616}{27} \zeta_2
+6 L \zeta_2
+2 L^2 \zeta_2
+\frac{224}{3} \ln (2) \zeta_2
-\frac{8}{5} \zeta_2^2
-\frac{184}{3} \zeta_3
-\frac{56}{9} L \zeta_3
\nonumber\\&
+x \Big(
        -\frac{1148}{27}
        -\frac{5246 L}{27}
        -\frac{472 L^2}{9}
        -12 L^3
        -\frac{2192}{9} \zeta_2
        -4 L \zeta_2
        +320 \ln (2) \zeta_2
        -\frac{352}{3} \zeta_3
\Big)
\nonumber\\&
+x^2 \Big(
        \frac{215}{27}
        +\frac{32696 L}{27}
        +\frac{9452 L^2}{27}
        +\frac{1772 L^3}{27}
        -\frac{4 L^4}{3}
        +\frac{47870}{27} \zeta_2
        +40 L \zeta_2
        -4 L^2 \zeta_2
\nonumber\\&
        -\frac{9344}{3} \ln (2) \zeta_2
        -\frac{8}{5} \zeta_2^2
        +\frac{13120}{9} \zeta_3
        +\frac{176}{9} L \zeta_3
\Big)
\Bigg]
\Bigg\} \,.
%
\\
\F_{V,2}^{(1)} &=
C_F  \Bigg\{
-4 L x
+ \ep  \Bigg[
x \Big(
        -16 L
        -2 L^2
        +4 \zeta_2
\Big)
+8 (-1+L) x^2
\Bigg]
+ \ep^2  \Bigg[
x \Big(
        -8 L^2
        -\frac{2 L^3}{3}
\nonumber\\&
        +\frac{1}{3} L \Big(
                -96
                +6 \zeta_2
        \Big)
        +16 \zeta_2
        +8 \zeta_3
\Big)
+
x^2 \Big(
        -40
        +32 L
        +4 L^2
        -8 \zeta_2
\Big)
\Bigg]
\Bigg\}\,.
%
\\
\F_{V,2}^{(2)} &=
C_F^2  \Bigg\{
%
 \frac{1}{\ep}  \Bigg[
8 L (1+L) x
\Bigg]
+ \Bigg[
x \Big(
        62 L
        +34 L^2
        +8 L^3
        +60 \zeta_2
        -48 L \zeta_2
        -192 \ln (2) \zeta_2
        +16 \zeta_3
\Big)
\nonumber\\&
+x^2 \Big(
        -232
        +232 L
        -200 L^2
        -\frac{64 L^3}{3}
        -\frac{4 L^4}{3}
        +752 \zeta_2
        +208 L \zeta_2
        +64 L^2 \zeta_2
        +384 \ln (2) \zeta_2
\nonumber\\&
        +\frac{1056}{5} \zeta_2^2
        -864 \zeta_3
        -448 L \zeta_3
\Big)
\Bigg]
+ \ep  \Bigg[
x \Big(
        -20
        +32 c_1
        +249 L
        +74 L^2
        +\frac{94 L^3}{3}
        +\frac{14 L^4}{3}
\nonumber\\&
        +492 \zeta_2
        -116 L \zeta_2
        -16 L^2 \zeta_2
        -1152 \ln (2) \zeta_2
        -\frac{1192}{5} \zeta_2^2
        +128 \zeta_3
        -192 L \zeta_3
\Big)
+x^2 \Big(
        -640
\nonumber\\&
        -64 c_1
        +392 L
        -308 L^2
        -\frac{728 L^3}{3}
        -\frac{40 L^4}{3}
        -\frac{4 L^5}{5}
        +2200 \zeta_2
        +1296 L \zeta_2
        +264 L^2 \zeta_2
\nonumber\\&
        +\frac{256}{3} L^3 \zeta_2
        +3072 \ln (2) \zeta_2
        -\frac{14176}{5} \zeta_2^2
        -\frac{4848}{5} L \zeta_2^2
        +2672 \zeta_3
        +1488 L \zeta_3
        -80 L^2 \zeta_3
        -1088 \zeta_2 \zeta_3
\nonumber\\&
        +4608 \zeta_5
\Big)
\Bigg]
\Bigg\}
+ C_F C_A  \Bigg\{
 \Bigg[
x \Big(
        12
        -\frac{346 L}{9}
        +\frac{2 L^2}{3}
        -\frac{244}{3} \zeta_2
        +96 \ln (2) \zeta_2
        +80 \zeta_3
\Big)
+x^2 \Big(
        -\frac{616}{3}
\nonumber\\&
        +\frac{232 L}{3}
        -72 L^2
        +\frac{32 L^3}{3}
        +\frac{2 L^4}{3}
        +656 \zeta_2
        +368 L \zeta_2
        +64 L^2 \zeta_2
        -192 \ln (2) \zeta_2
        +\frac{1104}{5} \zeta_2^2
\nonumber\\&
        -1456 \zeta_3
        -544 L \zeta_3
\Big)
\Bigg]
+ \ep  \Bigg[
x \Big(
        78
        -16 c_1
        -\frac{8057 L}{27}
        -\frac{250 L^2}{9}
        +\frac{2 L^3}{3}
        -\frac{1768}{9} \zeta_2
        -38 L \zeta_2
\nonumber\\&
        -8 L^2 \zeta_2
        +576 \ln (2) \zeta_2
        +\frac{1504}{5} \zeta_2^2
        -264 \zeta_3
        +48 L \zeta_3
\Big)
+x^2 \Big(
        \frac{2672}{9}
        +32 c_1
        -\frac{5912 L}{9}
        -4 L^2
\nonumber\\&
        -\frac{152 L^3}{3}
        +\frac{20 L^4}{3}
        +\frac{2 L^5}{5}
        +664 \zeta_2
        +656 L \zeta_2
        +160 L^2 \zeta_2
        +\frac{160}{3} L^3 \zeta_2
        -1536 \ln (2) \zeta_2
        -\frac{16032}{5} \zeta_2^2
\nonumber\\&
        -\frac{4104}{5} L \zeta_2^2
        +608 \zeta_3
        +1168 L \zeta_3
        -248 L^2 \zeta_3
        -224 \zeta_2 \zeta_3
        +5760 \zeta_5
\Big)
\Bigg]
\Bigg\}
\nonumber\\&
+ C_F n_l T_F  \Bigg\{
%
%
 \Bigg[
-\frac{32}{3} (-1+L) x^2
+x \Big(
        \frac{200 L}{9}
        +\frac{8 L^2}{3}
        -\frac{16}{3} \zeta_2
\Big)
\Bigg]
+ \ep  \Bigg[
x \Big(
        \frac{3844 L}{27}
        +\frac{296 L^2}{9}
\nonumber\\&
        +\frac{8 L^3}{3}
        -\frac{592}{9} \zeta_2
        +8 L \zeta_2
        -32 \zeta_3
\Big)
+
x^2 \Big(
        \frac{1472}{9}
        -\frac{1184 L}{9}
        -16 L^2
        +32 \zeta_2
\Big)
\Bigg]
\Bigg\}
\nonumber\\&
+ C_F T_F  \Bigg\{
%
%
%
 \Bigg[
x \Big(
        \frac{272}{3}
        +\frac{200 L}{9}
        +\frac{8 L^2}{3}
        -16 \zeta_2
\Big)
+
x^2 \Big(
        -\frac{544}{3}
        -\frac{992 L}{3}
        -64 L^2
        -512 \zeta_2
\Big)
\Bigg]
\nonumber\\&
+ \ep  \Bigg[
x \Big(
        \frac{1528}{9}
        +\frac{3844 L}{27}
        +\frac{296 L^2}{9}
        +\frac{8 L^3}{3}
        +\frac{256}{9} \zeta_2
        +8 L \zeta_2
        +128 \ln (2) \zeta_2
        -\frac{256}{3} \zeta_3
\Big)
\nonumber\\&
+
x^2 \Big(
        -\frac{2368}{9}
        -\frac{8672 L}{9}
        -336 L^2
        -\frac{128 L^3}{3}
        -\frac{4000}{3} \zeta_2
        +2304 \ln (2) \zeta_2
        -1088 \zeta_3
\Big)
\Bigg]
\Bigg\} \,.
\end{align}
We note that in the asymptotic limit, the magnetic part of the vector form factors vanish.

\subsubsection{Axial-vector form factor}
\begin{align}
\F_{A,1}^{(1),\ns} &=
C_F  \Bigg\{
\frac{1}{\ep}  \Bigg[
-2
-2 L
-4 L x^2
\Bigg]
+
 \Bigg[
-4
-3 L
-L^2
+2 \zeta_2
+2 (-2+3 L) x
+x^2 \Big(
        1
        -8 L
\nonumber\\&
        -2 L^2
        +4 \zeta_2
\Big)
\Bigg]
+ \ep  \Bigg[
-8
-8 L
-\frac{3 L^2}{2}
-\frac{L^3}{3}
+2 \zeta_2
+L \zeta_2
+4 \zeta_3
+x \Big(
        -10
        +6 L
        +3 L^2
\nonumber\\&
        -6 \zeta_2
\Big)
+x^2 \Big(
        12
        -27 L
        -4 L^2
        -\frac{2 L^3}{3}
        +8 \zeta_2
        +2 L \zeta_2
        +8 \zeta_3
\Big)
\Bigg]
+ \ep^2  \Bigg[
-16
-16 L
-4 L^2
\nonumber\\&
-\frac{L^3}{2}
-\frac{L^4}{12}
+6 \zeta_2
+\frac{3}{2} L \zeta_2
+\frac{1}{2} L^2 \zeta_2
+\frac{14}{5} \zeta_2^2
+\frac{20}{3} \zeta_3
+\frac{14}{3} L \zeta_3
+x \Big(
        -26
        +16 L
        +3 L^2
\nonumber\\&
        +L^3
        -8 \zeta_2
        -3 L \zeta_2
        -12 \zeta_3
\Big)
+x^2 \Big(
        25
        -46 L
        -\frac{27 L^2}{2}
        -\frac{4 L^3}{3}
        -\frac{L^4}{6}
        +\frac{55}{2} \zeta_2
        +4 L \zeta_2
\nonumber\\&
        +L^2 \zeta_2
        +\frac{28}{5} \zeta_2^2
        +16 \zeta_3
        +\frac{28}{3} L \zeta_3
\Big)
\Bigg]
\Bigg\}
%
\\
\F_{A,1}^{(2),\ns} &=
C_F^2  \Bigg\{
\frac{1}{\ep^2}  \Bigg[
2
+4 L
+2 L^2
+8 L (1+L) x^2
\Bigg]
+ \frac{1}{\ep}  \Bigg[
8
+14 L
+8 L^2
+2 L^3
-4 \zeta_2
-4 L \zeta_2
\nonumber\\&
-4 (1+L) (-2+3 L) x
+x^2 \Big(
        -2
        +30 L
        +32 L^2
        +8 L^3
        -8 \zeta_2
        -16 L \zeta_2
\Big)
\Bigg]
+ \Bigg[
46
+\frac{85 L}{2}
\nonumber\\&
+\frac{55 L^2}{2}
+\frac{20 L^3}{3}
+\frac{7 L^4}{6}
+39 \zeta_2
-4 L^2 \zeta_2
-48 \ln (2) \zeta_2
-\frac{118}{5} \zeta_2^2
-44 \zeta_3
-32 L \zeta_3
\nonumber\\&
+x \Big(
        -22
        +19 L
        -19 L^2
        -12 L^3
        +\frac{L^4}{3}
        -66 \zeta_2
        +44 L \zeta_2
        +8 L^2 \zeta_2
        +288 \ln (2) \zeta_2
        +\frac{144}{5} \zeta_2^2
\nonumber\\&
        -104 \zeta_3
        -80 L \zeta_3
\Big)
+x^2 \Big(
        \frac{1195}{2}
        -357 L
        +182 L^2
        +\frac{64 L^3}{3}
        +\frac{14 L^4}{3}
        -228 \zeta_2
        -216 L \zeta_2
        -52 L^2 \zeta_2
\nonumber\\&
        -576 \ln (2) \zeta_2
        -\frac{1308}{5} \zeta_2^2
        +488 \zeta_3
        +432 L \zeta_3
\Big)
\Bigg]
+ \ep  \Bigg[
4
+8 c_1
+\frac{479 L}{4}
+\frac{153 L^2}{2}
+\frac{137 L^3}{6}
\nonumber\\&
+\frac{11 L^4}{3}
+\frac{L^5}{2}
+163 \zeta_2
+17 L \zeta_2
-\frac{8}{3} L^3 \zeta_2
-24 \ln (2) \zeta_2
-160 \zeta_2^2
-\frac{106}{5} L \zeta_2^2
-\frac{346}{3} \zeta_3
-\frac{284}{3} L \zeta_3
\nonumber\\&
-\frac{112}{3} L^2 \zeta_3
-12 \zeta_2 \zeta_3
-18 \zeta_5
+x \Big(
        190
        -48 c_1
        -\frac{173 L}{2}
        -26 L^2
        -\frac{53 L^3}{3}
        -\frac{22 L^4}{3}
        +\frac{L^5}{5}
        -856 \zeta_2
\nonumber\\&
        +166 L \zeta_2
        -2 L^2 \zeta_2
        +\frac{8}{3} L^3 \zeta_2
        +1728 \ln (2) \zeta_2
        +\frac{1124}{5} \zeta_2^2
        -84 L \zeta_2^2
        -628 \zeta_3
        +228 L \zeta_3
        -52 L^2 \zeta_3 
\nonumber\\&
        +80 \zeta_2 \zeta_3
        +864 \zeta_5
\Big)
+x^2 \Big(
        -\frac{543}{4}
        +96 c_1
        +792 L
        -163 L^2
        +\frac{518 L^3}{3}
        +\frac{58 L^4}{3}
        +\frac{22 L^5}{15}
        +1708 \zeta_2
\nonumber\\&
        -510 L \zeta_2
        +136 L^2 \zeta_2
        -\frac{188}{3} L^3 \zeta_2
        -2688 \ln (2) \zeta_2
        +\frac{3596}{5} \zeta_2^2
        +\frac{2328}{5} L \zeta_2^2
        +832 \zeta_3
        -\frac{5008}{3} L \zeta_3
\nonumber\\&
        +\frac{452}{3} L^2 \zeta_3
        -328 \zeta_2 \zeta_3
        -5868 \zeta_5
\Big)
\Bigg]
\Bigg\}
+ C_F C_A  \Bigg\{
\frac{1}{\ep^2}  \Bigg[
\frac{11}{3}
+\frac{11 L}{3}
+\frac{22 L x^2}{3}
\Bigg]
+ \frac{1}{\ep}  \Bigg[
-\frac{49}{9}
-\frac{67 L}{9}
\nonumber\\&
+2 \zeta_2
+2 L \zeta_2
-2 \zeta_3
+x^2 \Big(
        4
        -\frac{188 L}{9}
        -4 L^2
        -\frac{4 L^3}{3}
        -4 \zeta_2
        -8 \zeta_3
\Big)
\Bigg]
+ \Bigg[
-\frac{1595}{27}
-\frac{2545 L}{54}
\nonumber\\&
-\frac{233 L^2}{18}
-\frac{11 L^3}{9}
-\frac{7}{9} \zeta_2
-\frac{22}{3} L \zeta_2
+2 L^2 \zeta_2
+24 \ln (2) \zeta_2
-\frac{3}{5} \zeta_2^2
+\frac{134}{3} \zeta_3
+26 L \zeta_3
+x \Big(
        -\frac{796}{9}
\nonumber\\&
        +\frac{241 L}{3}
        -9 L^2
        -\frac{L^4}{6}
        +158 \zeta_2
        +36 L \zeta_2
        +8 L^2 \zeta_2
        -144 \ln (2) \zeta_2
        +\frac{132}{5} \zeta_2^2
        -168 \zeta_3
        -56 L \zeta_3
\Big)
\nonumber\\&
+x^2 \Big(
        \frac{7523}{18}
        -\frac{9760 L}{27}
        +\frac{283 L^2}{9}
        -\frac{46 L^3}{9}
        -\frac{L^4}{3}
        -\frac{2528}{9} \zeta_2
        -\frac{164}{3} L \zeta_2
        -24 L^2 \zeta_2
        +288 \ln (2) \zeta_2
\nonumber\\&
        -\frac{544}{5} \zeta_2^2
        +\frac{1060}{3} \zeta_3
        +328 L \zeta_3
\Big)
\Bigg]
+ \ep  \Bigg[
-
\frac{28745}{162}
-4 c_1
-\frac{70165 L}{324}
-\frac{3337 L^2}{54}
-\frac{565 L^3}{54}
\nonumber\\&
-\frac{11 L^4}{12}
-\frac{71}{27} \zeta_2
-\frac{575}{18} L \zeta_2
-\frac{11}{2} L^2 \zeta_2
+\frac{4}{3} L^3 \zeta_2
+12 \ln (2) \zeta_2
+\frac{637}{5} \zeta_2^2
+\frac{88}{5}
 L \zeta_2^2
+\frac{1577}{9} \zeta_3
\nonumber\\&
+\frac{260}{3} L \zeta_3
+26 L^2 \zeta_3
-2 \zeta_2 \zeta_3
-157 \zeta_5
+x \Big(
        -\frac{10577}{27}
        +24 c_1
        +\frac{3799 L}{18}
        +\frac{112 L^2}{3}
        -9 L^3
        +\frac{L^4}{6}
\nonumber\\&
        -\frac{L^5}{10}
        +\frac{1744}{3} \zeta_2
        +153 L \zeta_2
        +42 L^2 \zeta_2
        +\frac{32}{3} L^3 \zeta_2
        -864 \ln (2) \zeta_2
        -\frac{2944}{5} \zeta_2^2
        -\frac{606}{5} L \zeta_2^2
        +184 \zeta_3
\nonumber\\&
        +20 L \zeta_3
        -10 L^2 \zeta_3
        -136 \zeta_2 \zeta_3
        +576 \zeta_5
\Big)
+x^2 \Big(
        -\frac{12007}{27}
        -48 c_1
        -\frac{19681 L}{162}
        -\frac{22517 L^2}{54}
\nonumber\\&
        +\frac{956 L^3}{27}
        -\frac{37 L^4}{6}
        -\frac{L^5}{15}
        -\frac{25085}{54} \zeta_2
        -\frac{4418}{9} L \zeta_2
        +43 L^2 \zeta_2
        -36 L^3 \zeta_2
        +1344 \ln (2) \zeta_2
        +\frac{8004}{5} \zeta_2^2
\nonumber\\&
        +548 L \zeta_2^2
        +\frac{4042}{9} \zeta_3
        -\frac{2720}{3} L \zeta_3
        +164 L^2 \zeta_3
        +628 \zeta_2 \zeta_3
        -2994 \zeta_5
\Big)
\Bigg]
\Bigg\}
+ C_F n_l T_F  \Bigg\{
\frac{1}{\ep^2}  \Bigg[
-\frac{4}{3}
\nonumber\\&
-\frac{4 L}{3}
-\frac{8 L x^2}{3}
\Bigg]
+ \frac{1}{\ep}  \Bigg[
\frac{20}{9}
+\frac{20 L}{9}
+\frac{40 L x^2}{9}
\Bigg]
+ \Bigg[
\frac{424}{27}
+\frac{418 L}{27}
+\frac{38 L^2}{9}
+\frac{4 L^3}{9}
-\frac{28}{9} \zeta_2
\nonumber\\&
+\frac{8}{3} L \zeta_2
-\frac{16}{3} \zeta_3
+x \Big(
        \frac{200}{9}
        -\frac{68 L}{3}
        -4 L^2
        +8 \zeta_2
\Big)
+x^2 \Big(
        -\frac{164}{9}
        +\frac{1352 L}{27}
        +\frac{88 L^2}{9}
        +\frac{8 L^3}{9}
\nonumber\\&
        -\frac{176}{9} \zeta_2
        +\frac{16}{3} L \zeta_2
        -\frac{32}{3} \zeta_3
\Big)
\Bigg]
+ \ep  \Bigg[
\frac{5204}{81}
+\frac{5813 L}{81}
+\frac{562 L^2}{27}
+\frac{94 L^3}{27}
+\frac{L^4}{3}
-\frac{176}{27} \zeta_2
+\frac{74}{9} L \zeta_2
\nonumber\\&
+2 L^2 \zeta_2
-\frac{96}{5} \zeta_2^2
-\frac{280}{9} \zeta_3
-\frac{16}{3} L \zeta_3
+x \Big(
        \frac{3808}{27}
        -\frac{862 L}{9}
        -\frac{80 L^2}{3}
        -4 L^3
        +\frac{232}{3} \zeta_2
        -12 L \zeta_2
\nonumber\\&
        +48 \zeta_3
\Big)
+x^2 \Big(
        -\frac{4234}{27}
        +\frac{20254 L}{81}
        +\frac{1838 L^2}{27}
        +\frac{224 L^3}{27}
        +\frac{2 L^4}{3}
        -\frac{3838}{27} \zeta_2
        +\frac{184}{9} L \zeta_2
        +4 L^2 \zeta_2
\nonumber\\&
        -\frac{192}{5} \zeta_2^2
        -\frac{896}{9} \zeta_3
        -\frac{32}{3} L \zeta_3
\Big)
\Bigg]
\Bigg\}
+ C_F T_F  \Bigg\{
 \Bigg[
\frac{1532}{27}
+\frac{530 L}{27}
+\frac{38 L^2}{9}
+\frac{4 L^3}{9}
-\frac{8}{3} \zeta_2
+4 L \zeta_2
\nonumber\\&
+x \Big(
        -\frac{160}{9}
        -\frac{164 L}{3}
        -20 L^2
        -136 \zeta_2
\Big)
+x^2 \Big(
        \frac{320}{9}
        +\frac{2980 L}{27}
        +\frac{520 L^2}{9}
        -\frac{16 L^3}{9}
        +\frac{1136}{3} \zeta_2
\nonumber\\&
        -8 L \zeta_2
\Big)
\Bigg]
+ \ep  \Bigg[
\frac{4138}{27}
+\frac{191 L}{3}
+\frac{562 L^2}{27}
+\frac{94 L^3}{27}
+\frac{L^4}{3}
-\frac{1616}{27} \zeta_2
+6 L \zeta_2
+2 L^2 \zeta_2
\nonumber\\&
+\frac{224}{3} \ln (2) \zeta_2
-\frac{8}{5} \zeta_2^2
-\frac{184}{3} \zeta_3
-\frac{56}{9} L \zeta_3
+x \Big(
        \frac{3628}{27}
        -\frac{1726 L}{9}
        -\frac{176 L^2}{3}
        -\frac{44 L^3}{3}
        -400 \zeta_2
\nonumber\\&
        -12 L \zeta_2
        +576 \ln (2) \zeta_2
        -256 \zeta_3
\Big)
+x^2 \Big(
        -\frac{10537}{27}
        +\frac{10904 L}{27}
        +\frac{2972 L^2}{27}
        +\frac{1196 L^3}{27}
        -\frac{4 L^4}{3}
\nonumber\\&
        +\frac{25118}{27} \zeta_2
        +40 L \zeta_2
        -4 L^2 \zeta_2
        -\frac{4736}{3} \ln (2) \zeta_2
        -\frac{8}{5} \zeta_2^2
        +\frac{6208}{9} \zeta_3
        +\frac{176}{9} L \zeta_3
\Big)
\Bigg]
\Bigg\}\,.
\end{align}
Here we observe that for $x=0$, the electric vector and axial-vector form factors are the same, as 
expected.
The bare singlet piece for the electric axial-vector form factor is given by 
\begin{align}
\hat{\F}_{A,1}^{(2),\sing} &= 
C_F T_F \Bigg\{
-\frac{6}{\ep} 
%
+ \Bigg[
-29
-12 L
+8 \zeta_2
+x \Big(
        8
        +24 L
        +8 L^2
        -48 \zeta_2
        -32 L \zeta_2
        +64 \zeta_3
\Big)
\nonumber\\&
+x^2 \Big(
        -4
        -16 L
        -56 L^2
        -\frac{16 L^3}{3}
        +32 \zeta_2
        +32 L \zeta_2
        +32 L^2 \zeta_2
        -\frac{64}{5} \zeta_2^2
        -128 \zeta_3
        -128 L \zeta_3
\Big)
\Bigg]
\nonumber\\&
+ \ep \Bigg[
-\frac{199}{2}
-58 L
-12 L^2
+26 \zeta_2
+16 L \zeta_2
+8 \zeta_3
+x \Big(
        -4
        +76 L
        +56 L^2
        +8 L^3
        -176 \zeta_2
\nonumber\\&
        -48 L \zeta_2
        -48 L^2 \zeta_2
        +384 \ln (2) \zeta_2
        +\frac{224}{5} \zeta_2^2
        -80 \zeta_3
        +32 L \zeta_3
\Big)
+x^2 \Big(
        -274
        +88 L
        -180 L^2
\nonumber\\&
        -\frac{136 L^3}{3}
        -\frac{10 L^4}{3}
        +632 \zeta_2
        -16 L \zeta_2
        +64 L^2 \zeta_2
        +32 L^3 \zeta_2
        -1536 \ln (2) \zeta_2
        -48 \zeta_2^2
        -\frac{448}{5} L \zeta_2^2
\nonumber\\&
        +400 \zeta_3
        +224 L \zeta_3
        -32 L^2 \zeta_3
        +160 \zeta_2 \zeta_3
        -144 \zeta_5
\Big)
\Bigg]
\Bigg\} \,.
\end{align}
The magnetic parts of the axial-vector form factor read
\begin{align}
\F_{A,2}^{(1),\ns} &=
C_F  \Bigg\{
-4 (2+3 L) x
-16 (1+L) x^2
+ \ep  \Bigg[
x \Big(
        -16
        -24 L
        -6 L^2
        +12 \zeta_2
\Big)
+x^2 \Big(
        -56
\nonumber\\&
        -8 L
        -8 L^2
        +16 \zeta_2
\Big)
\Bigg]
+ \ep^2  \Bigg[
x \Big(
        -32
        -48 L
        -12 L^2
        -2 L^3
        +20 \zeta_2
        +6 L \zeta_2
        +24 \zeta_3
\Big)
\nonumber\\&
+x^2 \Big(
        -136
        -16 L
        -4 L^2
        -\frac{8 L^3}{3}
        +8 L \zeta_2
        +32 \zeta_3
\Big)
\Bigg]
\Bigg\}\,.
%
\\
\F_{A,2}^{(2),\ns} &=
C_F^2  \Bigg\{
 \frac{1}{\ep}  \Bigg[
8 (1+L) (2+3 L) x
+32 (1+L)^2 x^2
\Bigg]
+ \Bigg[
x \Big(
        68
        +122 L
        +86 L^2
        +24 L^3
        +132 \zeta_2
\nonumber\\&
        -48 L \zeta_2
        -192 \ln (2) \zeta_2
        +48 \zeta_3
\Big)
+x^2 \Big(
        -80
        +416 L
        +48 L^2
        +32 L^3
        +\frac{4 L^4}{3}
        +16 L \zeta_2
\nonumber\\&
        +384 \ln (2) \zeta_2
        +\frac{32}{5} \zeta_2^2
        -160 \zeta_3
        -64 L \zeta_3
\Big)
\Bigg]
+ \ep  \Bigg[
x \Big(
        242
        +32 c_1
        +335 L
        +250 L^2
        +\frac{218 L^3}{3}
\nonumber\\&
        +14 L^4
        +812 \zeta_2
        -28 L \zeta_2
        -48 L^2 \zeta_2
        -1152 \ln (2) \zeta_2
        -\frac{1944}{5} \zeta_2^2
        +176 \zeta_3
        -96 L \zeta_3
\Big)
+x^2 \Big(
        316
\nonumber\\&
        -64 c_1
        +548 L
        +620 L^2
        +16 L^3
        +\frac{56 L^4}{3}
        +\frac{4 L^5}{5}
        -616 \zeta_2
        +384 L \zeta_2
        -24 L^2 \zeta_2
        -\frac{64}{3} L^3 \zeta_2
\nonumber\\&
        +3072 \ln (2) \zeta_2
        +\frac{96}{5} \zeta_2^2
        +\frac{496}{5} L \zeta_2^2
        -2000 \zeta_3
        -176 L \zeta_3
        -112 L^2 \zeta_3
        +576 \zeta_2 \zeta_3
        +768 \zeta_5
\Big)
\Bigg]
\Bigg\}
\nonumber\\& 
+ C_F C_A  \Bigg\{
 \Bigg[
x \Big(
        -\frac{968}{9}
        -\frac{458 L}{3}
        -22 L^2
        -84 \zeta_2
        +96 \ln (2) \zeta_2
        +48 \zeta_3
\Big)
+x^2 \Big(
        -\frac{2800}{9}
        -\frac{1936 L}{9}
\nonumber\\&
        -\frac{304 L^2}{3}
        -\frac{2 L^4}{3}
        +\frac{1184}{3} \zeta_2
        +144 L \zeta_2
        +32 L^2 \zeta_2
        -192 \ln (2) \zeta_2
        +\frac{528}{5} \zeta_2^2
        -560 \zeta_3
        -224 L \zeta_3
\Big)
\Bigg]
\nonumber\\&
+ \ep  \Bigg[
x \Big(
        -\frac{14872}{27}
        -16 c_1
        -\frac{6853 L}{9}
        -\frac{590 L^2}{3}
        -22 L^3
        -\frac{908}{3} \zeta_2
        -82 L \zeta_2
        +576 \ln (2) \zeta_2
\nonumber\\&
        +\frac{1248}{5} \zeta_2^2
        +232 \zeta_3
        +144 L \zeta_3
\Big)
+x^2 \Big(
        -\frac{61028}{27}
        +32 c_1
        -\frac{18980 L}{27}
        -\frac{5092 L^2}{9}
        -96 L^3
        -\frac{2 L^5}{5}
\nonumber\\&
        +\frac{12992}{9} \zeta_2
        +568 L \zeta_2
        +176 L^2 \zeta_2
        +\frac{128}{3} L^3 \zeta_2
        -1536 \ln (2) \zeta_2
        -\frac{7296}{5} \zeta_2^2
        -\frac{2424}{5} L \zeta_2^2
        +1312 \zeta_3
\nonumber\\&
        +304 L \zeta_3
        -40 L^2 \zeta_3
        -544 \zeta_2 \zeta_3
        +2304 \zeta_5
\Big)
\Bigg]
\Bigg\}
+ C_F n_l T_F  \Bigg\{
%
%
 \Bigg[
x \Big(
        \frac{304}{9}
        +\frac{136 L}{3}
        +8 L^2
        -16 \zeta_2
\Big)
\nonumber\\&
+
x^2 \Big(
        \frac{896}{9}
        +\frac{320 L}{9}
        +\frac{32 L^2}{3}
        -\frac{64}{3} \zeta_2
\Big)
\Bigg]
+ \ep  \Bigg[
x \Big(
        \frac{4664}{27}
        +\frac{2036 L}{9}
        +\frac{184 L^2}{3}
        +8 L^3
        -\frac{224}{3} \zeta_2
\nonumber\\&
        +24 L \zeta_2
        -96 \zeta_3
\Big)
+
x^2 \Big(
        \frac{18544}{27}
        +\frac{2704 L}{27}
        +\frac{368 L^2}{9}
        +\frac{32 L^3}{3}
        +\frac{128}{9} \zeta_2
        +32 L \zeta_2
        -128 \zeta_3
\Big)
\Bigg]
\Bigg\}
\nonumber\\&
+ C_F T_F  \Bigg\{
 \Bigg[
x \Big(
        -\frac{128}{9}
        +\frac{136 L}{3}
        +8 L^2
        +80 \zeta_2
\Big)
+
x^2 \Big(
        -\frac{1024}{9}
        -\frac{1408 L}{9}
        -\frac{160 L^2}{3}
        -320 \zeta_2
\Big)
\Bigg]
\nonumber\\&
+ \ep  \Bigg[
x \Big(
        -\frac{304}{27}
        +\frac{2036 L}{9}
        +\frac{184 L^2}{3}
        +8 L^3
        +\frac{1072}{3} \zeta_2
        +24 L \zeta_2
        -384 \ln (2) \zeta_2
        +192 \zeta_3
\Big)
\nonumber\\&
+
x^2 \Big(
        \frac{5200}{27}
        -\frac{5936 L}{27}
        -\frac{1360 L^2}{9}
        -32 L^3
        -\frac{6080}{9} \zeta_2
        +32 L \zeta_2
        +1280 \ln (2) \zeta_2
        -\frac{1600}{3} \zeta_3
\Big)
\Bigg]
\Bigg\} \,.
\end{align}
The bare singlet piece for the magnetic axial-vector form factor is given by 
\begin{align}
\hat{\F}_{A,2}^{(2),\sing} &= 
C_F T_F \Bigg\{
\Bigg[
x \Big(
        -80
        -48 L
        -8 L^2
        -32 \zeta_2
\Big)
+x^2 \Big(
        -160
        +\frac{4 L^4}{3}
        -256 \zeta_2
        -128 L \zeta_2
        +32 L^2 \zeta_2
\nonumber\\&
        +\frac{256}{5} \zeta_2^2
        +256 \zeta_3
        +192 L \zeta_3
\Big) 
\Bigg]
+ \ep \Bigg[
x \Big(
        -360
        -264 L
        -72 L^2
        -\frac{32 L^3}{3}
        +64 \zeta_2
        +16 L \zeta_2
\nonumber\\&
        -48 \zeta_3
\Big)
+x^2 \Big(
        -768
        -288 L
        -32 L^2
        -\frac{16 L^3}{3}
        -\frac{8 L^4}{3}
        +\frac{16 L^5}{15}
        -896 \zeta_2
        -192 L \zeta_2
\nonumber\\&
        -192 L^2 \zeta_2
        +16 L^3 \zeta_2
        +1536 \ln (2) \zeta_2
        +\frac{256}{5} \zeta_2^2
        -32 L \zeta_2^2
        -448 \zeta_3
        -512 L \zeta_3
\nonumber\\&
        +208 L^2 \zeta_3
        +64 \zeta_2 \zeta_3
        -704 \zeta_5
\Big) 
\Bigg]
\Bigg\} \,.
\end{align}
Similar to the vector form factor, the magnetic part of the axial-vector form factor also vanishes
for $x=0$. 

\subsubsection{Scalar form factor}
Next, we present the scalar form factor in the asymptotic limit.
\begin{align}
\F_{S}^{(1)} &=
C_F  \Bigg\{
\frac{1}{\ep}  \Bigg[
-2
-2 L
-4 L x^2
\Bigg]
+
 \Bigg[
-2
-L^2
+2 \zeta_2
+4 (-1+3 L) x
+x^2 \Big(
        1
        -2 L
        -2 L^2
\nonumber\\&
        +4 \zeta_2
\Big)
\Bigg]
+ \ep  \Bigg[
-4
-2 L
-\frac{L^3}{3}
-\zeta_2
+L \zeta_2
+4 \zeta_3
+x \Big(
        -4
        +12 L
        +6 L^2
        -12 \zeta_2
\Big)
+x^2 \Big(
        \frac{45}{2}
\nonumber\\&
        -24 L
        -L^2
        -\frac{2 L^3}{3}
        +2 \zeta_2
        +2 L \zeta_2
        +8 \zeta_3
\Big)
\Bigg]
+ \ep^2  \Bigg[
-8
-4 L
-L^2
-\frac{L^4}{12}
+\frac{1}{2} L^2 \zeta_2
+\frac{14}{5} \zeta_2^2
\nonumber\\&
+\frac{2}{3} \zeta_3
+\frac{14}{3} L \zeta_3
+\zeta_2
+x \Big(
        -8
        +28 L
        +6 L^2
        +2 L^3
        -14 \zeta_2
        -6 L \zeta_2
        -24 \zeta_3
\Big)
+x^2 \Big(
        \frac{193}{4}
        -34 L
\nonumber\\&
        -12 L^2
        -\frac{L^3}{3}
        -\frac{L^4}{6}
        +\frac{49}{2} \zeta_2
        +L \zeta_2
        +L^2 \zeta_2
        +\frac{28}{5} \zeta_2^2
        +4 \zeta_3
        +\frac{28}{3} L \zeta_3
\Big)
\Bigg]
\Bigg\} \,.
%
\\
\F_{S}^{(2)} &=
C_F^2  \Bigg\{
\frac{1}{\ep^2}  \Bigg[
2
+4 L
+2 L^2
+8 L (1+L) x^2
\Bigg]
+ \frac{1}{\ep}  \Bigg[
4
+4 L
+2 L^2
+2 L^3
-4 \zeta_2
-4 L \zeta_2
\nonumber\\&
-8 (1+L) (-1+3 L) x
+x^2 \Big(
        -2
        +10 L
        +8 L^2
        +8 L^3
        -8 \zeta_2
        -16 L \zeta_2
\Big)
\Bigg]
+ \Bigg[
29
+12 L
+6 L^2
\nonumber\\&
+\frac{2 L^3}{3}
+\frac{7 L^4}{6}
+6 \zeta_2
+12 L \zeta_2
-4 L^2 \zeta_2
-\frac{118}{5} \zeta_2^2
-56 \zeta_3
-32 L \zeta_3
+x \Big(
        32
        -72 L
        +4 L^2
        -24 L^3
\nonumber\\&
        -112 \zeta_2
        +64 L \zeta_2
        +8 L^2 \zeta_2
        +288 \ln (2) \zeta_2
        +\frac{136}{5} \zeta_2^2
        -120 \zeta_3
        -64 L \zeta_3
\Big)
+x^2 \Big(
        \frac{1069}{2}
        -380 L
\nonumber\\&
        +166 L^2
        -\frac{10 L^3}{3}
        +5 L^4
        -36 \zeta_2
        -132 L \zeta_2
        -36 L^2 \zeta_2
        -576 \ln (2) \zeta_2
        -\frac{724}{5} \zeta_2^2
        +72 \zeta_3
\nonumber\\&
        +224 L \zeta_3
\Big)
\Bigg]
+ \ep  \Bigg[
-\frac{113}{2}
+36 L
+14 L^2
+\frac{14 L^3}{3}
+\frac{L^4}{6}
+\frac{L^5}{2}
-40 \zeta_2
+24 L \zeta_2
+12 L^2 \zeta_2
\nonumber\\&
-\frac{8}{3} L^3 \zeta_2
+264 \ln (2) \zeta_2
-\frac{314}{5} \zeta_2^2
-\frac{106}{5} L \zeta_2^2
-\frac{478}{3} \zeta_3
-\frac{212}{3} L \zeta_3
-\frac{112}{3} L^2 \zeta_3
-12 \zeta_2 \zeta_3
-18 \zeta_5
\nonumber\\&
+x \Big(
        232
        -48 c_1
        -216 L
        -40 L^2
        +12 L^3
        -\frac{41 L^4}{3}
        -1016 \zeta_2
        +24 L \zeta_2
        +12 L^2 \zeta_2
        +8 L^3 \zeta_2
\nonumber\\&
        +1536 \ln (2) \zeta_2
        +\frac{1552}{5} \zeta_2^2
        -\frac{544}{5} L \zeta_2^2
        -616 \zeta_3
        +240 L \zeta_3
        -24 L^2 \zeta_3
        -64 \zeta_2 \zeta_3
        +672 \zeta_5
\Big)
\nonumber\\&
+x^2 \Big(
        -\frac{2149}{4}
        +96 c_1
        +895 L
        -221 L^2
        +\frac{472 L^3}{3}
        +\frac{37 L^4}{6}
        +\frac{5 L^5}{3}
        +2568 \zeta_2
        -366 L \zeta_2
\nonumber\\&
        +214 L^2 \zeta_2
        -52 L^3 \zeta_2
        -2304 \ln (2) \zeta_2
        -\frac{306}{5} \zeta_2^2
        +100 L \zeta_2^2
        +604 \zeta_3
        -\frac{4240}{3} L \zeta_3
        +\frac{128}{3} L^2 \zeta_3
\nonumber\\&
        -520 \zeta_2 \zeta_3
        -2916 \zeta_5
\Big)
\Bigg]
\Bigg\}
+ C_F C_A  \Bigg\{
\frac{1}{\ep^2}  \Bigg[
\frac{11}{3}
+\frac{11 L}{3}
+\frac{22 L x^2}{3}
\Bigg]
+ \frac{1}{\ep}  \Bigg[
-\frac{49}{9}
-\frac{67 L}{9}
+2 \zeta_2
\nonumber\\&
+2 L \zeta_2
-2 \zeta_3
+x^2 \Big(
        4
        -\frac{188 L}{9}
        -4 L^2
        -\frac{4 L^3}{3}
        -4 \zeta_2
        -8 \zeta_3
\Big)
\Bigg]
+ \Bigg[
-\frac{869}{27}
-\frac{242 L}{27}
-\frac{67 L^2}{9}
\nonumber\\&
-\frac{11 L^3}{9}
+\frac{182}{9} \zeta_2
-\frac{22}{3} L \zeta_2
+2 L^2 \zeta_2
-\frac{3}{5} \zeta_2^2
+\frac{98}{3} \zeta_3
+26 L \zeta_3
+x \Big(
        -\frac{580}{9}
        +\frac{416 L}{3}
        +12 L^2
\nonumber\\&
        +92 \zeta_2
        -144 \ln (2) \zeta_2
        -44 \zeta_3
\Big)
+x^2 \Big(
        \frac{3463}{36}
        -\frac{5705 L}{54}
        -\frac{37 L^2}{18}
        -\frac{43 L^3}{9}
        -\frac{L^4}{2}
        -\frac{611}{9} \zeta_2
        +\frac{58}{3} L \zeta_2
\nonumber\\&
        +10 L^2 \zeta_2
        +288 \ln (2) \zeta_2
        -\frac{122}{5} \zeta_2^2
        +\frac{364}{3} \zeta_3
        +96 L \zeta_3
\Big)
\Bigg]
+ \ep  \Bigg[
-\frac{6437}{162}
-\frac{2122 L}{81}
-\frac{341 L^2}{27}
\nonumber\\&
-\frac{134 L^3}{27}
-\frac{11 L^4}{12}
+\frac{1972}{27} \zeta_2
-\frac{103}{9} L \zeta_2
-\frac{11}{2} L^2 \zeta_2
+\frac{4}{3} L^3 \zeta_2
-132 \ln (2) \zeta_2
+65 \zeta_2^2
+\frac{88}{5} L \zeta_2^2
\nonumber\\&
+\frac{1055}{9} \zeta_3
+\frac{152}{3} L \zeta_3
+26 L^2 \zeta_3
-2 \zeta_2 \zeta_3
-157 \zeta_5
+x \Big(
        -\frac{2756}{27}
        +24 c_1
        +\frac{3700 L}{9}
        +\frac{476 L^2}{3}
\nonumber\\&
        +\frac{32 L^3}{3}
        -\frac{L^4}{6}
        +\frac{746}{3} \zeta_2
        +14 L \zeta_2
        +2 L^2 \zeta_2
        -768 \ln (2) \zeta_2
        -360 \zeta_2^2
        +220 \zeta_3
        -16 L \zeta_3
\Big)
\nonumber\\&
+x^2 \Big(
        \frac{40951}{216}
        -48 c_1
        -\frac{85217 L}{324}
        -\frac{24019 L^2}{108}
        +\frac{553 L^3}{54}
        -\frac{79 L^4}{12}
        -\frac{L^5}{6}
        -\frac{13060}{27} \zeta_2
        -\frac{188}{9} L \zeta_2
\nonumber\\&
        +28 L^2 \zeta_2
        +\frac{2}{3} L^3 \zeta_2
        +1152 \ln (2) \zeta_2
        +841 \zeta_2^2
        +\frac{798}{5} L \zeta_2^2
        -\frac{1916}{9} \zeta_3
        -\frac{908}{3} L \zeta_3
        +92 L^2 \zeta_3
\nonumber\\&
        +388 \zeta_2 \zeta_3
        -942 \zeta_5
\Big)
\Bigg]
\Bigg\}
+ C_F n_l T_F  \Bigg\{
\frac{1}{\ep^2}  \Bigg[
-\frac{4}{3}
-\frac{4 L}{3}
-\frac{8 L x^2}{3}
\Bigg]
+ \frac{1}{\ep}  \Bigg[
\frac{20}{9}
+\frac{20 L}{9}
+\frac{40 L x^2}{9}
\Bigg]
\nonumber\\&
+ \Bigg[
\frac{196}{27}
+\frac{112 L}{27}
+\frac{20 L^2}{9}
+\frac{4 L^3}{9}
+\frac{8}{9} \zeta_2
+\frac{8}{3} L \zeta_2
-\frac{16}{3} \zeta_3
+x \Big(
        \frac{128}{9}
        -\frac{112 L}{3}
        -8 L^2
        +16 \zeta_2
\Big)
\nonumber\\&
+x^2 \Big(
        -\frac{290}{9}
        +\frac{1064 L}{27}
        +\frac{52 L^2}{9}
        +\frac{8 L^3}{9}
        -\frac{104}{9} \zeta_2
        +\frac{16}{3} L \zeta_2
        -\frac{32}{3} \zeta_3
\Big)
\Bigg]
+ \ep  \Bigg[
\frac{1706}{81}
+\frac{1232 L}{81}
\nonumber\\&
+\frac{148 L^2}{27}
+\frac{40 L^3}{27}
+\frac{L^4}{3}
+\frac{328}{27} \zeta_2
+\frac{20}{9} L \zeta_2
+2 L^2 \zeta_2
-\frac{96}{5} \zeta_2^2
-\frac{64}{9} \zeta_3
-\frac{16}{3} L \zeta_3
+x \Big(
        \frac{1504}{27}
\nonumber\\&
        -\frac{1328 L}{9}
        -\frac{136 L^2}{3}
        -8 L^3
        +\frac{344}{3} \zeta_2
        -24 L \zeta_2
        +96 \zeta_3
\Big)
+x^2 \Big(
        -\frac{7375}{27}
        +\frac{16600 L}{81}
        +\frac{1496 L^2}{27}
\nonumber\\&
        +\frac{116 L^3}{27}
        +\frac{2 L^4}{3}
        -\frac{3154}{27} \zeta_2
        +\frac{76}{9} L \zeta_2
        +4 L^2 \zeta_2
        -\frac{192}{5} \zeta_2^2
        -\frac{464}{9} \zeta_3
        -\frac{32}{3} L \zeta_3
\Big)
\Bigg]
\Bigg\}
\nonumber\\&
+ C_F T_F  \Bigg\{
 \Bigg[
\frac{1628}{27}
+\frac{224 L}{27}
+\frac{20 L^2}{9}
+\frac{4 L^3}{9}
-\frac{68}{3} \zeta_2
+4 L \zeta_2
+x \Big(
        -\frac{160}{9}
        +\frac{128 L}{3}
        +8 L^2
        -\frac{L^4}{3}
\nonumber\\&
        +16 \zeta_2
        -8 L^2 \zeta_2
        -\frac{64}{5} \zeta_2^2
        -48 L \zeta_3
\Big)
+x^2 \Big(
        \frac{2192}{9}
        -\frac{1088 L}{27}
        -\frac{416 L^2}{9}
        -\frac{136 L^3}{9}
        -\frac{4 L^4}{3}
        +\frac{512}{3} \zeta_2
\nonumber\\&
        +40 L \zeta_2
        -32 L^2 \zeta_2
        -\frac{256}{5} \zeta_2^2
        -256 \zeta_3
        -192 L \zeta_3
\Big)
\Bigg]
+ \ep  \Bigg[
\frac{4214}{27}
+\frac{64 L}{9}
+\frac{148 L^2}{27}
+\frac{40 L^3}{27}
+\frac{L^4}{3}
\nonumber\\&
-\frac{4028}{27} \zeta_2
+2 L^2 \zeta_2
+\frac{512}{3} \ln (2) \zeta_2
-\frac{8}{5} \zeta_2^2
-\frac{328}{3} \zeta_3
-\frac{56}{9} L \zeta_3
+x \Big(
        \frac{2368}{27}
        +\frac{1696 L}{9}
        +\frac{344 L^2}{3}
\nonumber\\&
        +\frac{32 L^3}{3}
        +\frac{L^4}{3}
        -\frac{4 L^5}{15}
        -\frac{520}{3} \zeta_2
        -88 L \zeta_2
        -8 L^2 \zeta_2
        -4 L^3 \zeta_2
        +192 \ln (2) \zeta_2
        +\frac{96}{5} \zeta_2^2
        +8 L \zeta_2^2
\nonumber\\&
        +240 \zeta_3
        +112 L \zeta_3
        -52 L^2 \zeta_3
        -16 \zeta_2 \zeta_3
        +176 \zeta_5
\Big)
+x^2 \Big(
        \frac{2621}{27}
        +\frac{10172 L}{27}
        -\frac{10276 L^2}{27}
        -\frac{2116 L^3}{27}
\nonumber\\&
        -14 L^4
        -\frac{16 L^5}{15}
        +\frac{29510}{27} \zeta_2
        +148 L \zeta_2
        +148 L^2 \zeta_2
        -16 L^3 \zeta_2
        -\frac{7040}{3} \ln (2) \zeta_2
        -\frac{1504}{5} \zeta_2^2
        +32 L \zeta_2^2
\nonumber\\&
        +\frac{3904}{9} \zeta_3
        -\frac{5296}{9} L \zeta_3
        -208 L^2 \zeta_3
        -64 \zeta_2 \zeta_3
        +704 \zeta_5
\Big)
\Bigg]
\Bigg\}\,.
\end{align}
\subsubsection{Pseudo-scalar form factor}

\vspace*{1mm}
\noindent
The non-singlet part of the pseudo-scalar form factor can be obtained in this limit using Eq.~(\ref{eq:cwiFF})
as
\begin{align}
 \F_{P}^{(n),\ns} = \F_{A,1}^{(n),\ns} + \Bigg( - \frac{(1-x)^2}{4 x} \Bigg) \F_{A,2}^{(n),\ns} \quad 
\text{for } n=1,2.
\end{align}
The unrenormalized singlet piece is given by
\begin{align}
\hat{\F}_{P}^{(2),\sing} &= 
C_F T_F \Bigg\{
\Bigg[
x \Big(
        8 L^2
        -\frac{L^4}{3}
        +32 \zeta_2
        -8 L^2 \zeta_2
        -\frac{64}{5} \zeta_2^2
        -48 L \zeta_3
\Big)
+x^2 \Big(
        112
        -64 L
        +16 L^2
        +16 L^3
\nonumber\\&
        +32 \zeta_2
        +96 L \zeta_2
\Big)
\Bigg]
+ \ep \Bigg[
x \Big(
        16 L^2
        +8 L^3
        +L^4
        -\frac{4 L^5}{15}
        +16 L \zeta_2
        +8 L^2 \zeta_2
        -4 L^3 \zeta_2
        +\frac{224}{5} \zeta_2^2
\nonumber\\&
        +8 L \zeta_2^2
        +48 \zeta_3
        +208 L \zeta_3
        -52 L^2 \zeta_3
        -16 \zeta_2 \zeta_3
        +176 \zeta_5
\Big)
+x^2 \Big(
        -400
        +480 L
        -80 L^2
\nonumber\\&
        -\frac{56 L^3}{3}
        +\frac{32 L^4}{3}
        -64 \zeta_2
        -160 L \zeta_2
        -48 L^2 \zeta_2
        -\frac{176}{5} \zeta_2^2
        +32 \zeta_3
        -128 L \zeta_3
\Big) 
\Bigg]
\Bigg\} \,.
\end{align}
%
\subsection{Threshold region {\boldmath$q^2 \sim 4 m^2$}}

\vspace*{1mm}
\noindent
Now we provide the expansion of the form factors 
in the threshold region $q^2 \sim 4 m^2$ or $x\rightarrow-1$. 
We expand the form factors around $\beta=0$ up to ${\cal O}(\beta^2)$ and denote the form factors 
by $\tilde{F}$ in this limit.

\subsubsection{Vector form factor}
First, we present the electric component of the vector form factor 
\begin{align}
 \tilde{F}_{V,1}^{(1)} &=  C_F \Bigg\{
\frac{1}{\ep}  \Bigg[
\frac{8 \beta ^2}{3}
+ i \pi \Big(
-\frac{1}{\beta }-\beta
\Big)
\Bigg]
+  \Bigg[
\frac{6 \zeta_2}{\beta }
-6
+6 \beta  \zeta_2
-\frac{10 \beta ^2}{9}
+ i \pi \Big(
\frac{1}{\beta } \Big( -1+2 \ln (2)
\nonumber\\&
+2 \log (\beta ) \Big)
+\beta  (-1+2 \ln (2)+2 \log (\beta ))
\Big)
\Bigg]
+ \ep  \Bigg[
\frac{1}{\beta } \Big( 6 \zeta_2
-12 \zeta_2 \ln (2)
-12 \zeta_2 \log (\beta )
\Big)
+ 4
\nonumber\\&
+\beta  \Big(
        6 \zeta_2
        -12 \zeta_2 \ln (2)
        -12 \zeta_2 \log (\beta )
\Big)
+\beta ^2 \Big(
        \frac{344}{27}
        +\frac{4 \zeta_2}{3}
\Big)
+ i \pi \Big(
\frac{1}{\beta } \Big( -4
+\frac{3}{2} \zeta_2
+2 \ln (2)
\nonumber\\&
-2 \ln ^2(2)
+2 \log (\beta )
-4 \ln (2) \log (\beta )
-2 \log ^2(\beta )
\Big)
+\beta  \Big(
        -3
        +\frac{3}{2} \zeta_2
        +2 \ln (2)
        -2 \ln ^2(2)
\nonumber\\&
        +2 \log (\beta )
        -4 \ln (2) \log (\beta )
        -2 \log ^2(\beta )
\Big)
\Big)
\Bigg]
+ \ep^2  \Bigg[
 \frac{1}{\beta } \Big( 24 \zeta_2
+3 \zeta_2^2
-12 \zeta_2 \ln (2)
+12 \zeta_2 \ln ^2(2)
\nonumber\\&
-12 \zeta_2 \log (\beta )
+24 \zeta_2 \ln (2) \log (\beta )
+12 \zeta_2 \log ^2(\beta )
\Big)
-24
-3 \zeta_2
+\beta  \Big(
        18 \zeta_2
        +3 \zeta_2^2
        -12 \zeta_2 \ln (2)
\nonumber\\&
        +12 \zeta_2 \ln ^2(2)
        -12 \zeta_2 \log (\beta )
        +24 \zeta_2 \ln (2) \log (\beta )
        +12 \zeta_2 \log ^2(\beta )
\Big)
+\beta ^2 \Big(
        -\frac{472}{81}
        -\frac{5 \zeta_2}{9}
        -\frac{8 \zeta_3}{9}
\Big)
\nonumber\\&
+ i \pi \Big(
\frac{1}{\beta } \Big( 
-8
+\frac{3}{2} \zeta_2
+\frac{7}{3} \zeta_3
+8 \ln (2)
-3 \zeta_2 \ln (2)
-2 \ln ^2(2)
+\frac{4 \ln ^3(2)}{3}
+8 \log (\beta )
\nonumber\\&
-3 \zeta_2 \log (\beta )
-4 \ln (2) \log (\beta )
+4 \ln ^2(2) \log (\beta )
-2 \log ^2(\beta )
+4 \ln (2) \log ^2(\beta )
+\frac{4 \log ^3(\beta )}{3}
\Big)
\nonumber\\&
+ \beta  \Big(
        -4
        +\frac{3}{2} \zeta_2
        +\frac{7}{3} \zeta_3
        +6 \ln (2)
        -3 \zeta_2 \ln (2)
        -2 \ln ^2(2)
        +\frac{4 \ln ^3(2)}{3}
        +6 \log (\beta )
        -3 \zeta_2 \log (\beta )
\nonumber\\&
        -4 \ln (2) \log (\beta )
        +4 \ln ^2(2) \log (\beta )
        -2 \log ^2(\beta )
        +4 \ln (2) \log ^2(\beta )
        +\frac{4 \log ^3(\beta )}{3}
\Big)
\Big)
\Bigg]
\Bigg\} \,, 
\\
 \tilde{F}_{V,1}^{(2)} &=  C_F^2 \Bigg\{
\frac{1}{\ep^2}  \Bigg[
-\frac{3 \zeta_2}{\beta ^2}
-6 \zeta_2
-3 \beta ^2 \zeta_2
+ i \pi \Bigg(
-\frac{8 \beta }{3}
\Bigg)
\Bigg]
+ \frac{1}{\ep}  \Bigg[
\frac{1}{\beta ^2} \Big(
-6 \zeta_2
+12 \zeta_2 \ln (2)
\nonumber\\&
+12 \zeta_2 \log (\beta )
\Big)
-12 \zeta_2
+16 \beta  \zeta_2
+24 \zeta_2 \ln (2)
+24 \zeta_2 \log (\beta )
+\beta ^2 \Big(
        -16
        +3 \zeta_2
        +12 \zeta_2 \ln (2)
\nonumber\\&
        +12 \zeta_2 \log (\beta )
\Big)
+ i \pi \Bigg(
-\frac{6 \zeta_2}{\beta ^2}
+\frac{6}{\beta }
-12 \zeta_2
+\frac{8}{9} \beta  (5+6 \ln (2)+6 \log (\beta ))
-6 \beta ^2 \zeta_2
\Bigg)
\Bigg]
\nonumber\\&
+  \Bigg[
 \frac{1}{\beta ^2} \Big( -28 \zeta_2
+15 \zeta_2^2
+24 \zeta_2 \ln (2)
-24 \zeta_2 \ln ^2(2)
+24 \zeta_2 \log (\beta )
-48 \zeta_2 \ln (2) \log (\beta )
\nonumber\\&
-24 \zeta_2 \log ^2(\beta )
\Big)
-\frac{36 \zeta_2}{\beta }
+\frac{421}{15}
-\frac{1926}{25} \zeta_2
+30 \zeta_2^2
-\frac{81}{5} \zeta_3
+\frac{156}{5} \zeta_2 \ln (2)
-48 \zeta_2 \ln ^2(2)
\nonumber\\&
+\frac{56}{5} \zeta_2 \log (\beta )
-96 \zeta_2 \ln (2) \log (\beta )
-48 \zeta_2 \log ^2(\beta )
+\beta  \Big(
        \frac{256}{3} \zeta_2
        -32 \zeta_2 \ln (2)
        -32 \zeta_2 \log (\beta )
\Big)
\nonumber\\&
+\beta ^2 \Big(
        \frac{691}{35}
        -\frac{51302 \zeta_2}{3675}
        +15 \zeta_2^2
        +\frac{1192}{35} \zeta_3
        -\frac{108}{35} \zeta_2 \ln (2)
        -24 \zeta_2 \ln ^2(2)
        +\frac{412}{35} \zeta_2 \log (\beta )
\nonumber\\&
        -48 \zeta_2 \ln (2) \log (\beta )
        -24 \zeta_2 \log ^2(\beta )
\Big)
+ i \pi \Bigg(
\frac{1}{\beta ^2} \Big( -12 \zeta_2
+24 \zeta_2 \ln (2)
+24 \zeta_2 \log (\beta )
\Big)
\nonumber\\&
-\frac{1}{\beta } \Big( 12 (\ln (2)+\log (\beta )) \Big)
-\frac{28}{5} \zeta_2
+48 \zeta_2 \ln (2)
+48 \zeta_2 \log (\beta )
+\beta  \Big(
        -\frac{2498}{27}
        +\frac{8}{3} \zeta_2
\nonumber\\&
        +\frac{448 \ln (2)}{9}
        -\frac{16}{3} \ln ^2(2)
        +\frac{256 \log (\beta )}{9}
        -\frac{32}{3} \ln (2) \log (\beta )
        -\frac{16}{3} \log ^2(\beta )
\Big)
+\beta ^2 \Big(
        -\frac{281}{35} \zeta_2
\nonumber\\&
        +24 \zeta_2 \ln (2)
        +24 \zeta_2 \log (\beta )
\Big)
\Bigg)
\Bigg]
+ \ep  \Bigg[
\frac{1}{\beta ^2} \Big( -28 \zeta_2
+30 \zeta_2^2
-106 \zeta_2 \zeta_3
+112 \zeta_2 \ln (2)
\nonumber\\&
-60 \zeta_2^2 \ln (2)
-48 \zeta_2 \ln ^2(2)
+32 \zeta_2 \ln ^3(2)
+112 \zeta_2 \log (\beta )
-60 \zeta_2^2 \log (\beta )
-96 \zeta_2 \ln (2) \log (\beta )
\nonumber\\&
+96 \zeta_2 \ln ^2(2) \log (\beta )
-48 \zeta_2 \log ^2(\beta )
+96 \zeta_2 \ln (2) \log ^2(\beta )
+32 \zeta_2 \log ^3(\beta )
\Big)
+\frac{1}{\beta } \Big( 72 \ln (2) \zeta_2
\nonumber\\&
+72 \log (\beta ) \zeta_2
\Big)
-\frac{36163}{450}
-\frac{10 c_1}{3}
+\frac{157876}{375} \zeta_2
-\frac{306}{5} \zeta_2^2
-\frac{43756}{75} \zeta_3
-212 \zeta_2 \zeta_3
\nonumber\\&
-\frac{32552}{75} \zeta_2 \ln (2)
-120 \zeta_2^2 \ln (2)
+\frac{32}{5} \zeta_2 \ln ^2(2)
+64 \zeta_2 \ln ^3(2)
-\frac{7144}{25} \zeta_2 \log (\beta )
\nonumber\\&
-120 \zeta_2^2 \log (\beta )
-\frac{224}{5} \zeta_2 \ln (2) \log (\beta )
+192 \zeta_2 \ln ^2(2) \log (\beta )
-\frac{112}{5} \zeta_2 \log ^2(\beta )
\nonumber\\&
+192 \zeta_2 \ln (2) \log ^2(\beta )
+64 \zeta_2 \log ^3(\beta )
+\beta  \Big(
        \frac{14416}{9} \zeta_2
        +16 \zeta_2^2
        -\frac{3776}{3} \zeta_2 \ln (2)
        +32 \zeta_2 \ln ^2(2)
\nonumber\\&
        -\frac{2624}{3} \zeta_2 \log (\beta )
        +64 \zeta_2 \ln (2) \log (\beta )
        +32 \zeta_2 \log ^2(\beta )
\Big)
+\beta ^2 \Big(
        \frac{1880267}{22050}
        +\frac{16 c_1}{21}
        +\frac{61140199 \zeta_2}{385875}
\nonumber\\&
        +\frac{26498}{175} \zeta_2^2
        +\frac{286234 \zeta_3}{3675}
        -106 \zeta_2 \zeta_3
        +\frac{800012 \zeta_2 \ln (2)}{3675}
        -60 \zeta_2^2 \ln (2)
        +\frac{208}{7} \zeta_2 \ln ^2(2)
\nonumber\\&
        +32 \zeta_2 \ln ^3(2)
        +\frac{175684 \zeta_2 \log (\beta )}
        {1225}
        -60 \zeta_2^2 \log (\beta )
        -\frac{2368}{35} \zeta_2 \ln (2) \log (\beta )
        +96 \zeta_2 \ln ^2(2) \log (\beta )
\nonumber\\&
        -\frac{824}{35} \zeta_2 \log ^2(\beta )
        +96 \zeta_2 \ln (2) \log ^2(\beta )
        +32 \zeta_2 \log ^3(\beta )
\Big)
+ i \pi \Bigg(
\frac{1}{\beta ^2} \Big(
-56 \zeta_2
-18 \zeta_2^2
\nonumber\\&
+48 \zeta_2 \ln (2)
-48 \zeta_2 \ln ^2(2)
+48 \zeta_2 \log (\beta )
-96 \zeta_2 \ln (2) \log (\beta )
-48 \zeta_2 \log ^2(\beta )
\Big)
\nonumber\\&
+ \frac{1}{\beta } \Big(
36
-6 \zeta_2
+12 \ln ^2(2)
+24 \ln (2) \log (\beta )
+12 \log ^2(\beta )
\Big)
+\frac{3572}{25} \zeta_2
-36 \zeta_2^2
+\frac{112}{5} \zeta_2 \ln (2)
\nonumber\\&
-96 \zeta_2 \ln ^2(2)
+\frac{112}
{5} \zeta_2 \log (\beta )
-192 \zeta_2 \ln (2) \log (\beta )
-96 \zeta_2 \log ^2(\beta )
+\beta  \Big(
        -\frac{49163}{81}
        +\frac{1064}{9} \zeta_2
\nonumber\\&
        +\frac{64}{9} \zeta_3
        +\frac{19408 \ln (2)}{27}
        -\frac{16}{3} \zeta_2 \ln (2)
        -\frac{2656}{9} \ln ^2(2)
        +\frac{32 \ln ^3(2)}{9}
        +\frac{14416 \log (\beta )}{27}
\nonumber\\&
        -\frac{16}{3} \zeta_2 \log (\beta )
        -\frac{3776}{9} \ln (2) \log (\beta )
        +\frac{32}{3} \ln ^2(2) \log (\beta )
        -\frac{1312}{9} \log ^2(\beta )
        +\frac{32}{3} \ln (2) \log ^2(\beta )
\nonumber\\&
        +\frac{32 \log ^3(\beta )}{9}
\Big)
+\beta ^2 \Big(
        -\frac{62992 \zeta_2}{1225}
        -18 \zeta_2^2
        -\frac{76}{35} \zeta_2 \ln (2)
        -48 \zeta_2 \ln ^2(2)
        +\frac{674}{35} \zeta_2 \log (\beta )
\nonumber\\&
        -96 \zeta_2 \ln (2) \log (\beta )
        -48 \zeta_2 \log ^2(\beta )
\Big)
\Bigg)
\Bigg]
\Bigg\} 
+ C_F C_A \Bigg\{
\frac{1}{\ep^2}  \Bigg[
-\frac{44 \beta ^2}{9}
+ i \pi \Bigg(
\frac{11}{6 \beta }+\frac{11 \beta }{6}
\Bigg)
\Bigg]
\nonumber\\&
+ \frac{1}{\ep}  \Bigg[
-8 \beta  \zeta_2
+ \beta ^2 \Big(
        \frac{376}{27}
        +\frac{32 \zeta_2}{3}
\Big)
+ i \pi \Bigg(
-\frac{31}{18 \beta }
+\frac{1}{6} \beta  (-13-32 \ln (2)-16 \log (\beta ))
\Bigg)
\Bigg]
\nonumber\\& 
+  \Bigg[
\frac{1}{\beta } \Big( 
\frac{146}{3} \zeta_2
-44 \zeta_2 \ln (2)
-44 \zeta_2 \log (\beta )
\Big)
-\frac{379}{15}
+\frac{5482}{75} \zeta_2
-\frac{166}{5} \zeta_3
-\frac{312}{5} \zeta_2 \ln (2)
\nonumber\\&
-\frac{144}{5} \zeta_2 \log (\beta )
+\beta  \Big(
        -\frac{16}{3} \zeta_2
        +84 \zeta_2 \ln (2)
        +36 \zeta_2 \log (\beta )
\Big)
+\beta ^2 \Big(
        \frac{26779}{2835}
        +88 \zeta_2
        -\frac{346}{105} \zeta_3
\nonumber\\&
        -\frac{1952}{35} \zeta_2 \ln (2)
        -48 \zeta_2 \log (\beta )
\Big)
+ i \pi \Bigg(
\frac{1}{\beta } \Big( 
-\frac{394}{27}
+\frac{146 \ln (2)}{9}
-\frac{22 \ln ^2(2)}{3}
+\frac{146 \log (\beta )}{9}
\nonumber\\&
-\frac{44}{3} \ln (2) \log (\beta )
-\frac{22 \log ^2(\beta )}{3} 
\Big)
+\frac{72}{5} \zeta_2
+\beta  \Big(
        -\frac{139}{9}
        -\frac{40}{3} \zeta_2
        -\frac{152 \ln (2)}{9}
        +\frac{74 \ln ^2(2)}{3}
\nonumber\\&
        -\frac{16 \log (\beta )}{9}
        +28 \ln (2) \log (\beta )
        +6 \log ^2(\beta )
\Big)
+\frac{351}{14} \beta ^2 \zeta_2
\Bigg)
\Bigg]
+ \ep  \Bigg[
 \frac{1}{\beta } \Big(
\frac{2422}{9} \zeta_2
+99 \zeta_2^2
\nonumber\\&
-\frac{752}{3} \zeta_2 \ln (2)
+132 \zeta_2 \ln ^2(2)
-\frac{752}{3} \zeta_2 \log (\beta )
+264 \zeta_2 \ln (2) \log (\beta )
+132 \zeta_2 \log ^2(\beta )
\Big)
\nonumber\\&
-\frac{28301}{150}
+\frac{28 c_1}{5}
+\frac{228613 \zeta_2}{1125}
-\frac{7748}{25} \zeta_2^2
+\frac{8539}{75} \zeta_3
-\frac{1076}{5} \zeta_2 \ln (2)
-\frac{256}{5} \zeta_2 \ln ^2(2)
\nonumber\\&
-\frac{4144}{25} \zeta_2 \log (\beta )
+\frac{576}{5} \zeta_2 \ln (2) \log (\beta )
+\frac{288}{5} \zeta_2 \log ^2(\beta )
+\beta  \Big(
        \frac{976}{9} \zeta_2
        +59 \zeta_2^2
        +\frac{1612}{3} \zeta_2 \ln (2)
\nonumber\\&
        -508 \zeta_2 \ln ^2(2)
        +\frac{796}{3} \zeta_2 \log (\beta )
        -632 \zeta_2 \ln (2) \log (\beta )
        -172 \zeta_2 \log ^2(\beta )
\Big)
+\beta ^2 \Big(
        \frac{39802157}{198450}
\nonumber\\&
        -\frac{284 c_1}{105}
        +\frac{5048597 \zeta_2}{9450}
        -\frac{209161 \zeta_2^2}{1050}
        -\frac{244703 \zeta_3}{7350}
        -\frac{323804 \zeta_2 \ln (2)}{3675}
        +\frac{264}{7} \zeta_2 \ln ^2(2)
\nonumber\\&
        -\frac{1274}{5} \zeta_2 \log (\beta )
        +144 \zeta_2 \ln (2) \log (\beta )
        +96 \zeta_2 \log ^2(\beta )
\Big)
+ i \pi \Bigg(
\frac{1}{\beta } \Big( 
-\frac{13177}{162}
+\frac{38}{3} \zeta_2
+\frac{22}{3} \zeta_3
\nonumber\\&
+\frac{2422 \ln (2)}{27}
-11 \zeta_2 \ln (2)
-\frac{376}{9} \ln ^2(2)
+\frac{44 \ln ^3(2)}{3}
+\frac{2422 \log (\beta )}{27}
-11 \zeta_2 \log (\beta )
\nonumber\\&
-\frac{752}{9} \ln (2) \log (\beta )
+44 \ln ^2(2) \log (\beta )
-\frac{376}{9} \log ^2(\beta )
+44 \ln (2) \log ^2(\beta )
+\frac{44 \log ^3(\beta )}{3}
\Big)
\nonumber\\&
+\frac{2072}{25} \zeta_2
-\frac{288}{5} \zeta_2 \ln (2)
-\frac{288}{5} \zeta_2 \log (\beta )
+\beta  \Big(
        -\frac{2950}{27}
        -\frac{596}{9} \zeta_2
        -\frac{194}{3} \zeta_3
        -\frac{256 \ln (2)}{27}
\nonumber\\&
        +\frac{335}{3} \zeta_2 \ln (2)
        +150 \ln ^2(2)
        -\frac{764}{9} \ln ^3(2)
        +\frac{976 \log (\beta )}{27}
        +77 \zeta_2 \log (\beta )
        +\frac{1612}{9} \ln (2) \log (\beta )
\nonumber\\&
        -\frac{508}{3} \ln ^2(2) \log (\beta )
        +\frac{398 \log ^2(\beta )}{9}
        -\frac{316}{3} \ln (2) \log ^2(\beta )
        -\frac{172}{9} \log ^3(\beta )
\Big)
+\beta ^2 \Big(
        \frac{792}{5} \zeta_2
\nonumber\\&
        -\frac{786}{7} \zeta_2 \ln (2)
        -\frac{657}{7} \zeta_2 \log (\beta )
\Big)
\Bigg)
\Bigg]
\Bigg\} 
+ C_F T_F n_l \Bigg\{
\frac{1}{\ep^2}  \Bigg[
\frac{16 \beta ^2}{9}
+ i \pi \Bigg(
-\frac{2}{3 \beta }-\frac{2 \beta }{3}
\Bigg)
\Bigg]
\nonumber\\&
+ \frac{1}{\ep}  \Bigg[
-\frac{80 \beta ^2}{27}
+ i \pi \Bigg(
\frac{10}{9 \beta }+\frac{10 \beta }{9}
\Bigg)
\Bigg]
+  \Bigg[
\frac{1}{\beta } \Big(
-\frac{64}{3} \zeta_2
+16 \zeta_2 \ln (2)
+16 \zeta_2 \log (\beta )
\Big)
+4
\nonumber\\&
+\beta  \Big(
        -\frac{64}{3} \zeta_2
        +16 \zeta_2 \ln (2)
        +16 \zeta_2 \log (\beta )
\Big)
+\beta ^2 \Big(
        -\frac{868}{81}
        -\frac{64 \zeta_2}{9}
\Big)
+ i \pi \Bigg(
\frac{1}{\beta } \Big( 
\frac{206}{27}
-\frac{64 \ln (2)}{9}
\nonumber\\&
+\frac{8 \ln ^2(2)}{3}
-\frac{64 \log (\beta )}{9}
+\frac{16}{3} \ln (2) \log (\beta )
+\frac{8 \log ^2(\beta )}{3} 
\Big)
+ \beta  \Big(
        \frac{116}{27}
       -\frac{64 \ln (2)}{9}
       +\frac{8 \ln ^2(2)}{3}
\nonumber\\&
       -\frac{64 \log (\beta )}{9}
       +\frac{16}{3} \ln (2) \log (\beta )
       +\frac{8 \log ^2(\beta )}{3}
      \Big)
\Bigg)
\Bigg]
+ \ep  \Bigg[
\frac{1}{\beta } \Big(
-\frac{1112}{9} \zeta_2
-36 \zeta_2^2
+\frac{304}{3} \zeta_2 \ln (2)
\nonumber\\&
-48 \zeta_2 \ln ^2(2)
+\frac{304}{3} \zeta_2 \log (\beta )
-96 \zeta_2 \ln (2) \log (\beta )
-48 \zeta_2 \log ^2(\beta )
\Big)
+\frac{574}{9}
+36 \zeta_2
\nonumber\\&
+\beta  \Big(
        -\frac{680}{9} \zeta_2
        -36 \zeta_2^2
        +\frac{304}{3} \zeta_2 \ln (2)
        -48 \zeta_2 \ln ^2(2)
        +\frac{304}{3} \zeta_2 \log (\beta )
        -96 \zeta_2 \ln (2) \log (\beta )
\nonumber\\&
        -48 \zeta_2 \log ^2(\beta )
\Big)
+\beta ^2 \Big(
        -\frac{518}{81}
        -\frac{220 \zeta_2}{27}
        -\frac{128 \zeta_3}{9}
\Big)
+ i \pi \Bigg(
\frac{1}{\beta } \Big( 
\frac{3211}{81}
-\frac{16}{3} \zeta_2
-\frac{8}{3} \zeta_3
-\frac{1112 \ln (2)}{27}
\nonumber\\&
+4 \zeta_2 \ln (2)
+\frac{152 \ln ^2(2)}{9}
-\frac{16}{3} \ln ^3(2)
-\frac{1112 \log (\beta )}{27}
+4 \zeta_2 \log (\beta )
+\frac{304}{9} \ln (2) \log (\beta )
\nonumber\\&
-16 \ln ^2(2) \log (\beta )
+\frac{152 \log ^2(\beta )}{9}
-16 \ln (2) \log ^2(\beta )
-\frac{16}{3} \log ^3(\beta )
\Big)
+\beta  \Big(
        \frac{934}{81}
        -\frac{16}{3} \zeta_2
        -\frac{8}{3} \zeta_3
\nonumber\\&
        -\frac{680 \ln (2)}{27}
        +4 \zeta_2 \ln (2)
        +\frac{152 \ln ^2(2)}{9}
        -\frac{16}{3} \ln ^3(2)
        -\frac{680 \log (\beta )}{27}
        +4 \zeta_2 \log (\beta )
\nonumber\\&
        +\frac{304}{9} \ln (2) \log (\beta )
        -16 \ln ^2(2) \log (\beta )
        +\frac{152 \log ^2(\beta )}{9}
        -16 \ln (2) \log ^2(\beta )
        -\frac{16}{3} \log ^3(\beta )
\Big)
\Bigg)
\Bigg]
\Bigg\} 
\nonumber\\&
+ C_F T_F \Bigg\{
%
%
%
  \Bigg[
\frac{148}{3}
-\frac{416}{15} \zeta_2
+\beta ^2 \Big(
        -\frac{1372}{81}
        +\frac{52 \zeta_2}{5}
\Big)
+ i \pi \Bigg(
\frac{2 \zeta_2}{3 \beta }
+\beta  \Big(
        \frac{16}{15}
        +\frac{2 \zeta_2}{3}
\Big)
\Bigg)
\Bigg]
\nonumber\\&
+ \ep  \Bigg[
-\frac{4 \zeta_2^2}{\beta }
+\frac{2102}{15}
-\frac{28616}{225} \zeta_2
-\frac{1456}{15} \zeta_3
+\frac{832}{5} \zeta_2 \ln (2)
+\beta  \Big(
        -\frac{32 \zeta_2}{5}
        -4 \zeta_2^2
\Big)
\nonumber\\&
+\beta ^2 \Big(
        -\frac{47902}{1215}
        +\frac{475514 \zeta_2}{11025}
        +\frac{5104}{135} \zeta_3
        -\frac{6232}{105} \zeta_2 \ln (2)
        +\frac{24}{7} \zeta_2 \log (\beta )
\Big)
\nonumber\\&
+ i \pi \Bigg(
\frac{1}{\beta } \Big(
\frac{2}{3} \zeta_2
-\frac{4}{9} \zeta_3
-\frac{4}{3} \zeta_2 \ln (2)
-\frac{4}{3} \zeta_2 \log (\beta )
\Big)
+\beta  \Big(
        \frac{16}{15}
        +\frac{2}{3} \zeta_2
        -\frac{4}{9} \zeta_3
        -\frac{32 \ln (2)}{15}
\nonumber\\&
        -\frac{4}{3} \zeta_2 \ln (2)
        -\frac{32 \log (\beta )}{15}
        -\frac{4}{3} \zeta_2 \log (\beta )
\Big)
\Bigg)
\Bigg]
\Bigg\} \,. 
\end{align}
%
Next, we present the magnetic component of the vector form factor
%
\begin{align}
 \tilde{F}_{V,2}^{(1)} &=  C_F \Bigg\{
%
%
-2+\frac{4 \beta ^2}{3}
+ i \pi \Bigg(
\frac{1}{\beta }-\beta
\Bigg)
%
+ \ep  \Bigg[
-\frac{6 \zeta_2}{\beta }
-4
+6 \beta  \zeta_2
+\frac{28 \beta ^2}{9}
+ i \pi \Bigg(
-\frac{1}{\beta } \Big( 2 (-2
\nonumber\\&
+\ln (2)+\log (\beta )) \Big)
+\beta  (-5+2 \ln (2)+2 \log (\beta ))
\Bigg)
\Bigg]
+ \ep^2  \Bigg[
\frac{1}{\beta } \Big(
-24 \zeta_2
+12 \zeta_2 \ln (2)
\nonumber\\&
+12 \zeta_2 \log (\beta )
\Big)
-8
-\zeta_2
+\beta  \Big(
        30 \zeta_2
        -12 \zeta_2 \ln (2)
        -12 \zeta_2 \log (\beta )
\Big)
+\beta ^2 \Big(
        \frac{160}{27}
        +\frac{2 \zeta_2}{3}
\Big)
\nonumber\\&
+ i \pi \Bigg(
\frac{1}{\beta } \Big(
8
-\frac{3}{2} \zeta_2
-8 \ln (2)
+2 \ln ^2(2)
-8 \log (\beta )
+4 \ln (2) \log (\beta )
+2 \log ^2(\beta )
\Big)
\nonumber\\&
+\beta  \Big(
        -12
        +\frac{3}{2} \zeta_2
        +10 \ln (2)
        -2 \ln ^2(2)
        +10 \log (\beta )
        -4 \ln (2) \log (\beta )
        -2 \log ^2(\beta )
\Big)
\Bigg)
\Bigg]
\Bigg\} \,. 
\\
 \tilde{F}_{V,2}^{(2)} &=  C_F^2 \Bigg\{
 \frac{1}{\ep}  \Bigg[
\frac{6 \zeta_2}{\beta ^2}
+\beta ^2 \Big(
        -\frac{16}{3}
        -6 \zeta_2
\Big)
+ i \pi \Bigg(
\frac{2}{\beta }+\frac{10 \beta }{3}
\Bigg)
\Bigg]
+  \Bigg[
\frac{1}{\beta ^2} \Big(
28 \zeta_2
-24 \zeta_2 \ln (2)
\nonumber\\&
-24 \zeta_2 \log (\beta )
\Big)
+\frac{269}{15}
-\frac{4922}{75} \zeta_2
-\frac{12 \zeta_2}{\beta }
-60 \beta  \zeta_2
+\frac{41}{5} \zeta_3
+\frac{404}{5} \zeta_2 \ln (2)
+\frac{24}{5} \zeta_2 \log (\beta )
\nonumber\\&
+\beta ^2 \Big(
        \frac{3658}{315}
        +\frac{103928 \zeta_2}{3675}
        -\frac{471}{35} \zeta_3
        -\frac{1628}{35} \zeta_2 \ln (2)
        +\frac{512}{35} \zeta_2 \log (\beta )
\Big)
+ i \pi \Bigg(
\frac{12 \zeta_2}{\beta ^2}
-\frac{1}{\beta } \Big( 4 (\ln (2)
\nonumber\\&
+\log (\beta )) \Big)
-\frac{12}{5} \zeta_2
-\frac{2}{9} \beta  (-73+90 \ln (2)+90 \log (\beta ))
-\frac{181}{35} \beta ^2 \zeta_2
\Bigg)
\Bigg]
+ \ep  \Bigg[
\frac{1}{\beta ^2} \Big(
28 \zeta_2
-30 \zeta_2^2
\nonumber\\&
-112 \zeta_2 \ln (2)
+48 \zeta_2 \ln ^2(2)
-112 \zeta_2 \log (\beta )
+96 \zeta_2 \ln (2) \log (\beta )
+48 \zeta_2 \log ^2(\beta )
\Big)
\nonumber\\&
+\frac{1}{\beta } \Big(
24 \ln (2) \zeta_2
+24 \log (\beta ) \zeta_2
\Big)
+\frac{20713}{450}
-\frac{38 c_1}{3}
-\frac{529628 \zeta_2}{1125}
+\frac{1342}{5} \zeta_2^2
-\frac{15944}{75} \zeta_3
\nonumber\\&
+\frac{27752}{75} \zeta_2 \ln (2)
+\frac{448}{5} \zeta_2 \ln ^2(2)
+\frac{344}{25} \zeta_2 \log (\beta )
-\frac{96}{5} \zeta_2 \ln (2) \log (\beta )
-\frac{48}{5} \zeta_2 \log ^2(\beta )
\nonumber\\&
+\beta  \Big(
        -\frac{200}{3} \zeta_2
        +280 \zeta_2 \ln (2)
        +280 \zeta_2 \log (\beta )
\Big)
+\beta ^2 \Big(
        -\frac{4863529}{33075}
        +\frac{250 c_1}{21}
        +\frac{115201589 \zeta_2}{385875}
\nonumber\\&        
        -\frac{42283}{175} \zeta_2^2
        +\frac{330121 \zeta_3}{1225}
        -\frac{749248 \zeta_2 \ln (2)}{3675}
        -\frac{3672}{35} \zeta_2 \ln ^2(2)
        +\frac{245324 \zeta_2 \log (\beta )}{1225}
\nonumber\\&
        -\frac{1328}{35} \zeta_2 \ln (2) \log (\beta )
        -\frac{1024}{35} \zeta_2 \log ^2(\beta )
\Big)
+ i \pi \Bigg(
\frac{1}{\beta ^2} \Big(
56 \zeta_2
-48 \zeta_2 \ln (2)
-48 \zeta_2 \log (\beta )
\Big)
\nonumber\\&
+\frac{1}{\beta } \Big(
-4
-2 \zeta_2
+4 \ln ^2(2)
+8 \ln (2) \log (\beta )
+4 \log ^2(\beta )
\Big)
-\frac{172}{25} \zeta_2
+\frac{48}{5} \zeta_2 \ln (2)
+\frac{48}{5} \zeta_2 \log (\beta )
\nonumber\\&
+\beta  \Big(
        -\frac{241}{9}
        -\frac{50}{3} \zeta_2
        -\frac{8 \ln (2)}{9}
        +\frac{140 \ln ^2(2)}{3}
        -\frac{200 \log (\beta )}{9}
        +\frac{280}{3} \ln (2) \log (\beta )
        +\frac{140 \log ^2(\beta )}{3}
\Big)
\nonumber\\&
+\beta ^2 \Big(
        -\frac{147512 \zeta_2}{1225}
        +\frac{1924}{35} \zeta_2 \ln (2)
        +\frac{1174}{35} \zeta_2 \log (\beta )
\Big)
\Bigg)
\Bigg]
\Bigg\} 
+ C_F C_A \Bigg\{
%
%
%
%
%
%
  \Bigg[
-\frac{28 \zeta_2}{\beta }
-\frac{373}{45}
\nonumber\\&
+\frac{1156}{25} \zeta_2
-\frac{94}{5} \zeta_3
-\frac{328}{5} \zeta_2 \ln (2)
-\frac{96}{5} \zeta_2 \log (\beta )
+60 \beta  \zeta_2
+\beta ^2 \Big(
        -\frac{1234}{315}
        -\frac{1577}{25} \zeta_2
        +\frac{288}{35} \zeta_3
\nonumber\\&        
        +\frac{2344}{35} \zeta_2 \ln (2)
        +\frac{32}{5} \zeta_2 \log (\beta )
\Big)
+ i \pi \Bigg(
-\frac{1}{9 \beta } \Big( 4 (-25+21 \ln (2)+21 \log (\beta )) \Big)
+\frac{48}{5} \zeta_2
\nonumber\\&
+\frac{2}{3} \beta  (-49+38 \ln (2)+30 \log (\beta ))
-\frac{299}{70} \beta ^2 \zeta_2
\Bigg)
\Bigg]
+ \ep  \Bigg[
\frac{1}{\beta } \Big(
-\frac{682}{3} \zeta_2
+168 \zeta_2 \ln (2)
\nonumber\\&
+168 \zeta_2 \log (\beta )
\Big)
-\frac{181841}{1350}
+\frac{116 c_1}{15}
+\frac{104129}{375} \zeta_2
-\frac{6652}{25} \zeta_2^2
+\frac{6236}{75} \zeta_3
-\frac{4112}{15} \zeta_2 \ln (2)
\nonumber\\&
-\frac{224}{5} \zeta_2 \ln ^2(2)
-\frac{1856}{25} \zeta_2 \log (\beta )
+\frac{384}{5} \zeta_2 \ln (2) \log (\beta )
+\frac{192}{5} \zeta_2 \log ^2(\beta )
+\beta  \Big(
        \frac{1822}{3} \zeta_2
\nonumber\\&
        -520 \zeta_2 \ln (2)
        -424 \zeta_2 \log (\beta )
\Big)
+\beta ^2 \Big(
        \frac{16374871}{99225}
        -\frac{128 c_1}{21}
        -\frac{975547 \zeta_2}{5250}
        +\frac{82843}{350} \zeta_2^2
        -\frac{458029 \zeta_3}{3675}
\nonumber\\&
        -\frac{71626 \zeta_2 \ln (2)}{3675}
        +\frac{4056}{35} \zeta_2 \ln ^2(2)
        -\frac{638}{25} \zeta_2 \log (\beta )
        +\frac{112}{5} \zeta_2 \ln (2) \log (\beta )
        -\frac{64}{5} \zeta_2 \log ^2(\beta )
\Big)
\nonumber\\&
+ i \pi \Bigg(
\frac{1}{\beta } \Big(
\frac{4015}{54}
-\frac{15}{2} \zeta_2
-\frac{682 \ln (2)}{9}
+28 \ln ^2(2)
-\frac{682 \log (\beta )}{9}
+56 \ln (2) \log (\beta )
+28 \log ^2(\beta )
\Big)
\nonumber\\&
+\frac{928}{25} \zeta_2
-\frac{192}{5} \zeta_2 \ln (2)
-\frac{192}{5} \zeta_2 \log (\beta )
+\beta  \Big(
        -\frac{12113}{54}
        +\frac{79}{2} \zeta_2
        +\frac{2270 \ln (2)}{9}
        -108 \ln ^2(2)
\nonumber\\&        
        +\frac{1822 \log (\beta )}{9}
        -\frac{520}{3} \ln (2) \log (\beta )
        -\frac{212}{3} \log ^2(\beta )
\Big)
+\beta ^2 \Big(
        -\frac{456}{25} \zeta_2
        +\frac{1018}{35} \zeta_2 \ln (2)
\nonumber\\&        
        +\frac{373}{35} \zeta_2 \log (\beta )
\Big)
\Bigg)
\Bigg]
\Bigg\} 
+ C_F T_F n_l \Bigg\{
  \Bigg[
\frac{8 \zeta_2}{\beta }
+\frac{52}{9}
-8 \beta  \zeta_2
-\frac{40 \beta ^2}{9}
+ i \pi \Bigg(
\frac{1}{9 \beta } \Big( 2 (-25+12 \ln (2)
\nonumber\\&
+12 \log (\beta )) \Big)
-\frac{2}{9} \beta  (-31+12 \ln (2)+12 \log (\beta ))
\Bigg)
\Bigg]
+ \ep  \Bigg[
\frac{1}{\beta } \Big(
\frac{296}{3} \zeta_2
-48 \zeta_2 \ln (2)
\nonumber\\&
-48 \zeta_2 \log (\beta )
\Big)
+\frac{1010}{27}
+12 \zeta_2
+\beta  \Big(
        -\frac{368}{3} \zeta_2
        +48 \zeta_2 \ln (2)
        +48 \zeta_2 \log (\beta )
\Big)
+\beta ^2 \Big(
        -\frac{2228}{81}
\nonumber\\&
        -8 \zeta_2
\Big)
+ i \pi \Bigg(
\frac{1}{\beta }\Big(
-\frac{961}{27}
+2 \zeta_2
+\frac{296 \ln (2)}{9}
-8 \ln ^2(2)
+\frac{296 \log (\beta )}{9}
-16 \ln (2) \log (\beta )
\nonumber\\&
-8 \log ^2(\beta )
\Big)
+\beta  \Big(
        \frac{1405}{27}
        -2 \zeta_2
        -\frac{368 \ln (2)}{9}
        +8 \ln ^2(2)
        -\frac{368 \log (\beta )}{9}
        +16 \ln (2) \log (\beta )
\nonumber\\&
        +8 \log ^2(\beta )
\Big)
\Bigg)
\Bigg]
\Bigg\} 
+ C_F T_F \Bigg\{
  \Bigg[
-\frac{92}{9}
+\frac{32}{5} \zeta_2
+\beta ^2 \Big(
        \frac{248}{27}
        -\frac{36 \zeta_2}{5}
\Big)
\Bigg]
+ \ep  \Bigg[
\frac{82}{135}
+\frac{324}{25} \zeta_2
\nonumber\\&
+\frac{112}{5} \zeta_3
-\frac{192}{5} \zeta_2 \ln (2)
+\beta ^2 \Big(
        \frac{772}{405}
        -\frac{171322 \zeta_2}{11025}
        -\frac{96}{5} \zeta_3
        +\frac{1032}{35} \zeta_2 \ln (2)
        -\frac{24}{7} \zeta_2 \log (\beta )
\Big)
\nonumber\\&
+ i \pi \Bigg(
-\frac{2 \zeta_2}{3 \beta }
+\beta  \Big(
        \frac{16}{15}
        +\frac{2 \zeta_2}{3}
\Big)
\Bigg)
\Bigg]
\Bigg\} \,.
\end{align}
%

\subsubsection{Axial-vector form factor}
In the following, we provide the non-singlet part of the axial-vector form factors in the threshold 
region
\begin{align}
 \tilde{F}_{A,1}^{(1),\ns} &=  C_F \Bigg\{
\frac{1}{\ep}  \Bigg[
\frac{8 \beta ^2}{3}
+ i \pi \Bigg(
-\frac{1}{\beta }-\beta
\Bigg)
\Bigg]
+  \Bigg[
\frac{6 \zeta_2}{\beta }
-4
+6 \beta  \zeta_2
-\frac{22 \beta ^2}{9}
+ i \pi \Bigg(
\frac{1}{\beta } \Big( 2 (-1+\ln (2)
\nonumber\\&
+\log (\beta )) \Big)
+2 \beta  (\ln (2)+\log (\beta ))
\Bigg)
\Bigg]
+ \ep  \Bigg[
\frac{1}{\beta } \Big( 12 \zeta_2
-12 \zeta_2 \ln (2)
-12 \zeta_2 \log (\beta )
\Big)
\nonumber\\&
+\beta  \Big(
        -12 \ln (2) \zeta_2
        -12 \log (\beta ) \zeta_2
\Big)
+\beta ^2 \Big(
        \frac{404}{27}
        +\frac{4 \zeta_2}{3}
\Big)
+ i \pi \Bigg(
\frac{1}{\beta } \Big(
-4
+\frac{3}{2} \zeta_2
+4 \ln (2)
\nonumber\\&
-2 \ln ^2(2)
+4 \log (\beta )
-4 \ln (2) \log (\beta )
-2 \log ^2(\beta )
\Big)
+ \beta  \Big(
        -2
        +\frac{3}{2} \zeta_2
        -2 \ln ^2(2)
\nonumber\\&
        -4 \ln (2) \log (\beta )
        -2 \log ^2(\beta )
\Big)
\Bigg)
\Bigg]
+ \ep^2  \Bigg[
\frac{1}{\beta } \Big( 24 \zeta_2
+3 \zeta_2^2
-24 \zeta_2 \ln (2)
+12 \zeta_2 \ln ^2(2)
\nonumber\\&
-24 \zeta_2 \log (\beta )
+24 \zeta_2 \ln (2) \log (\beta )
+12 \zeta_2 \log ^2(\beta )
\Big)
-16
-2 \zeta_2
+\beta  \Big(
        12 \zeta_2
        +3 \zeta_2^2
\nonumber\\&
        +12 \zeta_2 \ln ^2(2)
        +24 \zeta_2 \ln (2) \log (\beta )
        +12 \zeta_2 \log ^2(\beta )
\Big)
+\beta ^2 \Big(
        -\frac{808}{81}
        -\frac{11 \zeta_2}{9}
        -\frac{8 \zeta_3}{9}
\Big)
\nonumber\\&
+ i \pi \Bigg(
\frac{1}{\beta } \Big(
-8
+3 \zeta_2
+\frac{7}{3} \zeta_3
+8 \ln (2)
-3 \zeta_2 \ln (2)
-4 \ln ^2(2)
+\frac{4 \ln ^3(2)}{3}
+8 \log (\beta )
\nonumber\\&
-3 \zeta_2 \log (\beta )
-8 \ln (2) \log (\beta )
+4 \ln ^2(2) \log (\beta )
-4 \log ^2(\beta )
+4 \ln (2) \log ^2(\beta )
+\frac{4 \log ^3(\beta )}{3}
\Big)
\nonumber\\&
+\beta  \Big(
        -4
        +\frac{7}{3} \zeta_3
        +4 \ln (2)
        -3 \zeta_2 \ln (2)
        +\frac{4 \ln ^3(2)}{3}
        +4 \log (\beta )
        -3 \zeta_2 \log (\beta )
        +4 \ln ^2(2) \log (\beta )
\nonumber\\&
        +4 \ln (2) \log ^2(\beta )
        +\frac{4 \log ^3(\beta )}{3}
\Big)
\Bigg)
\Bigg]
\Bigg\} \,.
\\
 \tilde{F}_{A,1}^{(2),\ns} &=  C_F^2 \Bigg\{
\frac{1}{\ep^2}  \Bigg[
-\frac{3 \zeta_2}{\beta ^2}
-6 \zeta_2
-3 \beta ^2 \zeta_2
+ i \pi \Bigg(
-\frac{8 \beta }{3}
\Bigg)
\Bigg]
+ \frac{1}{\ep}  \Bigg[
\frac{1}{\beta ^2} \Big(
-12 \zeta_2
+12 \zeta_2 \ln (2)
\nonumber\\&
+12 \zeta_2 \log (\beta )
\Big)
-12 \zeta_2
+16 \beta  \zeta_2
+24 \zeta_2 \ln (2)
+24 \zeta_2 \log (\beta )
+\beta ^2 \Big(
        -\frac{32}{3}
        +9 \zeta_2
        +12 \zeta_2 \ln (2)
\nonumber\\&
        +12 \zeta_2 \log (\beta )
\Big)
+ i \pi \Bigg(
-\frac{6 \zeta_2}{\beta ^2}
+\frac{4}{\beta }
-12 \zeta_2
+\frac{2}{9} \beta  (5+24 \ln (2)+24 \log (\beta ))
-6 \beta ^2 \zeta_2
\Bigg)
\Bigg]
\nonumber\\&
+  \Bigg[
\frac{1}{\beta ^2} \Big(
-24 \zeta_2
+15 \zeta_2^2
+48 \zeta_2 \ln (2)
-24 \zeta_2 \ln ^2(2)
+48 \zeta_2 \log (\beta )
-48 \zeta_2 \ln (2) \log (\beta )
\nonumber\\&
-24 \zeta_2 \log ^2(\beta )
\Big)
+\frac{46}{3}
-32 \zeta_2
-\frac{24 \zeta_2}{\beta }
+30 \zeta_2^2
-27 \zeta_3
+84 \zeta_2 \ln (2)
-48 \zeta_2 \ln ^2(2)
+8 \zeta_2 \log (\beta )
\nonumber\\&
-96 \zeta_2 \ln (2) \log (\beta )
-48 \zeta_2 \log ^2(\beta )
+\beta  \Big(
        -\frac{20}{3} \zeta_2
        -32 \zeta_2 \ln (2)
        -32 \zeta_2 \log (\beta )
\Big)
+\beta ^2 \Big(
        \frac{247}{9}
\nonumber\\&
        -\frac{1508}{75} \zeta_2
        +15 \zeta_2^2
        +50 \zeta_3
        -\frac{172}{5} \zeta_2 \ln (2)
        -24 \zeta_2 \ln ^2(2)
        -\frac{4}{5} \zeta_2 \log (\beta )
        -48 \zeta_2 \ln (2) \log (\beta )
\nonumber\\&
        -24 \zeta_2 \log ^2(\beta )
\Big)
+ i \pi \Bigg(
\frac{1}{\beta ^2} \Big(
-24 \zeta_2
+24 \zeta_2 \ln (2)
+24 \zeta_2 \log (\beta )
\Big)
-\frac{1}{\beta } \Big( 8 (-1+\ln (2)
\nonumber\\&
+\log (\beta )) \Big)
-4 \zeta_2
+48 \zeta_2 \ln (2)
+48 \zeta_2 \log (\beta )
+\beta  \Big(
        -\frac{956}{27}
        +\frac{8}{3} \zeta_2
        -\frac{20 \ln (2)}{9}
        -\frac{16}{3} \ln ^2(2)
\nonumber\\&
        -\frac{20 \log (\beta )}{9}
        -\frac{32}{3} \ln (2) \log (\beta )
        -\frac{16}{3} \log ^2(\beta )
\Big)
+\beta ^2 \Big(
        \frac{2}{5} \zeta_2
        +24 \zeta_2 \ln (2)
        +24 \zeta_2 \log (\beta )
\Big)
\Bigg)
\Bigg]
\nonumber\\&
+ \ep  \Bigg[
\frac{1}{\beta ^2}  \Big(
48 \zeta_2
+60 \zeta_2^2
-106 \zeta_2 \zeta_3
+96 \zeta_2 \ln (2)
-60 \zeta_2^2 \ln (2)
-96 \zeta_2 \ln ^2(2)
+32 \zeta_2 \ln ^3(2)
\nonumber\\&
+96 \zeta_2 \log (\beta )
-60 \zeta_2^2 \log (\beta )
-192 \zeta_2 \ln (2) \log (\beta )
+96 \zeta_2 \ln ^2(2) \log (\beta )
-96 \zeta_2 \log ^2(\beta )
\nonumber\\&
+96 \zeta_2 \ln (2) \log ^2(\beta )
+32 \zeta_2 \log ^3(\beta )
\Big)
+\frac{1}{\beta } \Big(
-48 \zeta_2
+48 \zeta_2 \ln (2)
+48 \zeta_2 \log (\beta )
\Big)
-\frac{313}{9}
\nonumber\\&
-\frac{38 c_1}{3}
+\frac{940}{3} \zeta_2
+\frac{282}{5} \zeta_2^2
-\frac{1202}{3} \zeta_3
-212 \zeta_2 \zeta_3
-\frac{424}{3} \zeta_2 \ln (2)
-120 \zeta_2^2 \ln (2)
+32 \zeta_2 \ln ^2(2)
\nonumber\\&
+64 \zeta_2 \ln ^3(2)
-304 \zeta_2 \log (\beta )
-120 \zeta_2^2 \log (\beta )
-32 \zeta_2 \ln (2) \log (\beta )
+192 \zeta_2 \ln ^2(2) \log (\beta )
\nonumber\\&
-16 \zeta_2 \log ^2(\beta )
+192 \zeta_2 \ln (2) \log ^2(\beta )
+64 \zeta_2 \log ^3(\beta )
+\beta  \Big(
        \frac{3964}{9} \zeta_2
        +16 \zeta_2^2
        +\frac{40}{3} \zeta_2 \ln (2)
\nonumber\\&
        +32 \zeta_2 \ln ^2(2)
        +\frac{40}{3} \zeta_2 \log (\beta )
        +64 \zeta_2 \ln (2) \log (\beta )
        +32 \zeta_2 \log ^2(\beta )
\Big)
+\beta ^2 \Big(
        -\frac{35237}{270}
        +\frac{28 c_1}{5}
\nonumber\\&
        -\frac{117812 \zeta_2}{1125}
        +\frac{3639}{25} \zeta_2^2
        +\frac{724}{15} \zeta_3
        -106 \zeta_2 \zeta_3
        -\frac{624}{5} \zeta_2 \ln (2)
        -60 \zeta_2^2 \ln (2)
        +\frac{488}{5} \zeta_2 \ln ^2(2)
\nonumber\\&
        +32 \zeta_2 \ln ^3(2)
        +\frac{3696}{25} \zeta_2 \log (\beta )
        -60 \zeta_2^2 \log (\beta )
        +\frac{16}
        {5} \zeta_2 \ln (2) \log (\beta )
        +96 \zeta_2 \ln ^2(2) \log (\beta )
\nonumber\\&
        +\frac{8}{5} \zeta_2 \log ^2(\beta )
        +96 \zeta_2 \ln (2) \log ^2(\beta )
        +32 \zeta_2 \log ^3(\beta )
\Big)
+ i \pi \Bigg(
\frac{1}{\beta ^2} \Big(
-48 \zeta_2
-18 \zeta_2^2
+96 \zeta_2 \ln (2)
\nonumber\\&
-48 \zeta_2 \ln ^2(2)
+96 \zeta_2 \log (\beta )
-96 \zeta_2 \ln (2) \log (\beta )
-48 \zeta_2 \log ^2(\beta )
\Big)
+\frac{1}{\beta } \Big(
32
-4 \zeta_2
-16 \ln (2)
\nonumber\\&
+8 \ln ^2(2)
-16 \log (\beta )
+16 \ln (2) \log (\beta )
+8 \log ^2(\beta )
\Big)
+152 \zeta_2
-36 \zeta_2^2
+16 \zeta_2 \ln (2)
\nonumber\\&
-96 \zeta_2 \ln ^2(2)
+16 \zeta_2 \log (\beta )
-192 \zeta_2 \ln (2) \log (\beta )
-96 \zeta_2 \log ^2(\beta )
+\beta  \Big(
        -\frac{18635}{81}
        -\frac{10}{9} \zeta_2
\nonumber\\&
        +\frac{64}{9} \zeta_3
        +\frac{5116 \ln (2)}{27}
        -\frac{16}{3} \zeta_2 \ln (2)
        +\frac{20 \ln ^2(2)}{9}
        +\frac{32 \ln ^3(2)}{9}
        +\frac{3964 \log (\beta )}{27}
        -\frac{16}{3} \zeta_2 \log (\beta )
\nonumber\\&
        +\frac{40}{9} \ln (2) \log (\beta )
        +\frac{32}{3} \ln ^2(2) \log (\beta )
        +\frac{20 \log ^2(\beta )}{9}
        +\frac{32}{3} \ln (2) \log ^2(\beta )
        +\frac{32 \log ^3(\beta )}{9}
\Big)
\nonumber\\&
+\beta ^2 \Big(
        -\frac{1848}{25} \zeta_2
        -18 \zeta_2^2
        -\frac{8}{5} \zeta_2 \ln (2)
        -48 \zeta_2 \ln ^2(2)
        -\frac{8}{5} \zeta_2 \log (\beta )
        -96 \zeta_2 \ln (2) \log (\beta )
\nonumber\\&
        -48 \zeta_2 \log ^2(\beta )
\Big)
\Bigg)
\Bigg]
\Bigg\} 
+ C_F C_A \Bigg\{
\frac{1}{\ep^2}  \Bigg[
-\frac{44 \beta ^2}{9}
+ i \pi \Bigg(
\frac{11}{6 \beta }+\frac{11 \beta }{6}
\Bigg)
\Bigg]
+ \frac{1}{\ep}  \Bigg[
-8 \beta  \zeta_2
+\beta ^2 \Big(
        \frac{376}{27}
\nonumber\\&
        +\frac{32 \zeta_2}{3}
\Big)
+ i \pi \Bigg(
-\frac{31}{18 \beta }
+\frac{1}{6} \beta  (-13-32 \ln (2)-16 \log (\beta ))
\Bigg)
\Bigg]
+  \Bigg[
\frac{1}{\beta } \Big(
\frac{194}{3} \zeta_2
-44 \zeta_2 \ln (2)
\nonumber\\&
-44 \zeta_2 \log (\beta )
\Big)
-\frac{202}{9}
+\frac{178}{3} \zeta_2
-18 \zeta_3
-72 \zeta_2 \ln (2)
-16 \zeta_2 \log (\beta )
+\beta  \Big(
        \frac{8}{3} \zeta_2
        +84 \zeta_2 \ln (2)
\nonumber\\&
        +36 \zeta_2 \log (\beta )
\Big)
+\beta ^2 \Big(
        \frac{1159}{405}
        +\frac{6263}{75} \zeta_2
        -\frac{346}{15} \zeta_3
        -\frac{216}{5} \zeta_2 \ln (2)
        -\frac{336}{5} \zeta_2 \log (\beta )
\Big)
\nonumber\\&
+ i \pi \Bigg(
\frac{1}{\beta } \Big(
-\frac{478}{27}
+\frac{194 \ln (2)}{9}
-\frac{22 \ln ^2(2)}{3}
+\frac{194 \log (\beta )}{9}
-\frac{44}{3} \ln (2) \log (\beta )
-\frac{22 \log ^2(\beta )}{3} 
\Big)
\nonumber\\&
+8 \zeta_2
+\beta  \Big(
        -\frac{143}{9}
        -\frac{40}{3} \zeta_2
        -\frac{80 \ln (2)}{9}
        +\frac{74 \ln ^2(2)}{3}
        +\frac{8 \log (\beta )}{9}
        +28 \ln (2) \log (\beta )
        +6 \log ^2(\beta )
\Big)
\nonumber\\&
+\frac{168}{5} \beta ^2 \zeta_2
\Bigg)
\Bigg]
+ \ep  \Bigg[
\frac{1}{\beta } \Big(
\frac{2704}{9} \zeta_2
+99 \zeta_2^2
-\frac{1040}{3} \zeta_2 \ln (2)
+132 \zeta_2 \ln ^2(2)
-\frac{1040}{3} \zeta_2 \log (\beta )
\nonumber\\&
+264 \zeta_2 \ln (2) \log (\beta )
+132 \zeta_2 \log ^2(\beta )
\Big)
-\frac{4103}{27}
+\frac{28 c_1}{3}
+\frac{2294}{9} \zeta_2
-\frac{1404}{5} \zeta_2^2
+73 \zeta_3
\nonumber\\&
-\frac{940}{3} \zeta_2 \ln (2)
-64 \zeta_2 \ln ^2(2)
-144 \zeta_2 \log (\beta )
+64 \zeta_2 \ln (2) \log (\beta )
+32 \zeta_2 \log ^2(\beta )
+\beta  \Big(
        \frac{1702}{9} \zeta_2
\nonumber\\&
        +59 \zeta_2^2
        +\frac{988}{3} \zeta_2 \ln (2)
        -508 \zeta_2 \ln ^2(2)
        +\frac{460}{3} \zeta_2 \log (\beta )
        -632 \zeta_2 \ln (2) \log (\beta )
        -172 \zeta_2 \log ^2(\beta )
\Big)
\nonumber\\&
+\beta ^2 \Big(
        \frac{361877}{1350}
        -4 c_1
        +\frac{2123531 \zeta_2}{3375}
        -\frac{17228}{75} \zeta_2^2
        -\frac{8419}{150} \zeta_3
        -\frac{6966}{25} \zeta_2 \ln (2)
        +\frac{576}{5} \zeta_2 \ln ^2(2)
\nonumber\\&
        -\frac{9536}{25} \zeta_2 \log (\beta )
        +\frac{1344}{5} \zeta_2 \ln (2) \log (\beta )
        +\frac{672}{5} \zeta_2 \log ^2(\beta )
\Big)
+ i \pi \Bigg(
\frac{1}{\beta } \Big(
-\frac{6338}{81}
+\frac{97}{6} \zeta_2
+\frac{22}{3} \zeta_3
\nonumber\\&
+\frac{2704 \ln (2)}{27}
-11 \zeta_2 \ln (2)
-\frac{520}{9} \ln ^2(2)
+\frac{44 \ln ^3(2)}{3}
+\frac{2704 \log (\beta )}{27}
-11 \zeta_2 \log (\beta )
\nonumber\\&
-\frac{1040}{9} \ln (2) \log (\beta )
+44 \ln ^2(2) \log (\beta )
-\frac{520}{9} \log ^2(\beta )
+44 \ln (2) \log ^2(\beta )
+\frac{44 \log ^3(\beta )}{3}
\Big)
\nonumber\\&
+72 \zeta_2
-32 \zeta_2 \ln (2)
-32 \zeta_2 \log (\beta )
+\beta  \Big(
        -\frac{9149}{54}
        -\frac{727}{18} \zeta_2
        -\frac{194}{3} \zeta_3
        +\frac{1238 \ln (2)}{27}
        +\frac{335}{3} \zeta_2 \ln (2)
\nonumber\\&
        +94 \ln ^2(2)
        -\frac{764}{9} \ln ^3(2)
        +\frac{1702 \log (\beta )}{27}
        +77 \zeta_2 \log (\beta )
        +\frac{988}{9} \ln (2) \log (\beta )
        -\frac{508}{3} \ln ^2(2) \log (\beta )
\nonumber\\&
        +\frac{230 \log ^2(\beta )}{9}
        -\frac{316}{3} \ln (2) \log ^2(\beta )
        -\frac{172}{9} \log ^3(\beta )
\Big)
+\beta ^2 \Big(
        \frac{4768}{25} \zeta_2
        -\frac{672}{5} \zeta_2 \ln (2)
\nonumber\\&
        -\frac{672}{5} \zeta_2 \log (\beta )
\Big)
\Bigg)
\Bigg]
\Bigg\} 
+ C_F T_F n_l \Bigg\{
\frac{1}{\ep^2}  \Bigg[
\frac{16 \beta ^2}{9}
+ i \pi \Bigg(
-\frac{2}{3 \beta }-\frac{2 \beta }{3}
\Bigg)
\Bigg]
+ \frac{1}{\ep}  \Bigg[
-\frac{80 \beta ^2}{27}
\nonumber\\&
+ i \pi \Bigg(
\frac{10}{9 \beta }+\frac{10 \beta }{9}
\Bigg)
\Bigg]
+  \Bigg[
\frac{1}{\beta } \Big(
-\frac{88}{3} \zeta_2
+16 \zeta_2 \ln (2)
+16 \zeta_2 \log (\beta )
\Big)
+\frac{56}{9}
+\beta  \Big(
        -\frac{40}{3} \zeta_2
\nonumber\\&
        +16 \zeta_2 \ln (2)
        +16 \zeta_2 \log (\beta )
\Big)
+\beta ^2 \Big(
        -\frac{940}{81}
        -\frac{64 \zeta_2}{9}
\Big)
+ i \pi \Bigg(
\frac{1}{\beta } \Big(
\frac{248}{27}
-\frac{88 \ln (2)}{9}
+\frac{8 \ln ^2(2)}{3}
\nonumber\\&
-\frac{88 \log (\beta )}{9}
+\frac{16}{3} \ln (2) \log (\beta )
+\frac{8 \log ^2(\beta )}{3}
\Big)
+\beta  \Big(
        \frac{38}{27}
-\frac{40 \ln (2)}{9}
+\frac{8 \ln ^2(2)}{3}
-\frac{40 \log (\beta )}{9}
\nonumber\\&
+\frac{16}{3} \ln (2) \log (\beta )
+\frac{8 \log ^2(\beta )}{3}
\Big)
\Bigg)
\Bigg]
+ \ep  \Bigg[
\frac{1}{\beta } \Big(
-\frac{1280}{9} \zeta_2
-36 \zeta_2^2
+\frac{448}{3} \zeta_2 \ln (2)
-48 \zeta_2 \ln ^2(2)
\nonumber\\&
+\frac{448}{3} \zeta_2 \log (\beta )
-96 \zeta_2 \ln (2) \log (\beta )
-48 \zeta_2 \log ^2(\beta )
\Big)
+\frac{1228}{27}
+24 \zeta_2
+\beta  \Big(
        -\frac{296}{9} \zeta_2
        -36 \zeta_2^2
\nonumber\\&
        +\frac{160}{3} \zeta_2 \ln (2)
        -48 \zeta_2 \ln ^2(2)
        +\frac{160}{3} \zeta_2 \log (\beta )
        -96 \zeta_2 \ln (2) \log (\beta )
        -48 \zeta_2 \log ^2(\beta )
\Big)
+\beta ^2 \Big(
        \frac{22}{9}
\nonumber\\&
        -\frac{4 \zeta_2}{27}
        -\frac{128 \zeta_3}{9}
\Big)
+ i \pi \Bigg(
\frac{1}{\beta } \Big(
\frac{3160}{81}
-\frac{22}{3} \zeta_2
-\frac{8}{3} \zeta_3
-\frac{1280 \ln (2)}{27}
+4 \zeta_2 \ln (2)
+\frac{224 \ln ^2(2)}{9}
\nonumber\\&
-\frac{16}{3} \ln ^3(2)
-\frac{1280 \log (\beta )}{27}
+4 \zeta_2 \log (\beta )
+\frac{448}{9} \ln (2) \log (\beta )
-16 \ln ^2(2) \log (\beta )
+\frac{224 \log ^2(\beta )}{9}
\nonumber\\&
-16 \ln (2) \log ^2(\beta )
-\frac{16}{3} \log ^3(\beta )
\Big)
+
\beta  \Big(
        \frac{733}{81}
        -\frac{10}{3} \zeta_2
        -\frac{8}{3} \zeta_3
        -\frac{296 \ln (2)}{27}
        +4 \zeta_2 \ln (2)
        +\frac{80 \ln ^2(2)}{9}
\nonumber\\&
        -\frac{16}{3} \ln ^3(2)
        -\frac{296 \log (\beta )}{27}
        +4 \zeta_2 \log (\beta )
        +\frac{160}{9} \ln (2) \log (\beta )
        -16 \ln ^2(2) \log (\beta )
        +\frac{80 \log ^2(\beta )}{9}
\nonumber\\&
        -16 \ln (2) \log ^2(\beta )
        -\frac{16}{3} \log ^3(\beta )
\Big)
\Bigg)
\Bigg]
\Bigg\} 
+ C_F T_F \Bigg\{
%
%
%
  \Bigg[
\frac{320}{9}
-\frac{64}{3} \zeta_2
+\beta ^2 \Big(
        -\frac{388}{81}
        +\frac{16 \zeta_2}{5}
\Big)
\nonumber\\&
+ i \pi \Bigg(
\frac{2 \zeta_2}{3 \beta }
+\frac{2}{3} \beta  \zeta_2
\Bigg)
\Bigg]
+ \ep  \Bigg[
-\frac{4 \zeta_2^2}{\beta }
+\frac{2944}{27}
-\frac{884}{9} \zeta_2
-\frac{224}{3} \zeta_3
+128 \zeta_2 \ln (2)
-4 \beta  \zeta_2^2
\nonumber\\&
+\beta ^2 \Big(
        -\frac{18106}{1215}
        +\frac{1256}{75} \zeta_2
        +\frac{2512}{135} \zeta_3
        -\frac{448}{15} \zeta_2 \ln (2)
\Big)
+ i \pi \Bigg(
\frac{1}{\beta } \Big(
\frac{4}{3} \zeta_2
-\frac{4}{9} \zeta_3
-\frac{4}{3} \zeta_2 \ln (2)
\nonumber\\&
-\frac{4}{3} \zeta_2 \log (\beta )
\Big)
+\beta  \Big(
        -\frac{4}{9} \zeta_3
        -\frac{4}{3} \zeta_2 \ln (2)
        -\frac{4}{3} \zeta_2 \log (\beta )
\Big)
\Bigg)
\Bigg]
\Bigg\} \,.
\end{align}
The unrenormalized singlet part is given by
\begin{align}
\tilde{\hat{F}}_{A,1}^{(2),\sing} &= 
C_F T_F \Bigg\{
-\frac{6}{\ep} 
%
-\frac{6}{\beta ^2}
-\frac{53}{3}
-\frac{76}{3} \zeta_2
+64 \zeta_2 \ln (2)
+\beta ^2 \Big(
        \frac{32}{3}
        +\frac{44}{15} \zeta_2
        -\frac{256}{5} \zeta_2 \ln (2)
\Big)
%
%
%
\nonumber\\&
+ \ep \Bigg[
\frac{1}{\beta ^2} \Big(
-15
+20 \zeta_2
-42 \zeta_3
+16 \ln (2)
+48 \zeta_2 \ln (2)
\Big)
-\frac{48 \zeta_2}{\beta }
-\frac{1169}{18}
-\frac{32 c_1}{3}
-\frac{1424}{9} \zeta_2
\nonumber\\&
+\frac{944}{5} \zeta_2^2
-\frac{98}{3} \zeta_3
+\frac{32 \ln (2)}{3}
-\frac{88}{3} \zeta_2 \ln (2)
+64 \zeta_2 \ln ^2(2)
+48 \beta  \zeta_2
+\beta ^2 \Big(
        \frac{2528}{45}
        +\frac{128 c_1}{15}
\nonumber\\&
        +\frac{15374}{225} \zeta_2
        -\frac{3776}{25} \zeta_2^2
        -\frac{14}{15} \zeta_3
        -\frac{176 \ln (2)}{15}
        -\frac{4936}{75} \zeta_2 \ln (2)
        -\frac{256}{5} \zeta_2 \ln ^2(2)
\Big)
\nonumber\\&
+ i \pi \Bigg(
\frac{1}{\beta ^2} \Big(
-8
+12 \zeta_2
\Big)
-\frac{1}{\beta } \Big( 16 (-1+\ln (2)) \Big)
-\frac{16}{3}
-16 \zeta_2
+\beta  \Big(        
-\frac{64}{3}
+16 \ln (2)
\Big)
\nonumber\\&
+\beta ^2 \Big(
        \frac{88}{15}
        +\frac{16 \zeta_2}{5}
\Big)
\Bigg) 
\Bigg]
\Bigg\},
\end{align}
%
%
\begin{align}
 \tilde{F}_{A,2}^{(1),{\rm ns}} &=  C_F \Bigg\{
%
%
%
%
-2-\frac{4 \beta ^2}{3}
+ i \pi \Bigg(
\frac{2}{\beta }-\beta
\Bigg)
%
+ \ep  \Bigg[
-\frac{12 \zeta_2}{\beta }
+4
+6 \beta  \zeta_2
-\frac{28 \beta ^2}{9}
+ i \pi \Bigg(
-\frac{1}{\beta } \Big( 4 (-1
\nonumber\\&
+\ln (2)+\log (\beta )) \Big)
+2 \beta  (-2+\ln (2)+\log (\beta ))
\Bigg)
\Bigg]
+ \ep^2  \Bigg[
\frac{1}{\beta } \Big(
-24 \zeta_2
+24 \zeta_2 \ln (2)
\nonumber\\&
+24 \zeta_2 \log (\beta )
\Big)
-8
-\zeta_2
+\beta  \Big(
        24 \zeta_2
        -12 \zeta_2 \ln (2)
        -12 \zeta_2 \log (\beta )
\Big)
+\beta ^2 \Big(
        -\frac{160}{27}
        -\frac{2 \zeta_2}{3}
\Big)
\nonumber\\&
+ i \pi \Bigg(
\frac{1}{\beta } \Big(
8
-3 \zeta_2
-8 \ln (2)
+4 \ln ^2(2)
-8 \log (\beta )
+8 \ln (2) \log (\beta )
+4 \log ^2(\beta )
\Big)
\nonumber\\&
+\beta  \Big(
        -8
        +\frac{3}{2} \zeta_2
        +8 \ln (2)
        -2 \ln ^2(2)
        +8 \log (\beta )
        -4 \ln (2) \log (\beta )
        -2 \log ^2(\beta )
\Big)
\Bigg)
\Bigg]
\Bigg\} \,.
\\
 \tilde{F}_{A,2}^{\ns{(2)}} &=  C_F^2 \Bigg\{
\frac{1}{\ep}  \Bigg[
\frac{12 \zeta_2}{\beta ^2}
+6 \zeta_2
+\beta ^2 \Big(
        -\frac{16}{3}
        -12 \zeta_2
\Big)
+ i \pi \Bigg(
\frac{2}{\beta }+\frac{26 \beta }{3}
\Bigg)
\Bigg]
+  \Bigg[
\frac{1}{\beta ^2} \Big(
24 \zeta_2
-48 \zeta_2 \ln (2)
\nonumber\\&
-48 \zeta_2 \log (\beta )
\Big)
-\frac{12 \zeta_2}{\beta }
+\frac{41}{3}
-18 \zeta_2
-45 \zeta_3
-60 \zeta_2 \ln (2)
-80 \zeta_2 \log (\beta )
+20 \beta  \zeta_2
+\beta ^2 \Big(
        \frac{218}{9}
\nonumber\\&
        -\frac{3292}{75} \zeta_2
        -7 \zeta_3
        -\frac{268}{5} \zeta_2 \ln (2)
        +\frac{264}{5} \zeta_2 \log (\beta )
\Big)
+ i \pi \Bigg(
\frac{24 \zeta_2}{\beta ^2}
-\frac{1}{\beta } \Big( 4 (3+\ln (2)+\log (\beta )) \Big)
\nonumber\\&
+40 \zeta_2
+\frac{2}{9} \beta  (-265+126 \ln (2)+30 \log (\beta ))
-\frac{132}{5} \beta ^2 \zeta_2
\Bigg)
\Bigg]
+ \ep  \Bigg[
\frac{1}{\beta ^2} \Big(
-48 \zeta_2
-60 \zeta_2^2
\nonumber\\&
-96 \zeta_2 \ln (2)
+96 \zeta_2 \ln ^2(2)
-96 \zeta_2 \log (\beta )
+192 \zeta_2 \ln (2) \log (\beta )
+96 \zeta_2 \log ^2(\beta )
\Big)
+\frac{1}{\beta } \Big( 
72 \zeta_2
\nonumber\\&
+24 \zeta_2 \ln (2)
+24 \zeta_2 \log (\beta )
\Big)
-\frac{823}{18}
-\frac{10 c_1}{3}
+\frac{776}{3} \zeta_2
-\frac{1572}{5} \zeta_2^2
-\frac{673}{3} \zeta_3
-\frac{1052}{3} \zeta_2 \ln (2)
\nonumber\\&
+112 \zeta_2 \ln ^2(2)
-128 \zeta_2 \log (\beta )
+320 \zeta_2 \ln (2) \log (\beta )
+160 \zeta_2 \log ^2(\beta )
+\beta  \Big(
        \frac{3608}{3} \zeta_2
\nonumber\\&        
        -968 \zeta_2 \ln (2)
        -584 \zeta_2 \log (\beta )
\Big)
+\beta ^2 \Big(
        \frac{6661}{135}
        +\frac{266 c_1}{15}
        -\frac{193438 \zeta_2}{1125}
        -\frac{6534}{25} \zeta_2^2
        +\frac{3811}{15} \zeta_3
\nonumber\\&        
        +\frac{9292}{15} \zeta_2 \ln (2)
        -\frac{1248}{5} \zeta_2 \ln ^2(2)
        +\frac{9704}{25} \zeta_2 \log (\beta )
        -\frac{1056}{5} \zeta_2 \ln (2) \log (\beta )
        -\frac{528}{5} \zeta_2 \log ^2(\beta )
\Big)
\nonumber\\&
+ i \pi \Bigg(
\frac{1}{\beta ^2} \Big(
48 \zeta_2
-96 \zeta_2 \ln (2)
-96 \zeta_2 \log (\beta )
\Big)
+\frac{1}{\beta } \Big(
-8
-2 \zeta_2
+24 \ln (2)
+4 \ln ^2(2)
+24 \log (\beta )
\nonumber\\&
+8 \ln (2) \log (\beta )
+4 \log ^2(\beta )
\Big)
+64 \zeta_2
-160 \zeta_2 \ln (2)
-160 \zeta_2 \log (\beta )
+\beta  \Big(
        -\frac{1031}{3}
        +\frac{302}{3} \zeta_2
\nonumber\\&        
        +\frac{5080 \ln (2)}{9}
        -\frac{740}{3} \ln ^2(2)
        +\frac{3608 \log (\beta )}{9}
        -\frac{968}{3} \ln (2) \log (\beta )
        -\frac{292}{3} \log ^2(\beta )
\Big)
\nonumber\\&
+\beta ^2 \Big(
        -\frac{4852}{25} \zeta_2
        +\frac{528}{5} \zeta_2 \ln (2)
        +\frac{528}{5} \zeta_2 \log (\beta )
\Big)
\Bigg)
\Bigg]
\Bigg\} 
+ C_F C_A \Bigg\{
\Bigg[
-\frac{44 \zeta_2}{\beta }
+\frac{151}{9}
+\frac{104}{3} \zeta_2
\nonumber\\&
-30 \zeta_3
-72 \zeta_2 \ln (2)
-32 \zeta_2 \log (\beta )
+84 \beta  \zeta_2
+\beta ^2 \Big(
        -\frac{2666}{45}
        -\frac{5138}{75} \zeta_2
        +\frac{252}{5} \zeta_3
        +\frac{656}{5} \zeta_2 \ln (2)
\nonumber\\&        
        +\frac{256}{5} \zeta_2 \log (\beta )
\Big)
+ i \pi \Bigg(
-\frac{1}{9 \beta } \Big( 4 (-32+33 \ln (2)+33 \log (\beta )) \Big)
+16 \zeta_2
+\frac{14}{9} \beta  (-23+18 \ln (2)
\nonumber\\&
+18 \log (\beta ))
-\frac{128}{5} \beta ^2 \zeta_2
\Bigg)
\Bigg]
+ \ep  \Bigg[
\frac{1}{\beta } \Big(
-\frac{776}{3} \zeta_2
+264 \zeta_2 \ln (2)
+264 \zeta_2 \log (\beta )
\Big)
-\frac{1433}{54}
+\frac{20 c_1}{3}
\nonumber\\&
+\frac{415}{9} \zeta_2
-\frac{1644}{5} \zeta_2^2
+46 \zeta_3
-\frac{104}{3} \zeta_2 \ln (2)
-32 \zeta_2 \ln ^2(2)
-96 \zeta_2 \log (\beta )
+128 \zeta_2 \ln (2) \log (\beta )
\nonumber\\&
+64 \zeta_2 \log ^2(\beta )
+\beta  \Big(
        \frac{1990}{3} \zeta_2
        -504 \zeta_2 \ln (2)
        -504 \zeta_2 \log (\beta )
\Big)
+\beta ^2 \Big(
        -\frac{145603}{2025}
        -\frac{40 c_1}{3}
\nonumber\\&        
        -\frac{202052 \zeta_2}{1125}
        +\frac{14536}{25} \zeta_2^2
        -\frac{2001}{25} \zeta_3
        +\frac{9148}{75} \zeta_2 \ln (2)
        +\frac{384}{5} \zeta_2 \ln ^2(2)
        +\frac{4736}{25} \zeta_2 \log (\beta )
\nonumber\\&        
        -\frac{1024}{5} \zeta_2 \ln (2) \log (\beta )
        -\frac{512}{5} \zeta_2 \log ^2(\beta )
\Big)
+ i \pi \Bigg(
\frac{1}{\beta } \Big(
\frac{1924}{27}
-11 \zeta_2
-\frac{776 \ln (2)}{9}
+44 \ln ^2(2)
\nonumber\\&
-\frac{776 \log (\beta )}{9}
+88 \ln (2) \log (\beta )
+44 \log ^2(\beta )
\Big)
+48 \zeta_2
-64 \zeta_2 \ln (2)
-64 \zeta_2 \log (\beta )
\nonumber\\&
+\beta  \Big(
        -\frac{10195}{54}
        +\frac{157}{6} \zeta_2
        +\frac{2182 \ln (2)}{9}
        -84 \ln ^2(2)
        +\frac{1990 \log (\beta )}{9}
        -168 \ln (2) \log (\beta )
\nonumber\\&        
        -84 \log ^2(\beta )
\Big)
+\beta ^2 \Big(
        -\frac{2368}{25} \zeta_2
        +\frac{512}{5} \zeta_2 \ln (2)
        +\frac{512}{5} \zeta_2 \log (\beta )
\Big)
\Bigg)
\Bigg]
\Bigg\} 
\nonumber\\&
+ C_F T_F n_l \Bigg\{
 \Bigg[
\frac{16 \zeta_2}{\beta }
-\frac{44}{9}
-8 \beta  \zeta_2
+8 \beta ^2
+ i \pi \Bigg(
\frac{1}{9 \beta } \Big( 16 (-4+3 \ln (2)+3 \log (\beta )) \Big)
-\frac{2}{9} \beta  (-25
\nonumber\\&
+12 \ln (2)+12 \log (\beta ))
\Bigg)
\Bigg]
+ \ep  \Bigg[
\frac{1}{\beta } \Big(
\frac{352}{3} \zeta_2
-96 \zeta_2 \ln (2)
-96 \zeta_2 \log (\beta )
\Big)
+\frac{338}{27}
+12 \zeta_2
\nonumber\\&
+\beta  \Big(
        -\frac{296}{3} \zeta_2
        +48 \zeta_2 \ln (2)
        +48 \zeta_2 \log (\beta )
\Big)
+\beta ^2 \Big(
        \frac{3044}{81}
        +8 \zeta_2
\Big)
+ i \pi \Bigg(
\frac{1}{\beta } \Big(
-\frac{944}{27}
+4 \zeta_2
\nonumber\\&
+\frac{352 \ln (2)}{9}
-16 \ln ^2(2)
+\frac{352 \log (\beta )}{9}
-32 \ln (2) \log (\beta )
-16 \log ^2(\beta )
\Big)
+\beta  \Big(
        \frac{889}{27}
        -2 \zeta_2
\nonumber\\&        
        -\frac{296 \ln (2)}{9}
        +8 \ln ^2(2)
        -\frac{296 \log (\beta )}{9}
        +16 \ln (2) \log (\beta )
        +8 \log ^2(\beta )
\Big)
\Bigg)
\Bigg]
\Bigg\} 
\nonumber\\&
+ C_F T_F \Bigg\{
 \Bigg[
\frac{196}{9}
-\frac{32}{3} \zeta_2
+\beta ^2 \Big(
        -\frac{632}{27}
        +\frac{64 \zeta_2}{5}
\Big)
+ i \pi \Bigg(
\frac{16 \beta }{15}
\Bigg)
\Bigg]
+ \ep  \Bigg[
\frac{1322}{27}
-\frac{412}{9} \zeta_2
-\frac{112}{3} \zeta_3
\nonumber\\&
+64 \zeta_2 \ln (2)
-\frac{32}{5} \beta  \zeta_2
+\beta ^2 \Big(
        -\frac{13348}{405}
        +\frac{9232}{225} \zeta_2
        +\frac{224}{5} \zeta_3
        -\frac{384}{5} \zeta_2 \ln (2)
\Big)
+ i \pi \Bigg(
-\frac{4 \zeta_2}{3 \beta }
\nonumber\\&
+\beta  \Big(
        \frac{32}{15}
        +\frac{2}{3} \zeta_2
        -\frac{32 \ln (2)}{15}
        -\frac{32 \log (\beta )}{15}
\Big)
\Bigg)
\Bigg]
\Bigg\} \,.
\end{align}
The unrenormalized singlet contribution is
\begin{align}
\tilde{\hat{F}}_{A,2}^{(2),\sing} &= 
C_F T_F \Bigg\{
\Bigg[
\frac{6}{\beta ^2}
+\frac{8}{3}
+\frac{136}{3} \zeta_2
-42 \zeta_3
+16 \ln (2)
-16 \zeta_2 \ln (2)
-48 \beta  \zeta_2
+\beta ^2 \Big(
        -30
        -\frac{904}{15} \zeta_2
\nonumber\\&
        +70 \zeta_3
        +\frac{16 \ln (2)}{3}
        +\frac{176}{5} \zeta_2 \ln (2)
\Big)
+ i \pi \Bigg(
-8
+12 \zeta_2
+\beta  (16-16 \ln (2))
+\beta ^2 \Big(
        -\frac{8}{3}
\nonumber\\&
        -20 \zeta_2
\Big)
\Bigg)
\Bigg]
+ \ep \Bigg[
\frac{1}{\beta ^2} \Big(
15
-20 \zeta_2
+42 \zeta_3
-16 \ln (2)
-48 \zeta_2 \ln (2)
\Big)
+\frac{48 \zeta_2}{\beta }
+\frac{364}{9}
-4 c_1
\nonumber\\&
+\frac{2684}{9} \zeta_2
-\frac{976}{5} \zeta_2^2
+\frac{182}{3} \zeta_3
+\frac{112 \ln (2)}{3}
-\frac{416}{3} \zeta_2 \ln (2)
-56 \zeta_3 \ln (2)
-16 \ln ^2(2)
\nonumber\\&
+200 \zeta_2 \ln ^2(2)
+\beta  \Big(
        -240 \zeta_2
        +288 \zeta_2 \ln (2)
        +96 \zeta_2 \log (\beta )
\Big)
+\beta ^2 \Big(
        -\frac{427}{5}
        +\frac{236 c_1}{45}
\nonumber\\&
        -\frac{101524}{225} \zeta_2
        +\frac{26288}{75} \zeta_2^2
        -\frac{1666}{15} \zeta_3
        -\frac{3392 \ln (2)}{45}
        +\frac{33536}{75} \zeta_2 \ln (2)
        +\frac{280}{3} \zeta_3 \ln (2)
        -\frac{16}{3} \ln ^2(2)
\nonumber\\&
        -\frac{1624}{5} \zeta_2 \ln ^2(2)
\Big)
+ i \pi \Bigg(
\frac{1}{\beta ^2} \Big(
8
-12 \zeta_2
\Big)
+\frac{1}{\beta } \Big( 16 (-1+\ln (2)) \Big)
-\frac{56}{3}
+28 \zeta_2
+28 \zeta_3
+16 \ln (2)
\nonumber\\&
-24 \zeta_2 \ln (2)
+\beta  \Big(
        \frac{160}{3}
        -80 \zeta_2
        -80 \ln (2)
        +48 \ln ^2(2)
        -32 \log (\beta )
        +32 \ln (2) \log (\beta )
\Big)
\nonumber\\&
+\beta ^2 \Big(
        \frac{1696}{45}
        -\frac{428}{15} \zeta_2
        -\frac{140}{3} \zeta_3
        +\frac{16 \ln (2)}{3}
        +40 \zeta_2 \ln (2)
\Big)
\Bigg) 
\Bigg]
\Bigg\} \,.
\end{align}

\subsubsection{Scalar form factor}
The scalar form factor is in this limit given by
\begin{align}
\tilde{F}_{S}^{(1)} &=  C_F \Bigg\{
\frac{1}{\ep}  \Bigg[
\frac{8 \beta ^2}{3}
+ i \pi \Bigg(
-\frac{1}{\beta }-\beta
\Bigg)
\Bigg]
+  \Bigg[
\frac{6 \zeta_2}{\beta }
-2
+6 \beta  \zeta_2
-\frac{76 \beta ^2}{9}
+ i \pi \Bigg(
\frac{1}{\beta } \Big( 2 (-1+\ln (2)
\nonumber\\&
+\log (\beta )) \Big)
+\beta  (3+2 \ln (2)+2 \log (\beta ))
\Bigg)
\Bigg]
+ \ep  \Bigg[
\frac{1}{\beta } \Big(
12 \zeta_2
-12 \zeta_2 \ln (2)
-12 \zeta_2 \log (\beta )
\Big)
+4
\nonumber\\&
+\beta  \Big(
        -18 \zeta_2
        -12 \zeta_2 \ln (2)
        -12 \zeta_2 \log (\beta )
\Big)
+\beta ^2 \Big(
        \frac{404}{27}
        +\frac{4 \zeta_2}{3}
\Big)
+ i \pi \Bigg(
\frac{1}{\beta } \Big(
-4
+\frac{3}{2} \zeta_2
+4 \ln (2)
\nonumber\\&
-2 \ln ^2(2)
+4 \log (\beta )
-4 \ln (2) \log (\beta )
-2 \log ^2(\beta )
\Big)
+\beta  \Big(
        4
        +\frac{3}{2} \zeta_2
        -6 \ln (2)
        -2 \ln ^2(2)
\nonumber\\&        
        -6 \log (\beta )
        -4 \ln (2) \log (\beta )
        -2 \log ^2(\beta )
\Big)
\Bigg)
\Bigg]
+ \ep^2  \Bigg[
\frac{1}{\beta } \Big(
24 \zeta_2
+3 \zeta_2^2
-24 \zeta_2 \ln (2)
+12 \zeta_2 \ln ^2(2)
\nonumber\\&
-24 \zeta_2 \log (\beta )
+24 \zeta_2 \ln (2) \log (\beta )
+12 \zeta_2 \log ^2(\beta )
\Big)
-8
-\zeta_2
+\beta  \Big(
        -24 \zeta_2
        +3 \zeta_2^2
        +36 \zeta_2 \ln (2)
\nonumber\\&        
        +12 \zeta_2 \ln ^2(2)
        +36 \zeta_2 \log (\beta )
        +24 \zeta_2 \ln (2) \log (\beta )
        +12 \zeta_2 \log ^2(\beta )
\Big)
+\beta ^2 \Big(
        -\frac{2752}{81}
        -\frac{38 \zeta_2}{9}
        -\frac{8 \zeta_3}{9}
\Big)
\nonumber\\&
+ i \pi \Bigg(
\frac{1}{\beta } \Big(
-8
+3 \zeta_2
+\frac{7}{3} \zeta_3
+8 \ln (2)
-3 \zeta_2 \ln (2)
-4 \ln ^2(2)
+\frac{4 \ln ^3(2)}{3}
+8 \log (\beta )
-3 \zeta_2 \log (\beta )
\nonumber\\&
-8 \ln (2) \log (\beta )
+4 \ln ^2(2) \log (\beta )
-4 \log ^2(\beta )
+4 \ln (2) \log ^2(\beta )
+\frac{4 \log ^3(\beta )}{3}
\Big)
+\beta  \Big(
        8
        -\frac{9}{2} \zeta_2
\nonumber\\&        
        +\frac{7}{3} \zeta_3
        -8 \ln (2)
        -3 \zeta_2 \ln (2)
        +6 \ln ^2(2)
        +\frac{4 \ln ^3(2)}{3}
        -8 \log (\beta )
        -3 \zeta_2 \log (\beta )
        +12 \ln (2) \log (\beta )
\nonumber\\&        
        +4 \ln ^2(2) \log (\beta )
        +6 \log ^2(\beta )
        +4 \ln (2) \log ^2(\beta )
        +\frac{4 \log ^3(\beta )}{3}
\Big)
\Bigg)
\Bigg]
\Bigg\} \,.
\\
\tilde{F}_{S}^{(2)} &=  C_F^2 \Bigg\{
\frac{1}{\ep^2}  \Bigg[
-\frac{3 \zeta_2}{\beta ^2}
-6 \zeta_2
-3 \beta ^2 \zeta_2
+ i \pi \Bigg(
-\frac{8 \beta }{3}
\Bigg)
\Bigg]
+ \frac{1}{\ep}  \Bigg[
\frac{1}{\beta ^2} \Big(
-12 \zeta_2
+12 \zeta_2 \ln (2)
\nonumber\\&
+12 \zeta_2 \log (\beta )
\Big)
+6 \zeta_2
+24 \zeta_2 \ln (2)
+24 \zeta_2 \log (\beta )
+16 \beta  \zeta_2
+\beta ^2 \Big(
        -\frac{16}{3}
        +27 \zeta_2
        +12 \zeta_2 \ln (2)
\nonumber\\&        
        +12 \zeta_2 \log (\beta )
\Big)
+ i \pi \Bigg(
-\frac{6 \zeta_2}{\beta ^2}
+\frac{2}{\beta }
-12 \zeta_2
+\frac{2}{9} \beta  (23+24 \ln (2)+24 \log (\beta ))
-6 \beta ^2 \zeta_2
\Bigg)
\Bigg]
\nonumber\\&
+  \Bigg[
\frac{1}{\beta ^2} \Big(
-24 \zeta_2
+15 \zeta_2^2
+48 \zeta_2 \ln (2)
-24 \zeta_2 \ln ^2(2)
+48 \zeta_2 \log (\beta )
-48 \zeta_2 \ln (2) \log (\beta )
\nonumber\\&
-24 \zeta_2 \log ^2(\beta )
\Big)
-\frac{12 \zeta_2}{\beta }
+5
+98 \zeta_2
+30 \zeta_2^2
-44 \zeta_3
-40 \zeta_2 \ln (2)
-48 \zeta_2 \ln ^2(2)
-88 \zeta_2 \log (\beta )
\nonumber\\&
-96 \zeta_2 \ln (2) \log (\beta )
-48 \zeta_2 \log ^2(\beta )
+\beta  \Big(
        -\frac{92}{3} \zeta_2
        -32 \zeta_2 \ln (2)
        -32 \zeta_2 \log (\beta )
\Big)
+\beta ^2 \Big(
        \frac{1684}{45}
\nonumber\\&        
        -\frac{2446}{25} \zeta_2
        +15 \zeta_2^2
        -\frac{58}{5} \zeta_3
        -\frac{964}{5} \zeta_2 \ln (2)
        -24 \zeta_2 \ln ^2(2)
        -\frac{684}{5} \zeta_2 \log (\beta )
        -48 \zeta_2 \ln (2) \log (\beta )
\nonumber\\&        
        -24 \zeta_2 \log ^2(\beta )
\Big)
+ i \pi \Bigg(
\frac{1}{\beta ^2} \Big(
-24 \zeta_2
+24 \zeta_2 \ln (2)
+24 \zeta_2 \log (\beta )
\Big)
-\frac{4 (\ln (2)+\log (\beta ))}{\beta }
\nonumber\\&
+44 \zeta_2
+48 \zeta_2 \ln (2)
+48 \zeta_2 \log (\beta )
+\beta  \Big(
        -\frac{632}{27}
        +\frac{8}{3} \zeta_2
        -\frac{92 \ln (2)}{9}
        -\frac{16}{3} \ln ^2(2)
        -\frac{92 \log (\beta )}{9}
\nonumber\\&        
        -\frac{32}{3} \ln (2) \log (\beta )
        -\frac{16}{3} \log ^2(\beta )
\Big)
+\beta ^2 \Big(
        \frac{342}{5} \zeta_2
        +24 \zeta_2 \ln (2)
        +24 \zeta_2 \log (\beta )
\Big)
\Bigg)
\Bigg]
\nonumber\\&
+ \ep  \Bigg[
\frac{1}{\beta ^2} \Big(
48 \zeta_2
+60 \zeta_2^2
-106 \zeta_2 \zeta_3
+96 \zeta_2 \ln (2)
-60 \zeta_2^2 \ln (2)
-96 \zeta_2 \ln ^2(2)
+32 \zeta_2 \ln ^3(2)
\nonumber\\&
+96 \zeta_2 \log (\beta )
-60 \zeta_2^2 \log (\beta )
-192 \zeta_2 \ln (2) \log (\beta )
+96 \zeta_2 \ln ^2(2) \log (\beta )
-96 \zeta_2 \log ^2(\beta )
\nonumber\\&
+96 \zeta_2 \ln (2) \log ^2(\beta )
+32 \zeta_2 \log ^3(\beta )
\Big)
+\frac{1}{\beta } \Big(
24 \ln (2) \zeta_2
+24 \log (\beta ) \zeta_2
\Big)
-\frac{451}{6}
-8 c_1
+\frac{2368}{3} \zeta_2
\nonumber\\&
-\frac{1278}{5} \zeta_2^2
-291 \zeta_3
-212 \zeta_2 \zeta_3
-740 \zeta_2 \ln (2)
-120 \zeta_2^2 \ln (2)
+176 \zeta_2 \ln ^2(2)
+64 \zeta_2 \ln ^3(2)
\nonumber\\&
-688 \zeta_2 \log (\beta )
-120 \zeta_2^2 \log (\beta )
+352 \zeta_2 \ln (2) \log (\beta )
+192 \zeta_2 \ln ^2(2) \log (\beta )
+176 \zeta_2 \log ^2(\beta )
\nonumber\\&
+192 \zeta_2 \ln (2) \log ^2(\beta )
+64 \zeta_2 \log ^3(\beta )
+\beta  \Big(
        \frac{3316}{9} \zeta_2
        +16 \zeta_2^2
        +\frac{184}{3} \zeta_2 \ln (2)
        +32 \zeta_2 \ln ^2(2)
\nonumber\\&        
        +\frac{184}{3} \zeta_2 \log (\beta )
        +64 \zeta_2 \ln (2) \log (\beta )
        +32 \zeta_2 \log ^2(\beta )
\Big)
+\beta ^2 \Big(
        -\frac{100966}{675}
        +\frac{28 c_1}{3}
        -\frac{61918}{125} \zeta_2
\nonumber\\&        
        -\frac{10939}{25} \zeta_2^2
        -\frac{15298}{75} \zeta_3
        -106 \zeta_2 \zeta_3
        -\frac{5392}{75} \zeta_2 \ln (2)
        -60 \zeta_2^2 \ln (2)
        +\frac{1144}{5} \zeta_2 \ln ^2(2)
        +32 \zeta_2 \ln ^3(2)
\nonumber\\&        
        +\frac{4776}{25} \zeta_2 \log (\beta )
        -60 \zeta_2^2 \log (\beta )
        +\frac{2736}
        {5} \zeta_2 \ln (2) \log (\beta )
        +96 \zeta_2 \ln ^2(2) \log (\beta )
        +\frac{1368}{5} \zeta_2 \log ^2(\beta )
\nonumber\\&        
        +96 \zeta_2 \ln (2) \log ^2(\beta )
        +32 \zeta_2 \log ^3(\beta )
\Big)
+ i \pi \Bigg(
\frac{1}{\beta ^2} \Big(
-48 \zeta_2
-18 \zeta_2^2
+96 \zeta_2 \ln (2)
-48 \zeta_2 \ln ^2(2)
\nonumber\\&
+96 \zeta_2 \log (\beta )
-96 \zeta_2 \ln (2) \log (\beta )
-48 \zeta_2 \log ^2(\beta )
\Big)
+\frac{1}{\beta } \Big(
8
-2 \zeta_2
+4 \ln ^2(2)
+8 \ln (2) \log (\beta )
\nonumber\\&
+4 \log ^2(\beta )
\Big)
+344 \zeta_2
-36 \zeta_2^2
-176 \zeta_2 \ln (2)
-96 \zeta_2 \ln ^2(2)
-176 \zeta_2 \log (\beta )
-192 \zeta_2 \ln (2) \log (\beta )
\nonumber\\&
-96 \zeta_2 \log ^2(\beta )
+\beta  \Big(
        -\frac{11021}{81}
        -\frac{46}{9} \zeta_2
        +\frac{64}{9} \zeta_3
        +\frac{4468 \ln (2)}{27}
        -\frac{16}{3} \zeta_2 \ln (2)
        +\frac{92 \ln ^2(2)}{9}
\nonumber\\&        
        +\frac{32 \ln ^3(2)}{9}
        +\frac{3316 \log (\beta )}{27}
        -\frac{16}{3} \zeta_2 \log (\beta )
        +\frac{184}{9} \ln (2) \log (\beta )
        +\frac{32}{3} \ln ^2(2) \log (\beta )
        +\frac{92 \log ^2(\beta )}{9}
\nonumber\\&        
        +\frac{32}{3} \ln (2) \log ^2(\beta )
        +\frac{32 \log ^3(\beta )}{9}
\Big)
+\beta ^2 \Big(
        -\frac{2388}{25} \zeta_2
        -18 \zeta_2^2
        -\frac{1368}{5} \zeta_2 \ln (2)
        -48 \zeta_2 \ln ^2(2)
\nonumber\\&        
        -\frac{1368}{5} \zeta_2 \log (\beta )
        -96 \zeta_2 \ln (2) \log (\beta )
        -48 \zeta_2 \log ^2(\beta )
\Big)
\Bigg)
\Bigg]
\Bigg\} 
+ C_F C_A \Bigg\{
\frac{1}{\ep^2}  \Bigg[
-\frac{44 \beta ^2}{9}
+ i \pi \Bigg(
\frac{11}{6 \beta }
\nonumber\\&
+\frac{11 \beta }{6}
\Bigg)
\Bigg]
+ \frac{1}{\ep}  \Bigg[
-8 \beta  \zeta_2
+\beta ^2 \Big(
        \frac{376}{27}
        +\frac{32 \zeta_2}{3}
\Big)
+ i \pi \Bigg(
-\frac{31}{18 \beta }
+\frac{1}{6} \beta  (-13-32 \ln (2)
\nonumber\\&
-16 \log (\beta ))
\Bigg)
\Bigg]
+  \Bigg[
\frac{1}{\beta } \Big(
\frac{194}{3} \zeta_2
-44 \zeta_2 \ln (2)
-44 \zeta_2 \log (\beta )
\Big)
+\frac{49}{9}
+38 \zeta_2
-20 \zeta_3
-64 \zeta_2 \ln (2)
\nonumber\\&
-16 \zeta_2 \log (\beta )
+\beta  \Big(
        -\frac{100}{3} \zeta_2
        +84 \zeta_2 \ln (2)
        +36 \zeta_2 \log (\beta )
\Big)
+\beta ^2 \Big(
        -\frac{16472}{405}
        +\frac{1679}{15} \zeta_2
        -\frac{472}{15} \zeta_3
\nonumber\\&        
        -\frac{336}{5} \zeta_2 \ln (2)
        -80 \zeta_2 \log (\beta )
\Big)
+ i \pi \Bigg(
\frac{1}{\beta } \Big(
-\frac{478}{27}
+\frac{194 \ln (2)}{9}
-\frac{22 \ln ^2(2)}{3}
+\frac{194 \log (\beta )}{9}
\nonumber\\&
-\frac{44}{3} \ln (2) \log (\beta )
-\frac{22 \log ^2(\beta )}{3}
\Big)
+8 \zeta_2
+\beta  \Big(
        \frac{43}{9}
        -\frac{40}{3} \zeta_2
        -\frac{188 \ln (2)}{9}
        +\frac{74 \ln ^2(2)}{3}
\nonumber\\&        
        -\frac{100 \log (\beta )}{9}
        +28 \ln (2) \log (\beta )
        +6 \log ^2(\beta )
\Big)
+40 \beta ^2 \zeta_2
\Bigg)
\Bigg]
+ \ep  \Bigg[
\frac{1}{\beta } \Big( 
\frac{2704}{9} \zeta_2
+99 \zeta_2^2
\nonumber\\&
-\frac{1040}{3} \zeta_2 \ln (2)
+132 \zeta_2 \ln ^2(2)
-\frac{1040}{3} \zeta_2 \log (\beta )
+264 \zeta_2 \ln (2) \log (\beta )
+132 \zeta_2 \log ^2(\beta )
\Big)
\nonumber\\&
-\frac{383}{54}
+8 c_1
+\frac{409}{3} \zeta_2
-264 \zeta_2^2
-\zeta_3
-220 \zeta_2 \ln (2)
-64 \zeta_2 \ln ^2(2)
-144 \zeta_2 \log (\beta )
\nonumber\\&
+64 \zeta_2 \ln (2) \log (\beta )
+32 \zeta_2 \log ^2(\beta )
+\beta  \Big(
        -\frac{692}{9} \zeta_2
        +59 \zeta_2^2
        +\frac{1636}{3} \zeta_2 \ln (2)
        -508 \zeta_2 \ln ^2(2)
\nonumber\\&        
        +\frac{1108}{3} \zeta_2 \log (\beta )
        -632 \zeta_2 \ln (2) \log (\beta )
        -172 \zeta_2 \log ^2(\beta )
\Big)
+\beta ^2 \Big(
        \frac{17767}{675}
        -\frac{32 c_1}{15}
        +\frac{450622}{675} \zeta_2
\nonumber\\&        
        -\frac{25112}{75} \zeta_2^2
        +\frac{2717}{150} \zeta_3
        -\frac{26306}{75} \zeta_2 \ln (2)
        +\frac{608}{5} \zeta_2 \ln ^2(2)
        -416 \zeta_2 \log (\beta )
        +320 \zeta_2 \ln (2) \log (\beta )
\nonumber\\&        
        +160 \zeta_2 \log ^2(\beta )
\Big)
+ i \pi \Bigg(
\frac{1}{\beta } \Big(
-\frac{6338}{81}
+\frac{97}{6} \zeta_2
+\frac{22}{3} \zeta_3
+\frac{2704 \ln (2)}{27}
-11 \zeta_2 \ln (2)
-\frac{520}{9} \ln ^2(2)
\nonumber\\&
+\frac{44 \ln ^3(2)}{3}
+\frac{2704 \log (\beta )}{27}
-11 \zeta_2 \log (\beta )
-\frac{1040}{9} \ln (2) \log (\beta )
+44 \ln ^2(2) \log (\beta )
\nonumber\\&
-\frac{520}{9} \log ^2(\beta )
+44 \ln (2) \log ^2(\beta )
+\frac{44 \log ^3(\beta )}{3}
\Big)
+72 \zeta_2
-32 \zeta_2 \ln (2)
-32 \zeta_2 \log (\beta )
\nonumber\\&
+\beta  \Big(
        -\frac{2317}{27}
        -\frac{422}{9} \zeta_2
        -\frac{194}{3} \zeta_3
        -\frac{1156 \ln (2)}{27}
        +\frac{335}{3} \zeta_2 \ln (2)
        +130 \ln ^2(2)
        -\frac{764}{9} \ln ^3(2)
\nonumber\\&        
        -\frac{692 \log (\beta )}{27}
        +77 \zeta_2 \log (\beta )
        +\frac{1636}{9} \ln (2) \log (\beta )
        -\frac{508}{3} \ln ^2(2) \log (\beta )
        +\frac{554 \log ^2(\beta )}{9}
\nonumber\\&        
        -\frac{316}{3} \ln (2) \log ^2(\beta )
        -\frac{172}{9} \log ^3(\beta )
\Big)
+\beta ^2 \Big(
        208 \zeta_2
        -160 \zeta_2 \ln (2)
        -160 \zeta_2 \log (\beta )
\Big)
\Bigg)
\Bigg]
\Bigg\} 
\nonumber\\&
+ C_F T_F n_l \Bigg\{
\frac{1}{\ep^2}  \Bigg[
\frac{16 \beta ^2}{9}
+ i \pi \Bigg(
-\frac{2}{3 \beta }-\frac{2 \beta }{3}
\Bigg)
\Bigg]
+ \frac{1}{\ep}  \Bigg[
-\frac{80 \beta ^2}{27}
+ i \pi \Bigg(
\frac{10}{9 \beta }+\frac{10 \beta }{9}
\Bigg)
\Bigg]
\nonumber\\&
+  \Bigg[
\frac{1}{\beta } \Big(
-\frac{88}{3} \zeta_2
+16 \zeta_2 \ln (2)
+16 \zeta_2 \log (\beta )
\Big)
-\frac{20}{9}
+\beta  \Big(
        \frac{32}{3} \zeta_2
        +16 \zeta_2 \ln (2)
        +16 \zeta_2 \log (\beta )
\Big)
\nonumber\\&
+\beta ^2 \Big(
        -\frac{400}{81}
        -\frac{64 \zeta_2}{9}
\Big)
+ i \pi \Bigg(
\frac{1}{\beta } \Big(
\frac{248}{27}
-\frac{88 \ln (2)}{9}
+\frac{8 \ln ^2(2)}{3}
-\frac{88 \log (\beta )}{9}
+\frac{16}{3} \ln (2) \log (\beta )
\nonumber\\&
+\frac{8 \log ^2(\beta )}{3}
\Big)
+\beta  \Big(
        -\frac{268}{27}
+\frac{32 \ln (2)}{9}
+\frac{8 \ln ^2(2)}{3}
+\frac{32 \log (\beta )}{9}
+\frac{16}{3} \ln (2) \log (\beta )
\nonumber\\&
+\frac{8 \log ^2(\beta )}{3}
\Big)
\Bigg)
\Bigg]
+ \ep  \Bigg[
\frac{1}{\beta } \Big(
-\frac{1280}{9} \zeta_2
-36 \zeta_2^2
+\frac{448}{3} \zeta_2 \ln (2)
-48 \zeta_2 \ln ^2(2)
+\frac{448}{3} \zeta_2 \log (\beta )
\nonumber\\&
-96 \zeta_2 \ln (2) \log (\beta )
-48 \zeta_2 \log ^2(\beta )
\Big)
+\frac{62}{27}
+12 \zeta_2
+\beta  \Big(
        \frac{1360}{9} \zeta_2
        -36 \zeta_2^2
        -\frac{272}{3} \zeta_2 \ln (2)
\nonumber\\&        
        -48 \zeta_2 \ln ^2(2)
        -\frac{272}{3} \zeta_2 \log (\beta )
        -96 \zeta_2 \ln (2) \log (\beta )
        -48 \zeta_2 \log ^2(\beta )
\Big)
+\beta ^2 \Big(
        \frac{800}{9}
        +\frac{968 \zeta_2}{27}
\nonumber\\&        
        -\frac{128 \zeta_3}{9}
\Big)
+ i \pi \Bigg(
\frac{1}{\beta } \Big(
\frac{3160}{81}
-\frac{22}{3} \zeta_2
-\frac{8}{3} \zeta_3
-\frac{1280 \ln (2)}{27}
+4 \zeta_2 \ln (2)
+\frac{224 \ln ^2(2)}{9}
-\frac{16}{3} \ln ^3(2)
\nonumber\\&
-\frac{1280 \log (\beta )}{27}
+4 \zeta_2 \log (\beta )
+\frac{448}{9} \ln (2) \log (\beta )
-16 \ln ^2(2) \log (\beta )
+\frac{224 \log ^2(\beta )}{9}
\nonumber\\&
-16 \ln (2) \log ^2(\beta )
-\frac{16}{3} \log ^3(\beta )
\Big)
+\beta  \Big(
        -\frac{3848}{81}
        +\frac{8}{3} \zeta_2
        -\frac{8}{3} \zeta_3
        +\frac{1360 \ln (2)}{27}
        +4 \zeta_2 \ln (2)
\nonumber\\&        
        -\frac{136}{9} \ln ^2(2)
        -\frac{16}{3} \ln ^3(2)
        +\frac{1360 \log (\beta )}{27}
        +4 \zeta_2 \log (\beta )
        -\frac{272}{9} \ln (2) \log (\beta )
        -16 \ln ^2(2) \log (\beta )
\nonumber\\&        
        -\frac{136}{9} \log ^2(\beta )
        -16 \ln (2) \log ^2(\beta )
        -\frac{16}{3} \log ^3(\beta )
\Big)
\Bigg)
\Bigg]
\Bigg\} 
+ C_F T_F \Bigg\{
%
%
%
\Bigg[
\frac{580}{9}
-\frac{212}{3} \zeta_2
-16 \ln (2)
\nonumber\\&
+64 \zeta_2 \ln (2)
+\beta ^2 \Big(
        -\frac{712}{81}
        +\frac{252}{5} \zeta_2
        -42 \zeta_3
        +\frac{32 \ln (2)}{3}
        -\frac{336}{5} \zeta_2 \ln (2)
\Big)
+ i \pi \Bigg(
\frac{2 \zeta_2}{3 \beta }
+8
+\frac{2}{3} \beta  \zeta_2
\nonumber\\&
+\beta ^2 \Big(
        -\frac{16}{3}
        +12 \zeta_2
\Big)
\Bigg)
\Bigg]
+ \ep  \Bigg[
-\frac{4 \zeta_2^2}{\beta }
+\frac{5186}{27}
-\frac{32 c_1}{3}
-\frac{2800}{9} \zeta_2
+\frac{944}{5} \zeta_2^2
-\frac{616}{3} \zeta_3
\nonumber\\&
-\frac{224 \ln (2)}{3}
+\frac{584}{3} \zeta_2 \ln (2)
+16 \ln ^2(2)
+64 \zeta_2 \ln ^2(2)
-4 \beta  \zeta_2^2
+\beta ^2 \Big(
        -\frac{35116}{1215}
        +\frac{68 c_1}{15}
\nonumber\\&        
        +\frac{18992}{75} \zeta_2
        -\frac{8656}{25} \zeta_2^2
        +\frac{5914}{135} \zeta_3
        +\frac{3584 \ln (2)}{45}
        -\frac{8696}{75} \zeta_2 \ln (2)
        -56 \zeta_3 \ln (2)
        -\frac{32}{3} \ln ^2(2)
\nonumber\\&        
        +\frac{744}{5} \zeta_2 \ln ^2(2)
\Big)
+ i \pi \Bigg(
\frac{1}{\beta } \Big(
\frac{4}{3} \zeta_2
-\frac{4}{9} \zeta_3
-\frac{4}{3} \zeta_2 \ln (2)
-\frac{4}{3} \zeta_2 \log (\beta )
\Big)
+\frac{112}{3}
-12 \zeta_2
-16 \ln (2)
\nonumber\\&
+\beta  \Big(
        -2 \zeta_2
        -\frac{4}{9} \zeta_3
        -\frac{4}{3} \zeta_2 \ln (2)
        -\frac{4}{3} \zeta_2 \log (\beta )
\Big)
+\beta ^2 \Big(
        -\frac{1792}{45}
        +40 \zeta_2
        +28 \zeta_3
        +\frac{32 \ln (2)}{3}
\nonumber\\&        
        -24 \zeta_2 \ln (2)
\Big)
\Bigg)
\Bigg]
\Bigg\} \,.
\end{align}

\subsubsection{Pseudo-scalar form factor}
The non-singlet part of the pseudo-scalar form factor can be obtained using the chiral 
Ward identity Eq.~(\ref{eq:cwiFF}) as follows
\begin{equation}
 \tilde{F}_{P}^{(n),\ns} = \tilde{F}_{A,1}^{(n),\ns} + \Big( 1 + \beta^2 + \beta^4 \Big) \tilde{F}_{A,2}^{(n),\ns} + {\cal O}(\beta^3)\,.
\end{equation}
The bare singlet piece is given by
\begin{align}
\tilde{\hat{F}}_{P}^{(2),\sing} &= 
C_F T_F \Bigg\{
\Bigg[
20 \zeta_2
-42 \zeta_3
+48 \zeta_2 \ln (2)
-48 \beta  \zeta_2
+\beta ^2 \Big(
        -\frac{40}{3}
        -12 \zeta_2
        +28 \zeta_3
        +16 \ln (2)
\nonumber\\&
        -32 \zeta_2 \ln (2)
\Big)
+ i \pi \Bigg(
12 \zeta_2
+\beta  (16-16 \ln (2))
+\beta ^2 \Big(
        -8
        -8 \zeta_2
\Big)
\Bigg)
\Bigg]
+ \ep \Bigg[
-\frac{44 c_1}{3}
+88 \zeta_2
\nonumber\\&
-\frac{32}{5} \zeta_2^2
+112 \zeta_3
-264 \zeta_2 \ln (2)
-56 \zeta_3 \ln (2)
+264 \zeta_2 \ln ^2(2)
+\beta  \Big(
        -144 \zeta_2
        +288 \zeta_2 \ln (2)
\nonumber\\&
        +96 \zeta_2 \log (\beta )
\Big)
+\beta ^2 \Big(
        \frac{256}{9}
        +\frac{88 c_1}{9}
        -108 \zeta_2
        +\frac{64}{15} \zeta_2^2
        -\frac{112}{3} \zeta_3
        -48 \ln (2)
        +\frac{680}{3} \zeta_2 \ln (2)
\nonumber\\&
        +\frac{112}{3} \zeta_3 \ln (2)
        -16 \ln ^2(2)
        -176 \zeta_2 \ln ^2(2)
\Big)
+ i \pi \Bigg(
-12 \zeta_2
+28 \zeta_3
-24 \zeta_2 \ln (2)
\nonumber\\&
+\beta  \Big(
        16
        -80 \zeta_2
        -48 \ln (2)
        +48 \ln ^2(2)
        -32 \log (\beta )
        +32 \ln (2) \log (\beta )
\Big)
\nonumber\\&
+\beta ^2 \Big(
        24
        -\frac{4}{3} \zeta_2
        -\frac{56}{3} \zeta_3
        +16 \ln (2)
        +16 \zeta_2 \ln (2)
\Big)
\Bigg) 
\Bigg]
\Bigg\} \,.
\end{align}
%

\section{Conclusion} \label{sec:conclu}

\vspace*{1mm}
\noindent
The massive form factors are basic building blocks to many observables
in heavy quark physics. The precision study of these objects will both shed 
light on the physical structure of the top quark itself and also on important
aspects of the mechanism to create the fermion masses. A
future electron-positron collider can achieve high precision and hence
an equal or better theory prediction is indispensable.  In a similar way this
also applies to the LHC for its high luminosity phase.  In the present
paper, we have computed the
heavy quark form factors for vector, axial-vector, scalar and
pseudo-scalar currents at two-loop level up to the ${\cal O}(\ep^2)$ contributions. 
These contributions
constitute  important ingredients to renormalize the three-loop form factors and do also contribute
to potential future 4-loop calculations.  In addition, they serve as a
cross-check of earlier results available in the literature. In the calculation we have used
both traditional techniques in solving the differential equations for the master integrals, as well as
a more recent automated method, based on coupled difference equations. Both methods play a role
in computing higher than second order corrections to the different form factors. 

\vspace{2ex}
\noindent
{\bf Acknowledgment.}~This work was supported in part by the Austrian
Science Fund (FWF) grant SFB F50 (F5009-N15), the European
Commission through contract PITN-GA-2012-316704 (HIGGSTOOLS), and ERC
Advanced Grant no.~320651, HEPGAME. The work of A.B was supported by
the DFG under Grant No. WO 1900/2. We thank D.~St\"ockinger for discussions.
We would like to thank Th.~Gehrmann \cite{Bernreuther:2004ih,
Bernreuther:2004th,Bernreuther:2005rw,Bernreuther:2005gw} and
J.~Gluza \cite{Gluza:2009yy} for providing their results in electronic form.
The Feynman diagrams have been drawn using {\tt Axodraw} \cite{Vermaseren:1994je}.

\newpage
\appendix

\section{Renormalization Constants} \label{app:renz}
In this appendix, we present corresponding renormalization constants up to relevant order in $\ep$.
The wave function renormalization constants in OS scheme up to two-loop \cite{Broadhurst:1991fy, Melnikov:2000zc,Marquard:2007uj,Marquard:2017XXX} are 
\begin{align}
Z_{2,OS}^{(1)} &=  C_F \Bigg[ 
- \frac{3}{\ep} 
- 4 
+ \ep \Bigg\{ -8 -\frac{3}{2} \zeta_2 \Bigg\}
+ \ep^2 \Bigg\{ - 16 -2 \zeta_2 + \zeta_3 \Bigg\}  
+ \ep^3 \Bigg\{ -32 -4 \zeta_2 
\nonumber\\&
-\frac{27}{40} \zeta_2^2 + \frac{4}{3} \zeta_3 \Bigg\} \Bigg] \,,
\\
Z_{2,OS}^{(2)} &= 
C_F^2 \Bigg[
\frac{1}{\ep^2} \Bigg\{ \frac{9}{2} \Bigg\}
+ \frac{1}{\ep} \Bigg\{ \frac{51}{4} \Bigg\}
+ \Bigg\{
\frac{433}{8}
-\frac{147}{2} \zeta_2
-24 \zeta_3
+96 \zeta_2 \ln (2)
\Bigg\}
+ \ep \Bigg\{
\frac{211}{16}
-16 c_1
\nonumber\\&
-\frac{1017}{4} \zeta_2
+\frac{1008}{5} \zeta_2^2
-297 \zeta_3
+552 \zeta_2 \ln (2)
\Bigg\}
+ \ep^2 \Bigg\{
\frac{4889}{32}
-92 c_1
-\frac{96 c_2}{5}
-\frac{8851}{8} \zeta_2
\nonumber\\&
+\frac{11703}{20} \zeta_2^2
-\frac{2069}{2} \zeta_3
+264 \zeta_2 \zeta_3
+2436 \zeta_5
+1968 \zeta_2 \ln (2)
\Bigg\}
\Bigg]
+ C_A C_F \Bigg[
\frac{1}{\ep^2} \Bigg\{ \frac{11}{2} \Bigg\}
\nonumber\\&
+ \frac{1}{\ep} \Bigg\{ -\frac{127}{12} \Bigg\}
+ \Bigg\{
-\frac{1705}{24}
+30 \zeta_2
+12 \zeta_3
-48 \zeta_2 \ln (2)
\Bigg\}
+ \ep \Bigg\{
-\frac{9907}{48}
+8 c_1
+\frac{769}{12} \zeta_2
\nonumber\\&
-\frac{504}{5} \zeta_2^2
+129 \zeta_3
-276 \zeta_2 \ln (2)
\Bigg\}
+ \ep^2 \Bigg\{
-\frac{79225}{96}
+46 c_1
+\frac{48 c_2}{5}
+\frac{6367}{24} \zeta_2
-\frac{14359}{40} \zeta_2^2
\nonumber\\&
+\frac{7595}{18} \zeta_3
-132 \zeta_2 \zeta_3
-1218 \zeta_5
-984 \zeta_2 \ln (2)
\Bigg\}
\Bigg]
+ C_F T_F n_l \Bigg[
\frac{1}{\ep^2} \Bigg\{ -2 \Bigg\}
+ \frac{1}{\ep} \Bigg\{ \frac{11}{3} \Bigg\}
\nonumber\\&
+ \Bigg\{
\frac{113}{6}
+8 \zeta_2
\Bigg\}
+ \ep \Bigg\{
\frac{851}{12}
+\frac{127}{3} \zeta_2
+16 \zeta_3
\Bigg\}
+ \ep^2 \Bigg\{
\frac{5753}{24}
+\frac{853}{6} \zeta_2
+\frac{597}{10} \zeta_2^2
+\frac{610}{9} \zeta_3
\Bigg\}
\Bigg]
\nonumber\\&
+ C_F T_F \Bigg[
%
 \frac{1}{\ep} 
+ \Bigg\{
\frac{947}{18}
-30 \zeta_2
\Bigg\}
+ \ep \Bigg\{
\frac{17971}{108}
-\frac{445}{3} \zeta_2
-\frac{340}{3} \zeta_3
+192 \zeta_2 \ln (2)
\Bigg\}
\nonumber\\&
+ \ep^2 \Bigg\{
\frac{422747}{648}
-32 c_1
-\frac{8605}{18} \zeta_2
+\frac{1683}{10} \zeta_2^2
-\frac{4810}{9} \zeta_3
+912 \zeta_2 \ln (2)
\Bigg\}
\Bigg] \,.
\end{align}
%
%
The heavy quark mass renormalization constants in OS scheme are \cite{Broadhurst:1991fy, Melnikov:2000zc,Marquard:2007uj,Marquard:2015qpa,Marquard:2016dcn}
\begin{align}
Z_{m,OS}^{(1)} &= C_F \Bigg[ 
- \frac{3}{\ep} 
- 4 
+ \ep \Bigg\{ -8 -\frac{3}{2} \zeta_2 \Bigg\}
+ \ep^2 \Bigg\{ - 16 -2 \zeta_2 + \zeta_3 \Bigg\}  
+ \ep^3 \Bigg\{ -32 -4 \zeta_2 
\nonumber\\&
-\frac{27}{40} \zeta_2^2 + \frac{4}{3} \zeta_3 \Bigg\} \Bigg] \,,
\\
Z_{m,OS}^{(2)} &= 
C_F^2 \Bigg[
\frac{1}{\ep^2} \Bigg\{ \frac{9}{2} \Bigg\}
+ \frac{1}{\ep} \Bigg\{ \frac{45}{4} \Bigg\}
+ \Bigg\{
\frac{199}{8}
-\frac{51}{2} \zeta_2
-12 \zeta_3
+48 \zeta_2 \ln (2)
\Bigg\}
+ \ep \Bigg\{
\frac{677}{16}
-8 c_1
-\frac{615}{4} \zeta_2
\nonumber\\&
+\frac{504}{5} \zeta_2^2
-135 \zeta_3
+288 \zeta_2 \ln (2)
\Bigg\}
+ \ep^2 \Bigg\{
\frac{1167}{32}
-48 c_1
-\frac{48 c_2}{5}
-\frac{4821}{8} \zeta_2
+\frac{7719}{20} \zeta_2^2
\nonumber\\&
-\frac{1203}{2} \zeta_3
+132 \zeta_2 \zeta_3
+1218 \zeta_5
+1056 \zeta_2 \ln (2)
\Bigg\}
\Bigg]
+ C_A C_F \Bigg[
\frac{1}{\ep^2} \Bigg\{ \frac{11}{2} \Bigg\}
+ \frac{1}{\ep} \Bigg\{ -\frac{97}{12} \Bigg\}
\nonumber\\&
+ \Bigg\{
-\frac{1111}{24}
+8 \zeta_2
+6 \zeta_3
-24 \zeta_2 \ln (2)
\Bigg\}
+ \ep \Bigg\{
-\frac{8581}{48}
+4 c_1
+\frac{271}{12} \zeta_2
-\frac{252}{5} \zeta_2^2
+52 \zeta_3
\nonumber\\&
-144 \zeta_2 \ln (2)
\Bigg\}
+ \ep^2 \Bigg\{
-\frac{58543}{96}
+24 c_1
+\frac{24 c_2}{5}
+\frac{1537}{24} \zeta_2
-\frac{9783}{40} \zeta_2^2
+\frac{3929}{18} \zeta_3
-66 \zeta_2 \zeta_3
\nonumber\\&
-609 \zeta_5
-528 \zeta_2 \ln (2)
\Bigg\}
\Bigg]
+ C_F T_F n_l \Bigg[
\frac{1}{\ep^2} \Bigg\{ -2 \Bigg\}
+ \frac{1}{\ep} \Bigg\{ \frac{5}{3} \Bigg\}
+ \Bigg\{
\frac{71}{6}
+8 \zeta_2
\Bigg\}
\nonumber\\&
+ \ep \Bigg\{
\frac{581}{12}
+\frac{97}{3} \zeta_2
+16 \zeta_3
\Bigg\}
+ \ep^2 \Bigg\{
\frac{4079}{24}
+\frac{643}{6} \zeta_2
+\frac{597}{10} \zeta_2^2
+\frac{478}{9} \zeta_3
\Bigg\}
\Bigg]
+ C_F T_F \Bigg[
\frac{1}{\ep^2} \Bigg\{ -2 \Bigg\}
\nonumber\\&
+ \frac{1}{\ep} \Bigg\{ \frac{5}{3} \Bigg\}
+ \Bigg\{
\frac{143}{6}
-16 \zeta_2
\Bigg\}
+ \ep \Bigg\{
\frac{1133}{12}
-\frac{227}{3} \zeta_2
-56 \zeta_3
+96 \zeta_2 \ln (2)
\Bigg\}
\nonumber\\&
+ \ep^2 \Bigg\{
\frac{8135}{24}
-16 c_1
-\frac{1553}{6} \zeta_2
+\frac{837}{10} \zeta_2^2
-\frac{2546}{9} \zeta_3
+480 \zeta_2 \ln (2)
\Bigg\}
\Bigg] \,.
\end{align}

\newpage

\section{The vector form factors up to two-loop} \label{app:vFF}

In this appendix, we present the vector form factors $F_{V,1}^{(n)}$
and $F_{V,2}^{(n)}$ up to two loops and ${\cal O}(\varepsilon)$. 

\begin{dmath*}
 F_{V,1}^{(1)} = C_F \Bigg[
\frac{1}{\ep}  \Bigg\{
-2
-2 \xi  H_0
\Bigg\}
+   \Bigg\{
-4
+\xi  \Big(
        4 H_{-1} H_0
        -H_0^2
        -4 H_{0,-1}
        +2 \zeta_2
\Big)
+\Big(
        -3-2 x-3 x^2\Big) \eta  H_0
\Bigg\}
\end{dmath*}
\vspace{-1.0cm}
\begin{dmath}
{\color{white}=}
+ \ep  \Bigg\{
-8
+\eta  \Big(
        2 \Big(
                3+2 x+3 x^2\Big) H_{-1} H_0
        +\frac{1}{2} \Big(
                -3-2 x-3 x^2\Big) H_0^2
        -2 \Big(
                3+2 x+3 x^2\Big) H_{0,-1}
        +2 \Big(
                1+x+2 x^2\Big) \zeta_2
\Big)
+\xi  \Big(
        \Big(
                -8
                -4 H_{-1}^2
                +\zeta_2
        \Big) H_0
        +2 H_{-1} H_0^2
        -\frac{1}{3} H_0^3
        +8 H_{-1} H_{0,-1}
        -4 H_{0,0,-1}
        -8 H_{0,-1,-1}
        -4 H_{-1} \zeta_2
        +4 \zeta_3
\Big)
\Bigg\}
+ \ep^2  \Bigg\{
-16
+\eta  \Big(
        \Big(
                -2 \Big(
                        3+2 x+3 x^2\Big) H_{-1}^2
                +\frac{1}{2} \Big(
                        3+2 x+3 x^2\Big) \zeta_2
        \Big) H_0
        +\Big(
                3+2 x+3 x^2\Big) H_{-1} H_0^2
        +\frac{1}{6} \Big(
                -3-2 x-3 x^2\Big) H_0^3
        +4 \Big(
                3+2 x+3 x^2\Big) H_{-1} H_{0,-1}
        -2 \Big(
                3+2 x+3 x^2\Big) H_{0,0,-1}
        -4 \Big(
                3+2 x+3 x^2\Big) H_{0,-1,-1}
        +2 \Big(
                3+5 x^2\Big) \zeta_2
        -2 \Big(
                3+2 x+3 x^2\Big) H_{-1} \zeta_2
        +\frac{4}{3} \Big(
                5+3 x+4 x^2\Big) \zeta_3
\Big)
+\xi  \Big(
        \Big(
                -16
                +\Big(
                        16
                        -2 \zeta_2
                \Big) H_{-1}
                +\frac{8}{3} H_{-1}^3
                +\frac{14}{3} \zeta_3
        \Big) H_0
        +\Big(
                -4
                -2 H_{-1}^2
                +\frac{1}{2} \zeta_2
        \Big) H_0^2
        +\frac{2}{3} H_{-1} H_0^3
        -\frac{1}{12} H_0^4
        +\Big(
                -16
                -8 H_{-1}^2
                -2 \zeta_2
        \Big) H_{0,-1}
        +8 H_{-1} H_{0,0,-1}
        +16 H_{-1} H_{0,-1,-1}
        -4 H_{0,0,0,-1}
        -8 H_{0,0,-1,-1}
        -16 H_{0,-1,-1,-1}
        +4 H_{-1}^2 \zeta_2
        +\frac{14}{5} \zeta_2^2
        -8 H_{-1} \zeta_3
\Big)
\Bigg\}
+ \ep^3  \Bigg\{
-32
+\eta  \Big(
        \Big(
                \frac{4}{3} \Big(
                        3+2 x+3 x^2\Big) H_{-1}^3
                +\Big(
                        -3-2 x-3 x^2\Big) H_{-1} \zeta_2
                +\frac{7}{3} \Big(
                        3+2 x+3 x^2\Big) \zeta_3
        \Big) H_0
        +\Big(
                \Big(
                        -3-2 x-3 x^2\Big) H_{-1}^2
                +\frac{1}{4} \Big(
                        3+2 x+3 x^2\Big) \zeta_2
                +\Big(
                        -1-x^2\Big) H_{-1} \zeta_2
        \Big) H_0^2
        +\frac{1}{3} \Big(
                3+2 x+3 x^2\Big) H_{-1} H_0^3
        +\frac{1}{24} \Big(
                -3-2 x-3 x^2\Big) H_0^4
        +\Big(
                -4 \Big(
                        3+2 x+3 x^2\Big) H_{-1}^2
                +\Big(
                        -3-2 x-3 x^2\Big) \zeta_2
        \Big) H_{0,-1}
        +4 \Big(
                3+2 x+3 x^2\Big) H_{-1} H_{0,0,-1}
        +8 \Big(
                3+2 x+3 x^2\Big) H_{-1} H_{0,-1,-1}
        -2 \Big(
                3+2 x+3 x^2\Big) H_{0,0,0,-1}
        -4 \Big(
                3+2 x+3 x^2\Big) H_{0,0,-1,-1}
        -8 \Big(
                3+2 x+3 x^2\Big) H_{0,-1,-1,-1}
        +4 \Big(
                3+5 x^2\Big) \zeta_2
        +2 \Big(
                3+2 x+3 x^2\Big) H_{-1}^2 \zeta_2
        +\frac{1}{20} \Big(
                75+56 x+93 x^2\Big) \zeta_2^2
        +\frac{4}{3} \Big(
                13+11 x^2\Big) \zeta_3
        -4 \Big(
                3+2 x+3 x^2\Big) H_{-1} \zeta_3
\Big)
+\xi  \Big(
        \Big(
                -32
                +\Big(
                        32
                        -\frac{28 \zeta_3}{3}
                \Big) H_{-1}
                +\Big(
                        -16
                        +2 \zeta_2
                \Big) H_{-1}^2
                -\frac{4}{3} H_{-1}^4
                +4 \zeta_2
                +\frac{47}{20} \zeta_2^2
        \Big) H_0
        +\Big(
                -8
                +8 H_{-1}
                +\frac{4}{3} H_{-1}^3
                +\frac{7}{3} \zeta_3
        \Big) H_0^2
        +\Big(
                \frac{1}{6} \Big(
                        -8
                        +\zeta_2
                \Big)
                -\frac{2}{3} H_{-1}^2
        \Big) H_0^3
        +\frac{1}{6} H_{-1} H_0^4
        -\frac{1}{60} H_0^5
        +\Big(
                -16 \zeta_2
                -\frac{28 \zeta_2^2}{5}
        \Big) H_{-1}
        +\Big(
                -32
                +\Big(
                        32
                        +4 \zeta_2
                \Big) H_{-1}
                +\frac{16}{3} H_{-1}^3
                +\frac{4}{3} \zeta_3
        \Big) H_{0,-1}
        +\Big(
                -16
                -8 H_{-1}^2
                -2 \zeta_2
        \Big) H_{0,0,-1}
        +\Big(
                -32
                -16 H_{-1}^2
                -4 \zeta_2
        \Big) H_{0,-1,-1}
        +8 H_{-1} H_{0,0,0,-1}
        +16 H_{-1} H_{0,0,-1,-1}
        +32 H_{-1} H_{0,-1,-1,-1}
        -4 H_{0,0,0,0,-1}
        -8 H_{0,0,0,-1,-1}
        -16 H_{0,0,-1,-1,-1}
        -32 H_{0,-1,-1,-1,-1}
        -\frac{8}{3} H_{-1}^3 \zeta_2
        +8 H_{-1}^2 \zeta_3
        -\frac{8}{3} \zeta_2 \zeta_3
        +12 \zeta_5
\Big)
\Bigg\}
\Bigg] \,.
\end{dmath}

\begin{dmath*}
 F_{V,1}^{(2)} =
%
C_F^2  \Bigg[
\frac{1}{\ep^2}  \Bigg\{
        2
        +4 \xi  H_0
        +2 \xi ^2 H_0^2
\Bigg\}
+ \frac{1}{\ep}  \Bigg\{
        8
        +\xi  \Big(
                -8 H_{-1} H_0
                +4 \Big(
                        2+x+x^2\Big) \eta  H_0^2
                +8 H_{0,-1}
                -4 \zeta_2
        \Big)
\end{dmath*}
\begin{dmath*}
{\color{white}=}
        +\xi ^2 \Big(
                -8 H_{-1} H_0^2
                +2 H_0^3
                +8 H_0 H_{0,-1}
                -4 H_0 \zeta_2
        \Big)
        +2 \Big(
                7+2 x+7 x^2\Big) \eta  H_0
\Bigg\}
+  \Bigg\{
                46
                +\eta ^2 \Big(
                        \Big(
                                -4 H_{-1} H_0^2 P_{103}
                                +8 H_0 H_{0,-1} P_{121}
                                -8 H_0 H_{0,1} P_{124}
                                +8 H_{0,0,1} P_{135}
                                -8 H_{0,0,-1} P_{136}
                                +\frac{1}{2} H_0^2 P_{161}
                                -4 P_{115} \zeta_3
                        \Big) x_+^2
                        +\Big(
                                \frac{16}{3} H_0^3 H_1 P_{176}
                                -4 H_0^2 H_{0,-1} P_{180}
                                -8 H_0 H_{0,0,-1} P_{193}
                                +48 H_{0,0,0,-1} P_{194}
                                -8 H_0^2 H_{0,1} P_{195}
                                -\frac{4}{3} H_0^3 P_{204}
                                +16 H_0 H_{0,0,1} P_{216}
                                -16 H_{0,0,0,1} P_{242}
                                +\frac{1}{6} H_0^4 P_{274}
                                +\Big(
                                        -4 x H_0 P_{134}
                                        +16 H_0 H_1 P_{174}
                                        -16 H_{0,1} P_{174}
                                        -8 H_{0,-1} P_{218}
                                        +4 H_0^2 P_{219}
                                \Big) \zeta_2
                                +\frac{2}{5} P_{308} \zeta_2^2
                        \Big) x_+^3
                \Big)
                +\xi  \Big(
                        8 H_{-1}
                        ^2 H_0
                        -16 H_{-1} H_{0,-1}
                        +16 H_{0,-1,-1}
                        +48 (-1+x) \ln (2) x_+ \zeta_2
                \Big)
                +\xi ^2 \Big(
                        16 H_{-1}^2 H_0^2
                        -8 H_{-1} H_0^3
                        +\Big(
                                32 H_0 H_1
                                -32 H_{0,-1}
                        \Big) H_{0,1}
                        +16 H_{0,1}^2
                        -32 H_{-1} H_0 H_{0,-1}
                        -32 H_0 H_1 H_{0,-1}
                        +8 H_{0,-1}^2
                        -64 H_1 H_{0,0,1}
                        +64 H_1 H_{0,0,-1}
                        -64 H_0 H_{0,1,1}
                        +32 H_0 H_{0,1,-1}
                        +32 H_0 H_{0,-1,1}
                        +16 H_0 H_{0,-1,-1}
                        +64 H_{0,0,1,1}
                        +32 H_{0,-1,0,1}
                        +16 H_{-1} H_0 \zeta_2
                        +16 H_1 \zeta_3
                \Big)
                +\eta  \Big(
                        x_+^3 \zeta_2 \Big(
                                39
                                +7 x
                                -750 x^2
                                +1074 x^3
                                -185 x^4
                                -57 x^5
                                -8 H_{-1} P_{51}
                        \Big)
                        +\frac{1}{2} \Big(
                                85-18 x+85 x^2\Big) H_0
                        +\Big(
                                2 H_{-1} H_0 P_{37}
                                -2 H_{0,-1} P_{37}
                                -32 \Big(
                                        2+x+10 x^2+x^3+2 x^4\Big) H_0 H_1
                                +32 \Big(
                                        2+x+10 x^2+x^3+2 x^4\Big) H_{0,1}
                        \Big) x_+^2
                        -16 H_0 P_{99} x_+^4 \zeta_3
                \Big)
                +4 H_0^2 H_1 P_{24} x_+^4
                +192 x \ln (2) x_+^2 \zeta_2
\Bigg\}
+ \ep  \Bigg\{
                \ln (2) x_+^4 \zeta_2 \Big(
                        -24 P_4
                        +48 H_1 P_{12}
                        +48 H_{-1} P_{12}
                \Big)
                +\xi  \Big(
                        \eta  \Big(
                                \ln (2) x_+^3 \zeta_2 \Big(
                                        -192 (-1+x) x^2 H_{0,1}
                                        -192 (-1+x) x^2 H_{0,-1}
                                \Big)
                                -\frac{4}{3} H_0^3 P_{66} x_+^3 \zeta_2
                        \Big)
                        -\frac{16}{3} H_{-1}^3 H_0
                        -32 x \Big(
                                2-29 x+2 x^2\Big) x_+^4 H_0 H_{0,1}^2
                        +16 H_{-1}^2 H_{0,-1}
                        -32 H_{-1} H_{0,-1,-1}
                        +32 H_{0,-1,-1,-1}
                        -4 (-1+x) x_+
                \Big)
                +\eta ^2 \Big(
                        16 H_0^2 H_{0,1,1} P_{53} x_+
                        +\Big(
                                H_{-1} H_0^2 P_{83}
                                +16 H_0 H_{0,1,1} P_{95}
                                +4 H_{-1}^2 H_0^2 P_{100}
                                +8 H_{0,1}^2 P_{119}
                                +16 H_0 H_{0,1,-1} P_{129}
                                +16 H_0 H_{0,-1,1} P_{129}
                                +16 H_{0,-1,0,1} P_{132}
                                +16 H_{0,0,1,-1} P_{133}
                                +16 H_{0,0,-1,1} P_{133}
                                -16 H_{0,0,-1,-1}
                                 P_{137}
                                +8 H_{0,-1}^2 P_{141}
                                +
                                \frac{1}{2} H_0^2 P_{163}
                                +\Big(
                                        32 H_0 P_{88}
                                        +32 H_{-1} H_0 P_{109}
                                        -16 H_{0,-1} P_{140}
                                \Big) H_{0,1}
                                +\Big(
                                        -16 H_{-1} H_0 P_{125}
                                        -8 H_0 P_{149}
                                \Big) H_{0,-1}
                                +\Big(
                                        -16 H_{-1} P_{133}
                                        -4 P_{156}
                                \Big) H_{0,0,1}
                                +\Big(
                                        16 H_{-1} P_{137}
                                        +2 P_{164}
                                \Big) H_{0,0,-1}
                                +\Big(
                                        4 H_{-1} P_{145}
                                        -\frac{2 P_{165}}{3}
                                \Big) \zeta_3
                        \Big) x_+^2
                        +\Big(
                                -16 H_0^2 H_{0,1,-1} P_{111}
                                -16 H_0^2 H_{0,-1,1} P_{111}
                                +\frac{16}{3} H_0^3 H_1^2 P_{186}
                                +32 H_{0,0,1,0,1} P_{189}
                                -16 H_0 H_{0,0,-1,-1} P_{199}
                                -32 H_{0,0,1,0,-1} P_{203}
                                -32 H_0 H_{0,0,1,-1} P_{213}
                                -32 H_0 H_{0,0,-1,1} P_{213}
                                +32 H_{0,0,-1,0,1} P_{220}
                                +8 H_0^2 H_{0,-1,-1} P_{230}
                                -32 H_{0,0,0,1,-1} P_{238}
                                -32 H_{0,0,0,-1,1} P_{238}
                                +\frac{1}{3} H_0^4 P_{241}
                                +32 H_{0,0,-1,0,-1} P_{249}
                                +96 H_{0,0,0,-1,-1} P_{250}
                                -32 H_0 H_{0,0,1,1} P_{256}
                                +32 H_{0,0,0,1,1} P_{287}
                                +\frac{1}{30} H_0^5 P_{290}
                                -16 H_{0,0,0,0,1} P_{305}
                                +16 H_{0,0,0,0,-1} P_{306}
                                +\frac{1}{6} H_0^3 P_{309}
                                +\Big(
                                        2 H_0^4 P_{175}
                                        -\frac{32}{3} H_{-1} H_0^3 P_{186}
                                \Big) H_1
                                +\Big(
                                        \frac{4}
                                        {3} H_0^3 P_{196}
                                        -
                                        \frac{2}{3} H_0^4 P_{231}
                                \Big) H_{-1}
                                +\Big(
                                        \frac{8}{3} H_0^3 P_{190}
                                        +16 H_{-1} H_0^2 P_{198}
                                        -4 H_0^2 P_{284}
                                \Big) H_{0,1}
                                +\Big(
                                        -16 x H_{-1} H_0^2 P_{98}
                                        +16 H_0^2 H_1 P_{178}
                                        -\frac{4}{3} H_0^3 P_{257}
                                        +4 H_0^2 P_{292}
                                \Big) H_{0,-1}
                                +\Big(
                                        32 H_{0,1} P_{170}
                                        -64 H_{-1} H_0 P_{179}
                                        +256 H_0 H_1 P_{185}
                                        -8 H_0 P_{228}
                                        -8 H_0^2 P_{232}
                                \Big) H_{0,0,1}
\end{dmath*}
\begin{dmath*}
{\color{white}=}
                                +\Big(
                                        32 H_{-1} H_0 P_{117}
                                        -8 H_0 P_{225}
                                        -32 H_{0,1} P_{227}
                                        +4 H_0^2 P_{239}
                                        -16 H_{0,-1} P_{278}
                                \Big) H_{0,0,-1}
                                +\Big(
                                        32 H_{-1} P_{182}
                                        -64 H_1 P_{248}
                                        +16 H_0 P_{293}
                                        +8 P_{300}
                                \Big) H_{0,0,0,1}
                                +\Big(
                                        -96 H_{-1} P_{112}
                                        +32 H_1 P_{285}
                                        -8 H_0 P_{297}
                                        -8 P_{303}
                                \Big) H_{0,0,0,-1}
                                +\Big(
                                        2 x H_0^2 P_{154}
                                        +H_0 P_{166}
                                        +32 H_0 H_1^2 P_{177}
                                        +64 H_{0,1,1} P_{186}
                                        -32 H_{0,1,-1} P_{223}
                                        -32 H_{0,-1,1} P_{224}
                                        -16 H_{0,0,1} P_{254}
                                        +16 H_{0,-1,-1} P_{275}
                                        +8 H_{0,0,-1} P_{294}
                                        +\Big(
                                                -64 H_{-1} H_0 P_{177}
                                                -8 H_0^2 P_{236}
                                        \Big) H_1
                                        +\Big(
                                                -8 H_0^2 P_{188}
                                                +8 H_0 P_{266}
                                        \Big) H_{-1}
                                        +\Big(
                                                64 H_{-1} P_{177}
                                                -32 H_1 P_{197}
                                                -8 P_{268}
                                        \Big) H_{0,1}
                                        +\Big(
                                                -32 H_{-1} P_{184}
                                                -8 P_{200}
                                                -16 H_0 P_{207}
                                                +32 H_1 P_{224}
                                        \Big) H_{0,-1}
                                        -4 P_{289} \zeta_3
                                \Big) \zeta_2
                                +\Big(
                                        -
                                        \frac{8}{5} H_{-1} P_{183}
                                        -\frac{8}{5} H_1 P_{296}
                                        +\frac{4 P_{307}}{5}
                                        -\frac{2}{5} H_0 P_{310}
                                \Big) \zeta_2^2
                                +\Big(
                                        -16 H_0 H_1 P_{209}
                                        -16 H_{0,-1} P_{245}
                                        +16 H_{0,1} P_{253}
                                        -\frac{4}{3} H_0^2 P_{291}
                                        -\frac{4}{3} H_0 P_{295}
                                \Big) \zeta_3
                                +6 P_{299} \zeta_5
                        \Big) x_+^3
                \Big)
                +\eta  \Big(
                        \ln (2) x_+^4 \zeta_2 \Big(
                                -3072 x^3 H_{0,1}
                                -3072 x^3 H_{0,-1}
                        \Big)
                        +\Big(
                                4 H_{-1} H_0 P_{40}
                                -2 H_{-1}^2 H_0 P_{41}
                                -4 H_{0,-1,-1} P_{41}
                                +\frac{1}{4} H_0 P_{43}
                                +32 \Big(
                                        2+x+10 x^2+x^3+2 x^4\Big) \zeta_2 H_1
                                +\Big(
                                        -4 H_0 P_{36}
                                        +192 \Big(
                                                2+x+10 x^2+x^3+2 x^4\Big) H_{-1} H_0
                                \Big) H_1
                                -32 \Big(
                                        2+x+10 x^2+x^3+2 x^4\Big) H_0 H_1^2
                                +\Big(
                                        4 P_{36}
                                        +64 \Big(
                                                2+x+10 x^2+x^3+2 x^4\Big) H_1
                                        -192 \Big(
                                                2+x+10 x^2+x^3+2 x^4\Big) H_{-1}
                                \Big) H_{0,1}
                                +\Big(
                                        -4 P_{40}
                                        +4 H_{-1} P_{41}
                                        -192 \Big(
                                                2+x+10 x^2+x^3+2 x^4\Big) H_1
                                \Big) H_{0,-1}
                                -64 \Big(
                                        2+x+10 x^2+x^3+2 x^4\Big) H_{0,1,1}
                                +192 \Big(
                                        2+x+10 x^2+x^3+2 x^4\Big) H_{0,1,-1}
                                +192 \Big(
                                        2+x+10 x^2+x^3+2 x^4\Big) H_{0,-1,1}
                        \Big) x_+^2
                        +\Big(
                                -2 H_0^2 H_1 P_{71}
                                +\Big(
                                        163
                                        +251 x
                                        -3084 x^2
                                        +4532 x^3
                                        -847 x^4
                                        -183 x^5
                                        +8 H_{-1}
                                        ^2 P_{63}
                                        -2 H_{-1} P_{72}
                                \Big) \zeta_2
                        \Big) x_+^3
                        +\Big(
                                384 H_0 H_1 H_{0,0,-1} P_{92}
                                -48 H_0 H_{0,-1}^2 P_{97}
                                +32 H_0 H_{0,-1,0,1} P_{122}
                                -32 H_{0,-1} H_{0,0,1} P_{127}
                                -
                                \frac{4}{3} H_0^3 H_1 P_{151}
                                +\Big(
                                        32 H_0^2 H_1 P_{104}
                                        -32 H_0 H_{0,-1} P_{106}
                                \Big) H_{0,1}
                                +\Big(
                                        -16 H_0 H_{0,1} P_{128}
                                        -8 H_0 H_1 P_{142}
                                \Big) \zeta_2
                                -32 H_{-1} H_0 P_{99} \zeta_3
                        \Big) x_+^4
                \Big)
                +\xi ^2 \Big(
                        -\frac{64}{3} H_{-1}^3 H_0^2
                        +16 H_{-1}^2 H_0^3
                        +\Big(
                                128 H_{-1} H_0 H_{0,-1}
                                -96 H_{0,-1}^2
                        \Big) H_1
                        +\Big(
                                \Big(
                                        -128 H_{-1} H_0
                                        +64 H_{0,-1}
                                \Big) H_1
                                +64 H_0 H_1^2
                                +128 H_{-1} H_{0,-1}
                        \Big) H_{0,1}
                        +\Big(
                                32 H_1
                                -64 H_{-1}
                        \Big) H_{0,1}^2
                        +64 H_{-1}^2 H_0 H_{0,-1}
                        -64 H_0 H_1^2 H_{0,-1}
                        -32 H_{-1} H_{0,-1}^2
                        +\Big(
                                256 H_{-1} H_1
                                -128 H_1^2
                        \Big) H_{0,0,1}
                        +\Big(
                                -256 H_{-1} H_1
                                +128 H_1^2
                        \Big) H_{0,0,-1}
                        +\Big(
                                256 H_{-1} H_0
                                -192 H_0 H_1
                                -256 H_{0,-1}
                        \Big) H_{0,1,1}
                        +\Big(
                                -128 H_{-1} H_0
                                +64 H_0 H_1
                                -64 H_{0,1}
                                +192 H_{0,-1}
                        \Big) H_{0,1,-1}
                        +\Big(
                                -128 H_{-1} H_0
                                +64 H_0 H_1
                                +192 H_{0,1}
                                +192 H_{0,-1}
                        \Big) H_{0,-1,1}
                        +\Big(
                                -64 H_{-1} H_0
                                +64 H_0 H_1
                                -320 H_{0,1}
                                +32 H_{0,-1}
                        \Big) H_{0,-1,-1}
                        +\Big(
                                256 H_1
                                -256 H_{-1}
                        \Big) H_{0,0,1,1}
                        -256 H_1 H_{0,0,1,-1}
                        -256 H_1 H_{0,0,-1,1}
                        +256 H_1 H_{0,0,-1,-1}
                        +192 H_0 H_{0,1,1,1}
                        -64 H_0 H_{0,1,-1,-1}
                        -128 H_{-1} H_{0,-1,0,1}
                        -64 H_0 H_{0,-1,1,-1}
                        -64 H_0 H_{0,-1,-1,1}
                        +32 H_0 H_{0,-1,-1,-1}
                        -384 H_{0,0,1,1,1}
                        +768 H_{0,0,1,1,-1}
                        +256 H_{0,0,1,-1,1}
                        -256 H_{0,0,-1,1,1}
                        -64 H_{0,1,0,1,1}
                        +256 H_{0,1,0,1,-1}
                        -256 H_{0,-1,0,1,1}
                        +128 H_{0,-1,0,1,-1}
                        +128 H_{0,-1,0,-1,1}
                        -256 H_{0,-1,1,0,1}
                        +320 H_{0,-1,-1,0,1}
                        -32 H_{-1}^2 H_0 \zeta_2
                        +\Big(
                                -64 H_{-1} H_1
                                +32 H_1^2
                        \Big) \zeta_3
                \Big)
                +28 x x_+^2
                +8 c_1 \Big(
                        1-4 x+x^2\Big) x_+^2
                +\Big(
                        -8 H_{-1} H_0^2 H_1 P_{12}
                        -32 H_0 H_{0,-1,-1} P_{19}
                        +16 H_1 H_{0,0,1} P_{20}
                        -16 H_{0,0,1,1} P_{20}
                        -16 H_0 H_1 H_{0,-1} P_{21}
                        -16 H_0 H_1 H_{0,1} P_{27}
                        +4 H_0^2 H_1^2 P_{29}
                        +16 H_1 H_{0,0,-1} P_{30}
                        +\Big(
                                -48 H_{-1} H_1 P_{12}
                                +48 H_{-1,1} P_{12}
                        \Big) \zeta_2
                        -4 H_1 P_{39} \zeta_3
                \Big) x_+^4
\Bigg\}
\Bigg]
\end{dmath*}
\begin{dmath*}
{\color{white}=}
%
+ C_F C_A  \Bigg[
\frac{1}{\ep^2}  \Bigg\{
        \frac{11}{3}
        +\frac{11 \xi  H_0}{3}
\Bigg\}
+ \frac{1}{\ep}  \Bigg\{
        -\frac{49}{9}
        +\xi  \Big(
                \eta  \Big(
                        -\frac{4}{3} x^2 H_0^3
                        -2 \Big(
                                -1+3 x^2\Big) H_0 \zeta_2
                \Big)
                -\frac{67}{9} H_0
                +4 H_{-1} H_0
                -4 H_0 H_1
                +4 H_{0,1}
                -4 H_{0,-1}
        \Big)
        +\eta  \Big(
                -4 x^2 H_0^2
                -2 \Big(
                        -1+3 x^2\Big) \zeta_2
        \Big)
        +\xi ^2 \Big(
                -4 H_0 H_{0,1}
                +4 H_0 H_{0,-1}
                +8 H_{0,0,1}
                -8 H_{0,0,-1}
                -2 \zeta_3
        \Big)
\Bigg\}
+  \Bigg\{
                \frac{1}{27} \Big(
                        -1595-3514 x-1595 x^2\Big) x_+^2
                +\eta ^2 x_+^3 \Big(
                        -32 x H_0 H_{0,0,1} P_{94}
                        -\frac{1}{6} x H_0^4 P_{123}
                        +\frac{4}{3} H_0^3 H_1 P_{169}
                        +4 H_0^2 H_{0,-1} P_{211}
                        -8 H_{0,0,0,-1}
                         P_{237}
                        +8 H_{0,0,0,1} P_{251}
                        +\Big(
                                16 H_0 H_1 P_{172}
                                +2 H_0^2 P_{173}
                                -4 H_{0,1} P_{192}
                                +4 H_{0,-1} P_{217}
                        \Big) \zeta_2
                        +
                        \frac{1}{5} P_{168} \zeta_2^2
                        -2 H_0 P_{244} \zeta_3
                \Big)
                +\xi ^2 \Big(
                        \Big(
                                -8 H_1
                                +8 H_{-1}
                        \Big) \zeta_3
                        +\Big(
                                16 H_{-1} H_0
                                -16 H_0 H_1
                                -8 H_{0,-1}
                        \Big) H_{0,1}
                        -4 H_{0,1}^2
                        -16 H_{-1} H_0 H_{0,-1}
                        +16 H_0 H_1 H_{0,-1}
                        +12 H_{0,-1}^2
                        +\Big(
                                32 H_1
                                -32 H_{-1}
                        \Big) H_{0,0,1}
                        +\Big(
                                -32 H_1
                                +32 H_{-1}
                        \Big) H_{0,0,-1}
                        +24 H_0 H_{0,1,1}
                        -8 H_0 H_{0,1,-1}
                        -8 H_0 H_{0,-1,1}
                        -8 H_0 H_{0,-1,-1}
                        -32 H_{0,0,1,1}
                        +32 H_{0,0,1,-1}
                        +32 H_{0,0,-1,1}
                        -32 H_{0,0,-1,-1}
                \Big)
                +\eta  \Big(
                        \Big(
                                \frac{1}{54} H_0 P_3
                                +4 H_0 H_1 P_{11}
                                -4 H_{0,1} P_{11}
                                +\frac{4}{9} H_{-1} H_0 P_{35}
                                -\frac{4}{9} H_{0,-1} P_{35}
                        \Big) x_+^2
                        +\Big(
                                4 H_0^2 H_1 P_{49}
                                +8 H_{0,0,1} P_{57}
                                +\frac{2}{3} H_{-1} H_0^2 P_{67}
                                -\frac{4}{3} H_{0,0,-1} P_{73}
                                +\frac{1}{18} H_0^2 P_{77}
                                +\Big(
                                        \frac{4}{3} H_{-1} P_{56}
                                        +\frac{P_{79}}{9}
                                \Big) \zeta_2
                                -\frac{2}{3} P_{50} \zeta_3
                        \Big) x_+^3
                        +\Big(
                                \frac{1}{9} H_0^3 P_{84}
                                +4 H_0^2 H_{0,1} P_{96}
                                -\frac{2}{3} H_0 P_{143} \zeta_2
                        \Big) x_+^4
                \Big)
                +\xi  \Big(
                        \eta  \Big(
                                \frac{4}{3} \Big(
                                        3+x^2\Big) H_{-1}
                                 H_0^3
                                +16 H_{-1} H_0 \zeta_2
                        \Big)
                        -
                        \frac{104}{3} H_{-1}^2 H_0
                        +24 H_{-1} H_0 H_1
                        -4 H_0 H_1^2
                        +\Big(
                                8 H_1
                                -24 H_{-1}
                        \Big) H_{0,1}
                        +\Big(
                                -24 H_1
                                +\frac{208 H_{-1}}{3}
                        \Big) H_{0,-1}
                        -8 H_{0,1,1}
                        +24 H_{0,1,-1}
                        +24 H_{0,-1,1}
                        -\frac{208}{3} H_{0,-1,-1}
                        +4 H_1 \zeta_2
                        -24 (-1+x) \ln (2) x_+ \zeta_2
                \Big)
                +\Big(
                        24 H_0 H_{0,1} P_9
                        -4 H_0 H_{0,-1} P_{17}
                \Big) x_+^4
                -8 (-1+x) H_0 H_{0,0,-1} P_8 x_+^5
                -96 x \ln (2) x_+^2 \zeta_2
\Bigg\}
+ \ep  \Bigg\{
                \frac{1}{162} \Big(
                        -28745-70126 x-28745 x^2\Big) x_+^2
                +\ln (2) x_+^4 \zeta_2 \Big(
                        12 P_4
                        -24 H_1 P_{12}
                        -24 H_{-1} P_{12}
                \Big)
                +\eta  \Big(
                        \ln (2) x_+^4 \zeta_2 \Big(
                                1536 x^3 H_{0,1}
                                +1536 x^3 H_{0,-1}
                        \Big)
                        +\Big(
                                \frac{1}{324} H_0 P_1
                                +4 H_0 H_1^2 P_{11}
                                +8 H_{0,1,1} P_{11}
                                -24 H_{0,1,-1} P_{11}
                                -24 H_{0,-1,1} P_{11}
                                +\frac{4}{9} H_{-1}^2 H_0 P_{34}
                                +\frac{8}{9} H_{0,-1,-1} P_{34}
                                +\frac{1}{27} H_{-1} H_0 P_{45}
                                +\Big(
                                        -24 H_{-1} H_0 P_{11}
                                        +4 H_0 P_{15}
                                \Big) H_1
                                +\Big(
                                        -8 H_1 P_{11}
                                        +24 H_{-1} P_{11}
                                        -4 P_{15}
                                \Big) H_{0,1}
                                +\Big(
                                        \frac{P_2}{27}
                                        +24 H_1 P_{11}
                                        -\frac{8}{9} H_{-1} P_{34}
                                \Big) H_{0,-1}
                                -4 H_1 P_{11} \zeta_2
                        \Big) x_+^2
                        +\Big(
                                -8 H_{0,0,1,1} P_{58}
                                +2 H_0^2 H_1^2 P_{59}
                                +8 H_{0,0,1,-1} P_{62}
                                +8 H_{0,0,-1,1} P_{62}
                                -2 H_{-1}^2 H_0^2 P_{68}
                                -8 H_{0,0,-1,-1} P_{70}
                                +\frac{1}{9} H_{-1} H_0^2 P_{74}
                                +\frac{1}{54} H_0^2 P_{78}
                                +\Big(
                                        4 H_{-1} H_0^2 P_{55}
                                        +2 H_0^2 P_{64}
                                \Big) H_1
                                +\Big(
                                        8 H_1 P_{58}
                                        -4 P_{60}
                                        -8 H_{-1} P_{62}
                                \Big) H_{0,0,1}
                                +\Big(
                                        8 H_1 P_{54}
                                        +8 H_{-1} P_{70}
                                        +\frac{2 P_{80}
                                        }{9}
                                \Big) H_{0,0,-1}
                                +\Big(
                                        -48 H_{-1} H_1 P_{48}
                                        -12 H_{-1}^2 P_{52}
                                        -
                                        \frac{2}{9} H_{-1} P_{76}
                                        +\frac{P_{81}}{27}
                                \Big) \zeta_2
                                +\Big(
                                        \frac{P_{46}}{9}
                                        -2 H_{-1} P_{65}
                                        -2 H_1 P_{69}
                                \Big) \zeta_3
                        \Big) x_+^3
                        +\Big(
                                \frac{1}{12} H_0^4 P_{85}
                                +32 H_0 H_{0,0,1,-1} P_{87}
                                +32 H_0 H_{0,0,-1,1} P_{87}
                                +48 H_0 H_{0,-1}^2 P_{93}
                                +16 H_0 H_{0,1}^2 P_{101}
                                -16 H_0 H_{0,-1,0,1} P_{110}
                                +4 H_0 H_{0,0,1} P_{116}
                                -16 H_0 H_{0,0,-1,-1} P_{120}
                                +\frac{2}{3} H_0^3 H_1 P_{146}
                                +2 H_0^2 H_{0,-1} P_{150}
                                +4 H_{0,0,0,1} P_{152}
                                +\frac{2}{3} H_{-1} H_0^3 P_{153}
\end{dmath*}
\begin{dmath*}
{\color{white}=}
                                -4 H_{0,0,0,-1} P_{155}
                                +\frac{1}{54} H_0^3 P_{162}
                                +\Big(
                                        8 H_{-1} H_0^2 P_{96}
                                        -16 H_0^2 H_1 P_{104}
                                        -16 H_0 H_{0,-1} P_{108}
                                        -2 H_0^2 P_{147}
                                \Big) H_{0,1}
                                +\Big(
                                        -32 H_0 H_1 P_{105}
                                        -4 H_0 P_{148}
                                \Big) H_{0,0,-1}
                                +\Big(
                                        \frac{1}{18} H_0 P_{82}
                                        +2 H_{-1} H_0 P_{126}
                                        -2 H_{0,-1} P_{130}
                                        -4 H_{0,1} P_{131}
                                        +4 H_0 H_1 P_{139}
                                        +\frac{1}{2} H_0^2 P_{159}
                                \Big) \zeta_2
                                +\frac{1}{5} P_{160} \zeta_2^2
                                +\frac{4}{3} H_0 P_{144} \zeta_3
                        \Big) x_+^4
                \Big)
                +\xi  \Big(
                        \eta  \Big(
                                \ln (2) x_+^3 \zeta_2 \Big(
                                        96 (-1+x) x^2 H_{0,1}
                                        +96 (-1+x) x^2 H_{0,-1}
                                \Big)
                                -
                                \frac{4}{3} \Big(
                                        3+5 x^2\Big) H_{-1}^2 H_0^3
                                -8 \Big(
                                        1+3 x^2\Big) H_{-1}^2 H_0 \zeta_2
                        \Big)
                        +80 H_{-1}^3 H_0
                        -72 H_{-1}^2 H_0 H_1
                        +24 H_{-1} H_0 H_1^2
                        -\frac{8}{3} H_0 H_1^3
                        +\Big(
                                -48 H_{-1} H_1
                                +8 H_1^2
                                +72 H_{-1}^2
                        \Big) H_{0,1}
                        +\Big(
                                144 H_{-1} H_1
                                -24 H_1^2
                                -240 H_{-1}^2
                        \Big) H_{0,-1}
                        +\Big(
                                -16 H_1
                                +48 H_{-1}
                        \Big) H_{0,1,1}
                        +\Big(
                                48 H_1
                                -144 H_{-1}
                        \Big) H_{0,1,-1}
                        +\Big(
                                48 H_1
                                -144 H_{-1}
                        \Big) H_{0,-1,1}
                        +\Big(
                                -144 H_1
                                +480 H_{-1}
                        \Big) H_{0,-1,-1}
                        +16 H_{0,1,1,1}
                        -48 H_{0,1,1,-1}
                        -48 H_{0,1,-1,1}
                        +144 H_{0,1,-1,-1}
                        -48 H_{0,-1,1,1}
                        +144 H_{0,-1,1,-1}
                        +144 H_{0,-1,-1,1}
                        -480 H_{0,-1,-1,-1}
                        +4 H_1^2 \zeta_2
                \Big)
                +\eta ^2 \Big(
                        \Big(
                                -\frac{1}{30} x H_0^5 P_{138}
                                +128 H_0 H_{0,0,1,1} P_{187}
                                -8 H_0^2 H_{0,1,1} P_{202}
                                +8 H_0^2 H_{0,1,-1} P_{205}
                                +8 H_0^2 H_{0,-1,1} P_{205}
                                +\frac{1}{3} H_{-1} H_0^4 P_{212}
                                -\frac{4}{3} H_0^3 H_1^2 P_{215}
                                -\frac{8}{3} H_0^3 H_{0,1} P_{229}
                                -16 H_{0,0,1,0,1} P_{246}
                                -8 H_0^2 H_{0,-1,-1} P_{247}
                                -32 H_{0,0,-1,0,-1} P_{252}
                                +16 H_{0,0,-1,0,1} P_{255}
                                +16 H_{0,0,1,0,-1}
                                 P_{262}
                                +32 H_{0,0,0,1,-1} P_{283}
                                +32 H_{0,0,0,-1,1} P_{283}
                                -16 H_{0,0,0,1,1} P_{298}
                                +8 H_{0,0,0,0,1} P_{301}
                                -8 H_{0,0,0,0,-1} P_{302}
                                -16 H_{0,0,0,-1,-1} P_{304}
                                +\Big(
                                        \frac{1}{3} H_0^4 P_{86}
                                        +\frac{8}{3} H_{-1} H_0^3 P_{215}
                                \Big) H_1
                                +\Big(
                                        -8 H_{-1} H_0^2 P_{118}
                                        -8 H_0^2 H_1 P_{201}
                                        +\frac{4}{3} H_0^3 P_{288}
                                \Big) H_{0,-1}
                                +\Big(
                                        32 H_{-1} H_0 P_{171}
                                        -32 H_0 H_1 P_{210}
                                        +16 H_{0,1} P_{221}
                                        -16 H_{0,-1} P_{263}
                                        +4 H_0^2 P_{269}
                                \Big) H_{0,0,1}
                                +\Big(
                                        -16 H_{0,1} P_{233}
                                        +32 H_{0,-1} P_{243}
                                        -4 H_0^2 P_{281}
                                \Big) H_{0,0,-1}
                                +\Big(
                                        -16 H_{-1} P_{214}
                                        +16 H_1 P_{259}
                                        -8 H_0 P_{286}
                                \Big) H_{0,0,0,1}
                                +\Big(
                                        16 H_{-1} P_{226}
                                        -16 H_1 P_{240}
                                        +8 H_0 P_{280}
                                \Big) H_{0,0,0,-1}
                                +\Big(
                                        -8 x^2 H_{-1} H_0^2 P_{47}
                                        -8 H_0 H_1^2 P_{206}
                                        -8 H_{0,1,1} P_{234}
                                        +\frac{4}{3} H_0^3 P_{261}
                                        -16 H_{0,0,-1} P_{264}
                                        +8 H_{0,1,-1} P_{270}
                                        +8 H_{0,-1,1} P_{271}
                                        -8 H_{0,-1,-1} P_{276}
                                        +8 H_{0,0,1} P_{282}
                                        +\Big(
                                                4 x H_0^2 P_{107}
                                                +16 H_{-1} H_0 P_{206}
                                        \Big) H_1
                                        +\Big(
                                                32 x H_1 P_{102}
                                                -32 x H_{-1} P_{102}
                                                -4 H_0 P_{279}
                                        \Big) H_{0,1}
                                        +\Big(
                                                16 H_{-1} P_{208}
                                                -16 H_1 P_{235}
                                                +4 H_0 P_{277}
                                        \Big) H_{0,-1}
                                        +2 P_{267} \zeta_3
                                \Big) \zeta_2
                                +\Big(
                                        \frac{8}{5} H_{-1} P_{222}
                                        -\frac{2}{5} H_0 P_{260}
                                        -
                                        \frac{4}{5} H_1 P_{265}
                                \Big) \zeta_2^2
                                +\Big(
                                        16 H_0 H_1 P_{181}
                                        -16 H_{-1} H_0 P_{191}
                                        -4 H_{0,1} P_{258}
                                        +2 H_0^2 P_{272}
                                        +4 H_{0,-1} P_{273}
                                \Big) \zeta_3
                                +P_{167} \zeta_5
                        \Big) x_+^3
                        -8 H_{-1} H_0^2 x_+^2 \zeta_2
                \Big)
                +\xi ^2 \Big(
                        \Big(
                                -64 H_{-1} H_0 H_{0,-1}
                                +48 H_{0,-1}^2
                        \Big) H_1
                        +\Big(
                                -32 H_{-1}^2 H_0
                                +\Big(
                                        64 H_{-1} H_0
                                        -32 H_{0,-1}
                                \Big) H_1
                                -32 H_0 H_1^2
                                +32 H_{-1} H_{0,-1}
                        \Big) H_{0,1}
                        +\Big(
                                -16 H_1
                                +16 H_{-1}
                        \Big) H_{0,1}^2
                        +32 H_{-1}^2 H_0 H_{0,-1}
                        +32 H_0 H_1^2 H_{0,-1}
                        -48 H_{-1} H_{0,-1}^2
                        +\Big(
                                -128 H_{-1} H_1
                                +64 H_1^2
                                +64 H_{-1}^2
                        \Big) H_{0,0,1}
                        +\Big(
                                128 H_{-1} H_1
                                -64 H_1^2
                                -64 H_{-1}^2
                        \Big) H_{0,0,-1}
                        +\Big(
                                -96 H_{-1} H_0
                                +96 H_0 H_1
                                +16 H_{0,1}
                                +80 H_{0,-1}
                        \Big) H_{0,1,1}
                        +\Big(
                                32 H_{-1} H_0
                                -32 H_0 H_1
                                +16 H_{0,1}
                                -48 H_{0,-1}
                        \Big) H_{0,1,-1}
                        +\Big(
                                32 H_{-1} H_0
                                -32 H_0 H_1
                                -48 H_{0,1}
                                -48 H_{0,-1}
                        \Big) H_{0,-1,1}
                        +\Big(
                                32 H_{-1} H_0
                                -32 H_0 H_1
                                +16 H_{0,1}
                                -48 H_{0,-1}
                        \Big) H_{0,-1,-1}
                        +\Big(
                                -128 H_1
                                +128 H_{-1}
                        \Big) H_{0,0,1,1}
                        +\Big(
                                128 H_1
                                -128 H_{-1}
                        \Big) H_{0,0,1,-1}
                        +\Big(
                                128 H_1
                                -128 H_{-1}
                        \Big) H_{0,0,-1,1}
                        +\Big(
                                -128 H_1
                                +128 H_{-1}
                        \Big) H_{0,0,-1,-1}
                        -112 H_0 H_{0,1,1,1}
                        +16 H_0 H_{0,1,1,-1}
                        +16 H_0 H_{0,1,-1,1}
\end{dmath*}
\begin{dmath*}
{\color{white}=}
                        +16 H_0 H_{0,1,-1,-1}
                        +16 H_0 H_{0,-1,1,1}
                        +16 H_0 H_{0,-1,1,-1}
                        +16 H_0 H_{0,-1,-1,1}
                        +16 H_0 H_{0,-1,-1,-1}
                        +128 H_{0,0,1,1,1}
                        -256 H_{0,0,1,1,-1}
                        -128 H_{0,0,1,-1,1}
                        +128 H_{0,0,1,-1,-1}
                        +128 H_{0,0,-1,1,-1}
                        +128 H_{0,0,-1,-1,1}
                        +256 H_{0,0,-1,-1,-1}
                        -64 H_{0,1,0,1,-1}
                        +64 H_{0,-1,0,1,1}
                        +192 H_{0,-1,0,-1,-1}
                        +64 H_{0,-1,1,0,1}
                        +\Big(
                                32 H_{-1} H_1
                                -16 H_1^2
                                -16 H_{-1}^2
                        \Big) \zeta_3
                \Big)
                -4 c_1 \Big(
                        1-4 x+x^2\Big) x_+^2
                +\Big(
                        8 H_{-1} H_0 H_{0,-1} P_5
                        -8 H_0 H_{0,1,1} P_{10}
                        -8 H_{0,1}^2 P_{14}
                        -24 H_{0,-1,0,1} P_{16}
                        -16 H_0 H_{0,1,-1} P_{22}
                        -16 H_0 H_{0,-1,1} P_{22}
                        +8 H_0 H_1 H_{0,-1} P_{23}
                        +16 H_0 H_{0,-1} P_{25}
                        +8 H_0 H_{0,-1,-1} P_{31}
                        -4 H_{0,-1}^2 P_{33}
                        +\Big(
                                -16 H_{-1} H_0 P_{13}
                                +4 H_0 P_{18}
                                +8 H_0 H_1 P_{26}
                                +16 H_{0,-1} P_{28}
                        \Big) H_{0,1}
                        -24 H_{-1,1} P_{12} \zeta_2
                \Big) x_+^4
                -16 (-1+x) H_{-1} H_0 H_{0,0,-1} P_8 x_+^5
\Bigg\}
\Bigg]
\\
+ C_F n_l T_F  \Bigg[
\frac{1}{\ep^2}  \Bigg\{
        -\frac{4}{3}
        -\frac{4 \xi  H_0}{3}
\Bigg\}
+ \frac{1}{\ep}  \Bigg\{
        \frac{20}{9}
        +\frac{20 \xi  H_0}{9}
\Bigg\}
+  \Bigg\{
                \frac{424}{27}
                +\eta  \Big(
                        \frac{2}{27} \Big(
                                209+6 x+209 x^2\Big) H_0
                        -\frac{8}{9} \Big(
                                19+6 x+19 x^2\Big) H_{-1} H_0
                        +\frac{2}{9} \Big(
                                19+6 x+19 x^2\Big) H_0^2
                        +\frac{8}{9} \Big(
                                19+6 x+19 x^2\Big) H_{0,-1}
                        -\frac{4}{9} \Big(
                                7+6 x+31 x^2\Big) \zeta_2
                \Big)
                +\xi  \Big(
                        \frac{16}{3} H_{-1}^2 H_0
                        -\frac{8}{3} H_{-1} H_0^2
                        +\frac{4}{9} H_0^3
                        -\frac{32}{3} H_{-1} H_{0,-1}
                        +\frac{16}{3} H_{0,0,-1}
                        +\frac{32}{3} H_{0,-1,-1}
                        +\Big(
                                \frac{8}{3} H_0
                                +\frac{16}{3} H_{-1}
                        \Big) \zeta_2
                        -\frac{16 \zeta_3}{3}
                \Big)
\Bigg\}
+ \ep  \Bigg\{
                \frac{5204}{81}
                +\xi  \Big(
                        \Big(
                                -\frac{16 H_0}{3}
                                +32 H_{-1}
                        \Big) \zeta_3
                        -\frac{32}{3} H_{-1}^3 H_0
                        +8 H_{-1}^2 H_0^2
                        -\frac{8}{3} H_{-1} H_0^3
                        +32 H_{-1}^2 H_{0,-1}
                        -32 H_{-1} H_{0,0,-1}
                        -64 H_{-1} H_{0,-1,-1}
                        +16 H_{0,0,0,-1}
                        +32 H_{0,0,-1,-1}
                        +64 H_{0,-1,-1,-1}
                        +\Big(
                                -8 H_{-1} H_0
                                +2 H_0^2
                                -16 H_{-1}^2
                                +24 H_{0,-1}
                        \Big) \zeta_2
                        -\frac{96 \zeta_2^2}{5}
                \Big)
                +\eta  \Big(
                        \frac{1}{81} \Big(
                                5813-1218 x+5813 x^2\Big) H_0
                        +\frac{8}{9} \Big(
                                47+18 x+47 x^2\Big) H_{-1}^2 H_0
                        +\frac{2}{27} \Big(
                                281+6 x+281 x^2\Big) H_0^2
                        +\frac{2}{27} \Big(
                                47+18 x+47 x^2\Big) H_0^3
                        +\frac{1}{3} \Big(
                                1+x^2\Big) H_0^4
                        +\Big(
                                -\frac{8}{27} \Big(
                                        281+6 x+281 x^2\Big) H_0
                                -\frac{4}{9} \Big(
                                        47+18 x+47 x^2\Big) H_0^2
                        \Big) H_{-1}
                        +\Big(
                                \frac{8}{27} \Big(
                                        281+6 x+281 x^2\Big)
                                -\frac{16}{9} \Big(
                                        47+18 x+47 x^2\Big) H_{-1}
                        \Big) H_{0,-1}
                        +\frac{8}{9} \Big(
                                47+18 x+47 x^2\Big) H_{0,0,-1}
                        +\frac{16}{9} \Big(
                                47+18 x+47 x^2\Big) H_{0,-1,-1}
                        +\Big(
                                -\frac{8}{27} \Big(
                                        22+3 x+259 x^2\Big)
                                +\frac{2}{9} \Big(
                                        37+18 x+37 x^2\Big) H_0
                                +\frac{8}
                                {9} \Big(
                                        47+18 x+47 x^2\Big) H_{-1}
                        \Big) \zeta_2
                        -
                        \frac{8}{9} \Big(
                                35+18 x+59 x^2\Big) \zeta_3
                \Big)
\Bigg\}
\Bigg]
\\
+ C_F T_F  \Bigg[
%
%
%
 \Bigg\{
                \frac{4}{27} \Big(
                        383+178 x+383 x^2\Big) x_+^2
                +\eta  \Big(
                        \frac{2}{27} H_0 P_{42} x_+^2
                        +\Big(
                                \frac{8}{3} x H_0^3 P_6
                                +4 H_0 P_{90} \zeta_2
                        \Big) x_+^4
                \Big)
                -\frac{4}{9} (-1+x) \xi  \eta  H_0^3 P_6 x_+^3
                +\Big(
                        \frac{2}{9} H_0^2 P_{32}
                        -\frac{8}{3} P_7 \zeta_2
                \Big) x_+^4
\Bigg\}
+ \ep  \Bigg\{
                \frac{2}{27} \Big(
                        2069+1990 x+2069 x^2\Big) x_+^2
                +\frac{32}{3} (-1+x)^2 \Big(
                        7+72 x+7 x^2\Big) \ln (2) x_+^4 \zeta_2
                +\xi  \eta  x_+^3 \Big(
                        -\frac{8}{9} (-1+x) H_{-1} H_0^3 P_6
                        -\frac{1}{3} (-1+x) H_0^4 P_6
                        -\frac{16}{3} (-1+x) H_0^2 H_{0,1} P_6
                        +\frac{16}{3} (-1+x) H_0^2 H_{0,-1} P_6
                        +\frac{64}{3} (-1+x) H_0 H_{0,0,1} P_6
                        -\frac{32}{3} (-1+x) H_0 H_{0,0,-1} P_6
                        -32 (-1+x) H_{0,0,0,1} P_6
                        +\frac{16}{3} (-1+x) H_{0,0,0,-1} P_6
                \Big)
\end{dmath*}
\begin{dmath*}
{\color{white}=}
                +\eta  \Big(
                        \Big(
                                -\frac{4}{27} H_{-1} H_0 P_{42}
                                +\frac{4}{27} H_{0,-1} P_{42}
                                +\frac{1}{27} H_0 P_{44}
                        \Big) x_+^2
                        +\Big(
                                -\frac{2}{27} H_0^2 P_{61}
                                +\frac{4 P_{75} \zeta_2}{27}
                        \Big) x_+^3
                        +\Big(
                                \frac{16}{3} x H_{-1} H_0^3 P_6
                                +2 x H_0^4 P_6
                                +32 x H_0^2 H_{0,1} P_6
                                -32 x H_0^2 H_{0,-1} P_6
                                -128 x H_0 H_{0,0,1} P_6
                                +64 x H_0 H_{0,0,-1} P_6
                                +192 x H_{0,0,0,1} P_6
                                -32 x H_{0,0,0,-1} P_6
                                -\frac{2}{27} H_0^3 P_{157}
                                +\Big(
                                        \frac{8}{3} H_{-1} H_0 P_{89}
                                        +2 H_0^2 P_{90}
                                        -\frac{8}{3} H_{0,-1} P_{113}
                                        -\frac{2}
                                        {9} H_0 P_{158}
                                \Big) \zeta_2
                                -
                                \frac{8}{5} P_{91} \zeta_2^2
                                -\frac{8}{9} H_0 P_{114} \zeta_3
                        \Big) x_+^4
                \Big)
                +\Big(
                        -\frac{4}{9} H_{-1} H_0^2 P_{32}
                        +\frac{8}{9} H_0^2 H_1 P_{32}
                        -\frac{16}{9} H_0 H_{0,1} P_{32}
                        +\frac{16}{9} H_{0,0,1} P_{32}
                        +\frac{8}{9} H_{0,0,-1} P_{32}
                        -\frac{8}{3} H_{-1} P_{32} \zeta_2
                        -\frac{8}{9} P_{38} \zeta_3
                \Big) x_+^4
\Bigg\}
\Bigg] \,.
\end{dmath*}
%
The polynomials are defined as
\begin{align}
P_1 &= -70165 x^4-177188 x^3-77966 x^2-177188 x-70165,\nonumber\\
P_2 &= -3623 x^4-23056 x^3+6062 x^2-23056 x-3623,\nonumber\\
P_3 &= -2545 x^4-6812 x^3-4646 x^2-6812 x-2545,\nonumber\\
P_4 &= x^4-56 x^3-42 x^2-56 x+1,\nonumber\\
P_5 &= x^4-16 x^3+26 x^2-16 x+1,\nonumber\\
P_6 &= x^4-2 x^3+12 x^2-2 x+1,\nonumber\\
P_7 &= x^4+37 x^3-124 x^2+37 x+1,\nonumber\\
P_8 &= 2 x^4-9 x^3+122 x^2-9 x+2,\nonumber\\
P_9 &= 2 x^4+9 x^3+18 x^2+9 x+2,\nonumber\\
P_{10} &= 3 x^4-94 x^3+322 x^2-94 x+3,\nonumber\\
P_{11} &= 3 x^4-8 x^3+26 x^2-8 x+3,\nonumber\\
P_{12} &= 3 x^4+16 x^3+62 x^2+16 x+3,\nonumber\\
P_{13} &= 4 x^4-13 x^3+140 x^2-13 x+4,\nonumber\\
P_{14} &= 4 x^4+27 x^3+10 x^2+27 x+4,\nonumber\\
P_{15} &= 5 x^4-90 x^3+138 x^2-90 x+5,\nonumber\\
P_{16} &= 5 x^4+26 x^3+46 x^2+26 x+5,\nonumber\\
P_{17} &= 5 x^4+79 x^3-146 x^2+79 x+5,\nonumber\\
P_{18} &= 7 x^4-42 x^3+234 x^2-42 x+7,\nonumber\\
P_{19} &= 7 x^4-3 x^3+349 x^2-3 x+7,\nonumber\\
P_{20} &= 7 x^4+7 x^3-48 x^2+7 x+7,\nonumber\\
P_{21} &= 7 x^4+11 x^3+284 x^2+11 x+7,\nonumber\\
P_{22} &= 7 x^4+85 x^3-96 x^2+85 x+7,\nonumber\\
P_{23} &= 7 x^4+104 x^3-142 x^2+104 x+7,\nonumber\\
P_{24} &= 8 x^4+15 x^3+194 x^2+15 x+8,\nonumber\\
P_{25} &= 10 x^4+46 x^3+29 x^2+46 x+10,\nonumber\\
P_{26} &= 11 x^4-40 x^3+342 x^2-40 x+11,\nonumber\\
P_{27} &= 11 x^4+22 x^3+58 x^2+22 x+11,\nonumber\\
P_{28} &= 11 x^4+72 x^3+44 x^2+72 x+11,\nonumber\\
P_{29} &= 15 x^4+37 x^3+164 x^2+37 x+15,\nonumber\\
P_{30} &= 17 x^4+38 x^3+630 x^2+38 x+17,\nonumber\\
P_{31} &= 18 x^4+227 x^3-272 x^2+227 x+18,\nonumber\\
P_{32} &= 19 x^4-2 x^3+206 x^2-2 x+19,\nonumber\\
P_{33} &= 19 x^4+211 x^3-246 x^2+211 x+19,\nonumber\\
P_{34} &= 28 x^4-187 x^3+110 x^2-187 x+28,\nonumber\\
P_{35} &= 31 x^4+59 x^3+164 x^2+59 x+31,\nonumber\\
P_{36} &= 43 x^4+20 x^3+130 x^2+20 x+43,\nonumber\\
P_{37} &= 55 x^4-48 x^3+562 x^2-48 x+55,\nonumber\\
P_{38} &= 69 x^4+408 x^3-698 x^2+408 x+69,\nonumber\\
P_{39} &= 79 x^4+142 x^3+1698 x^2+142 x+79,\nonumber\\
P_{40} &= 81 x^4-104 x^3+766 x^2-104 x+81,\nonumber\\
P_{41} &= 257 x^4-48 x^3+2078 x^2-48 x+257,\nonumber\\
P_{42} &= 265 x^4-124 x^3+3494 x^2-124 x+265,\nonumber\\
P_{43} &= 479 x^4+704 x^3-30 x^2+704 x+479,\nonumber\\
P_{44} &= 1719 x^4+452 x^3+20170 x^2+452 x+1719,\nonumber\\
P_{45} &= 3623 x^4+23056 x^3-6062 x^2+23056 x+3623,\nonumber\\
P_{46} &= -1045 x^5-8139 x^4+3518 x^3+1634 x^2+8151 x+1577,\nonumber\\
P_{47} &= x^5+5 x^4-9 x^3+60 x^2-46 x+19,\nonumber\\
P_{48} &= 2 x^5+8 x^4+25 x^3-21 x^2-5 x-1,\nonumber\\
P_{49} &= 2 x^5+13 x^4-5 x^3-11 x^2-25 x-6,\nonumber\\
P_{50} &= 5 x^5-285 x^4+1592 x^3-1840 x^2+99 x-67,\nonumber\\
P_{51} &= 5 x^5-30 x^4+53 x^3-61 x^2+24 x-7,\nonumber\\
P_{52} &= 5 x^5-26 x^4+87 x^3-119 x^2+2 x-13,\nonumber\\
P_{53} &= 5 x^5+8 x^4+6 x^3-22 x^2-16 x-13,\nonumber\\
P_{54} &= 5 x^5+171 x^4-494 x^3+398 x^2-243 x-29,\nonumber\\
P_{55} &= 9 x^5+23 x^4+2 x^3+94 x^2+49 x+15,\nonumber\\
P_{56} &= 10 x^5-105 x^4+139 x^3-203 x^2+57 x-26,\nonumber\\
P_{57} &= 10 x^5+29 x^4+59 x^3-43 x^2-17 x-6,\nonumber\\
P_{58} &= 11 x^5-81 x^4+480 x^3-448 x^2+105 x-3,\nonumber\\
P_{59} &= 11 x^5-21 x^4+284 x^3-316 x^2-3 x-19,\nonumber\\
P_{60} &= 13 x^5+61 x^4-316 x^3+460 x^2-101 x+11,\nonumber\\
P_{61} &= 16 x^5-120 x^4-287 x^3-3083 x^2-21 x-281,\nonumber\\
P_{62} &= 25 x^5-45 x^4+614 x^3-518 x^2+117 x-1,\nonumber\\
P_{63} &= 26 x^5-45 x^4+305 x^3-313 x^2+39 x-28,\nonumber\\
P_{64} &= 27 x^5-37 x^4+236 x^3-92 x^2-3 x-3,\nonumber\\
P_{65} &= 41 x^5+423 x^4-482 x^3+1106 x^2+45 x+115,\nonumber\\
P_{66} &= 42 x^5+136 x^4+87 x^3+89 x^2-4 x+2,\nonumber\\
P_{67} &= 53 x^5-12 x^4+851 x^3-571 x^2+222 x+17,\nonumber\\
P_{68} &= 56 x^5+43 x^4+695 x^3-383 x^2+191 x+22,\nonumber\\
P_{69} &= 57 x^5+183 x^4+430 x^3-606 x^2-315 x-101,\nonumber\\
P_{70} &= 58 x^5+9 x^4+779 x^3-467 x^2+225 x+20,\nonumber\\
P_{71} &= 77 x^5+139 x^4+406 x^3+298 x^2+53 x+51,\nonumber\\
P_{72} &= 83 x^5-217 x^4+2594 x^3-646 x^2+59 x+47,\nonumber\\
P_{73} &= 83 x^5+432 x^4-499 x^3+779 x^2-222 x-13,\nonumber\\
P_{74} &= 97 x^5+1533 x^4+3340 x^3-4976 x^2-357 x-533,\nonumber\\
P_{75} &= 139 x^5+2247 x^4-9395 x^3+6025 x^2-2388 x-404,\nonumber\\
P_{76} &= 187 x^5+399 x^4-5192 x^3+9388 x^2-843 x+349,\nonumber\\
P_{77} &= 217 x^5+81 x^4-860 x^3+616 x^2-621 x-233,\nonumber\\
P_{78} &= 254 x^5-25098 x^4-1801 x^3-10009 x^2-10761 x-3337,\nonumber\\
P_{79} &= 347 x^5-1737 x^4+5228 x^3-3040 x^2+1737 x-7,\nonumber\\
P_{80} &= 1343 x^5+3651 x^4-5788 x^3+7424 x^2-4827 x-907,\nonumber\\
P_{81} &= 4774 x^5-4044 x^4+30811 x^3-3449 x^2+12363 x-71,\nonumber\\
P_{82} &= -1367 x^6+1822 x^5-25085 x^4+19772 x^3-1829 x^2+2110 x-575,\nonumber\\
P_{83} &= -399 x^6-124 x^5-2335 x^4-1448 x^3+2075 x^2-124 x+307,\nonumber\\
P_{84} &= -89 x^6+154 x^5-1547 x^4+1262 x^3-29 x^2+4 x-11,\nonumber\\
P_{85} &= -29 x^6+188 x^5-335 x^4+974 x^3+67 x^2+2 x-11,\nonumber\\
P_{86} &= -9 x^6+55 x^5-422 x^4+254 x^3-175 x^2-63 x-24,\nonumber\\
P_{87} &= x^6-31 x^5+213 x^4-428 x^3+213 x^2-31 x+1,\nonumber\\
P_{88} &= x^6-12 x^5+14 x^4-38 x^3+14 x^2-12 x+1,\nonumber\\
P_{89} &= x^6+4 x^5-5 x^4+128 x^3-5 x^2+4 x+1,\nonumber\\
P_{90} &= x^6+4 x^5+3 x^4+48 x^3+3 x^2+4 x+1,\nonumber\\
P_{91} &= x^6+4 x^5+6 x^4+18 x^3+6 x^2+4 x+1,\nonumber\\
P_{92} &= x^6+4 x^5+11 x^4+11 x^2+4 x+1,\nonumber\\
P_{93} &= x^6+6 x^5-4 x^4+21 x^3-4 x^2+6 x+1,\nonumber\\
P_{94} &= x^6+7 x^5-23 x^4+57 x^3-50 x^2+28 x-4,\nonumber\\
P_{95} &= 2 x^6-11 x^5-114 x^4+182 x^3-114 x^2-11 x+2,\nonumber\\
P_{96} &= 2 x^6-5 x^5+92 x^4-162 x^3+92 x^2-5 x+2,\nonumber\\
P_{97} &= 2 x^6+6 x^5+33 x^4-68 x^3+33 x^2+6 x+2,\nonumber\\
P_{98} &= 2 x^6+9 x^5-21 x^4+102 x^3-88 x^2+31 x-3,\nonumber\\
P_{99} &= 2 x^6+11 x^5-9 x^4+48 x^3-9 x^2+11 x+2,\nonumber\\
P_{100} &= 2 x^6+13 x^5+246 x^4-274 x^3+248 x^2+17 x+4,\nonumber\\
P_{101} &= 3 x^6-4 x^5+115 x^4-302 x^3+115 x^2-4 x+3,\nonumber\\
P_{102} &= 3 x^6+4 x^5+47 x^4-116 x^3+137 x^2-32 x+5,\nonumber\\
P_{103} &= 3 x^6+5 x^5+153 x^4-202 x^3+155 x^2+9 x+5,\nonumber\\
P_{104} &= 3 x^6+13 x^5+22 x^4+12 x^3+22 x^2+13 x+3,\nonumber\\
P_{105} &= 3 x^6+24 x^5-37 x^4+196 x^3-37 x^2+24 x+3,\nonumber\\
P_{106} &= 4 x^6+6 x^5+141 x^4-220 x^3+141 x^2+6 x+4,\nonumber\\
P_{107} &= 4 x^6+41 x^5-173 x^4+608 x^3-580 x^2+193 x-29,\nonumber\\
P_{108} &= 5 x^6-24 x^5+291 x^4-698 x^3+291 x^2-24 x+5,\nonumber\\
P_{109} &= 5 x^6+14 x^5-19 x^4+64 x^3-19 x^2+14 x+5,\nonumber\\
P_{110} &= 5 x^6+59 x^5-175 x^4+508 x^3-175 x^2+59 x+5,\nonumber\\
P_{111} &= 6 x^6-57 x^5+167 x^4-223 x^3+17 x^2-30 x-8,\nonumber\\
P_{112} &= 6 x^6-53 x^5+267 x^4-239 x^3+73 x^2+6 x+4,\nonumber\\
P_{113} &= 7 x^6+28 x^5+x^4+536 x^3+x^2+28 x+7,\nonumber\\
P_{114} &= 7 x^6+28 x^5+13 x^4+416 x^3+13 x^2+28 x+7,\nonumber\\
P_{115} &= 7 x^6+40 x^5-163 x^4+280 x^3-159 x^2+48 x+11,\nonumber\\
P_{116} &= 7 x^6+132 x^5+71 x^4-2312 x^3+1121 x^2-396 x-15,\nonumber\\
P_{117} &= 9 x^6-84 x^5+355 x^4-341 x^3+94 x^2-3 x+2,\nonumber\\
P_{118} &= 9 x^6-57 x^5+232 x^4-176 x^3+97 x^2+15 x+8,\nonumber\\
P_{119} &= 9 x^6+11 x^5+139 x^4-254 x^3+139 x^2+11 x+9,\nonumber\\
P_{120} &= 10 x^6-25 x^5+470 x^4-874 x^3+470 x^2-25 x+10,\nonumber\\
P_{121} &= 11 x^6-8 x^5+339 x^4-556 x^3+339 x^2-8 x+11,\nonumber\\
P_{122} &= 11 x^6+35 x^5+195 x^4-324 x^3+195 x^2+35 x+11,\nonumber\\
P_{123} &= 12 x^6+37 x^5+49 x^4+42 x^3+42 x^2+11 x-1,\nonumber\\
P_{124} &= 13 x^6+3 x^5+231 x^4-430 x^3+231 x^2+3 x+13,\nonumber\\
P_{125} &= 13 x^6+14 x^5+111 x^4-20 x^3+111 x^2+14 x+13,\nonumber\\
P_{126} &= 13 x^6+34 x^5-61 x^4+908 x^3-873 x^2+138 x-63,\nonumber\\
P_{127} &= 13 x^6+57 x^5+76 x^4-56 x^3+76 x^2+57 x+13,\nonumber\\
P_{128} &= 14 x^6+52 x^5+153 x^4-76 x^3+153 x^2+52 x+14,\nonumber\\
P_{129} &= 15 x^6-9 x^5+517 x^4-982 x^3+517 x^2-9 x+15,\nonumber\\
P_{130} &= 15 x^6+706 x^5-55 x^4+2888 x^3-159 x^2+198 x-41,\nonumber\\
P_{131} &= 17 x^6-150 x^5+1153 x^4-1852 x^3+883 x^2-174 x+11,\nonumber\\
P_{132} &= 17 x^6+25 x^5+231 x^4-418 x^3+231 x^2+25 x+17,\nonumber\\
P_{133} &= 17 x^6+46 x^5-109 x^4+348 x^3-109 x^2+46 x+17,\nonumber\\
P_{134} &= 18 x^6-x^5-85 x^4+150 x^3+76 x^2+99 x-1,\nonumber\\
P_{135} &= 18 x^6+7 x^5+290 x^4-502 x^3+290 x^2+7 x+18,\nonumber\\
P_{136} &= 19 x^6-21 x^5+525 x^4-910 x^3+523 x^2-25 x+17,\nonumber\\
P_{137} &= 24 x^6+15 x^5-24 x^4+234 x^3-26 x^2+11 x+22,\nonumber\\
P_{138} &= 24 x^6+75 x^5+91 x^4+118 x^3+50 x^2+29 x-3,\nonumber\\
P_{139} &= 25 x^6-126 x^5+1233 x^4-1820 x^3+859 x^2-166 x+11,\nonumber\\
P_{140} &= 25 x^6+19 x^5+479 x^4-854 x^3+479 x^2+19 x+25,\nonumber\\
P_{141} &= 27 x^6-20 x^5+835 x^4-1428 x^3+835 x^2-20 x+27,\nonumber\\
P_{142} &= 28 x^6+129 x^5-244 x^4+892 x^3-334 x^2+27 x+14,\nonumber\\
P_{143} &= 29 x^6-352 x^5+839 x^4-1502 x^3+317 x^2-46 x+11,\nonumber\\
P_{144} &= 29 x^6+665 x^5-2515 x^4+4174 x^3-715 x^2+269 x+65,\nonumber\\
P_{145} &= 37 x^6-94 x^5+223 x^4-108 x^3+231 x^2-78 x+45,\nonumber\\
P_{146} &= 37 x^6-10 x^5+1175 x^4-1924 x^3+735 x^2-386 x-27,\nonumber\\
P_{147} &= 39 x^6+20 x^5+1021 x^4-2324 x^3+865 x^2-376 x-21,\nonumber\\
P_{148} &= 44 x^6+265 x^5+704 x^4-2880 x^3+1696 x^2-529 x-20,\nonumber\\
P_{149} &= 45 x^6-80 x^5+239 x^4-792 x^3+239 x^2-80 x+45,\nonumber\\
P_{150} &= 50 x^6+119 x^5+610 x^4-2060 x^3+1254 x^2-435 x-2,\nonumber\\
P_{151} &= 53 x^6+173 x^5+357 x^4+892 x^3-935 x^2-17 x-11,\nonumber\\
P_{152} &= 59 x^6-326 x^5+1675 x^4+1888 x^3-1503 x^2+446 x+9,\nonumber\\
P_{153} &= 63 x^6+9 x^5+1145 x^4+762 x^3-1073 x^2+385 x-11,\nonumber\\
P_{154} &= 72 x^6+381 x^5-659 x^4+1986 x^3-786 x^2+9 x+21,\nonumber\\
P_{155} &= 81 x^6-429 x^5+863 x^4+3222 x^3-2399 x^2+667 x+43,\nonumber\\
P_{156} &= 93 x^6-130 x^5+491 x^4-716 x^3-21 x^2-194 x-35,\nonumber\\
P_{157} &= 105 x^6-30 x^5+1065 x^4-56 x^3-431 x^2-14 x-47,\nonumber\\
P_{158} &= 125 x^6-94 x^5+1295 x^4-220 x^3-201 x^2-78 x-27,\nonumber\\
P_{159} &= 133 x^6-332 x^5+4847 x^4-2420 x^3+159 x^2+24 x-11,\nonumber\\
P_{160} &= 145 x^6-738 x^5+15433 x^4-15038 x^3+12285 x^2-404 x+637,\nonumber\\
P_{161} &= 229 x^6-60 x^5+1517 x^4-968 x^3+247 x^2+4 x+55,\nonumber\\
P_{162} &= 443 x^6+2186 x^5-12793 x^4+14416 x^3+2471 x^2-766 x-565,\nonumber\\
P_{163} &= 649 x^6-416 x^5+4063 x^4-2272 x^3+975 x^2-80 x+153,\nonumber\\
P_{164} &= 759 x^6-516 x^5+4247 x^4-4888 x^3-163 x^2-516 x+53,\nonumber\\
P_{165} &= 1043 x^6-224 x^5+10999 x^4-14384 x^3+2761 x^2+400 x+173,\nonumber\\
P_{166} &= -337 x^7-125 x^6-3325 x^5+2351 x^4-431 x^3-1203 x^2-19 x+17,\nonumber\\
P_{167} &= -117 x^7-1119 x^6+4661 x^5-17313 x^4+15395 x^3-6031 x^2+297 x-157,\nonumber\\
P_{168} &= -71 x^7-353 x^6+621 x^5-2761 x^4+2243 x^3-991 x^2+131 x-3,\nonumber\\
P_{169} &= x^7-7 x^6+69 x^5-267 x^4+239 x^3-89 x^2-5 x-5,\nonumber\\
P_{170} &= x^7-6 x^6+115 x^5-150 x^4+206 x^3-75 x^2+30 x+7,\nonumber\\
P_{171} &= x^7-5 x^6+61 x^5-93 x^4+121 x^3-41 x^2+17 x+3,\nonumber\\
P_{172} &= x^7-2 x^6+37 x^5-130 x^4+123 x^3-42 x^2-x-2,\nonumber\\
P_{173} &= x^7-x^6+33 x^5-99 x^4+113 x^3-23 x^2+7 x+1,\nonumber\\
P_{174} &= x^7+x^6+29 x^5-39 x^4+53 x^3-19 x^2+5 x+1,\nonumber\\
P_{175} &= x^7+2 x^6+16 x^5+41 x^4+15 x^3+24 x^2+22 x+7,\nonumber\\
P_{176} &= x^7+2 x^6+17 x^5-16 x^4+30 x^3-7 x^2+4 x+1,\nonumber\\
P_{177} &= x^7+4 x^6-7 x^5+30 x^4-16 x^3+17 x^2+2 x+1,\nonumber\\
P_{178} &= x^7+6 x^6-26 x^5-17 x^4-39 x^3-14 x^2-30 x-9,\nonumber\\
P_{179} &= x^7+7 x^6-31 x^5+215 x^4-173 x^3+61 x^2+11 x+5,\nonumber\\
P_{180} &= x^7+9 x^6-57 x^5+183 x^4-197 x^3+47 x^2-15 x-3,\nonumber\\
P_{181} &= x^7+9 x^6-33 x^5+113 x^4-120 x^3+28 x^2-12 x-2,\nonumber\\
P_{182} &= x^7+14 x^6-91 x^5+692 x^4-552 x^3+191 x^2+46 x+19,\nonumber\\
P_{183} &= x^7+35 x^6-220 x^5+1700 x^4-846 x^3+830 x^2+331 x+121,\nonumber\\
P_{184} &= 2 x^7+3 x^6+19 x^5-76 x^4+62 x^3-29 x^2-9 x-4,\nonumber\\
P_{185} &= 2 x^7+6 x^6+13 x^5-13 x^4+20 x^3-8 x^2-3 x-1,\nonumber\\
P_{186} &= 2 x^7+7 x^6-2 x^5+37 x^4-9 x^3+22 x^2+5 x+2,\nonumber\\
P_{187} &= 2 x^7+7 x^6+5 x^5+14 x^4-7 x^3-4 x-1,\nonumber\\
P_{188} &= 2 x^7+8 x^6-x^5+83 x^4+x^3+61 x^2+28 x+10,\nonumber\\
P_{189} &= 2 x^7+23 x^6-172 x^5+363 x^4-433 x^3+122 x^2-53 x-12,\nonumber\\
P_{190} &= 2 x^7+24 x^6-157 x^5+447 x^4-587 x^3+57 x^2-84 x-22,\nonumber\\
P_{191} &= 3 x^7+76 x^5-247 x^4+240 x^3-81 x^2-3 x-4,\nonumber\\
P_{192} &= 3 x^7-11 x^6+143 x^5-527 x^4+485 x^3-173 x^2-7 x-9,\nonumber\\
P_{193} &= 3 x^7-9 x^6+183 x^5-689 x^4+703 x^3-173 x^2+15 x-1,\nonumber\\
P_{194} &= 3 x^7+3 x^6+68 x^5-246 x^4+260 x^3-58 x^2+3 x-1,\nonumber\\
P_{195} &= 3 x^7+4 x^6+63 x^5-182 x^4+210 x^3-43 x^2+8 x+1,\nonumber\\
P_{196} &= 3 x^7+6 x^6-400 x^5+317 x^4+389 x^3-590 x^2+19,\nonumber\\
P_{197} &= 3 x^7+11 x^6-9 x^5+67 x^4-25 x^3+39 x^2+7 x+3,\nonumber\\
P_{198} &= 3 x^7+14 x^6-33 x^5+224 x^4-168 x^3+73 x^2+10 x+5,\nonumber\\
P_{199} &= 3 x^7+59 x^6-477 x^5+1807 x^4-1849 x^3+447 x^2-77 x-9,\nonumber\\
P_{200} &= 3 x^7+177 x^6+507 x^5+365 x^4-1153 x^3+413 x^2-93 x+37,\nonumber\\
P_{201} &= 4 x^7+3 x^6+77 x^5-316 x^4+260 x^3-117 x^2-27 x-12,\nonumber\\
P_{202} &= 4 x^7+33 x^6-113 x^5+462 x^4-490 x^3+93 x^2-45 x-8,\nonumber\\
P_{203} &= 4 x^7+35 x^6-212 x^5+557 x^4-655 x^3+142 x^2-77 x-18,\nonumber\\
P_{204} &= 5 x^7-7 x^6-94 x^5+45 x^4+7 x^3-5 x^2-10 x-5,\nonumber\\
P_{205} &= 5 x^7+3 x^6+103 x^5-269 x^4+241 x^3-123 x^2-15 x-9,\nonumber\\
P_{206} &= 5 x^7+5 x^6+89 x^5-239 x^4+267 x^3-69 x^2+7 x-1,\nonumber\\
P_{207} &= 5 x^7+9 x^6+32 x^5-314 x^4+188 x^3-122 x^2-63 x-23,\nonumber\\
P_{208} &= 5 x^7+12 x^6+32 x^5-61 x^4+75 x^3-22 x^2-6 x-3,\nonumber\\
P_{209} &= 5 x^7+15 x^6+37 x^5-73 x^4+59 x^3-47 x^2-21 x-7,\nonumber\\
P_{210} &= 5 x^7+27 x^6-51 x^5+247 x^4-219 x^3+71 x^2-15 x-1,\nonumber\\
P_{211} &= 6 x^7+9 x^6+87 x^5-190 x^4+218 x^3-67 x^2+3 x-2,\nonumber\\
P_{212} &= 6 x^7+17 x^6+41 x^5+84 x^4+84 x^3+79 x^2+55 x+18,\nonumber\\
P_{213} &= 7 x^7-5 x^6+293 x^5-1119 x^4+1133 x^3-283 x^2+11 x-5,\nonumber\\
P_{214} &= 7 x^7+2 x^6+174 x^5-21 x^4+245 x^3-14 x^2+94 x+25,\nonumber\\
P_{215} &= 7 x^7+11 x^6+99 x^5-225 x^4+281 x^3-59 x^2+13 x+1,\nonumber\\
P_{216} &= 7 x^7+13 x^6+107 x^5-367 x^4+409 x^3-77 x^2+5 x-1,\nonumber\\
P_{217} &= 7 x^7+15 x^6+49 x^5-143 x^4+129 x^3-59 x^2-21 x-9,\nonumber\\
P_{218} &= 7 x^7+15 x^6+53 x^5-131 x^4+145 x^3-43 x^2-9 x-5,\nonumber\\
P_{219} &= 7 x^7+19 x^6+46 x^5-20 x^4+62 x^3-16 x^2-x-1,\nonumber\\
P_{220} &= 8 x^7-5 x^6+351 x^5-878 x^4+1004 x^3-261 x^2+59 x+10,\nonumber\\
P_{221} &= 10 x^7-15 x^6+345 x^5-1186 x^4+1158 x^3-365 x^2+3 x-14,\nonumber\\
P_{222} &= 10 x^7-5 x^6+294 x^5-496 x^4+755 x^3-109 x^2+116 x+27,\nonumber\\
P_{223} &= 10 x^7+29 x^6+29 x^5-40 x^4+54 x^3-19 x^2-23 x-8,\nonumber\\
P_{224} &= 10 x^7+29 x^6+35 x^5-142 x^4+156 x^3-25 x^2-23 x-8,\nonumber\\
P_{225} &= 11 x^7-15 x^6+131 x^5+1575 x^4-3873 x^3+2265 x^2-13 x+47,\nonumber\\
P_{226} &= 11 x^7-3 x^6+317 x^5-487 x^4+711 x^3-157 x^2+99 x+21,\nonumber\\
P_{227} &= 11 x^7+12 x^6+294 x^5-665 x^4+777 x^3-214 x^2+36 x+5,\nonumber\\
P_{228} &= 11 x^7+13 x^6+403 x^5-1715 x^4+2781 x^3-1669 x^2-43 x-37,\nonumber\\
P_{229} &= 11 x^7+15 x^6+174 x^5-413 x^4+420 x^3-169 x^2-12 x-10,\nonumber\\
P_{230} &= 11 x^7+55 x^6-183 x^5+653 x^4-639 x^3+193 x^2-49 x-9,\nonumber\\
P_{231} &= 12 x^7+35 x^6+69 x^5+120 x^4+118 x^3+101 x^2+67 x+22,\nonumber\\
P_{232} &= 12 x^7+71 x^6-253 x^5+1158 x^4-1200 x^3+223 x^2-89 x-18,\nonumber\\
P_{233} &= 13 x^7-48 x^6+636 x^5-1993 x^4+1965 x^3-656 x^2+36 x-17,\nonumber\\
P_{234} &= 13 x^7+19 x^6+193 x^5-457 x^4+555 x^3-123 x^2+23 x+1,\nonumber\\
P_{235} &= 13 x^7+26 x^6+138 x^5-441 x^4+455 x^3-128 x^2-20 x-11,\nonumber\\
P_{236} &= 13 x^7+32 x^6+136 x^5-289 x^4+317 x^3-116 x^2-20 x-9,\nonumber\\
P_{237} &= 13 x^7+75 x^6-197 x^5+655 x^4-543 x^3+277 x^2-27 x+3,\nonumber\\
P_{238} &= 13 x^7+119 x^6-727 x^5+2363 x^4-2517 x^3+617 x^2-185 x-35,\nonumber\\
P_{239} &= 13 x^7+121 x^6-755 x^5+2461 x^4-2755 x^3+545 x^2-247 x-55,\nonumber\\
P_{240} &= 13 x^7+128 x^6-498 x^5+2121 x^4-1897 x^3+658 x^2-32 x+19,\nonumber\\
P_{241} &= 13 x^7+132 x^6+185 x^5+187 x^4-215 x^3-10 x^2+17 x+11,\nonumber\\
P_{242} &= 14 x^7+31 x^6+166 x^5-587 x^4+657 x^3-116 x^2-x-4,\nonumber\\
P_{243} &= 14 x^7+40 x^6+71 x^5-165 x^4+207 x^3-41 x^2-22 x-8,\nonumber\\
P_{244} &= 15 x^7+9 x^6+319 x^5-967 x^4+981 x^3-309 x^2-3 x-13,\nonumber\\
P_{245} &= 15 x^7+49 x^6-2 x^5+34 x^4-20 x^3+12 x^2-43 x-13,\nonumber\\
P_{246} &= 16 x^7-13 x^6+487 x^5-1568 x^4+1540 x^3-507 x^2+x-20,\nonumber\\
P_{247} &= 16 x^7+11 x^6+319 x^5-862 x^4+890 x^3-299 x^2+x-12,\nonumber\\
P_{248} &= 17 x^7+49 x^6+127 x^5-89 x^4+201 x^3-47 x^2-x-1,\nonumber\\
P_{249} &= 17 x^7+63 x^6-63 x^5+167 x^4-209 x^3+33 x^2-81 x-23,\nonumber\\
P_{250} &= 17 x^7+69 x^6-116 x^5+434 x^4-448 x^3+106 x^2-75 x-19,\nonumber\\
P_{251} &= 17 x^7+70 x^6-54 x^5+189 x^4-77 x^3+134 x^2-22 x-1,\nonumber\\
P_{252} &= 18 x^7+30 x^6+241 x^5-635 x^4+677 x^3-211 x^2-12 x-12,\nonumber\\
P_{253} &= 18 x^7+43 x^6+185 x^5-710 x^4+752 x^3-155 x^2-25 x-12,\nonumber\\
P_{254} &= 18 x^7+53 x^6+81 x^5-204 x^4+190 x^3-91 x^2-59 x-20,\nonumber\\
P_{255} &= 19 x^7-46 x^6+778 x^5-2375 x^4+2347 x^3-798 x^2+34 x-23,\nonumber\\
P_{256} &= 19 x^7+65 x^6+47 x^5+109 x^4-67 x^3-17 x^2-47 x-13,\nonumber\\
P_{257} &= 19 x^7+91 x^6-279 x^5+593 x^4-859 x^3+89 x^2-205 x-57,\nonumber\\
P_{258} &= 19 x^7+95 x^6-173 x^5+7 x^4-77 x^3+123 x^2-125 x-29,\nonumber\\
P_{259} &= 19 x^7+113 x^6-261 x^5+1317 x^4-1093 x^3+421 x^2-17 x+13,\nonumber\\
P_{260} &= 20 x^7-181 x^6+1831 x^5-5450 x^4+5282 x^3-1951 x^2+109 x-44,\nonumber\\
P_{261} &= 20 x^7+56 x^6+131 x^5-118 x^4+265 x^3-26 x^2+7 x+1,\nonumber\\
P_{262} &= 21 x^7-4 x^6+554 x^5-1661 x^4+1633 x^3-574 x^2-8 x-25,\nonumber\\
P_{263} &= 21 x^7+50 x^6+188 x^5-565 x^4+593 x^3-168 x^2-38 x-17,\nonumber\\
P_{264} &= 21 x^7+55 x^6+131 x^5-375 x^4+445 x^3-81 x^2-25 x-11,\nonumber\\
P_{265} &= 22 x^7-179 x^6+1753 x^5-4744 x^4+5262 x^3-1383 x^2+401 x+52,\nonumber\\
P_{266} &= 22 x^7-x^6+405 x^5-124 x^4-50 x^3+221 x^2-x+40,\nonumber\\
P_{267} &= 23 x^7+65 x^6+25 x^5-129 x^4+283 x^3+85 x^2+x-1,\nonumber\\
P_{268} &= 23 x^7+92 x^6-436 x^5+1149 x^4-1213 x^3+298 x^2-22 x-19,\nonumber\\
P_{269} &= 24 x^7+11 x^6+543 x^5-1222 x^4+1236 x^3-533 x^2-5 x-22,\nonumber\\
P_{270} &= 25 x^7+49 x^6+259 x^5-685 x^4+699 x^3-249 x^2-43 x-23,\nonumber\\
P_{271} &= 25 x^7+49 x^6+271 x^5-889 x^4+903 x^3-261 x^2-43 x-23,\nonumber\\
P_{272} &= 25 x^7+94 x^6-4 x^5+567 x^4-301 x^3+194 x^2+20 x+13,\nonumber\\
P_{273} &= 27 x^7+17 x^6+525 x^5-1825 x^4+1671 x^3-635 x^2-83 x-49,\nonumber\\
P_{274} &= 27 x^7+83 x^6+117 x^5+117 x^4+121 x^3+53 x^2+19 x+7,\nonumber\\
P_{275} &= 29 x^7+57 x^6+247 x^5-901 x^4+887 x^3-257 x^2-63 x-31,\nonumber\\
P_{276} &= 31 x^7+63 x^6+253 x^5-899 x^4+885 x^3-263 x^2-69 x-33,\nonumber\\
P_{277} &= 31 x^7+69 x^6+263 x^5-783 x^4+825 x^3-233 x^2-51 x-25,\nonumber\\
P_{278} &= 31 x^7+85 x^6+183 x^5-763 x^4+749 x^3-193 x^2-91 x-33,\nonumber\\
P_{279} &= 31 x^7+121 x^6-53 x^5+505 x^4-379 x^3+143 x^2-67 x-13,\nonumber\\
P_{280} &= 32 x^7-57 x^6+1229 x^5-3252 x^4+3154 x^3-1299 x^2+15 x-46,\nonumber\\
P_{281} &= 34 x^7-7 x^6+899 x^5-2460 x^4+2446 x^3-909 x^2+x-36,\nonumber\\
P_{282} &= 34 x^7+101 x^6+153 x^5-116 x^4+270 x^3-43 x^2-35 x-12,\nonumber\\
P_{283} &= 35 x^7-5 x^6+918 x^5-2502 x^4+2572 x^3-868 x^2+35 x-25,\nonumber\\
P_{284} &= 37 x^7+78 x^6-64 x^5+1211 x^4-2191 x^3+1360 x^2+50 x+31,\nonumber\\
P_{285} &= 37 x^7+104 x^6+326 x^5-271 x^4+495 x^3-166 x^2-8 x-5,\nonumber\\
P_{286} &= 38 x^7+29 x^6+807 x^5-1678 x^4+1748 x^3-757 x^2+x-28,\nonumber\\
P_{287} &= 40 x^7+167 x^6-262 x^5+911 x^4-897 x^3+272 x^2-161 x-38,\nonumber\\
P_{288} &= 42 x^7+71 x^6+557 x^5-1378 x^4+1490 x^3-477 x^2-23 x-26,\nonumber\\
P_{289} &= 43 x^7+77 x^6+485 x^5-1245 x^4+1567 x^3-255 x^2+61 x+3,\nonumber\\
P_{290} &= 43 x^7+135 x^6+177 x^5+261 x^4+145 x^3+113 x^2+39 x+15,\nonumber\\
P_{291} &= 46 x^7+141 x^6+284 x^5+193 x^4+325 x^3+86 x^2+81 x+28,\nonumber\\
P_{292} &= 47 x^7+3 x^6+543 x^5+259 x^4-1821 x^3+1571 x^2+15 x+23,\nonumber\\
P_{293} &= 50 x^7+209 x^6-257 x^5+2042 x^4-1776 x^3+447 x^2-95 x-12,\nonumber\\
P_{294} &= 51 x^7+121 x^6+375 x^5-1683 x^4+1613 x^3-425 x^2-151 x-61,\nonumber\\
P_{295} &= 55 x^7+216 x^6-2082 x^5+6547 x^4-5893 x^3+1446 x^2+24 x+71,\nonumber\\
P_{296} &= 56 x^7+101 x^6+925 x^5-3030 x^4+2512 x^3-1295 x^2-323 x-130,\nonumber\\
P_{297} &= 65 x^7+345 x^6-1061 x^5+5123 x^4-4969 x^3+1171 x^2-279 x-43,\nonumber\\
P_{298} &= 67 x^7+74 x^6+1200 x^5-3387 x^4+3527 x^3-1100 x^2-14 x-47,\nonumber\\
P_{299} &= 77 x^7+151 x^6+1175 x^5-2699 x^4+3217 x^3-805 x^2+71 x-3,\nonumber\\
P_{300} &= 91 x^7+153 x^6+833 x^5-2047 x^4+3597 x^3-1845 x^2+15 x-29,\nonumber\\
P_{301} &= 96 x^7+179 x^6+1307 x^5-2262 x^4+2682 x^3-1007 x^2+x-36,\nonumber\\
P_{302} &= 98 x^7+115 x^6+1791 x^5-3676 x^4+3984 x^3-1571 x^2+17 x-54,\nonumber\\
P_{303} &= 111 x^7+60 x^6+836 x^5-3631 x^4+6545 x^3-2672 x^2+84 x-53,\nonumber\\
P_{304} &= 119 x^7+177 x^6+1763 x^5-4297 x^4+4773 x^3-1423 x^2+27 x-51,\nonumber\\
P_{305} &= 139 x^7+512 x^6-90 x^5+3131 x^4-2207 x^3+750 x^2-116 x-7,\nonumber\\
P_{306} &= 141 x^7+544 x^6-346 x^5+4169 x^4-3273 x^3+986 x^2-160 x-13,\nonumber\\
P_{307} &= 146 x^7+1297 x^6-1553 x^5+8418 x^4-9754 x^3+3745 x^2-563 x-200,\nonumber\\
P_{308} &= 181 x^7+479 x^6+1355 x^5-2119 x^4+2973 x^3-745 x^2-113 x-59,\nonumber\\
P_{309} &= 543 x^7+587 x^6+2403 x^5+3623 x^4-3279 x^3+277 x^2-195 x+137,\nonumber\\
P_{310} &= 649 x^7+1613 x^6+5451 x^5-6985 x^4+11899 x^3-1941 x^2+493 x+53 \,.
\end{align}

\newpage

\begin{dmath*}
 F_{V,2}^{(1)} = C_F \Bigg[
%
%
  \Bigg\{
-4 x \eta  H_0
\Bigg\}
+ \ep  \Bigg\{
\eta  \Big(
        -16 x H_0
        +8 x H_{-1} H_0
        -2 x H_0^2
        -8 x H_{0,-1}
        +4 x \zeta_2
\Big)
\Bigg\}
~~~~~~~~~~~
\end{dmath*}
\vspace{-1.0cm}
\begin{dmath}
{\color{white}=}
+ \ep^2  \Bigg\{
\eta  \Big(
        \Big(
                -32 x
                +32 x H_{-1}
                -8 x H_{-1}^2
                +2 x \zeta_2
        \Big) H_0
        +\Big(
                -8 x
                +4 x H_{-1}
        \Big) H_0^2
        -\frac{2}{3} x H_0^3
        +\Big(
                -32 x
                +16 x H_{-1}
        \Big) H_{0,-1}
        -8 x H_{0,0,-1}
        -16 x H_{0,-1,-1}
        +16 x \zeta_2
        -8 x H_{-1} \zeta_2
        +8 x \zeta_3
\Big)
\Bigg\}
+ \ep^3  \Bigg\{
\eta  \Big(
        \Big(
                -64 x
                +\Big(
                        64 x
                        -4 x \zeta_2
                \Big) H_{-1}
                -32 x H_{-1}^2
                +\frac{16}{3} x H_{-1}^3
                +8 x \zeta_2
                +\frac{28}{3} x \zeta_3
        \Big) H_0
        +\Big(
                -16 x
                +16 x H_{-1}
                -4 x H_{-1}^2
                +x \zeta_2
        \Big) H_0^2
        +\Big(
                -\frac{8 x}{3}
                +\frac{4 x H_{-1}}{3}
        \Big) H_0^3
        -\frac{1}{6} x H_0^4
        +\Big(
                -32 x \zeta_2
                -16 x \zeta_3
        \Big) H_{-1}
        +\Big(
                -64 x
                +64 x H_{-1}
                -16 x H_{-1}^2
                -4 x \zeta_2
        \Big) H_{0,-1}
        +\Big(
                -32 x
                +16 x H_{-1}
        \Big) H_{0,0,-1}
        +\Big(
                -64 x
                +32 x H_{-1}
        \Big) H_{0,-1,-1}
        -8 x H_{0,0,0,-1}
        -16 x H_{0,0,-1,-1}
        -32 x H_{0,-1,-1,-1}
        +32 x \zeta_2
        +8 x H_{-1}^2 \zeta_2
        +\frac{28}{5} x \zeta_2^2
        +32 x \zeta_3
\Big)
\Bigg\}
\Bigg] \,.
\end{dmath}

\begin{dmath*}
 F_{V,2}^{(2)} =
%
C_F^2  \Bigg[
%
%
 \frac{1}{\ep}  \Bigg\{
        8 x \eta  H_0
        +8 x \xi  \eta  H_0^2
\Bigg\}
+  \Bigg\{
                \xi ^2 \eta ^3 \Big(
                        -\frac{4}{3} (-1+x)^2 x^2 H_0^4
                        -192 (-1+x)^2 x^2 H_{0,-1} \zeta_2
                \Big)
\end{dmath*}
\vspace{-1.0cm}
\begin{dmath*}
{\color{white}=}
                +\eta ^3 x_+^2 \Big(
                        384 x^2 H_{0,0,0,-1} P_{358}
                        +32 x^2 H_0^2 H_{0,-1} P_{362}
                        +\frac{16}{3} x^4 H_0^4
                        +\Big(
                                32 x^2 H_0^2 P_{334}
                                -1536 x^4 H_{0,-1}
                        \Big) \zeta_2
                        +\frac{96}{5} x^2 P_{372} \zeta_2^2
                \Big)
                +\xi  \Big(
                        \eta ^3 x_+ \Big(
                                4 (-1+x) x^3 H_0^4
                                -1728 (-1+x) x^3 H_{0,-1} \zeta_2
                        \Big)
                        +\eta  \Big(
                                \Big(
                                        128 (-1+x) x H_0 H_1
                                        -128 (-1+x) x H_{0,1}
                                \Big) x_+
                                +\Big(
                                        -64 (-1+x) x^2 H_0^3 H_1
                                        +1344 (-1+x) x^2 H_{0,0,0,1}
                                        +\Big(
                                                -384 (-1+x) x^2 H_0 H_1
                                                +384 (-1+x) x^2 H_{0,1}
                                        \Big) \zeta_2
                                        +448 (-1+x) x^2 H_0 \zeta_3
                                \Big) x_+^3
                        \Big)
                \Big)
                +\eta ^2 \Big(
                        \Big(
                                32 x H_{0,0,1}
                                 P_{313}
                                -8 x H_{-1} H_0^2 P_{314}
                                +48 x H_0 H_{0,-1} P_{315}
                                -80 x H_{0,0,-1} P_{317}
                                -16 x H_0^2 H_1 P_{327}
                                +2 x H_0^2 P_{450}
                                +48 x H_{-1} P_{355} \zeta_2
                                +16 x P_{316} \zeta_3
                        \Big) x_+^2
                        +\Big(
                                -
                                \frac{8}{3} x H_0^3 P_{471}
                                -16 x H_0 P_{473} \zeta_2
                        \Big) x_+^3
                \Big)
                +\eta  \Big(
                        -4 x \Big(
                                -15-173 x+323 x^2+x^3\Big) x_+^3 \zeta_2
                        +62 x H_0
                        +\Big(
                                24 x \Big(
                                        13-38 x+13 x^2\Big) H_{-1} H_0
                                +128 x^2 H_0 H_1
                                -128 x^2 H_{0,1}
                                -24 x \Big(
                                        13-38 x+13 x^2\Big) H_{0,-1}
                        \Big) x_+^2
                        +\Big(
                                -64 x^3 H_0^3 H_1
                                -32 x^2 \Big(
                                        11-41 x+11 x^2\Big) H_0^2 H_{0,1}
                                +512 x^2 \Big(
                                        2-11 x+2 x^2\Big) H_0 H_{0,0,1}
                                -64 x^2 \Big(
                                        19-73 x+19 x^2\Big) H_0 H_{0,0,-1}
                                +9408 x^3 H_{0,0,0,1}
                                +\Big(
                                        -384 x^3 H_0 H_1
                                        +384 x^3 H_{0,1}
                                \Big) \zeta_2
                                +448 x^3 H_0 \zeta_3
                        \Big) x_+^4
                \Big)
                +1824 x^2 H_0 H_{0,1} x_+^4
                -192 x \log (2) x_+^2 \zeta_2
\Bigg\}
+ \ep  \Bigg\{
                \xi ^2 \eta ^3 \Big(
                        -
                        \frac{8}{3} (-1+x)^2 x^2 H_{-1} H_0^4
                        -4096 (-1+x)^2 x^2 H_{0,0,-1,0,-1}
                        +\Big(
                                -384 (-1+x)^2 x^2 H_{-1} H_{0,-1}
                                +384 (-1+x)^2 x^2 H_{0,1,-1}
                        \Big) \zeta_2
                \Big)
                +\xi  \Big(
                        \eta ^2 x_+ \Big(
                                -64 (-1+x) x \Big(
                                        3+22 x+3 x^2\Big) H_1 H_{0,0,1}
                                -32 (-1+x) x \Big(
                                        23-20 x+23 x^2\Big) H_{-1} H_{0,0,-1}
                                +64 (-1+x) x \Big(
                                        3+22 x+3 x^2\Big) H_{0,0,1,1}
                                +32 (-1+x) x \Big(
                                        23-20 x+23 x^2\Big) H_{0,0,-1,-1}
                        \Big)
\end{dmath*}
\begin{dmath*}
{\color{white}=}
                        +\eta ^3 x_+ \Big(
                                8 (-1+x) x^3 H_{-1} H_0^4
                                +\frac{4}{15} (-1+x) x^2 \Big(
                                        3-4 x+3 x^2\Big) H_0^5
                                +256 (-1+x) x^2 \Big(
                                        29-61 x+29 x^2\Big) H_0 H_{0,0,1,-1}
                                +256 (-1+x) x^2 \Big(
                                        29-61 x+29 x^2\Big) H_0 H_{0,0,-1,1}
                                -8192 (-1+x) x^3 H_{0,0,-1,0,-1}
                                +\Big(
                                        32 (-1+x) x^2 \Big(
                                                19-31 x+19 x^2\Big) H_0^2 H_1
                                        -3456 (-1+x) x^3 H_{-1} H_{0,-1}
                                        -2688 (-1+x) x^3 H_{0,1,-1}
                                \Big) \zeta_2
                        \Big)
                        +\eta  \Big(
                                \Big(
                                        -768 (-1+x) x H_{-1} H_0 H_1
                                        +128 (-1+x) x H_0 H_1^2
                                        +\Big(
                                                -256 (-1+x) x H_1
                                                +768 (-1+x) x H_{-1}
                                        \Big) H_{0,1}
                                        +768 (-1+x) x H_1 H_{0,-1}
                                        +256 (-1+x) x H_{0,1,1}
                                        -768 (-1+x) x H_{0,1,-1}
                                        -768 (-1+x) x H_{0,-1,1}
                                        -128 (-1+x) x H_1 \zeta_2
                                \Big) x_+
                                +\Big(
                                        -128 (-1+x) x^2 H_{-1} H_0^3 H_1
                                        +64 (-1+x) x^2 H_0^3 H_1^2
                                        -128 (-1+x) x^2 H_0^2 H_1 H_{0,1}
                                        -1024 (-1+x) x^2 H_0 H_1 H_{0,0,1}
                                        +1024 (-1+x) x^2 H_0 H_1 H_{0,0,-1}
                                        +2688 (-1+x) x^2 H_{-1} H_{0,0,0,1}
                                        -1024 (-1+x) x^2 H_0 H_{0,0,1,1}
                                        +\Big(
                                                -768 (-1+x) x^2 H_{-1} H_0 H_1
                                                +384 (-1+x) x^2 H_0 H_1^2
                                                +\Big(
                                                        -768 (-1+x) x^2 H_1
                                                        +768 (-1+x) x^2 H_{-1}
                                                \Big) H_{0,1}
                                                +768 (-1+x) x^2 H_{0,1,1}
                                        \Big) \zeta_2
                                        +\Big(
                                                896 (-1+x) x^2 H_{-1} H_0
                                                +256 (-1+x) x^2 H_0 H_1
                                        \Big) \zeta_3
                                \Big) x_+^3
                        \Big)
                \Big)
                +\eta ^3 \Big(
                        \log (2) x_+^2 \zeta_2 \Big(
                                1536 (-2+x) x^3 (-1+2 x) H_{0,1}
                                +1536 (-2+x) x^3 (-1+2 x) H_{0,-1}
                        \Big)
                        +64 x^2 \Big(
                                3-5 x+3 x^2\Big) H_0^2 H_{0,1,1}
                        +\Big(
                                384 x^2 H_0 H_{0,-1}^2 P_{333}
                                +384 x^2 H_0^2 H_{0,1,-1} P_{335}
                                +384 x^2 H_0^2 H_{0,-1,1} P_{335}
                                -256 x^2 H_0 H_{0,1}^2 P_{349}
                                +
                                \frac{8}{3} x^2 H_0^4 H_1 P_{364}
                                +1536 x^2 H_{0,0,0,1,-1} P_{383}
                                +1536 x^2 H_{0,0,0,-1,1} P_{383}
                                -768 x^2 H_{0,0,0,-1,-1} P_{396}
                                -128 x^2 H_0 H_{0,-1,0,1} P_{400}
                                -64 x^2 H_0^2 H_{0,-1,-1} P_{404}
                                -128 x^2 H_{0,0,1,0,1} P_{415}
                                +128 x^2 H_{0,0,1,0,-1} P_{423}
                                +128 x^2 H_{0,0,-1,0,1} P_{436}
                                +64 x^2 H_{0,0,0,0,1} P_{455}
                                -64 x^2 H_{0,0,0,0,-1} P_{460}
                                +\frac{2}{3} x H_0^4 P_{492}
                                -\frac{2}{3} x H_0^3 P_{506}
                                +\frac{32}{3} x^4 H_{-1} H_0^4
                                +\frac{4}{15} x^3 \Big(
                                        3-4 x+3 x^2\Big) H_0^5
                                +\Big(
                                        256 x^2 H_0 H_{0,-1} P_{380}
                                        -\frac{64}{3} x^2 H_0^3 P_{391}
                                        +16 x H_0^2 P_{485}
                                \Big) H_{0,1}
                                +\Big(
                                        64 x^2 H_{-1} H_0^2 P_{362}
                                        -64 x^2 H_0^2 H_1 P_{363}
                                        +\frac{32}{3} x^2 H_0^3 P_{414}
                                        -8 x H_0^2 P_{495}
                                \Big) H_{0,-1}
                                +\Big(
                                        -128 x^2 H_{0,-1} P_{375}
                                        +128 x^2 H_{0,1} P_{405}
                                        +32 x^2 H_0^2 P_{430}
                                        -64 x H_0 P_{478}
                                \Big) H_{0,0,1}
                                +\Big(
                                        -512 x^2 H_{0,-1} P_{318}
                                        -128 x^2 H_{0,1} P_{427}
                                        -32 x^2 H_0^2 P_{441}
                                        -16 x H_0 P_{487}
                                \Big) H_{0,0,-1}
                                +\Big(
                                        -64 x^2 H_0 P_{443}
                                        -32 x P_{491}
                                \Big) H_{0,0,0,1}
                                +\Big(
                                        768 x^2 H_{-1}
                                         P_{358}
                                        +384 x^2 H_1 P_{378}
                                        +64 x^2 H_0 P_{452}
                                        +16 x P_{502}
                                \Big) H_{0,0,0,-1}
                                +1024 x^3 \Big(
                                        29-61 x+29 x^2\Big) H_0 H_{0,0,1,-1}
                                +1024 x^3 \Big(
                                        29-61 x+29 x^2\Big) H_0 H_{0,0,-1,1}
                                +128 x^2 \Big(
                                        5-11 x+5 x^2
                                \Big)
\Big(11-34 x+11 x^2\Big) H_0 H_{0,0,-1,-1}
                                -384 x^2 \Big(
                                        5-6 x+5 x^2
                                \Big)
\Big(7-13 x+7 x^2\Big) H_{0,0,0,1,1}
                                -4096 x^4 H_{0,0,-1,0,-1}
                                +\Big(
                                        -64 x^2 H_{0,0,1} P_{312}
                                        +384 x^2 H_{0,-1,1} P_{325}
                                        +512 x^2 H_{0,0,-1} P_{332}
                                        +64 x^2 H_{-1} H_0^2 P_{334}
                                        +384 x^2 H_{0,-1,-1} P_{353}
                                        +128 x^2 H_0 H_{0,1} P_{354}
                                        +
                                        \frac{32}{3} x^2 H_0^3 P_{369}
                                        -8 x H_0^2 P_{484}
                                        +4 x H_0 P_{507}
                                        +128 x^3 \Big(
                                                19-31 x+19 x^2\Big) H_0^2 H_1
                                        +\Big(
                                                -384 x^2 H_1 P_{325}
                                                -128 x^2 H_0 P_{331}
                                                -16 x P_{499}
                                                -3072 x^4 H_{-1}
                                        \Big) H_{0,-1}
                                        -3072 x^4 H_{0,1,-1}
                                        -64 x^2 P_{386} \zeta_3
                                \Big) \zeta_2
                                +\Big(
                                        \frac{192}{5} x^2 H_{-1} P_{372}
                                        -\frac{96}{5} x^2 H_1 P_{426}
                                        -\frac{16}{5} x^2 H_0 P_{462}
                                        +\frac{8 x P_{496}}{5}
                                \Big) \zeta_2^2
                                +\Big(
                                        -16 x^2 H_0^2 P_{356}
                                        +128 x^2 H_{0,-1} P_{370}
                                        +64 x^2 H_{0,1} P_{394}
                                        +16 x H_0 P_{479}
                                \Big) \zeta_3
                                +96 x^2 P_{419} \zeta_5
                        \Big) x_+^2
                \Big)
\end{dmath*}
\begin{dmath*}
{\color{white}=}
                +\eta ^2 \Big(
                        \log (2) x_+^2 \zeta_2 \Big(
                                -384 x H_1 P_{324}
                                -384 x H_{-1} P_{324}
                                -384 x P_{344}
                        \Big)
                        +\Big(
                                8 x H_{-1}^2 H_0^2 P_{311}
                                +16 x H_0^2 H_1^2 P_{319}
                                -64 x H_{0,-1,0,1} P_{328}
                                +128 x H_{0,0,1,-1} P_{336}
                                +128 x H_{0,0,-1,1} P_{336}
                                +64 x H_0 H_{0,1,1} P_{337}
                                -64 x H_0 H_{0,1,-1} P_{339}
                                -64 x H_0 H_{0,-1,1} P_{339}
                                +16 x H_{0,-1}^2 P_{340}
                                +32 x H_0 H_{0,-1,-1} P_{371}
                                -4 x H_{-1} H_0^2 P_{459}
                                +2 x H_0^2 P_{466}
                                +\Big(
                                        64 x H_{-1} H_0^2 P_{324}
                                        +8 x H_0^2 P_{401}
                                \Big) H_1
                                +\Big(
                                        -128 x H_0 H_1 P_{322}
                                        +64 x H_{-1} H_0 P_{326}
                                        +64 x H_{0,-1} P_{330}
                                        +64 x H_0 P_{343}
                                \Big) H_{0,1}
                                +\Big(
                                        -384 x H_{-1} H_0 P_{323}
                                        +64 x H_0 H_1 P_{338}
                                        -16 x H_0 P_{413}
                                \Big) H_{0,-1}
                                +\Big(
                                        -128 x H_{-1} P_{336}
                                        -16 x P_{420}
                                        -64 x^2 \Big(
                                                3+22 x+3 x^2\Big) H_1
                                \Big) H_{0,0,1}
                                +\Big(
                                        -128 x H_1 P_{346}
                                        +8 x P_{464}
                                        +128 x^2 \Big(
                                                23-20 x+23 x^2\Big) H_{-1}
                                \Big) H_{0,0,-1}
                                +64 x^2 \Big(
                                        3+22 x+3 x^2\Big) H_{0,0,1,1}
                                -128 x^2 \Big(
                                        23-20 x+23 x^2\Big) H_{0,0,-1,-1}
                                +\Big(
                                        384 x H_{-1} H_1 P_{324}
                                        -384 x H_{-1,1} P_{324}
                                        -48 x H_{-1}^2 P_{373}
                                        +24 x H_{-1} P_{425}
                                        +4 x P_{428}
                                \Big) \zeta_2
                                +\Big(
                                        32 x H_1 P_{347}
                                        +32 x H_{-1} P_{390}
                                        -16 x P_{440}
                                \Big) \zeta_3
                        \Big) x_+^2
                        +\Big(
                                -
                                \frac{16}{3} x H_0^3 H_1 P_{470}
                                +\frac{8}{3} x H_{-1} H_0^3 P_{472}
                                +\Big(
                                        16 x H_{-1} H_0 P_{476}
                                        -32 (-2+x) x \Big(
                                                5+39 x-129 x^2+101 x^3\Big) H_{0,1}
                                \Big) \zeta_2
                        \Big) x_+^3
                \Big)
                +\eta  \Big(
                        \Big(
                                3 x \Big(
                                        83+206 x+83 x^2\Big) H_0
                                +32 x \Big(
                                        41-60 x+41 x^2\Big) H_{-1} H_0
                                -24 x \Big(
                                        51-122 x+51 x^2\Big) H_{-1}^2 H_0
                                +\Big(
                                        -176 x \Big(
                                                3+2 x+3 x^2\Big) H_0
                                        -768 x^2 H_{-1} H_0
                                \Big) H_1
                                +128 x^2 H_0 H_1^2
                                +\Big(
                                        176 x \Big(
                                                3+2 x+3 x^2\Big)
                                        -256 x^2 H_1
                                        +768 x^2 H_{-1}
                                \Big) H_{0,1}
                                +\Big(
                                        -32 x \Big(
                                                41-60 x+41 x^2\Big)
                                        +768 x^2 H_1
                                        +48 x \Big(
                                                51-122 x+51 x^2\Big) H_{-1}
                                \Big) H_{0,-1}
                                +256 x^2 H_{0,1,1}
                                -768 x^2 H_{0,1,-1}
                                -768 x^2 H_{0,-1,1}
                                -48 x \Big(
                                        51-122 x+51 x^2\Big) H_{0,-1,-1}
                                -128 x^2 H_1 \zeta_2
                        \Big) x_+^2
                        +\Big(
                                -128 x^3 H_{-1} H_0^3 H_1
                                +64 x^3 H_0^3 H_1^2
                                +\Big(
                                        -64 x^2 \Big(
                                                11-41 x+11 x^2\Big) H_{-1} H_0^2
                                        +640 x^3 H_0^2 H_1
                                \Big) H_{0,1}
                                +\Big(
                                        1024 x^2 \Big(
                                                2-11 x+2 x^2\Big) H_{-1} H_0
                                        -4096 x^3 H_0 H_1
                                \Big) H_{0,0,1}
                                +\Big(
                                        -128 x^2 \Big(
                                                19-73 x+19 x^2\Big) H_{-1} H_0
                                        +4096 x^3 H_0 H_1
                                \Big) H_{0,0,-1}
                                +\Big(
                                        -1536 (-2+x) x^2 (-1+2 x) H_1
                                        +18816 x^3 H_{-1}
                                \Big) H_{0,0,0,1}
                                -1024 x^3 H_0 H_{0,0,1,1}
                                +\Big(
                                        -32 x H_0 H_1 P_{329}
                                        -768 x^3 H_{-1} H_0 H_1
                                        +384 x^3 H_0 H_1^2
                                        +\Big(
                                                -768 x^3 H_1
                                                +768 x^3 H_{-1}
                                        \Big) H_{0,1}
                                        +768 x^3 H_{0,1,1}
                                \Big) \zeta_2
                                +\Big(
                                        896 x^3 H_{-1} H_0
                                        +1024 x^3 H_0 H_1
                                \Big) \zeta_3
                        \Big) x_+^4
                \Big)
                -20 x x_+^2
                +32 c_1 x x_+^2
                -1056 x^2 H_{0,1}^2 x_+^4
\Bigg\}
\Bigg]
\\
+ C_F C_A  \Bigg[
%
%
%
%
  \Bigg\{
                \eta ^2 x_+^2 \Big(
                        -4 x H_{-1} H_0^2 P_{382}
                        +8 x H_0 H_{0,-1} P_{392}
                        -8 x H_{0,0,-1} P_{397}
                        -\frac{2}{3} x H_0^2 P_{402}
                        +\Big(
                                -24 x H_{-1} P_{355}
                                -\frac{4 x P_{451}}{3}
                        \Big) \zeta_2
                \Big)
                +\xi ^2 \eta ^3 \Big(
                        \frac{2}{3} (-1+x)^2 x^2 H_0^4
                        +96 (-1+x)^2 x^2 H_{0,-1} \zeta_2
                \Big)
                +\xi  \eta ^3 x_+ \Big(
                        -2 (-1+x) x^3 H_0^4
                        +864 (-1+x) x^3 H_{0,-1} \zeta_2
                \Big)
                +\eta ^3 x_+^2 \Big(
                        16 x^2 H_0^2 H_{0,-1}
                         P_{389}
                        +96 x^2 H_{0,0,0,-1} P_{398}
                        -
                        \frac{4}{3} x^2 H_0^3 P_{474}
                        -\frac{8}{3} x^4 H_0^4
                        +\Big(
                                8 x^2 H_0^2 P_{367}
                                -8 x^2 H_0 P_{475}
                                +768 x^4 H_{0,-1}
                        \Big) \zeta_2
                        +\frac{24}{5} x^2 P_{417} \zeta_2^2
                \Big)
                +\eta  \Big(
                        \Big(
                                -\frac{2}{9} x \Big(
                                        173+670 x+173 x^2\Big) H_0
                                +\frac{8}{3} x \Big(
                                        5-8 x+5 x^2\Big) H_{-1} H_0
                                -\frac{8}{3} x \Big(
                                        5-8 x+5 x^2\Big) H_{0,-1}
                        \Big) x_+^2
                        +\Big(
                                -32 x^2 \Big(
                                        17-47 x+17 x^2\Big) \zeta_3 H_0
                                +32 x^2 \Big(
                                        3-8 x+3 x^2\Big) H_0^3 H_1
                                -80 x^2 \Big(
                                        5-8 x+5 x^2\Big) H_0^2 H_{0,1}
                                +128 x^2 \Big(
                                        8-5 x+8 x^2\Big) H_0 H_{0,0,1}
                                -32 x^2 \Big(
                                        41-62 x+41 x^2\Big) H_0 H_{0,0,-1}
                                -96 x^2 \Big(
                                        13+4 x+13 x^2\Big) H_{0,0,0,1}
\end{dmath*}
\begin{dmath*}
{\color{white}=}
                                +\Big(
                                        192 x^2 \Big(
                                                3-8 x+3 x^2\Big) H_0 H_1
                                        -192 x^2 \Big(
                                                3-8 x+3 x^2\Big) H_{0,1}
                                \Big) \zeta_2
                        \Big) x_+^4
                \Big)
                +12 x x_+^2
                +\Big(
                        -48 (-1+x) x H_0 H_1
                        +48 (-1+x) x H_{0,1}
                \Big) x_+^3
                +\Big(
                        16 x \Big(
                                2+x+2 x^2\Big) H_0^2 H_1
                        -32 x \Big(
                                2+7 x+2 x^2\Big) H_0 H_{0,1}
                        +32 x \Big(
                                2+13 x+2 x^2\Big) H_{0,0,1}
                        +16 x \Big(
                                5-71 x+5 x^2\Big) \zeta_3
                \Big) x_+^4
                +96 x \log (2) x_+^2 \zeta_2
\Bigg\}
+ \ep  \Bigg\{
                \xi ^2 \Big(
                        \eta ^3 \Big(
                                \frac{4}{3} (-1+x)^2 x^2 H_{-1} H_0^4
                                +192 (-1+x)^2 x^2 H_{-1} H_{0,-1} \zeta_2
                        \Big)
                        +64 (-1+x)^2 x \eta ^2 H_{-1} H_0 H_{0,-1}
                \Big)
                +\eta ^2 \Big(
                        \log (2) x_+^2 \zeta_2 \Big(
                                192 x H_1 P_{324}
                                +192 x H_{-1} P_{324}
                                +192 x P_{344}
                        \Big)
                        +\Big(
                                -32 x H_0^2 H_1^2 P_{320}
                                +128 x H_{0,0,1,1} P_{341}
                                -128 x H_0 H_{0,1,1} P_{342}
                                +64 x H_{0,-1,0,1} P_{345}
                                -64 x H_{0,0,1,-1} P_{357}
                                -64 x H_{0,0,-1,1} P_{357}
                                +8 x H_{-1} H_0^2 P_{384}
                                +16 x H_{0,0,-1,-1} P_{411}
                                +8 x H_{0,-1}^2 P_{418}
                                -16 x H_0 H_{0,-1,-1} P_{421}
                                -\frac{2}{9} x H_0^2 P_{469}
                                -32 x^2 \Big(
                                        31-64 x+31 x^2\Big) H_{-1}^2 H_0^2
                                +\Big(
                                        -32 x H_{-1} H_0^2 P_{324}
                                        -32 x^2 \Big(
                                                -4+23 x-16 x^2+6 x^3\Big) H_0^2
                                \Big) H_1
                                +\Big(
                                        64 x H_0 H_1 P_{348}
                                        -32 x H_{-1} H_0 P_{352}
                                        +32 x H_0 P_{361}
                                        -32 x H_{0,-1} P_{379}
                                \Big) H_{0,1}
                                +\Big(
                                        -8 x H_0 P_{408}
                                        +768 x^3 H_{-1} H_0
                                \Big) H_{0,-1}
                                +\Big(
                                        -64 x P_{321}
                                        -128 x H_1 P_{341}
                                        +64 x H_{-1} P_{357}
                                \Big) H_{0,0,1}
                                +\Big(
                                        64 x H_1 P_{376}
                                        +16 x P_{381}
                                        -16 x H_{-1} P_{411}
                                \Big) H_{0,0,-1}
                                +\Big(
                                        -192 x H_{-1} H_1 P_{324}
                                        +192 x H_{-1,1} P_{324}
                                        +24 x H_{-1}^2 P_{373}
                                        -16 x H_{-1} P_{374}
                                        -
                                        \frac{4 x P_{468}}{9}
                                \Big) \zeta_2
                                +\Big(
                                        -16 x H_1 P_{377}
                                        -16 x H_{-1} P_{395}
                                        -8 x P_{429}
                                \Big) \zeta_3
                        \Big) x_+^2
                \Big)
                +\eta ^3 \Big(
                        \log (2) x_+^2 \zeta_2 \Big(
                                -768 (-2+x) x^3 (-1+2 x) H_{0,1}
                                -768 (-2+x) x^3 (-1+2 x) H_{0,-1}
                        \Big)
                        +\Big(
                                96 x^2 H_0^2 H_{0,1,-1} P_{366}
                                +96 x^2 H_0^2 H_{0,-1,1} P_{366}
                                +96 x^2 H_0 H_{0,-1}^2 P_{368}
                                -64 x^2 H_0 H_{0,1}^2 P_{407}
                                +32 x^2 H_0^2 H_{0,1,1} P_{410}
                                -128 x^2 H_{0,0,-1,0,-1} P_{416}
                                -128 x^2 H_0 H_{0,0,1,-1} P_{431}
                                -128 x^2 H_0 H_{0,0,-1,1} P_{431}
                                -32 x^2 H_0^2 H_{0,-1,-1} P_{432}
                                +384 x^2 H_{0,0,0,1,-1} P_{433}
                                +384 x^2 H_{0,0,0,-1,1} P_{433}
                                -64 x^2 H_0 H_{0,-1,0,1} P_{434}
                                -192 x^2 H_{0,0,0,1,1} P_{439}
                                -192 x^2 H_{0,0,0,-1,-1} P_{444}
                                -64 x^2 H_{0,0,1,0,1} P_{445}
                                +64 x^2 H_0 H_{0,0,-1,-1} P_{447}
                                +64 x^2 H_{0,0,1,0,-1} P_{449}
                                +64 x^2 H_{0,0,-1,0,1} P_{457}
                                +32 x^2 H_{0,0,0,0,1} P_{461}
                                -32 x^2 H_{0,0,0,0,-1} P_{465}
                                -\frac{1}{3} x^2 H_0^4 P_{477}
                                +\frac{2}{3} x H_0^3 P_{481}
                                -\frac{2}{15} x^3 \Big(
                                        3-4 x+3 x^2\Big) H_0^5
                                +\Big(
                                        \frac{4}{3}
                                         x^2 H_0^4 P_{387}
                                        +
                                        \frac{16}{3} x H_0^3 P_{480}
                                \Big) H_1
                                +\Big(
                                        \frac{4}{3} x H_0^3 P_{500}
                                        -\frac{16}{3} x^4 H_0^4
                                \Big) H_{-1}
                                +\Big(
                                        -\frac{16}{3} x^2 H_0^3 P_{435}
                                        +64 x^2 H_0 H_{0,-1} P_{437}
                                        -16 x H_0^2 P_{482}
                                \Big) H_{0,1}
                                +\Big(
                                        -32 x^2 H_0^2 H_1 P_{388}
                                        +32 x^2 H_{-1} H_0^2 P_{389}
                                        +\frac{16}{3} x^2 H_0^3 P_{442}
                                        -4 x H_0^2 P_{488}
                                \Big) H_{0,-1}
                                +\Big(
                                        -64 x^2 H_{0,-1} P_{399}
                                        +64 x^2 H_{0,1} P_{438}
                                        +16 x^2 H_0^2 P_{448}
                                        -32 x H_0 P_{483}
                                \Big) H_{0,0,1}
                                +\Big(
                                        256 x^2 H_{0,-1} P_{318}
                                        -64 x^2 H_{0,1} P_{454}
                                        -16 x^2 H_0^2 P_{458}
                                        +8 x H_0 P_{497}
                                \Big) H_{0,0,-1}
                                +\Big(
                                        -32 x^2 H_0 P_{456}
                                        +32 x P_{494}
                                \Big) H_{0,0,0,1}
                                +\Big(
                                        192 x^2 H_{-1} P_{398}
                                        +192 x^2 H_1 P_{422}
                                        +32 x^2 H_0 P_{463}
                                        -8 x P_{504}
                                \Big) H_{0,0,0,-1}
                                +\Big(
                                        32 x^2 H_{0,0,1} P_{312}
                                        -256 x^2 H_{0,0,-1} P_{332}
                                        -192 x^2 H_{0,-1,-1} P_{353}
                                        +192 x^2 H_{0,-1,1} P_{359}
                                        +192 x^2 H_{0,1,-1} P_{360}
                                        +\frac{8}{3} x^2 H_0^3 P_{393}
                                        +4 x H_0^2 P_{498}
                                        -2 x H_0 P_{505}
                                        +\Big(
                                                -16 x^2 H_0^2 P_{424}
                                                +32 x H_0 P_{489}
                                        \Big) H_1
                                        +\Big(
                                                16 x^2 H_0^2 P_{367}
                                                +8 x H_0 P_{486}
                                        \Big) H_{-1}
                                        +\Big(
                                                32 x^2 H_0 P_{403}
                                                -32 x P_{490}
                                        \Big) H_{0,1}
                                        +\Big(
                                                -192 x^2 H_1 P_{359}
                                                +32 x^2 H_0 P_{385}
                                                +8 x P_{501}
                                                +1536 x^4 H_{-1}
                                        \Big) H_{0,-1}
                                        -32 x^2 P_{365} \zeta_3
                                \Big) \zeta_2
                                +\Big(
                                        \frac{48}{5} x^2 H_{-1} P_{417}
                                        -\frac{48}{5} x^2 H_1 P_{453}
                                        -\frac{8}{5} x^2 H_0 P_{467}
                                        +\frac{4 x P_{503}}{5}
                                \Big) \zeta_2^2
\end{dmath*}
\begin{dmath*}
{\color{white}=}
                                +\Big(
                                        64 x^2 H_{0,-1} P_{406}
                                        -8 x^2 H_0^2 P_{409}
                                        +32 x^2 H_{0,1} P_{412}
                                        -16 x H_0 P_{493}
                                \Big) \zeta_3
                                +48 x^2 P_{446} \zeta_5
                        \Big) x_+^2
                \Big)
                +\xi  \Big(
                        \eta ^2 x_+ \Big(
                                -4 (-1+x) x \Big(
                                        31-64 x+31 x^2\Big) H_{-1}^2 H_0^2
                                +448 (-1+x) x^2 H_{-1} H_0 H_{0,-1}
                        \Big)
                        +\eta  x_+^3 \Big(
                                64 (-1+x) x^2 H_0^2 H_1 H_{0,1}
                                +512 (-1+x) x^2 H_0 H_{0,0,1,1}
                        \Big)
                        +\eta ^3 x_+ \Big(
                                -4 (-1+x) x^3 H_{-1} H_0^4
                                -\frac{2}{15} (-1+x) x^2 \Big(
                                        3-4 x+3 x^2\Big) H_0^5
                                +1728 (-1+x) x^3 H_{-1} H_{0,-1} \zeta_2
                        \Big)
                        +\Big(
                                -576 (-1+x) x H_0 H_{0,1,-1}
                                -576 (-1+x) x H_0 H_{0,-1,1}
                        \Big) x_+^3
                \Big)
                +\eta  \Big(
                        \Big(
                                -\frac{7}{27} x \Big(
                                        1151+3922 x+1151 x^2\Big) H_0
                                -\frac{4}{9} x \Big(
                                        749-2246 x+749 x^2\Big) H_{-1} H_0
                                +8 x \Big(
                                        7-16 x+7 x^2\Big) H_{-1}^2 H_0
                                +32 x \Big(
                                        11-19 x+11 x^2\Big) H_0 H_1
                                -32 x \Big(
                                        11-19 x+11 x^2\Big) H_{0,1}
                                +\Big(
                                        \frac{4}{9} x \Big(
                                                749-2246 x+749 x^2\Big)
                                        -16 x \Big(
                                                7-16 x+7 x^2\Big) H_{-1}
                                \Big) H_{0,-1}
                                +16 x \Big(
                                        7-16 x+7 x^2\Big) H_{0,-1,-1}
                        \Big) x_+^2
                        +\Big(
                                64 x^2 \Big(
                                        3-8 x+3 x^2\Big) H_{-1} H_0^3 H_1
                                -32 x^2 \Big(
                                        3-8 x+3 x^2\Big) H_0^3 H_1^2
                                +\Big(
                                        -160 x^2 \Big(
                                                5-8 x+5 x^2\Big) H_{-1} H_0^2
                                        -320 x^3 H_0^2 H_1
                                \Big) H_{0,1}
                                +\Big(
                                        256 x^2 \Big(
                                                8-5 x+8 x^2\Big) H_{-1} H_0
                                        +256 x^2 \Big(
                                                10-19 x+10 x^2\Big) H_0 H_1
                                \Big) H_{0,0,1}
                                +\Big(
                                        -64 x^2 \Big(
                                                41-62 x+41 x^2\Big) H_{-1} H_0
                                        -256 x^2 \Big(
                                                10-19 x+10 x^2\Big) H_0 H_1
                                \Big) H_{0,0,-1}
                                +\Big(
                                        -384 x^2 \Big(
                                                16-35 x+16 x^2\Big) H_1
                                        -192 x^2 \Big(
                                                13+4 x+13 x^2\Big) H_{-1}
                                \Big) H_{0,0,0,1}
                                +512 x^3 H_0 H_{0,0,1,1}
                                +\Big(
                                        384 x^2 \Big(
                                                3-8 x+3 x^2\Big) H_{-1} H_0 H_1
                                        -192 x^2 \Big(
                                                3-8 x+3 x^2\Big) H_0 H_1^2
                                        +\Big(
                                                384 x^2 \Big(
                                                        3-8 x+3 x^2\Big) H_1
                                                -384 x^2 \Big(
                                                        3-8 x+3 x^2\Big) H_{-1}
                                        \Big) H_{0,1}
                                        -384 x^2 \Big(
                                                3-8 x+3 x^2\Big) H_{0,1,1}
                                \Big) \zeta_2
                                +\Big(
                                        -64 x^2 \Big(
                                                17-47 x+17 x^2\Big) H_{-1} H_0
                                        -64 x^2 \Big(
                                                10-19 x+10 x^2\Big) H_0 H_1
                                \Big) \zeta_3
                        \Big) x_+^4
                \Big)
                +78 x x_+^2
                -16 c_1 x x_+^2
                +\Big(
                        288 (-1+x) x H_{-1} H_0 H_1
                        -48 (-1+x) x H_0 H_1^2
                        +\Big(
                                96 (-1+x) x H_1
                                -288 (-1+x) x H_{-1}
                        \Big) H_{0,1}
                        -288 (-1+x) x H_1 H_{0,-1}
                        -96 (-1+x) x H_{0,1,1}
                        +288 (-1+x) x H_{0,1,-1}
                        +288 (-1+x) x H_{0,-1,1}
                        +48 (-1+x) x H_1 \zeta_2
                \Big) x_+^3
                +\Big(
                        32 (-2+x) x (-1+2 x) H_{0,1}^2
                        -32 x \Big(
                                11-62 x+11 x^2\Big) H_0 H_1 H_{0,-1}
                        -2880 x^2 H_0 H_{0,1,-1}
                        -2880 x^2 H_0 H_{0,-1,1}
                \Big) x_+^4
\Bigg\}
\Bigg]
\\
+ C_F n_l T_F  \Bigg[
%
%
 \eta  \Bigg\{
                \frac{200}{9} x H_0
                -\frac{32}{3} x H_{-1} H_0
                +\frac{8}{3} x H_0^2
                +\frac{32}{3} x H_{0,-1}
                -\frac{16 x \zeta_2}{3}
\Bigg\}
+ \ep \eta  \Bigg\{
                \frac{3844}{27} x H_0
                +32 x H_{-1}^2 H_0
                +\frac{296}{9} x H_0^2
                +\frac{8}{3} x H_0^3
                +\Big(
                        -\frac{1184}{9} x H_0
                        -16 x H_0^2
                \Big) H_{-1}
                +\Big(
                        \frac{1184 x}{9}
                        -64 x H_{-1}
                \Big) H_{0,-1}
                +32 x H_{0,0,-1}
                +64 x H_{0,-1,-1}
                +\Big(
                        -\frac{592 x}{9}
                        +8 x H_0
                        +32 x H_{-1}
                \Big) \zeta_2
                -32 x \zeta_3
\Bigg\}
\Bigg]
\\
+ C_F T_F  \Bigg[
%
%
  \Bigg\{
                \xi  x_+^3 \Big(
                        -\frac{8}{3} (-1+x) x H_0^2
                        +16 (-1+x) x \zeta_2
                \Big)
                +\eta  \Big(
                        \frac{8}{9} x \Big(
                                25-322 x+25 x^2\Big) x_+^2 H_0
                        +\Big(
                                -32 x^3 H_0^3
                                -192 x^3 H_0 \zeta_2
                        \Big) x_+^4
                \Big)
                +\frac{272 x x_+^2}{3}
                +\Big(
                        -\frac{160}{3} x^2 H_0^2
                        -576 x^2 \zeta_2
                \Big) x_+^4
\Bigg\}
\end{dmath*}
\begin{dmath*}
{\color{white}=}
+ \ep  \Bigg\{
                \eta  \Big(
                        \Big(
                                \frac{4}{27} x \Big(
                                        961-4138 x+961 x^2\Big) H_0
                                -\frac{16}{9} x \Big(
                                        25-322 x+25 x^2\Big) H_{-1} H_0
                                +\frac{16}{9} x \Big(
                                        25-322 x+25 x^2\Big) H_{0,-1}
                        \Big) x_+^2
                        +\Big(
                                -\frac{16}{9} x \Big(
                                        -16+684 x-981 x^2+41 x^3\Big) \zeta_2
                                -\frac{8}{9} x \Big(
                                        -37+273 x+24 x^2+12 x^3\Big) H_0^2
                        \Big) x_+^3
                        +\Big(
                                -\frac{8}{9} x H_0^3 P_{351}
                                -64 x^3 H_{-1} H_0^3
                                -24 x^3 H_0^4
                                -384 x^3 H_0^2 H_{0,1}
                                +384 x^3 H_0^2 H_{0,-1}
                                +1536 x^3 H_0 H_{0,0,1}
                                -768 x^3 H_0 H_{0,0,-1}
                                -2304 x^3 H_{0,0,0,1}
                                +384 x^3 H_{0,0,0,-1}
                                +\Big(
                                        -\frac{8}{3} x H_0 P_{350}
                                        -384 x^3 H_{-1} H_0
                                        -96 x^3 H_0^2
                                        +1536 x^3 H_{0,-1}
                                \Big) \zeta_2
                                +\frac{96}{5} x^3 \zeta_2^2
                                +384 x^3 H_0 \zeta_3
                        \Big) x_+^4
                \Big)
                +\xi  \Big(
                        \Big(
                                \frac{16}{3} (-1+x) x H_{-1} H_0^2
                                -\frac{32}{3} (-1+x) x H_0^2 H_1
                                +\frac{64}{3} (-1+x) x H_0 H_{0,1}
                                -\frac{64}{3} (-1+x) x H_{0,0,1}
                                -\frac{32}{3} (-1+x) x H_{0,0,-1}
                                +32 (-1+x) x H_{-1} \zeta_2
                        \Big) x_+^3
                        -128 (-1+x) x \log (2) x_+^3 \zeta_2
                \Big)
                +\frac{1528 x x_+^2}{9}
                +\Big(
                        \frac{320}{3} x^2 H_{-1} H_0^2
                        -\frac{640}
                        {3} x^2 H_0^2 H_1
                        +\frac{1280}{3} x^2 H_0 H_{0,1}
                        -\frac{1280}{3} x^2 H_{0,0,1}
                        -\frac{640}{3} x^2 H_{0,0,-1}
                        +640 x^2 H_{-1} \zeta_2
                        -\frac{64}{3} x \Big(
                                4+67 x+4 x^2\Big) \zeta_3
                \Big) x_+^4
                +2816 x^2 \log (2) x_+^4 \zeta_2
\Bigg\}
\Bigg] \,.
\end{dmath*}
The polynomials are given by
\begin{align}
P_{311} &= x^4-96 x^3+178 x^2-96 x+1,\nonumber\\
P_{312} &= x^4-69 x^3+106 x^2-69 x+1,\nonumber\\
P_{313} &= x^4-67 x^3+140 x^2-67 x+1,\nonumber\\
P_{314} &= x^4-64 x^3+122 x^2-64 x+1,\nonumber\\
P_{315} &= x^4-54 x^3+102 x^2-54 x+1,\nonumber\\
P_{316} &= x^4-52 x^3+62 x^2-52 x+1,\nonumber\\
P_{317} &= x^4-52 x^3+98 x^2-52 x+1,\nonumber\\
P_{318} &= x^4-29 x^3+51 x^2-29 x+1,\nonumber\\
P_{319} &= x^4-27 x^3+64 x^2-27 x+1,\nonumber\\
P_{320} &= x^4-20 x^3+39 x^2-20 x+1,\nonumber\\
P_{321} &= x^4-18 x^3+61 x^2-30 x+7,\nonumber\\
P_{322} &= x^4-2 x^3+12 x^2-2 x+1,\nonumber\\
P_{323} &= x^4-x^3+6 x^2-x+1,\nonumber\\
P_{324} &= x^4+5 x^3-10 x^2+5 x+1,\nonumber\\
P_{325} &= x^4+15 x^3-26 x^2+15 x+1,\nonumber\\
P_{326} &= x^4+22 x^3-18 x^2+22 x+1,\nonumber\\
P_{327} &= x^4+47 x^3-88 x^2+47 x+1,\nonumber\\
P_{328} &= x^4+53 x^3-112 x^2+53 x+1,\nonumber\\
P_{329} &= x^4+93 x^3-222 x^2+91 x+9,\nonumber\\
P_{330} &= x^4+115 x^3-228 x^2+115 x+1,\nonumber\\
P_{331} &= 2 x^4-44 x^3+61 x^2-44 x+2,\nonumber\\
P_{332} &= 2 x^4-33 x^3+53 x^2-33 x+2,\nonumber\\
P_{333} &= 2 x^4-16 x^3+27 x^2-16 x+2,\nonumber\\
P_{334} &= 2 x^4-10 x^3+9 x^2-10 x+2,\nonumber\\
P_{335} &= 2 x^4-10 x^3+17 x^2-10 x+2,\nonumber\\
P_{336} &= 2 x^4+27 x^3-28 x^2+27 x+2,\nonumber\\
P_{337} &= 2 x^4+29 x^3-42 x^2+29 x+2,\nonumber\\
P_{338} &= 2 x^4+75 x^3-130 x^2+75 x+2,\nonumber\\
P_{339} &= 2 x^4+137 x^3-246 x^2+137 x+2,\nonumber\\
P_{340} &= 3 x^4-410 x^3+770 x^2-410 x+3,\nonumber\\
P_{341} &= 3 x^4-35 x^3+65 x^2-35 x+3,\nonumber\\
P_{342} &= 3 x^4-32 x^3+59 x^2-32 x+3,\nonumber\\
P_{343} &= 3 x^4-26 x^3-12 x^2-26 x+3,\nonumber\\
P_{344} &= 3 x^4-4 x^3+4 x^2-4 x+3,\nonumber\\
P_{345} &= 3 x^4+2 x^3-8 x^2+2 x+3,\nonumber\\
P_{346} &= 3 x^4+80 x^3-140 x^2+80 x+3,\nonumber\\
P_{347} &= 3 x^4+202 x^3-388 x^2+202 x+3,\nonumber\\
P_{348} &= 4 x^4-55 x^3+104 x^2-55 x+4,\nonumber\\
P_{349} &= 4 x^4-8 x^3+9 x^2-8 x+4,\nonumber\\
P_{350} &= 5 x^4-172 x^3-54 x^2-12 x-3,\nonumber\\
P_{351} &= 5 x^4-124 x^3-24 x^2+36 x-3,\nonumber\\
P_{352} &= 5 x^4-88 x^3+162 x^2-88 x+5,\nonumber\\
P_{353} &= 5 x^4-53 x^3+70 x^2-53 x+5,\nonumber\\
P_{354} &= 5 x^4-41 x^3+61 x^2-41 x+5,\nonumber\\
P_{355} &= 5 x^4-12 x^3+26 x^2-12 x+5,\nonumber\\
P_{356} &= 5 x^4-9 x^3+20 x^2-9 x+5,\nonumber\\
P_{357} &= 6 x^4-83 x^3+152 x^2-83 x+6,\nonumber\\
P_{358} &= 6 x^4-40 x^3+67 x^2-40 x+6,\nonumber\\
P_{359} &= 7 x^4-49 x^3+78 x^2-49 x+7,\nonumber\\
P_{360} &= 7 x^4-41 x^3+58 x^2-41 x+7,\nonumber\\
P_{361} &= 7 x^4-34 x^3+84 x^2-34 x+7,\nonumber\\
P_{362} &= 7 x^4-31 x^3+50 x^2-31 x+7,\nonumber\\
P_{363} &= 7 x^4-7 x^3+2 x^2-7 x+7,\nonumber\\
P_{364} &= 7 x^4+9 x^3-26 x^2+9 x+7,\nonumber\\
P_{365} &= 7 x^4+21 x^3-11 x^2+21 x+7,\nonumber\\
P_{366} &= 8 x^4-31 x^3+44 x^2-31 x+8,\nonumber\\
P_{367} &= 8 x^4-31 x^3+60 x^2-31 x+8,\nonumber\\
P_{368} &= 8 x^4-19 x^3+24 x^2-19 x+8,\nonumber\\
P_{369} &= 8 x^4-12 x^3+11 x^2-12 x+8,\nonumber\\
P_{370} &= 8 x^4-4 x^3+9 x^2-4 x+8,\nonumber\\
P_{371} &= 9 x^4+398 x^3-698 x^2+398 x+9,\nonumber\\
P_{372} &= 11 x^4-77 x^3+109 x^2-77 x+11,\nonumber\\
P_{373} &= 11 x^4-56 x^3+118 x^2-56 x+11,\nonumber\\
P_{374} &= 11 x^4-35 x^3+267 x^2-91 x+40,\nonumber\\
P_{375} &= 11 x^4+29 x^3-84 x^2+29 x+11,\nonumber\\
P_{376} &= 12 x^4-79 x^3+136 x^2-79 x+12,\nonumber\\
P_{377} &= 12 x^4+43 x^3-112 x^2+43 x+12,\nonumber\\
P_{378} &= 13 x^4-49 x^3+74 x^2-49 x+13,\nonumber\\
P_{379} &= 13 x^4-38 x^3+54 x^2-38 x+13,\nonumber\\
P_{380} &= 14 x^4-62 x^3+97 x^2-62 x+14,\nonumber\\
P_{381} &= 14 x^4-55 x^3+325 x^2-107 x+39,\nonumber\\
P_{382} &= 15 x^4-140 x^3+254 x^2-140 x+15,\nonumber\\
P_{383} &= 15 x^4-67 x^3+106 x^2-67 x+15,\nonumber\\
P_{384} &= 15 x^4+17 x^3-115 x^2+69 x-10,\nonumber\\
P_{385} &= 16 x^4-139 x^3+200 x^2-139 x+16,\nonumber\\
P_{386} &= 17 x^4-123 x^3+167 x^2-123 x+17,\nonumber\\
P_{387} &= 17 x^4-111 x^3+182 x^2-111 x+17,\nonumber\\
P_{388} &= 17 x^4-95 x^3+154 x^2-95 x+17,\nonumber\\
P_{389} &= 17 x^4-71 x^3+106 x^2-71 x+17,\nonumber\\
P_{390} &= 17 x^4-9 x^3+50 x^2-9 x+17,\nonumber\\
P_{391} &= 18 x^4-82 x^3+133 x^2-82 x+18,\nonumber\\
P_{392} &= 19 x^4-148 x^3+254 x^2-148 x+19,\nonumber\\
P_{393} &= 20 x^4-129 x^3+212 x^2-129 x+20,\nonumber\\
P_{394} &= 21 x^4-221 x^3+398 x^2-221 x+21,\nonumber\\
P_{395} &= 21 x^4-119 x^3+230 x^2-119 x+21,\nonumber\\
P_{396} &= 22 x^4-72 x^3+115 x^2-72 x+22,\nonumber\\
P_{397} &= 23 x^4-156 x^3+254 x^2-156 x+23,\nonumber\\
P_{398} &= 24 x^4-73 x^3+100 x^2-73 x+24,\nonumber\\
P_{399} &= 25 x^4-182 x^3+318 x^2-182 x+25,\nonumber\\
P_{400} &= 25 x^4-165 x^3+286 x^2-165 x+25,\nonumber\\
P_{401} &= 25 x^4-24 x^3+54 x^2+40 x-39,\nonumber\\
P_{402} &= 25 x^4+8 x^3-240 x^2+64 x-1,\nonumber\\
P_{403} &= 26 x^4-71 x^3+112 x^2-71 x+26,\nonumber\\
P_{404} &= 27 x^4-107 x^3+170 x^2-107 x+27,\nonumber\\
P_{405} &= 27 x^4-71 x^3+92 x^2-71 x+27,\nonumber\\
P_{406} &= 28 x^4-149 x^3+225 x^2-149 x+28,\nonumber\\
P_{407} &= 28 x^4-137 x^3+216 x^2-137 x+28,\nonumber\\
P_{408} &= 29 x^4-38 x^3+210 x^2-38 x+29,\nonumber\\
P_{409} &= 31 x^4-144 x^3+214 x^2-144 x+31,\nonumber\\
P_{410} &= 33 x^4-154 x^3+238 x^2-154 x+33,\nonumber\\
P_{411} &= 39 x^4-368 x^3+686 x^2-368 x+39,\nonumber\\
P_{412} &= 39 x^4-34 x^3-8 x^2-34 x+39,\nonumber\\
P_{413} &= 41 x^4-274 x^3+130 x^2-274 x+41,\nonumber\\
P_{414} &= 41 x^4-121 x^3+186 x^2-121 x+41,\nonumber\\
P_{415} &= 43 x^4-143 x^3+204 x^2-143 x+43,\nonumber\\
P_{416} &= 44 x^4-223 x^3+342 x^2-223 x+44,\nonumber\\
P_{417} &= 46 x^4-135 x^3+224 x^2-135 x+46,\nonumber\\
P_{418} &= 47 x^4-316 x^3+534 x^2-316 x+47,\nonumber\\
P_{419} &= 48 x^4-250 x^3+391 x^2-250 x+48,\nonumber\\
P_{420} &= 49 x^4-232 x^3-42 x^2-168 x-15,\nonumber\\
P_{421} &= 51 x^4-344 x^3+590 x^2-344 x+51,\nonumber\\
P_{422} &= 51 x^4-223 x^3+342 x^2-223 x+51,\nonumber\\
P_{423} &= 53 x^4-205 x^3+312 x^2-205 x+53,\nonumber\\
P_{424} &= 53 x^4-199 x^3+306 x^2-199 x+53,\nonumber\\
P_{425} &= 53 x^4-174 x^3+356 x^2+6 x+15,\nonumber\\
P_{426} &= 57 x^4-325 x^3+490 x^2-325 x+57,\nonumber\\
P_{427} &= 57 x^4-245 x^3+386 x^2-245 x+57,\nonumber\\
P_{428} &= 59 x^4+1366 x^3-2290 x^2+710 x+123,\nonumber\\
P_{429} &= 61 x^4-128 x^3+134 x^2-76 x+33,\nonumber\\
P_{430} &= 67 x^4-367 x^3+614 x^2-367 x+67,\nonumber\\
P_{431} &= 67 x^4-231 x^3+322 x^2-231 x+67,\nonumber\\
P_{432} &= 69 x^4-301 x^3+454 x^2-301 x+69,\nonumber\\
P_{433} &= 70 x^4-291 x^3+438 x^2-291 x+70,\nonumber\\
P_{434} &= 71 x^4-243 x^3+338 x^2-243 x+71,\nonumber\\
P_{435} &= 72 x^4-295 x^3+436 x^2-295 x+72,\nonumber\\
P_{436} &= 73 x^4-317 x^3+498 x^2-317 x+73,\nonumber\\
P_{437} &= 80 x^4-335 x^3+508 x^2-335 x+80,\nonumber\\
P_{438} &= 81 x^4-388 x^3+610 x^2-388 x+81,\nonumber\\
P_{439} &= 81 x^4-386 x^3+606 x^2-386 x+81,\nonumber\\
P_{440} &= 89 x^4-453 x^3+539 x^2-143 x-8,\nonumber\\
P_{441} &= 91 x^4-419 x^3+684 x^2-419 x+91,\nonumber\\
P_{442} &= 103 x^4-491 x^3+750 x^2-491 x+103,\nonumber\\
P_{443} &= 107 x^4-591 x^3+986 x^2-591 x+107,\nonumber\\
P_{444} &= 112 x^4-519 x^3+784 x^2-519 x+112,\nonumber\\
P_{445} &= 113 x^4-520 x^3+810 x^2-520 x+113,\nonumber\\
P_{446} &= 120 x^4-464 x^3+701 x^2-464 x+120,\nonumber\\
P_{447} &= 125 x^4-474 x^3+686 x^2-474 x+125,\nonumber\\
P_{448} &= 125 x^4-449 x^3+634 x^2-449 x+125,\nonumber\\
P_{449} &= 127 x^4-560 x^3+858 x^2-560 x+127,\nonumber\\
P_{450} &= 127 x^4-314 x^3+256 x^2-54 x+17,\nonumber\\
P_{451} &= 143 x^4-620 x^3+732 x^2-460 x+61,\nonumber\\
P_{452} &= 151 x^4-795 x^3+1318 x^2-795 x+151,\nonumber\\
P_{453} &= 151 x^4-559 x^3+862 x^2-559 x+151,\nonumber\\
P_{454} &= 159 x^4-673 x^3+1018 x^2-673 x+159,\nonumber\\
P_{455} &= 163 x^4-827 x^3+1350 x^2-827 x+163,\nonumber\\
P_{456} &= 181 x^4-633 x^3+886 x^2-633 x+181,\nonumber\\
P_{457} &= 191 x^4-805 x^3+1218 x^2-805 x+191,\nonumber\\
P_{458} &= 209 x^4-856 x^3+1266 x^2-856 x+209,\nonumber\\
P_{459} &= 215 x^4-306 x^3-148 x^2+522 x-187,\nonumber\\
P_{460} &= 221 x^4-1153 x^3+1914 x^2-1153 x+221,\nonumber\\
P_{461} &= 245 x^4-907 x^3+1302 x^2-907 x+245,\nonumber\\
P_{462} &= 303 x^4-1763 x^3+2306 x^2-1763 x+303,\nonumber\\
P_{463} &= 305 x^4-1143 x^3+1646 x^2-1143 x+305,\nonumber\\
P_{464} &= 379 x^4-1402 x^3+372 x^2-574 x-23,\nonumber\\
P_{465} &= 379 x^4-1397 x^3+1986 x^2-1397 x+379,\nonumber\\
P_{466} &= 497 x^4-686 x^3+638 x^2-294 x+37,\nonumber\\
P_{467} &= 513 x^4-1705 x^3+2998 x^2-1705 x+513,\nonumber\\
P_{468} &= 1277 x^4-3170 x^3+1683 x^2-2680 x+442,\nonumber\\
P_{469} &= 1666 x^4-3184 x^3-189 x^2+430 x+125,\nonumber\\
P_{470} &= x^5+64 x^4+185 x^3-443 x^2+238 x+11,\nonumber\\
P_{471} &= 8 x^5+46 x^4-23 x^3+x^2-x-3,\nonumber\\
P_{472} &= 9 x^5+237 x^4-112 x^3-160 x^2+331 x+31,\nonumber\\
P_{473} &= 14 x^5+43 x^4-2 x^3+22 x^2-4 x+3,\nonumber\\
P_{474} &= 19 x^5-176 x^4+382 x^3-356 x^2+91 x-8,\nonumber\\
P_{475} &= 19 x^5-138 x^4+314 x^3-356 x^2+159 x-46,\nonumber\\
P_{476} &= 23 x^5-137 x^4+208 x^3+160 x^2-43 x+45,\nonumber\\
P_{477} &= 31 x^5-220 x^4+488 x^3-560 x^2+153 x-20,\nonumber\\
P_{478} &= x^6+33 x^5-265 x^4+592 x^3-557 x^2+199 x+5,\nonumber\\
P_{479} &= 2 x^6+159 x^5-718 x^4+1124 x^3-616 x^2+69 x-12,\nonumber\\
P_{480} &= 3 x^6-158 x^5+520 x^4-742 x^3+507 x^2-172 x+18,\nonumber\\
P_{481} &= 3 x^6+360 x^5-1283 x^4+1288 x^3-5 x^2-12 x+1,\nonumber\\
P_{482} &= 4 x^6-121 x^5+529 x^4-846 x^3+594 x^2-177 x+21,\nonumber\\
P_{483} &= 4 x^6-51 x^5-262 x^4+760 x^3-684 x^2+203 x-26,\nonumber\\
P_{484} &= 5 x^6+323 x^5-654 x^4+738 x^3-185 x^2-29 x+2,\nonumber\\
P_{485} &= 6 x^6+7 x^5-316 x^4+906 x^3-906 x^2+303 x+8,\nonumber\\
P_{486} &= 7 x^6+44 x^5-207 x^4+712 x^3-739 x^2+324 x-45,\nonumber\\
P_{487} &= 9 x^6+210 x^5+697 x^4-2900 x^3+3103 x^2-1134 x-17,\nonumber\\
P_{488} &= 9 x^6+232 x^5-1443 x^4+3096 x^3-2669 x^2+1032 x-113,\nonumber\\
P_{489} &= 10 x^6-187 x^5+555 x^4-742 x^3+472 x^2-143 x+11,\nonumber\\
P_{490} &= 10 x^6-181 x^5+543 x^4-742 x^3+484 x^2-149 x+11,\nonumber\\
P_{491} &= 13 x^6-114 x^5+763 x^4-1462 x^3+1305 x^2-512 x-17,\nonumber\\
P_{492} &= 16 x^6-14 x^5+33 x^4-180 x^3+64 x^2-6 x+7,\nonumber\\
P_{493} &= 18 x^6-301 x^5+883 x^4-980 x^3+490 x^2-79 x-3,\nonumber\\
P_{494} &= 21 x^6-358 x^5+281 x^4+484 x^3-777 x^2+250 x-33,\nonumber\\
P_{495} &= 23 x^6-342 x^5+199 x^4+1128 x^3-1799 x^2+726 x+17,\nonumber\\
P_{496} &= 38 x^6-1856 x^5+4771 x^4-9562 x^3+6836 x^2-2070 x-149,\nonumber\\
P_{497} &= 39 x^6+12 x^5-1857 x^4+4240 x^3-3727 x^2+1204 x-135,\nonumber\\
P_{498} &= 45 x^6-704 x^5+1622 x^4-1268 x^3+327 x^2+36 x-2,\nonumber\\
P_{499} &= 47 x^6+66 x^5-173 x^4-816 x^3+697 x^2-378 x+29,\nonumber\\
P_{500} &= 65 x^6-604 x^5+891 x^4+712 x^3-1837 x^2+972 x-103,\nonumber\\
P_{501} &= 77 x^6-464 x^5+909 x^4-1576 x^3+703 x^2-288 x+15,\nonumber\\
P_{502} &= 105 x^6-168 x^5+2339 x^4-5364 x^3+4403 x^2-1524 x-31,\nonumber\\
P_{503} &= 147 x^6-3672 x^5+12192 x^4-15040 x^3+9997 x^2-3256 x+376,\nonumber\\
P_{504} &= 155 x^6-1264 x^5-351 x^4+4144 x^3-5011 x^2+1488 x-169,\nonumber\\
P_{505} &= 159 x^6-750 x^5+1921 x^4-2228 x^3+469 x^2-294 x+19,\nonumber\\
P_{506} &= 165 x^6-710 x^5-709 x^4+1724 x^3-665 x^2+306 x-47,\nonumber\\
P_{507} &= 327 x^6-458 x^5+1457 x^4-1020 x^3-499 x^2+158 x-29
\end{align}

\newpage

\section{The axial-vector form factor up to two-loop} \label{app:avFF}
In this appendix, we present the axial-vector form factors $F_{A,1}^{(n)}$ and $F_{A,2}^{(n)}$ up to two loops and ${\cal O}(\varepsilon)$. 

\begin{dmath*}
 F_{A,1}^{(1)} = C_F \Bigg[
\frac{1}{\ep}  \Bigg\{
-2
-2 \xi  H_0
\Bigg\}
+   \Bigg\{
-4
+\xi  \Big(
        4 H_{-1} H_0
        -H_0^2
        -4 H_{0,-1}
        +2 \zeta_2
\Big)
+\Big(
        -3+2 x-3 x^2\Big) \eta  H_0
\Bigg\}
\end{dmath*}
\vspace{-0.6cm}
\begin{dmath}
{\color{white}=}
+ \ep  \Bigg\{
-8
+\eta  \Big(
        2 \Big(
                3-2 x+3 x^2\Big) H_{-1} H_0
        +\frac{1}{2} \Big(
                -3+2 x-3 x^2\Big) H_0^2
        -2 \Big(
                3-2 x+3 x^2\Big) H_{0,-1}
        +2 \Big(
                1-x+2 x^2\Big) \zeta_2
\Big)
+\xi  \Big(
        \Big(
                -8
                -4 H_{-1}^2
                +\zeta_2
        \Big) H_0
        +2 H_{-1} H_0^2
        -\frac{1}{3} H_0^3
        +8 H_{-1} H_{0,-1}
        -4 H_{0,0,-1}
        -8 H_{0,-1,-1}
        -4 H_{-1} \zeta_2
        +4 \zeta_3
\Big)
\Bigg\}
+ \ep^2  \Bigg\{
-16
+\eta  \Big(
        \Big(
                -2 \Big(
                        3-2 x+3 x^2\Big) H_{-1}^2
                +\frac{1}{2} \Big(
                        3-2 x+3 x^2\Big) \zeta_2
        \Big) H_0
        +\Big(
                3-2 x+3 x^2\Big) H_{-1} H_0^2
        +\frac{1}{6} \Big(
                -3+2 x-3 x^2\Big) H_0^3
        +4 \Big(
                3-2 x+3 x^2\Big) H_{-1} H_{0,-1}
        -2 \Big(
                3-2 x+3 x^2\Big) H_{0,0,-1}
        -4 \Big(
                3-2 x+3 x^2\Big) H_{0,-1,-1}
        +2 \Big(
                3+5 x^2\Big) \zeta_2
        -2 \Big(
                3-2 x+3 x^2\Big) H_{-1} \zeta_2
        +\frac{4}{3} \Big(
                5-3 x+4 x^2\Big) \zeta_3
\Big)
+\xi  \Big(
        \Big(
                -16
                +\Big(
                        16
                        -2 \zeta_2
                \Big) H_{-1}
                +\frac{8}{3} H_{-1}^3
                +\frac{14}{3} \zeta_3
        \Big) H_0
        +\Big(
                -4
                -2 H_{-1}^2
                +\frac{1}{2} \zeta_2
        \Big) H_0^2
        +\frac{2}{3} H_{-1} H_0^3
        -\frac{1}{12} H_0^4
        +\Big(
                -16
                -8 H_{-1}^2
                -2 \zeta_2
        \Big) H_{0,-1}
        +8 H_{-1} H_{0,0,-1}
        +16 H_{-1} H_{0,-1,-1}
        -4 H_{0,0,0,-1}
        -8 H_{0,0,-1,-1}
        -16 H_{0,-1,-1,-1}
        +4 H_{-1}^2 \zeta_2
        +\frac{14}{5} \zeta_2^2
        -8 H_{-1} \zeta_3
\Big)
\Bigg\}
\\
+ \ep^3  \Bigg\{
-32
+\eta  \Big(
        \Big(
                \frac{4}{3} \Big(
                        3-2 x+3 x^2\Big) H_{-1}^3
                +\Big(
                        -3+2 x-3 x^2\Big) H_{-1} \zeta_2
                +\frac{7}{3} \Big(
                        3-2 x+3 x^2\Big) \zeta_3
        \Big) H_0
        +\Big(
                \Big(
                        -3+2 x-3 x^2\Big) H_{-1}^2
                +\frac{1}{4} \Big(
                        3-2 x+3 x^2\Big) \zeta_2
                +\Big(
                        -1-x^2\Big) H_{-1} \zeta_2
        \Big) H_0^2
        +\frac{1}{3} \Big(
                3-2 x+3 x^2\Big) H_{-1} H_0^3
        +\frac{1}{24} \Big(
                -3+2 x-3 x^2\Big) H_0^4
        +\Big(
                -4 \Big(
                        3-2 x+3 x^2\Big) H_{-1}^2
                +\Big(
                        -3+2 x-3 x^2\Big) \zeta_2
        \Big) H_{0,-1}
        +4 \Big(
                3-2 x+3 x^2\Big) H_{-1} H_{0,0,-1}
        +8 \Big(
                3-2 x+3 x^2\Big) H_{-1} H_{0,-1,-1}
        -2 \Big(
                3-2 x+3 x^2\Big) H_{0,0,0,-1}
        -4 \Big(
                3-2 x+3 x^2\Big) H_{0,0,-1,-1}
        -8 \Big(
                3-2 x+3 x^2\Big) H_{0,-1,-1,-1}
        +4 \Big(
                3+5 x^2\Big) \zeta_2
        +2 \Big(
                3-2 x+3 x^2\Big) H_{-1}^2 \zeta_2
        +\frac{1}{20} \Big(
                75-56 x+93 x^2\Big) \zeta_2^2
        +\frac{4}{3} \Big(
                13+11 x^2\Big) \zeta_3
        -4 \Big(
                3-2 x+3 x^2\Big) H_{-1} \zeta_3
\Big)
+\xi  \Big(
        \Big(
                -32
                +\Big(
                        32
                        -\frac{28 \zeta_3}{3}
                \Big) H_{-1}
                +\Big(
                        -16
                        +2 \zeta_2
                \Big) H_{-1}^2
                -\frac{4}{3} H_{-1}^4
                +4 \zeta_2
                +\frac{47}{20} \zeta_2^2
        \Big) H_0
        +\Big(
                -8
                +8 H_{-1}
                +\frac{4}{3} H_{-1}^3
                +\frac{7}{3} \zeta_3
        \Big) H_0^2
        +\Big(
                \frac{1}{6} \Big(
                        -8
                        +\zeta_2
                \Big)
                -\frac{2}{3} H_{-1}^2
        \Big) H_0^3
        +\frac{1}{6} H_{-1} H_0^4
        -\frac{1}{60} H_0^5
        +\Big(
                -16 \zeta_2
                -\frac{28 \zeta_2^2}{5}
        \Big) H_{-1}
        +\Big(
                -32
                +\Big(
                        32
                        +4 \zeta_2
                \Big) H_{-1}
                +\frac{16}{3} H_{-1}^3
                +\frac{4}{3} \zeta_3
        \Big) H_{0,-1}
        +\Big(
                -16
                -8 H_{-1}^2
                -2 \zeta_2
        \Big) H_{0,0,-1}
        +\Big(
                -32
                -16 H_{-1}^2
                -4 \zeta_2
        \Big) H_{0,-1,-1}
        +8 H_{-1} H_{0,0,0,-1}
        +16 H_{-1} H_{0,0,-1,-1}
        +32 H_{-1} H_{0,-1,-1,-1}
        -4 H_{0,0,0,0,-1}
        -8 H_{0,0,0,-1,-1}
        -16 H_{0,0,-1,-1,-1}
        -32 H_{0,-1,-1,-1,-1}
        -\frac{8}{3} H_{-1}^3 \zeta_2
        +8 H_{-1}^2 \zeta_3
        -\frac{8}{3} \zeta_2 \zeta_3
        +12 \zeta_5
\Big)
\Bigg\}
\Bigg] \,.
\end{dmath}


\begin{dmath*}
 F_{A,1}^{(2),\ns} =
%
C_F^2  \Bigg[
\frac{1}{\ep^2}  \Bigg\{
        2
        +4 \xi  H_0
        +2 \xi ^2 H_0^2
\Bigg\}
+ \frac{1}{\ep}  \Bigg\{
        8
        +\xi  \Big(
                -8 H_{-1} H_0
                +4 \Big(
                        2-x+x^2\Big) \eta  H_0^2
                +8 H_{0,-1}
                -4 \zeta_2
        \Big)
\end{dmath*}
\vspace{-1.0cm}
\begin{dmath*}
{\color{white}=}
        +\xi ^2 \Big(
                -8 H_{-1} H_0^2
                +2 H_0^3
                +8 H_0 H_{0,-1}
                -4 H_0 \zeta_2
        \Big)
        +2 \Big(
                7-2 x+7 x^2\Big) \eta  H_0
\Bigg\}
+  \Bigg\{
                46
                +\eta  \Big(
                        \frac{1}{2} \Big(
                                85-6 x+85 x^2\Big) H_0
                        +2 \Big(
                                55-82 x+55 x^2\Big) H_{-1} H_0
                        -32 \Big(
                                2-x+2 x^2\Big) H_0 H_1
                        +32 \Big(
                                2-x+2 x^2\Big) H_{0,1}
                        -2 \Big(
                                55-82 x+55 x^2\Big) H_{0,-1}
                \Big)
                +\xi  \Big(
                        8 H_{-1}^2 H_0
                        -16 H_{-1} H_{0,-1}
                        +16 H_{0,-1,-1}
                        +48 (-1+x) \log (2) x_+ \zeta_2
                \Big)
                +\eta ^2 \Big(
                        -4 H_{-1} H_0^2 P_{528}
                        +4 H_0^2 H_1 P_{552}
                        +8 H_0 H_{0,-1} P_{559}
                        -8 H_0 H_{0,1} P_{563}
                        +8 H_{0,0,1} P_{574}
                        -8 H_{0,0,-1} P_{576}
                        +\frac{1}{2} H_0^2 P_{603}
                        +\Big(
                                \frac{16}{3} H_0^3 H_1 P_{614}
                                +\Big(
                                        16 H_0 H_1 P_{612}
                                        -16 H_{0,1} P_{612}
                                \Big) \zeta_2
                        \Big) x_+
                        +\Big(
                                39
                                -146 x
                                -46 x^3
                                +57 x^4
                                +8 H_{-1} P_{536}
                        \Big) \zeta_2
                        -4 P_{547} \zeta_3
                \Big)
                +\eta ^3 \Big(
                        \frac{1}{6} H_0^4 P_{643}
                        +4 H_0^2 H_{0,-1} P_{658}
                        +8 H_0 H_{0,0,-1} P_{670}
                        +8 H_0^2 H_{0,1} P_{673}
                        -48 H_{0,0,0,-1} P_{674}
                        +\frac{4}{3} H_0^3 P_{690}
                        -16 H_0 H_{0,0,1} P_{698}
                        +16 H_{0,0,0,1} P_{727}
                        +\Big(
                                4 x H_0 P_{632}
                                -4 H_0^2 P_{699}
                                +8 H_{0,-1} P_{701}
                        \Big) \zeta_2
                        -
                        \frac{2}{5} P_{811} \zeta_2^2
                        -16 H_0 P_{667} \zeta_3
                \Big)
                +\xi ^2 \Big(
                        16 H_{-1}^2 H_0^2
                        -8 H_{-1} H_0^3
                        +\Big(
                                32 H_0 H_1
                                -32 H_{0,-1}
                        \Big) H_{0,1}
                        +16 H_{0,1}^2
                        -32 H_{-1} H_0 H_{0,-1}
                        -32 H_0 H_1 H_{0,-1}
                        +8 H_{0,-1}^2
                        -64 H_1 H_{0,0,1}
                        +64 H_1 H_{0,0,-1}
                        -64 H_0 H_{0,1,1}
                        +32 H_0 H_{0,1,-1}
                        +32 H_0 H_{0,-1,1}
                        +16 H_0 H_{0,-1,-1}
                        +64 H_{0,0,1,1}
                        +32 H_{0,-1,0,1}
                        +16 H_{-1} H_0 \zeta_2
                        +16 H_1 \zeta_3
                \Big)
                +192 x \log (2) x_+^2 \zeta_2
\Bigg\}
+ \ep  \Bigg\{
                \xi  \Big(
                        \eta ^2 \log (2) \zeta_2 \Big(
                                192 x^2 H_{0,1}
                                +192 x^2 H_{0,-1}
                        \Big)
                        -\frac{16}{3} H_{-1}^3 H_0
                        + \Big( 32 (-1+x^2) x \Big(
                                10-21 x+10 x^2\Big) \eta ^3 \Big) H_0 H_{0,1}^2
                        +16 H_{-1}
                        ^2 H_{0,-1}
                        -32 H_{-1} H_{0,-1,-1}
                        +32 H_{0,-1,-1,-1}
                        -4 (-1+x) x_+
                \Big)
                +\eta  \Big(
                        \frac{1}{4} \Big(
                                479+70 x+479 x^2\Big) H_0
                        +4 \Big(
                                81-146 x+81 x^2\Big) H_{-1} H_0
                        -2 \Big(
                                257-286 x+257 x^2\Big) H_{-1}^2 H_0
                        +\Big(
                                -4 \Big(
                                        43-18 x+43 x^2\Big) H_0
                                +192 \Big(
                                        2-x+2 x^2\Big) H_{-1} H_0
                        \Big) H_1
                        -32 \Big(
                                2-x+2 x^2\Big) H_0 H_1^2
                        +\Big(
                                4 \Big(
                                        43-18 x+43 x^2\Big)
                                +64 \Big(
                                        2-x+2 x^2\Big) H_1
                                -192 \Big(
                                        2-x+2 x^2\Big) H_{-1}
                        \Big) H_{0,1}
                        +\Big(
                                -4 \Big(
                                        81-146 x+81 x^2\Big)
                                -192 \Big(
                                        2-x+2 x^2\Big) H_1
                                +4 \Big(
                                        257-286 x+257 x^2\Big) H_{-1}
                        \Big) H_{0,-1}
                        -64 \Big(
                                2-x+2 x^2\Big) H_{0,1,1}
                        +192 \Big(
                                2-x+2 x^2\Big) H_{0,1,-1}
                        +192 \Big(
                                2-x+2 x^2\Big) H_{0,-1,1}
                        -4 \Big(
                                257-286 x+257 x^2\Big) H_{0,-1,-1}
                        +32 \Big(
                                2-x+2 x^2\Big) H_1 \zeta_2
                \Big)
                +\eta ^2 \Big(
                        H_{-1} H_0^2 P_{509}
                        +4 H_{-1}^2 H_0^2 P_{522}
                        +16 H_0 H_{0,1,1} P_{525}
                        -32 H_0 H_{0,-1,-1} P_{542}
                        -16 H_{0,0,1,1} P_{544}
                        +8 H_{0,1}^2 P_{553}
                        +16 H_0 H_{0,1,-1} P_{567}
                        +16 H_0 H_{0,-1,1} P_{567}
                        +4 H_0^2 H_1^2 P_{568}
                        +16 H_{0,-1,0,1} P_{573}
                        -16 H_{0,0,-1,-1} P_{581}
                        +8 H_{0,-1}^2 P_{585}
                        +\frac{1}{2} H_0^2 P_{605}
                        +\log (2) \zeta_2 \Big(
                                -24 P_{512}
                                +48 \Big(
                                        3-2 x^2+3 x^4\Big) H_1
                                +48 \Big(
                                        3-2 x^2+3 x^4\Big) H_{-1}
                        \Big)
                        +\Big(
                                2 H_0^2 P_{596}
                                -8 \Big(
                                        3-2 x^2+3 x^4\Big) H_{-1} H_0^2
                        \Big) H_1
                        +\Big(
                                32 H_0 P_{520}
                                -16 H_0 H_1 P_{560}
                                -16 H_{0,-1} P_{583}
                                +160 \Big(
                                        1-4 x^2+x^4\Big) H_{-1} H_0
                        \Big) H_{0,1}
                        +\Big(
                                -16 H_0 H_1 P_{543}
                                -16 H_{-1} H_0 P_{564}
                                -8 H_0 P_{590}
                        \Big) H_{0,-1}
                        +\Big(
                                16 H_1 P_{544}
                                -4 P_{600}
                                -16 \Big(
                                        17-78 x^2+17 x^4\Big) H_{-1}
                        \Big) H_{0,0,1}
                        +\Big(
                                16 H_1 P_{571}
                                +16 H_{-1} P_{581}
                                +2 P_{606}
                        \Big) H_{0,0,-1}
                        +16 \Big(
                                17-78 x^2+17 x^4\Big) H_{0,0,1,-1}
                        +16 \Big(
                                17-78 x^2+17 x^4\Big) H_{0,0,-1,1}
\end{dmath*}
\begin{dmath*}
{\color{white}=}
                        +\Big(
                                -
                                \frac{32}{3} H_{-1} H_0^3 H_1 P_{618}
                                +\frac{16}{3} H_0^3 H_1^2 P_{618}
                                +\Big(
                                        -64 H_{-1} H_0 H_1 P_{616}
                                        +32 H_0 H_1^2 P_{616}
                                        +64 H_{0,1,1} P_{618}
                                        +\Big(
                                                64 H_{-1} P_{616}
                                                -32 H_1 P_{620}
                                        \Big) H_{0,1}
                                \Big) \zeta_2
                        \Big) x_+
                        +\Big(
                                163
                                -514 x
                                +82 x^2
                                -74 x^3
                                +183 x^4
                                -8 H_{-1}^2 P_{584}
                                +2 H_{-1} P_{598}
                                -48 \Big(
                                        3-2 x^2+3 x^4\Big) H_{-1} H_1
                                +48 \Big(
                                        3-2 x^2+3 x^4\Big) H_{-1,1}
                        \Big) \zeta_2
                        +\Big(
                                4 H_{-1} P_{587}
                                -4 H_1 P_{597}
                                -\frac{2 P_{607}}{3}
                        \Big) \zeta_3
                \Big)
                +\xi ^2 \Big(
                        -\frac{64}{3} H_{-1}^3 H_0^2
                        +16 H_{-1}^2 H_0^3
                        +\Big(
                                128 H_{-1} H_0 H_{0,-1}
                                -96 H_{0,-1}^2
                        \Big) H_1
                        +\Big(
                                \Big(
                                        -128 H_{-1} H_0
                                        +64 H_{0,-1}
                                \Big) H_1
                                +64 H_0 H_1^2
                                +128 H_{-1} H_{0,-1}
                        \Big) H_{0,1}
                        +\Big(
                                32 H_1
                                -64 H_{-1}
                        \Big) H_{0,1}^2
                        +64 H_{-1}^2 H_0 H_{0,-1}
                        -64 H_0 H_1^2 H_{0,-1}
                        -32 H_{-1} H_{0,-1}^2
                        +\Big(
                                256 H_{-1} H_1
                                -128 H_1^2
                        \Big) H_{0,0,1}
                        +\Big(
                                -256 H_{-1} H_1
                                +128 H_1^2
                        \Big) H_{0,0,-1}
                        +\Big(
                                256 H_{-1} H_0
                                -192 H_0 H_1
                                -256 H_{0,-1}
                        \Big) H_{0,1,1}
                        +\Big(
                                -128 H_{-1} H_0
                                +64 H_0 H_1
                                -64 H_{0,1}
                                +192 H_{0,-1}
                        \Big) H_{0,1,-1}
                        +\Big(
                                -128 H_{-1} H_0
                                +64 H_0 H_1
                                +192 H_{0,1}
                                +192 H_{0,-1}
                        \Big) H_{0,-1,1}
                        +\Big(
                                -64 H_{-1} H_0
                                +64 H_0 H_1
                                -320 H_{0,1}
                                +32 H_{0,-1}
                        \Big) H_{0,-1,-1}
                        +\Big(
                                256 H_1
                                -256 H_{-1}
                        \Big) H_{0,0,1,1}
                        -256 H_1 H_{0,0,1,-1}
                        -256 H_1 H_{0,0,-1,1}
                        +256 H_1 H_{0,0,-1,-1}
                        +192 H_0 H_{0,1,1,1}
                        -64 H_0 H_{0,1,-1,-1}
                        -128 H_{-1} H_{0,-1,0,1}
                        -64 H_0 H_{0,-1,1,-1}
                        -64 H_0 H_{0,-1,-1,1}
                        +32 H_0 H_{0,-1,-1,-1}
                        -384 H_{0,0,1,1,1}
                        +768 H_{0,0,1,1,-1}
                        +256 H_{0,0,1,-1,1}
                        -256 H_{0,0,-1,1,1}
                        -64 H_{0,1,0,1,1}
                        +256 H_{0,1,0,1,-1}
                        -256 H_{0,-1,0,1,1}
                        +128 H_{0,-1,0,1,-1}
                        +128 H_{0,-1,0,-1,1}
                        -256 H_{0,-1,1,0,1}
                        +320 H_{0,-1,-1,0,1}
                        -32 H_{-1}^2 H_0 \zeta_2
                        +\Big(
                                -64 H_{-1} H_1
                                +32 H_1^2
                        \Big) \zeta_3
                \Big)
                +\eta ^3 \Big(
                        16 H_0^2 H_{0,1,-1} P_{623}
                        +16 H_0^2 H_{0,-1,1} P_{623}
                        +
                        \frac{1}{6} H_0^3 P_{639}
                        +\frac{1}{30} H_0^5 P_{641}
                        +\frac{1}{3} H_0^4 P_{644}
                        -48 H_0 H_{0,-1}^2 P_{663}
                        -32 H_{0,0,1,0,1} P_{669}
                        +16 H_0 H_{0,0,-1,-1} P_{682}
                        +32 H_{0,0,1,0,-1} P_{686}
                        -16 H_0^2 H_{0,1,1} P_{688}
                        +32 H_0 H_{0,0,1,-1} P_{696}
                        +32 H_0 H_{0,0,-1,1} P_{696}
                        -32 H_{0,0,-1,0,1} P_{703}
                        +32 H_0 H_{0,-1,0,1} P_{712}
                        -8 H_0^2 H_{0,-1,-1} P_{715}
                        +32 H_{0,0,0,1,-1} P_{726}
                        +32 H_{0,0,0,-1,1} P_{726}
                        -32 H_{0,0,-1,0,-1} P_{738}
                        -96 H_{0,0,0,-1,-1} P_{739}
                        +32 H_0 H_{0,0,1,1} P_{744}
                        -32 H_{0,0,0,1,1} P_{780}
                        +16 H_{0,0,0,0,1} P_{807}
                        -16 H_{0,0,0,0,-1} P_{808}
                        +\Big(
                                -\frac{2}{3} H_0^4 P_{675}
                                -\frac{4}{3} H_0^3 P_{790}
                        \Big) H_1
                        +\Big(
                                -\frac{4}{3} H_0^3 P_{680}
                                +\frac{2}{3} H_0^4 P_{716}
                        \Big) H_{-1}
                        +\Big(
                                -\frac{8}{3} H_0^3 P_{668}
                                +32 H_0^2 H_1 P_{677}
                                -16 H_{-1} H_0^2 P_{679}
                                -32 H_0 H_{0,-1} P_{683}
                                +4 H_0^2 P_{776}
                        \Big) H_{0,1}
                        +128 x^2 \Big(
                                10-21 x+10 x^2\Big) H_0 H_{0,1}^2
                        +\Big(
                                16 x H_{-1} H_0^2 P_{617}
                                -16 H_0^2 H_1 P_{655}
                                +\frac{4}{3} H_0^3 P_{747}
                                -4 H_0^2 P_{786}
                        \Big) H_{0,-1}
                        +\Big(
                                -32 H_{0,1} P_{647}
                                +64 H_{-1} H_0 P_{654}
                                -256 H_0 H_1 P_{662}
                                +8 H_0 P_{714}
                                +8 H_0^2 P_{717}
                                -32 H_{0,-1} P_{722}
                        \Big) H_{0,0,1}
                        +\Big(
                                -32 H_{-1} H_0 P_{629}
                                +128 H_0 H_1 P_{676}
                                +8 H_0 P_{709}
                                +32 H_{0,1} P_{710}
                                -4 H_0^2 P_{725}
                                +16 H_{0,-1} P_{768}
                        \Big) H_{0,0,-1}
                        +\Big(
                                -32 H_{-1} P_{657}
                                +64 H_1 P_{736}
                                -16 H_0 P_{788}
                                -8 P_{800}
                        \Big) H_{0,0,0,1}
                        +\Big(
                                96 H_{-1} P_{624}
                                -32 H_1 P_{777}
                                +8 H_0 P_{795}
                                +8 P_{803}
                        \Big) H_{0,0,0,-1}
                        +\Big(
                                -2 x H_0^2 P_{635}
                                +32 H_{0,1,-1} P_{707}
                                +32 H_{0,-1,1} P_{708}
                                +16 H_{0,0,1} P_{742}
                                -16 H_{0,-1,-1} P_{765}
                                +
                                \frac{4}{3} H_0^3 P_{782}
                                -8 H_{0,0,-1} P_{789}
                                +H_0 P_{812}
                                +\Big(
                                        8 H_0^2 P_{719}
                                        -8 H_0 P_{763}
                                \Big) H_1
                                +\Big(
                                        8 H_0^2 P_{665}
                                        -8 H_0 P_{754}
                                \Big) H_{-1}
                                +\Big(
                                        -16 H_0 P_{728}
                                        +8 P_{755}
                                \Big) H_{0,1}
                                +\Big(
                                        32 H_{-1} P_{666}
                                        +8 P_{681}
                                        +16 H_0 P_{694}
                                        -32 H_1 P_{708}
                                \Big) H_{0,-1}
                                +4 P_{783} \zeta_3
                        \Big) \zeta_2
                        +\Big(
                                \frac{8}{5} H_{-1} P_{659}
                                +\frac{8}{5} H_1 P_{792}
                                -\frac{4 P_{810}}{5}
                                +\frac{2}{5} H_0 P_{814}
                        \Big) \zeta_2^2
                        +\Big(
                                -32 H_{-1} H_0 P_{667}
                                +16 H_0 H_1 P_{691}
                                +16 H_{0,-1} P_{731}
                                -16 H_{0,1} P_{741}
                                +\frac{4}{3} H_0^2 P_{785}
                                +\frac{4}{3} H_0 P_{791}
                        \Big) \zeta_3
                        -6 P_{798} \zeta_5
                \Big)
                +8 x x_+^2
                +8 c_1 \Big(
                        1-4 x+x^2\Big) x_+^2
\Bigg\}
\Bigg]
\end{dmath*}
\begin{dmath*}
{\color{white}=}
%
+ C_F C_A  \Bigg[
\frac{1}{\ep^2}  \Bigg\{
        \frac{11}{3}
        +\frac{11 \xi  H_0}{3}
\Bigg\}
+ \frac{1}{\ep}  \Bigg\{
        -\frac{49}{9}
        +\xi  \Big(
                \eta  \Big(
                        -\frac{4}{3} x^2 H_0^3
                        -2 \Big(
                                -1+3 x^2\Big) H_0 \zeta_2
                \Big)
                -\frac{67}{9} H_0
                +4 H_{-1} H_0
                -4 H_0 H_1
                +4 H_{0,1}
                -4 H_{0,-1}
        \Big)
        +\eta  \Big(
                -4 x^2 H_0^2
                -2 \Big(
                        -1+3 x^2\Big) \zeta_2
        \Big)
        +\xi ^2 \Big(
                -4 H_0 H_{0,1}
                +4 H_0 H_{0,-1}
                +8 H_{0,0,1}
                -8 H_{0,0,-1}
                -2 \zeta_3
        \Big)
\Bigg\}
+  \Bigg\{
                -\frac{1595}{27}
                +\eta  \Big(
                        -\frac{5}{54} \Big(
                                509-114 x+509 x^2\Big) H_0
                        +\frac{4}{9} \Big(
                                31-105 x+31 x^2\Big) H_{-1} H_0
                        +4 \Big(
                                3-2 x+3 x^2\Big) H_0 H_1
                        -4 \Big(
                                3-2 x+3 x^2\Big) H_{0,1}
                        -\frac{4}{9} \Big(
                                31-105 x+31 x^2\Big) H_{0,-1}
                \Big)
                +\xi ^2 \Big(
                        \Big(
                                -8 H_1
                                +8 H_{-1}
                        \Big) \zeta_3
                        +\Big(
                                16 H_{-1} H_0
                                -16 H_0 H_1
                                -8 H_{0,-1}
                        \Big) H_{0,1}
                        -4 H_{0,1}^2
                        -16 H_{-1} H_0 H_{0,-1}
                        +16 H_0 H_1 H_{0,-1}
                        +12 H_{0,-1}^2
                        +\Big(
                                32 H_1
                                -32 H_{-1}
                        \Big) H_{0,0,1}
                        +\Big(
                                -32 H_1
                                +32 H_{-1}
                        \Big) H_{0,0,-1}
                        +24 H_0 H_{0,1,1}
                        -8 H_0 H_{0,1,-1}
                        -8 H_0 H_{0,-1,1}
                        -8 H_0 H_{0,-1,-1}
                        -32 H_{0,0,1,1}
                        +32 H_{0,0,1,-1}
                        +32 H_{0,0,-1,1}
                        -32 H_{0,0,-1,-1}
                \Big)
                +\xi  \Big(
                        \eta  \Big(
                                \frac{4}{3} \Big(
                                        3+x^2\Big) H_{-1} H_0^3
                                +16 H_{-1} H_0 \zeta_2
                        \Big)
                        -\frac{104}{3} H_{-1}^2 H_0
                        +24 H_{-1} H_0 H_1
                        -4 H_0 H_1^2
                        +\Big(
                                8 H_1
                                -24 H_{-1}
                        \Big) H_{0,1}
                        +\Big(
                                -24 H_1
                                +\frac{208 H_{-1}}{3}
                        \Big) H_{0,-1}
                        -8 H_{0,1,1}
                        +24 H_{0,1,-1}
                        +24 H_{0,-1,1}
                        -\frac{208}{3} H_{0,-1,-1}
                        +4 H_1 \zeta_2
                        -24 (-1+x) \log (2) x_+ \zeta_2
                \Big)
                +\eta ^2 \Big(
                        -4 H_0^2 H_1 P_{526}
                        -4 H_0 H_{0,-1} P_{538}
                        -8 H_{0,0,1} P_{555}
                        -\frac{2}{3} H_{-1} H_0^2 P_{591}
                        +\frac{4}{3} H_{0,0,-1} P_{599}
                        - \frac{1}{18} (1+x) ( 233-299 x+65 x^2+217 x^3 ) H_0^2
                        + 8 (1+x)^2 \Big(
                                6-11 x+6 x^2\Big) H_0 H_{0,1}
                        +\Big(
                                \frac{4}{3} H_0^3 H_1 P_{611}
                                +\Big(
                                        16 H_0 H_1 P_{613}
                                        -4 H_{0,1} P_{619}
                                \Big) \zeta_2
                        \Big) x_+
                        +\Big(
                                \frac{P_{510}}{9}
                                -\frac{4}{3} H_{-1} P_{554}
                        \Big) \zeta_2
                        +\frac{2 P_{534} \zeta_3}{3}
                \Big)
                +\eta ^3 \Big(
                        32 x H_0 H_{0,0,1} P_{615}
                        +\frac{1}{6} x H_0^4 P_{630}
                        +\frac{1}{9} H_0^3 P_{640}
                        +8 H_0 H_{0,0,-1} P_{660}
                        +4 H_0^2 H_{0,1} P_{661}
                        -4 H_0^2 H_{0,-1} P_{695}
                        +8 H_{0,0,0,-1} P_{723}
                        -8 H_{0,0,0,1} P_{737}
                        +\Big(
                                -2 H_0^2 P_{650}
                                -4 H_{0,-1} P_{700}
                                -
                                \frac{2}{3} H_0 P_{764}
                        \Big) \zeta_2
                        +\frac{1}{5} P_{797} \zeta_2^2
                        +2 H_0 P_{730} \zeta_3
                \Big)
                -96 x \log (2) x_+^2 \zeta_2
\Bigg\}
+ \ep  \Bigg\{
                \eta  \Big(
                        -\frac{5}{324} \Big(
                                14033+462 x+14033 x^2\Big) H_0
                        +\frac{1}{27} \Big(
                                3623+534 x+3623 x^2\Big) H_{-1} H_0
                        +\frac{4}{9} \Big(
                                28+279 x+28 x^2\Big) H_{-1}^2 H_0
                        +\frac{1}{9} \Big(
                                -533-659 x-677 x^2+97 x^3\Big) x_+ H_{-1} H_0^2
                        +\Big(
                                4 \Big(
                                        5-32 x+5 x^2\Big) H_0
                                -24 \Big(
                                        3-2 x+3 x^2\Big) H_{-1} H_0
                        \Big) H_1
                        +4 \Big(
                                3-2 x+3 x^2\Big) H_0 H_1^2
                        +\Big(
                                -4 \Big(
                                        5-32 x+5 x^2\Big)
                                -8 \Big(
                                        3-2 x+3 x^2\Big) H_1
                                +24 \Big(
                                        3-2 x+3 x^2\Big) H_{-1}
                        \Big) H_{0,1}
                        +\Big(
                                \frac{1}{27} \Big(
                                        -3623-534 x-3623 x^2\Big)
                                +24 \Big(
                                        3-2 x+3 x^2\Big) H_1
                                -\frac{8}{9} \Big(
                                        28+279 x+28 x^2\Big) H_{-1}
                        \Big) H_{0,-1}
                        +8 \Big(
                                3-2 x+3 x^2\Big) H_{0,1,1}
                        -24 \Big(
                                3-2 x+3 x^2\Big) H_{0,1,-1}
                        -24 \Big(
                                3-2 x+3 x^2\Big) H_{0,-1,1}
                        +\frac{8}{9} \Big(
                                28+279 x+28 x^2\Big) H_{0,-1,-1}
                        -4 \Big(
                                3-2 x+3 x^2\Big) H_1 \zeta_2
                \Big)
                +\xi  \Big(
                        \eta ^2 \log (2) \zeta_2 \Big(
                                -96 x^2 H_{0,1}
                                -96 x^2 H_{0,-1}
                        \Big)
                        +\eta  \Big(
                                -\frac{4}{3} \Big(
                                        3+5 x^2\Big) H_{-1}^2 H_0^3
                                -8 \Big(
                                        1+3 x^2\Big) H_{-1}^2 H_0 \zeta_2
                        \Big)
                        +80 H_{-1}^3 H_0
                        -72 H_{-1}^2 H_0 H_1
                        +24 H_{-1} H_0 H_1^2
                        -\frac{8}{3} H_0 H_1^3
                        +\Big(
                                -48 H_{-1} H_1
                                +8 H_1^2
                                +72 H_{-1}^2
                        \Big) H_{0,1}
                        +\Big(
                                144 H_{-1} H_1
                                -24 H_1^2
                                -240 H_{-1}^2
                        \Big) H_{0,-1}
                        +\Big(
                                -16 H_1
                                +48 H_{-1}
                        \Big) H_{0,1,1}
                        +\Big(
                                48 H_1
                                -144 H_{-1}
                        \Big) H_{0,1,-1}
                        +\Big(
                                48 H_1
                                -144 H_{-1}
                        \Big) H_{0,-1,1}
                        +\Big(
                                -144 H_1
                                +480 H_{-1}
                        \Big) H_{0,-1,-1}
                        +16 H_{0,1,1,1}
                        -48 H_{0,1,1,-1}
\end{dmath*}
\begin{dmath*}
{\color{white}=}
                        -48 H_{0,1,-1,1}
                        +144 H_{0,1,-1,-1}
                        -48 H_{0,-1,1,1}
                        +144 H_{0,-1,1,-1}
                        +144 H_{0,-1,-1,1}
                        -480 H_{0,-1,-1,-1}
                        +4 H_1^2 \zeta_2
                \Big)
                +\eta ^2 \Big(
                        \frac{1}{54} H_0^2 P_{511}
                        -24 H_0 H_{0,1,1} P_{513}
                        -8 H_{0,1}^2 P_{533}
                        -16 H_0 H_{0,1,-1} P_{548}
                        -16 H_0 H_{0,-1,1} P_{548}
                        +8 H_{0,0,1,1} P_{556}
                        -2 H_0^2 H_1^2 P_{558}
                        -8 H_{0,-1,0,1} P_{569}
                        +8 H_0 H_{0,-1,-1} P_{575}
                        -4 H_{0,-1}^2 P_{578}
                        -8 H_{0,0,1,-1} P_{582}
                        -8 H_{0,0,-1,1} P_{582}
                        +2 H_{-1}^2 H_0^2 P_{592}
                        +8 H_{0,0,-1,-1} P_{594}
                        +\log (2) \zeta_2 \Big(
                                12 P_{512}
                                -24 \Big(
                                        3-2 x^2+3 x^4\Big) H_1
                                -24 \Big(
                                        3-2 x^2+3 x^4\Big) H_{-1}
                        \Big)
                        +\Big(
                                -2 H_0^2 P_{586}
                                -4 \Big(
                                        -15+2 x^2+9 x^4\Big) H_{-1} H_0^2
                        \Big) H_1
                        +\Big(
                                -16 H_{-1} H_0 P_{531}
                                +4 H_0 P_{541}
                                +8 H_0 H_1 P_{557}
                                +16 H_{0,-1} P_{561}
                        \Big) H_{0,1}
                        +\Big(
                                8 H_{-1} H_0 P_{514}
                                +8 H_0 H_1 P_{549}
                                +8 H_0 P_{580}
                        \Big) H_{0,-1}
                        +\Big(
                                -8 H_1 P_{556}
                                +4 P_{565}
                                +8 H_{-1} P_{582}
                        \Big) H_{0,0,1}
                        +\Big(
                                -8 H_1 P_{539}
                                -8 H_{-1} P_{594}
                                -\frac{2 P_{609}}{9}
                        \Big) H_{0,0,-1}
                        +\Big(
                                \frac{8}{3} H_{-1} H_0^3 H_1 P_{627}
                                -\frac{4}{3} H_0^3 H_1^2 P_{627}
                                +\Big(
                                        16 H_{-1} H_0 H_1 P_{622}
                                        -8 H_0 H_1^2 P_{622}
                                        -8 H_{0,1,1} P_{631}
                                        +\Big(
                                                32 x \Big(
                                                        3-8 x+14 x^2+3 x^4\Big) H_1
                                                -32 x \Big(
                                                        3-8 x+14 x^2+3 x^4\Big) H_{-1}
                                        \Big) H_{0,1}
                                \Big) \zeta_2
                        \Big) x_+
                        +\Big(
                                \frac{P_{508}}{27}
                                +12 H_{-1}^2 P_{535}
                                +\frac{2}{9} H_{-1} P_{602}
                                +48 \Big(
                                        1-x^2+2 x^4\Big) H_{-1} H_1
                                -24 \Big(
                                        3-2 x^2+3 x^4\Big) H_{-1,1}
                        \Big) \zeta_2
                        +\Big(
                                2 H_{-1} P_{589}
                                +2 H_1 P_{593}
                                +\frac{P_{608}}{9}
                        \Big) \zeta_3
                \Big)
                +\xi ^2 \Big(
                        \Big(
                                -64 H_{-1} H_0 H_{0,-1}
                                +48 H_{0,-1}^2
                        \Big) H_1
                        +\Big(
                                -32 H_{-1}^2 H_0
                                +\Big(
                                        64 H_{-1} H_0
                                        -32 H_{0,-1}
                                \Big) H_1
                                -32 H_0 H_1^2
                                +32 H_{-1} H_{0,-1}
                        \Big) H_{0,1}
                        +\Big(
                                -16 H_1
                                +16 H_{-1}
                        \Big) H_{0,1}^2
                        +32 H_{-1}^2 H_0 H_{0,-1}
                        +32 H_0 H_1^2 H_{0,-1}
                        -48 H_{-1} H_{0,-1}^2
                        +\Big(
                                -128 H_{-1} H_1
                                +64 H_1^2
                                +64 H_{-1}^2
                        \Big) H_{0,0,1}
                        +\Big(
                                128 H_{-1} H_1
                                -64 H_1^2
                                -64 H_{-1}^2
                        \Big) H_{0,0,-1}
                        +\Big(
                                -96 H_{-1} H_0
                                +96 H_0 H_1
                                +16 H_{0,1}
                                +80 H_{0,-1}
                        \Big) H_{0,1,1}
                        +\Big(
                                32 H_{-1} H_0
                                -32 H_0 H_1
                                +16 H_{0,1}
                                -48 H_{0,-1}
                        \Big) H_{0,1,-1}
                        +\Big(
                                32 H_{-1} H_0
                                -32 H_0 H_1
                                -48 H_{0,1}
                                -48 H_{0,-1}
                        \Big) H_{0,-1,1}
                        +\Big(
                                32 H_{-1} H_0
                                -32 H_0 H_1
                                +16 H_{0,1}
                                -48 H_{0,-1}
                        \Big) H_{0,-1,-1}
                        +\Big(
                                -128 H_1
                                +128 H_{-1}
                        \Big) H_{0,0,1,1}
                        +\Big(
                                128 H_1
                                -128 H_{-1}
                        \Big) H_{0,0,1,-1}
                        +\Big(
                                128 H_1
                                -128 H_{-1}
                        \Big) H_{0,0,-1,1}
                        +\Big(
                                -128 H_1
                                +128 H_{-1}
                        \Big) H_{0,0,-1,-1}
                        -112 H_0 H_{0,1,1,1}
                        +16 H_0 H_{0,1,1,-1}
                        +16 H_0 H_{0,1,-1,1}
                        +16 H_0 H_{0,1,-1,-1}
                        +16 H_0 H_{0,-1,1,1}
                        +16 H_0 H_{0,-1,1,-1}
                        +16 H_0 H_{0,-1,-1,1}
                        +16 H_0 H_{0,-1,-1,-1}
                        +128 H_{0,0,1,1,1}
                        -256 H_{0,0,1,1,-1}
                        -128 H_{0,0,1,-1,1}
                        +128 H_{0,0,1,-1,-1}
                        +128 H_{0,0,-1,1,-1}
                        +128 H_{0,0,-1,-1,1}
                        +256 H_{0,0,-1,-1,-1}
                        -64 H_{0,1,0,1,-1}
                        +64 H_{0,-1,0,1,1}
                        +192 H_{0,-1,0,-1,-1}
                        +64 H_{0,-1,1,0,1}
                        +\Big(
                                32 H_{-1} H_1
                                -16 H_1^2
                                -16 H_{-1}^2
                        \Big) \zeta_3
                \Big)
                +\eta ^3 \Big(
                        \frac{1}{30} x H_0^5 P_{634}
                        +\frac{1}{12} H_0^4 P_{642}
                        +32 H_0 H_{0,0,1,-1} P_{646}
                        +32 H_0 H_{0,0,-1,1} P_{646}
                        +48 H_0 H_{0,-1}^2 P_{651}
                        -80 H_0 H_{0,-1,0,1} P_{656}
                        -128 H_0 H_{0,0,1,1} P_{664}
                        +16 H_0 H_{0,1}^2 P_{671}
                        +8 H_0^2 H_{0,1,1} P_{685}
                        -8 H_0^2 H_{0,1,-1} P_{689}
                        -8 H_0^2 H_{0,-1,1} P_{689}
                        -16 H_0 H_{0,0,-1,-1} P_{704}
                        +16 H_{0,0,1,0,1} P_{733}
                        +8 H_0^2 H_{0,-1,-1} P_{734}
                        +32 H_{0,0,-1,0,-1} P_{740}
                        -16 H_{0,0,-1,0,1} P_{743}
                        -16 H_{0,0,1,0,-1} P_{750}
                        -32 H_{0,0,0,1,-1} P_{774}
                        -32 H_{0,0,0,-1,1} P_{774}
                        +16 H_{0,0,0,1,1} P_{796}
                        -8 H_{0,0,0,0,1} P_{801}
                        +8 H_{0,0,0,0,-1} P_{802}
                        +16 H_{0,0,0,-1,-1} P_{805}
                        +\frac{1}{54} H_0^3 P_{813}
                        +\Big(
                                \frac{1}{3} H_0^4 P_{626}
                                +\frac{2}{3} H_0^3 P_{775}
                        \Big) H_1
                        +\Big(
                                \frac{1}{3} H_0^4 P_{645}
                                +\frac{2}{3} H_0^3 P_{794}
                        \Big) H_{-1}
                        +\Big(
                                8 H_{-1} H_0^2 P_{661}
                                -16 H_0^2 H_1 P_{677}
                                -16 H_0 H_{0,-1} P_{687}
                                +\frac{8}{3} H_0^3 P_{713}
                                -2 H_0^2 P_{779}
                        \Big) H_{0,1}
                        +\Big(
                                8 H_{-1} H_0^2 P_{625}
                                +8 H_0^2 H_1 P_{684}
                                -\frac{4}{3} H_0^3 P_{781}
                                +2 H_0^2 P_{787}
                        \Big) H_{0,-1}
                        +\Big(
                                -32 H_{-1} H_0 P_{649}
                                +32 H_0 H_1 P_{693}
                                +4 H_0 P_{702}
                                -16 H_{0,1} P_{706}
                                +16 H_{0,-1} P_{751}
                                -4 H_0^2 P_{757}
                        \Big) H_{0,0,1}
\end{dmath*}
\begin{dmath*}
{\color{white}=}
                        +\Big(
                                16 H_{-1} H_0 P_{660}
                                -32 H_0 H_1 P_{678}
                                +16 H_{0,1} P_{718}
                                -32 H_{0,-1} P_{729}
                                +4 H_0^2 P_{772}
                                -4 H_0 P_{784}
                        \Big) H_{0,0,-1}
                        +\Big(
                                16 H_{-1} P_{697}
                                -16 H_1 P_{745}
                                +8 H_0 P_{778}
                                +4 P_{793}
                        \Big) H_{0,0,0,1}
                        +\Big(
                                -16 H_{-1} P_{711}
                                +16 H_1 P_{724}
                                -8 H_0 P_{771}
                                -4 P_{799}
                        \Big) H_{0,0,0,-1}
                        +\Big(
                                \frac{1}{18} H_0 P_{638}
                                -\frac{4}{3} H_0^3 P_{749}
                                +16 H_{0,0,-1} P_{752}
                                -8 H_{0,1,-1} P_{759}
                                -8 H_{0,-1,1} P_{760}
                                +8 H_{0,-1,-1} P_{770}
                                -8 H_{0,0,1} P_{773}
                                +\frac{1}{2} H_0^2 P_{806}
                                +\Big(
                                        -4 x H_0^2 P_{621}
                                        +4 H_0 P_{758}
                                \Big) H_1
                                +\Big(
                                        8 H_0^2 P_{653}
                                        +2 H_0 P_{721}
                                \Big) H_{-1}
                                +\Big(
                                        -4 P_{735}
                                        +4 H_0 P_{769}
                                \Big) H_{0,1}
                                +\Big(
                                        -16 H_{-1} P_{692}
                                        +16 H_1 P_{720}
                                        -2 P_{732}
                                        -4 H_0 P_{767}
                                \Big) H_{0,-1}
                                -2 P_{756} \zeta_3
                        \Big) \zeta_2
                        +\Big(
                                -\frac{8}{5} H_{-1} P_{705}
                                +\frac{2}{5} H_0 P_{748}
                                +\frac{4}{5} H_1 P_{753}
                                +\frac{P_{809}}{5}
                        \Big) \zeta_2^2
                        +\Big(
                                -16 H_0 H_1 P_{652}
                                +16 H_{-1} H_0 P_{672}
                                +4 H_{0,1} P_{746}
                                -2 H_0^2 P_{761}
                                -4 H_{0,-1} P_{762}
                                +\frac{4}{3} H_0 P_{766}
                        \Big) \zeta_3
                        +P_{804} \zeta_5
                \Big)
                -\frac{28745}{162} (1+x)^2 x_+^2
                -4 c_1 \Big(
                        1-4 x+x^2\Big) x_+^2
\Bigg\}
\Bigg]
\\
+ C_F n_l T_F  \Bigg[
\frac{1}{\ep^2}  \Bigg\{
        -\frac{4}{3}
        -\frac{4 \xi  H_0}{3}
\Bigg\}
+ \frac{1}{\ep}  \Bigg\{
        \frac{20}{9}
        +\frac{20 \xi  H_0}{9}
\Bigg\}
+  \Bigg\{
                \frac{424}{27}
                +\eta  \Big(
                        \frac{2}{27} \Big(
                                209-78 x+209 x^2\Big) H_0
                        -\frac{8}{9} \Big(
                                19-6 x+19 x^2\Big) H_{-1} H_0
                        +\frac{2}{9} \Big(
                                19-6 x+19 x^2\Big) H_0^2
                        +\frac{8}{9} \Big(
                                19-6 x+19 x^2\Big) H_{0,-1}
                        -\frac{4}{9} \Big(
                                7-6 x+31 x^2\Big) \zeta_2
                \Big)
                +\xi  \Big(
                        \frac{16}
                        {3} H_{-1}^2 H_0
                        -
                        \frac{8}{3} H_{-1} H_0^2
                        +\frac{4}{9} H_0^3
                        -\frac{32}{3} H_{-1} H_{0,-1}
                        +\frac{16}{3} H_{0,0,-1}
                        +\frac{32}{3} H_{0,-1,-1}
                        +\Big(
                                \frac{8}{3} H_0
                                +\frac{16}{3} H_{-1}
                        \Big) \zeta_2
                        -\frac{16 \zeta_3}{3}
                \Big)
\Bigg\}
+ \ep  \Bigg\{
                \frac{5204}{81}
                +\xi  \Big(
                        \Big(
                                -\frac{16 H_0}{3}
                                +32 H_{-1}
                        \Big) \zeta_3
                        -\frac{32}{3} H_{-1}^3 H_0
                        +8 H_{-1}^2 H_0^2
                        -\frac{8}{3} H_{-1} H_0^3
                        +32 H_{-1}^2 H_{0,-1}
                        -32 H_{-1} H_{0,0,-1}
                        -64 H_{-1} H_{0,-1,-1}
                        +16 H_{0,0,0,-1}
                        +32 H_{0,0,-1,-1}
                        +64 H_{0,-1,-1,-1}
                        +\Big(
                                -8 H_{-1} H_0
                                +2 H_0^2
                                -16 H_{-1}^2
                                +24 H_{0,-1}
                        \Big) \zeta_2
                        -\frac{96 \zeta_2^2}{5}
                \Big)
                +\eta  \Big(
                        \frac{1}{81} \Big(
                                5813-1014 x+5813 x^2\Big) H_0
                        +\frac{8}{9} \Big(
                                47-18 x+47 x^2\Big) H_{-1}^2 H_0
                        +\frac{2}{27} \Big(
                                281-78 x+281 x^2\Big) H_0^2
                        +\frac{2}
                        {27} \Big(
                                47-18 x+47 x^2\Big) H_0^3
                        +
                        \frac{1}{3} \Big(
                                1+x^2\Big) H_0^4
                        +\Big(
                                -\frac{8}{27} \Big(
                                        281-78 x+281 x^2\Big) H_0
                                -\frac{4}{9} \Big(
                                        47-18 x+47 x^2\Big) H_0^2
                        \Big) H_{-1}
                        +\Big(
                                \frac{8}{27} \Big(
                                        281-78 x+281 x^2\Big)
                                -\frac{16}{9} \Big(
                                        47-18 x+47 x^2\Big) H_{-1}
                        \Big) H_{0,-1}
                        +\frac{8}{9} \Big(
                                47-18 x+47 x^2\Big) H_{0,0,-1}
                        +\frac{16}{9} \Big(
                                47-18 x+47 x^2\Big) H_{0,-1,-1}
                        +\Big(
                                -\frac{8}{27} \Big(
                                        22-39 x+259 x^2\Big)
                                +\frac{2}{9} \Big(
                                        37-18 x+37 x^2\Big) H_0
                                +\frac{8}{9} \Big(
                                        47-18 x+47 x^2\Big) H_{-1}
                        \Big) \zeta_2
                        -\frac{8}{9} \Big(
                                35-18 x+59 x^2\Big) \zeta_3
                \Big)
\Bigg\}
\Bigg]
\\
+ C_F T_F  \Bigg[
%
%
%
  \Bigg\{
                \frac{4}{27} \Big(
                        383+646 x+383 x^2\Big) x_+^2
                +\eta  x_+^2 \Big(
                        \frac{4}{9} H_0^3 P_{517}
                        +\frac{2}{27} H_0 P_{604}
                        +4 H_0 P_{518} \zeta_2
                \Big)
                +\Big(
                        \frac{2}{9} H_0^2 P_{577}
                        -\frac{8}{3} P_{521} \zeta_2
                \Big) x_+^4
\Bigg\}
+ \ep  \Bigg\{
                \frac{2}{27} \Big(
                        2069+4378 x+2069 x^2\Big) x_+^2
                +\eta  \Big(
                        \frac{1}{27} \Big(
                                1719-2918 x+1719 x^2\Big) H_0
                        +\Big(
                                \frac{8}{9} H_{-1} H_0^3 P_{517}
                                +\frac{1}{3} H_0^4 P_{517}
                                +\frac{16}{3} H_0^2 H_{0,1} P_{517}
                                -\frac{64}{3} H_0 H_{0,0,1} P_{517}
                                +\frac{32}{3} H_0 H_{0,0,-1} P_{517}
                                +32 H_{0,0,0,1} P_{517}
                                -\frac{16}
                                {3} H_{0,0,0,-1} P_{517}
                                -\frac{2}{27} H_0^2 P_{570}
                                -\frac{4}{27} H_{-1} H_0 P_{604}
\end{dmath*}
\begin{dmath}
{\color{white}=}
                                -\frac{8}{5} x \Big(
                                        2+2 x^2+x^3\Big) \zeta_2^2
                                +\Big(
                                        -\frac{16}{3} H_0^2 P_{517}
                                        +\frac{4 P_{604}}{27}
                                \Big) H_{0,-1}
                                +\Big(
                                        \frac{8}{3} H_{-1} H_0 P_{516}
                                        +2 H_0^2 P_{518}
                                        -\frac{8}{3} H_{0,-1} P_{545}
                                        +\frac{4 P_{601}}{27}
                                \Big) \zeta_2
                                -\frac{8}{9} H_0 P_{546} \zeta_3
                        \Big) x_+^2
                        +\Big(
                                -\frac{2}{27} H_0^3 P_{636}
                                -\frac{2}{9} H_0 P_{637} \zeta_2
                        \Big) x_+^3
                \Big)
                +\Big(
                        -\frac{4}{9} H_{-1} H_0^2 P_{577}
                        +\frac{8}{9} H_0^2 H_1 P_{577}
                        -\frac{16}{9} H_0 H_{0,1} P_{577}
                        +\frac{16}{9} H_{0,0,1} P_{577}
                        +\frac{8}{9} H_{0,0,-1} P_{577}
                        -\frac{8}{3} H_{-1} P_{577} \zeta_2
                        -\frac{8}{9} P_{595} \zeta_3
                \Big) x_+^4
                +\frac{32}{3} \log (2) P_{550} x_+^4 \zeta_2
                +\frac{8}{5} (-1+x) \xi  \eta  x_+ \zeta_2^2
\Bigg\}
\Bigg] \,.
\end{dmath}
%
%
\begin{dmath*}
 \hat F_{A,1}^{(2),\sing} =
%
C_F T_F   \Bigg[
\frac{1}{\ep}  \Bigg\{
-6
\Bigg\}
+ \Bigg\{
        \eta ^3 \Big(
                -32 x \zeta_2 \Big(
                        1-x+4 x^2-x^3+x^4\Big) H_0
                -\frac{16}{3} x^2 \Big(
                        1+4 x-3 x^2+2 x^3\Big) H_0^3
        \Big)
\end{dmath*}
\vspace{-1.0cm}
\begin{dmath*}
{\color{white}=}
        +\eta  \Big(
                -4 \Big(
                        3-2 x+3 x^2\Big) H_0
                +\Big(
                        \frac{64}{3} x^2 H_0^3 H_1
                        +64 x^2 H_0^2 H_{0,1}
                        -512 x^2 H_0 H_{0,0,1}
                        +256 x^2 H_0 H_{0,0,-1}
                        +1024 x^2 H_{0,0,0,1}
                        -768 x^2 H_{0,0,0,-1}
                        +\Big(
                                32 x^2 H_0^2
                                +128 x^2 H_0 H_1
                                -128 x^2 H_{0,1}
                        \Big) \zeta_2
                        -\frac{64}{5} x^2 \zeta_2^2
                        -128 x^2 H_0 \zeta_3
                \Big) x_+^2
        \Big)
        +\Big(
                -32 (-1+x) H_{-1} H_0
                +16 (-1+x) H_0 H_1
                -16 (-1+x) H_{0,1}
                +32 (-1+x) H_{0,-1}
        \Big) x_+
        +\Big(
                -29
                -34 x
                -29 x^2
                +8 x (1+3 x) H_0^2
                -64 x H_0^2 H_1
                +192 x H_0 H_{0,1}
                -128 x H_0 H_{0,-1}
                -256 x H_{0,0,1}
                +256 x H_{0,0,-1}
                +64 x \zeta_3
        \Big) x_+^2
        -48 x \xi  \eta  \zeta_2
        + 8 \xi ^2  \zeta_2
\Bigg\}
+ \ep  \Bigg\{
        \log (2) x_+^2 \zeta_2 \Big(
                384 x
                -384 x H_1
                -384 x H_{-1}
        \Big)
        +\eta  \Big(
                \log (2) x_+^2 \zeta_2 \Big(
                        768 x^2 H_{0,1}
                        +768 x^2 H_{0,-1}
                \Big)
                -2 \Big(
                        29+3 x+15 x^2+25 x^3\Big) x_+ H_0
                +8 \Big(
                        19-42 x+19 x^2\Big) H_{-1} H_0
                -32 (-2+x) (-1+2 x) H_0 H_1
                -8 \Big(
                        19-42 x+19 x^2\Big) H_{0,-1}
                +\Big(
                        \Big(
                                \frac{128}{3} x^2 H_{-1} H_0^3
                                +16 x^2 H_0^4
                        \Big) H_1
                        -\frac{64}{3} x^2 H_0^3 H_1^2
                        +\Big(
                                16 P_{532}
                                +128 x^2 H_{-1} H_0^2
                                +\frac{64}{3} x^2 H_0^3
                                +128 x^2 H_0^2 H_1
                                +256 x^2 H_0 H_{0,-1}
                        \Big) H_{0,1}
                        -512 x^2 H_0 H_{0,1}^2
                        +\Big(
                                \frac{256}{3} x^2 H_0^3
                                -256 x^2 H_0^2 H_1
                        \Big) H_{0,-1}
                        -384 x^2 H_0 H_{0,-1}^2
                        +\Big(
                                \Big(
                                        -64 x \Big(
                                                5-14 x+3 x^2\Big)
                                        -1024 x^2 H_{-1}
                                \Big) H_0
                                -448 x^2 H_0^2
                                +1280 x^2 H_{0,1}
                                -1792 x^2 H_{0,-1}
                        \Big) H_{0,0,1}
                        +\Big(
                                \Big(
                                        -128 x \Big(
                                                -5+3 x+3 x^2\Big)
                                        +512 x^2 H_{-1}
                                \Big) H_0
                                +128 x^2 H_0^2
                                -512 x^2 H_{0,1}
                                +1792 x^2 H_{0,-1}
                        \Big) H_{0,0,-1}
                        +256 x^2 H_0^2 H_{0,1,1}
                        -128 x^2 H_0^2 H_{0,1,-1}
                        -128 x^2 H_0^2 H_{0,-1,1}
                        -128 x^2 H_0^2 H_{0,-1,-1}
                        +\Big(
                                1536 x^2 H_0
                                -512 x^2 H_1
                                +2048 x^2 H_{-1}
                        \Big) H_{0,0,0,1}
                        +\Big(
                                -1536 x^2 H_0
                                +1280 x^2 H_1
                                -1536 x^2 H_{-1}
                        \Big) H_{0,0,0,-1}
                        +2048 x^2 H_0 H_{0,0,1,-1}
                        +2048 x^2 H_0 H_{0,0,-1,1}
                        -1024 x^2 H_0 H_{0,0,-1,-1}
                        +1024 x^2 H_0 H_{0,-1,0,1}
                        -1664 x^2 H_{0,0,0,0,1}
                        +2560 x^2 H_{0,0,0,0,-1}
                        -3328 x^2 H_{0,0,0,1,1}
                        +256 x^2 H_{0,0,0,1,-1}
                        +256 x^2 H_{0,0,0,-1,1}
                        -2304 x^2 H_{0,0,0,-1,-1}
                        -1280 x^2 H_{0,0,1,0,1}
                        +768 x^2 H_{0,0,1,0,-1}
                        +512 x^2 H_{0,0,-1,0,1}
                        -1280 x^2 H_{0,0,-1,0,-1}
                        +\Big(
                                64 x^2 H_{-1} H_0^2
                                +32 x^2 H_0^3
                                +256 x^2 H_{-1} H_0 H_1
                                -128 x^2 H_0 H_1^2
                                +\Big(
                                        -128 x^2 H_0
                                        +256 x^2 H_1
                                        -256 x^2 H_{-1}
                                \Big) H_{0,1}
                                +\Big(
                                        384 x^2 H_0
                                        -1024 x^2 H_1
                                \Big) H_{0,-1}
                                +384 x^2 H_{0,0,1}
                                -1536 x^2 H_{0,0,-1}
                                -256 x^2 H_{0,1,1}
                                +256 x^2 H_{0,1,-1}
                                +1024 x^2 H_{0,-1,1}
                                -768 x^2 H_{0,-1,-1}
                                +160 x^2 \zeta_3
                        \Big) \zeta_2
                        +\Big(
                                -
                                \frac{448}{5} x^2 H_0
                                +\frac{64}{5} x^2 H_1
                                -\frac{128}{5} x^2 H_{-1}
                        \Big) \zeta_2^2
\end{dmath*}
\vspace{-1.0cm}
\begin{dmath}
{\color{white}=}
                        +\Big(
                                -32 x \Big(
                                        -1-9 x+5 x^2\Big) H_0
                                -256 x^2 H_{-1} H_0
                                -32 x^2 H_0^2
                                -1344 x^2 H_{0,1}
                                +448 x^2 H_{0,-1}
                        \Big) \zeta_3
                        -144 x^2 \zeta_5
                \Big) x_+^2
        \Big)      
        +\eta ^3 \Big(
                \frac{32}{3} x H_{-1} H_0^3 P_{524}
                -64 x H_0^2 H_{0,1} P_{530}
                +64 x H_0^2 H_{0,-1} P_{537}
                +\frac{32}{3} x H_0^3 H_1 P_{551}
                -64 x H_{0,0,0,-1} P_{566}
                +32 x H_{0,0,0,1} P_{588}
                -\frac{8}{3} x H_0^3 P_{628}
                -\frac{2}{3} x^2 \Big(
                        5+26 x-19 x^2+12 x^3\Big) H_0^4
                +\Big(
                        -64 x H_{0,-1} P_{515}
                        -64 x H_{0,1} P_{523}
                        +64 x H_{-1} H_0 P_{524}
                        +64 x H_0 H_1 P_{529}
                        +16 H_0 P_{648}
                \Big) \zeta_2
                -\frac{16}{5} x P_{579} \zeta_2^2
        \Big)
        +\eta ^2 \Big(
                16 H_0 H_{0,1} P_{527}
                -16 H_{-1} H_0^2 P_{540}
                +32 H_{0,0,-1} P_{562}
                +32 \Big(
                        -1+2 x^2-4 x^3+x^4\Big) H_0^2 H_1
                -32 \Big(
                        3-x-x^3+3 x^4\Big) H_0 H_{0,-1}
                -16 x \Big(
                        21-26 x+5 x^2+8 x^3\Big) H_{0,0,1}
                +\Big(
                        4 H_0^2 P_{633}
                        +2 P_{610} \zeta_2
                \Big) x_+
                -16 H_{-1} P_{519} \zeta_2
                -8 P_{572} \zeta_3
        \Big)
        +\Big(
                128 (-1+x) H_{-1}
                ^2 H_0
                -96 (-1+x) H_{-1} H_0 H_1
                +16 (-1+x) H_0 H_1^2
                +\Big(
                        -32 (-1+x) H_1
                        +96 (-1+x) H_{-1}
                \Big) H_{0,1}
                +\Big(
                        96 (-1+x) H_1
                        -256 (-1+x) H_{-1}
                \Big) H_{0,-1}
                +32 (-1+x) H_{0,1,1}
                -96 (-1+x) H_{0,1,-1}
                -96 (-1+x) H_{0,-1,1}
                +256 (-1+x) H_{0,-1,-1}
                -16 (-1+x) H_1 \zeta_2
        \Big) x_+
        +\Big(
                \frac{1}{2} \Big(
                        -199-278 x-199 x^2\Big)
                +\Big(
                        672 x H_1
                        -96 x H_{-1}
                \Big) \zeta_3
                +32 x H_{-1}^2 H_0^2
                +\Big(
                        -8 (-1+x)^2
                        +64 x H_{-1} H_0^2
                \Big) H_1
                -96 x H_0^2 H_1^2
                +\Big(
                        -256 x H_{-1} H_0
                        +384 x H_0 H_1
                        +640 x H_{0,-1}
                \Big) H_{0,1}
                -64 x H_{0,1}^2
                +\Big(
                        128 x H_{-1} H_0
                        +128 x H_0 H_1
                \Big) H_{0,-1}
                -384 x H_{0,-1}^2
                +\Big(
                        -384 x H_1
                        +384 x H_{-1}
                \Big) H_{0,0,1}
                +\Big(
                        -384 x H_1
                        -384 x H_{-1}
                \Big) H_{0,0,-1}
                -256 x H_0 H_{0,1,1}
                -384 x H_0 H_{0,1,-1}
                -384 x H_0 H_{0,-1,1}
                +640 x H_0 H_{0,-1,-1}
                +384 x H_{0,0,1,1}
                -384 x H_{0,0,1,-1}
                -384 x H_{0,0,-1,1}
                +384 x H_{0,0,-1,-1}
                -384 x H_{0,-1,0,1}
                +\Big(
                        384 x H_{-1} H_1
                        +192 x H_{-1}^2
                        -384 x H_{-1,1}
                \Big) \zeta_2
        \Big) x_+^2
        -16 x (3+5 x) H_0^2 x_+^3 \zeta_2
\Bigg\}
\Bigg] \,.
\end{dmath}
%
%
The polynomials are defined as
\begin{align}
P_{508} &= -4774 x^4+21306 x^3-28365 x^2+14928 x-71\,,\nonumber\\
P_{509} &= -399 x^4+486 x^3-660 x^2-182 x+307\,,\nonumber\\
P_{510} &= -347 x^4+2262 x^3-3174 x^2+1698 x-7\,,\nonumber\\
P_{511} &= -254 x^4+4452 x^3-4779 x^2+462 x-3337\,,\nonumber\\
P_{512} &= x^4-60 x^3+110 x^2-60 x+1\,,\nonumber\\
P_{513} &= x^4-30 x^3+54 x^2-30 x+1\,,\nonumber\\
P_{514} &= x^4-16 x^3+26 x^2-16 x+1\,,\nonumber\\
P_{515} &= x^4-6 x^3-16 x^2+14 x-9\,,\nonumber\\
P_{516} &= x^4+2 x^3-10 x^2+2 x+1\,,\nonumber\\
P_{517} &= x^4+2 x^3-4 x^2+2 x+1\,,\nonumber\\
P_{518} &= x^4+2 x^3-2 x^2+2 x+1\,,\nonumber\\
P_{519} &= x^4+10 x^3-42 x^2+2 x+5\,,\nonumber\\
P_{520} &= x^4+10 x^3-35 x^2+10 x+1\,,\nonumber\\
P_{521} &= x^4+55 x^3+68 x^2+55 x+1\,,\nonumber\\
P_{522} &= 2 x^4-21 x^3+142 x^2-21 x+4\,,\nonumber\\
P_{523} &= 2 x^4-7 x^3+14 x^2-7 x+2\,,\nonumber\\
P_{524} &= 2 x^4-5 x^3-4 x^2+7 x-4\,,\nonumber\\
P_{525} &= 2 x^4-x^3-70 x^2-x+2\,,\nonumber\\
P_{526} &= 2 x^4+5 x^3-14 x^2+5 x+6\,,\nonumber\\
P_{527} &= 2 x^4+11 x^3-18 x^2+11 x+2\,,\nonumber\\
P_{528} &= 3 x^4-15 x^3+76 x^2-15 x+5\,,\nonumber\\
P_{529} &= 3 x^4-9 x^3+14 x^2-5 x+1\,,\nonumber\\
P_{530} &= 3 x^4-4 x^3-2 x^2+8 x-3\,,\nonumber\\
P_{531} &= 4 x^4-27 x^3+44 x^2-27 x+4\,,\nonumber\\
P_{532} &= 4 x^4-3 x^3-10 x^2-3 x+4\,,\nonumber\\
P_{533} &= 4 x^4+9 x^3-22 x^2+9 x+4\,,\nonumber\\
P_{534} &= 5 x^4-252 x^3+408 x^2-252 x+67\,,\nonumber\\
P_{535} &= 5 x^4-27 x^3+34 x^2-27 x+13\,,\nonumber\\
P_{536} &= 5 x^4-27 x^3+36 x^2-27 x+7\,,\nonumber\\
P_{537} &= 5 x^4-9 x^3+4 x^2+7 x-3\,,\nonumber\\
P_{538} &= 5 x^4+49 x^3-98 x^2+49 x+5\,,\nonumber\\
P_{539} &= 5 x^4+128 x^3-254 x^2+128 x+29\,,\nonumber\\
P_{540} &= 6 x^4-10 x^3-7 x^2+12 x-5\,,\nonumber\\
P_{541} &= 7 x^4-102 x^3+162 x^2-102 x+7\,,\nonumber\\
P_{542} &= 7 x^4-39 x^3+135 x^2-39 x+7\,,\nonumber\\
P_{543} &= 7 x^4-23 x^3+96 x^2-23 x+7\,,\nonumber\\
P_{544} &= 7 x^4-11 x^3-52 x^2-11 x+7\,,\nonumber\\
P_{545} &= 7 x^4+14 x^3-34 x^2+14 x+7\,,\nonumber\\
P_{546} &= 7 x^4+14 x^3-22 x^2+14 x+7\,,\nonumber\\
P_{547} &= 7 x^4+30 x^3-142 x^2+30 x+11\,,\nonumber\\
P_{548} &= 7 x^4+47 x^3-96 x^2+47 x+7\,,\nonumber\\
P_{549} &= 7 x^4+64 x^3-126 x^2+64 x+7\,,\nonumber\\
P_{550} &= 7 x^4+82 x^3+110 x^2+82 x+7\,,\nonumber\\
P_{551} &= 8 x^4-19 x^3+14 x^2+5 x-4\,,\nonumber\\
P_{552} &= 8 x^4-19 x^3+46 x^2-19 x+8\,,\nonumber\\
P_{553} &= 9 x^4-13 x^3+40 x^2-13 x+9\,,\nonumber\\
P_{554} &= 10 x^4-81 x^3+108 x^2-81 x+26\,,\nonumber\\
P_{555} &= 10 x^4-3 x^3-6 x^2-3 x+6\,,\nonumber\\
P_{556} &= 11 x^4-98 x^3+170 x^2-98 x+3\,,\nonumber\\
P_{557} &= 11 x^4-72 x^3+118 x^2-72 x+11\,,\nonumber\\
P_{558} &= 11 x^4-46 x^3+66 x^2-46 x+19\,,\nonumber\\
P_{559} &= 11 x^4-42 x^3+136 x^2-42 x+11\,,\nonumber\\
P_{560} &= 11 x^4-14 x^3-30 x^2-14 x+11\,,\nonumber\\
P_{561} &= 11 x^4+20 x^3-52 x^2+20 x+11\,,\nonumber\\
P_{562} &= 12 x^4-12 x^3-7 x^2+10 x+1\,,\nonumber\\
P_{563} &= 13 x^4-29 x^3+64 x^2-29 x+13\,,\nonumber\\
P_{564} &= 13 x^4-24 x^3+50 x^2-24 x+13\,,\nonumber\\
P_{565} &= 13 x^4+126 x^3-218 x^2+142 x-11\,,\nonumber\\
P_{566} &= 14 x^4-41 x^3+68 x^2-29 x+8\,,\nonumber\\
P_{567} &= 15 x^4-53 x^3+188 x^2-53 x+15\,,\nonumber\\
P_{568} &= 15 x^4-17 x^3-8 x^2-17 x+15\,,\nonumber\\
P_{569} &= 15 x^4+10 x^3-38 x^2+10 x+15\,,\nonumber\\
P_{570} &= 16 x^4-136 x^3-7 x^2+344 x-281\,,\nonumber\\
P_{571} &= 17 x^4-46 x^3+190 x^2-46 x+17\,,\nonumber\\
P_{572} &= 17 x^4-26 x^3+24 x^2+10 x-1\,,\nonumber\\
P_{573} &= 17 x^4-23 x^3+56 x^2-23 x+17\,,\nonumber\\
P_{574} &= 18 x^4-39 x^3+82 x^2-39 x+18\,,\nonumber\\
P_{575} &= 18 x^4+125 x^3-256 x^2+125 x+18\,,\nonumber\\
P_{576} &= 19 x^4-69 x^3+196 x^2-69 x+17\,,\nonumber\\
P_{577} &= 19 x^4-14 x^3+14 x^2-14 x+19\,,\nonumber\\
P_{578} &= 19 x^4+109 x^3-230 x^2+109 x+19\,,\nonumber\\
P_{579} &= 20 x^4-53 x^3+36 x^2+15 x-14\,,\nonumber\\
P_{580} &= 20 x^4+23 x^3-58 x^2+23 x+20\,,\nonumber\\
P_{581} &= 24 x^4-27 x^3-42 x^2-27 x+22\,,\nonumber\\
P_{582} &= 25 x^4-108 x^3+178 x^2-108 x+1\,,\nonumber\\
P_{583} &= 25 x^4-53 x^3+148 x^2-53 x+25\,,\nonumber\\
P_{584} &= 26 x^4-81 x^3+102 x^2-81 x+28\,,\nonumber\\
P_{585} &= 27 x^4-102 x^3+320 x^2-102 x+27\,,\nonumber\\
P_{586} &= 27 x^4-78 x^3+106 x^2-62 x+3\,,\nonumber\\
P_{587} &= 37 x^4-108 x^3+66 x^2-108 x+45\,,\nonumber\\
P_{588} &= 38 x^4-127 x^3+158 x^2-55 x+2\,,\nonumber\\
P_{589} &= 41 x^4+216 x^3-322 x^2+216 x-115\,,\nonumber\\
P_{590} &= 45 x^4-8 x^3-186 x^2-8 x+45\,,\nonumber\\
P_{591} &= 53 x^4-141 x^3+228 x^2-141 x-17\,,\nonumber\\
P_{592} &= 56 x^4-103 x^3+162 x^2-103 x-22\,,\nonumber\\
P_{593} &= 57 x^4-8 x^3-82 x^2-8 x+101\,,\nonumber\\
P_{594} &= 58 x^4-135 x^3+214 x^2-135 x-20\,,\nonumber\\
P_{595} &= 69 x^4+564 x^3+790 x^2+564 x+69\,,\nonumber\\
P_{596} &= 77 x^4-132 x^3+230 x^2-68 x-51\,,\nonumber\\
P_{597} &= 79 x^4-182 x^3+362 x^2-182 x+79\,,\nonumber\\
P_{598} &= 83 x^4-70 x^3-180 x^2+310 x-47\,,\nonumber\\
P_{599} &= 83 x^4+153 x^3-360 x^2+153 x+13\,,\nonumber\\
P_{600} &= 93 x^4+28 x^3-330 x^2+92 x-35\,,\nonumber\\
P_{601} &= 139 x^4+3248 x^3-7 x^2-3040 x-404\,,\nonumber\\
P_{602} &= 187 x^4-1314 x^3+810 x^2+234 x-349\,,\nonumber\\
P_{603} &= 229 x^4-266 x^3+244 x^2-70 x+55\,,\nonumber\\
P_{604} &= 265 x^4-208 x^3+14 x^2-208 x+265\,,\nonumber\\
P_{605} &= 649 x^4-902 x^3+1050 x^2-246 x+153\,,\nonumber\\
P_{606} &= 759 x^4-550 x^3-828 x^2+118 x+53\,,\nonumber\\
P_{607} &= 1043 x^4-618 x^3-1024 x^2+762 x+173\,,\nonumber\\
P_{608} &= 1045 x^4+5364 x^3-8166 x^2+2628 x+1577\,,\nonumber\\
P_{609} &= 1343 x^4+2430 x^3-4194 x^2+1530 x+907\,,\nonumber\\
P_{610} &= x^5-15 x^4-70 x^3+18 x^2-43 x+13\,,\nonumber\\
P_{611} &= x^5-5 x^4+18 x^3-26 x^2+x-5\,,\nonumber\\
P_{612} &= x^5-5 x^4+34 x^3-30 x^2+7 x+1\,,\nonumber\\
P_{613} &= x^5-2 x^4+10 x^3-12 x^2+x-2\,,\nonumber\\
P_{614} &= x^5-2 x^4+18 x^3-14 x^2+4 x+1\,,\nonumber\\
P_{615} &= x^5+4 x^4-15 x^3+26 x^2-16 x+4\,,\nonumber\\
P_{616} &= x^5+4 x^4-14 x^3+18 x^2-2 x+1\,,\nonumber\\
P_{617} &= 2 x^5+5 x^4-34 x^3+52 x^2-36 x+5\,,\nonumber\\
P_{618} &= 2 x^5+5 x^4-12 x^3+20 x^2-x+2\,,\nonumber\\
P_{619} &= 3 x^5-9 x^4+38 x^3-50 x^2+3 x-9\,,\nonumber\\
P_{620} &= 3 x^5+9 x^4-26 x^3+38 x^2-3 x+3\,,\nonumber\\
P_{621} &= 4 x^5+19 x^4-72 x^3+96 x^2-76 x+19\,,\nonumber\\
P_{622} &= 5 x^5-x^4+26 x^3-18 x^2+5 x-1\,,\nonumber\\
P_{623} &= 6 x^5-51 x^4+56 x^3-43 x^2+6 x+8\,,\nonumber\\
P_{624} &= 6 x^5-47 x^4+80 x^3-51 x^2+6 x-4\,,\nonumber\\
P_{625} &= 7 x^5-24 x^4+56 x^3-32 x^2+7 x-8\,,\nonumber\\
P_{626} &= 7 x^5-22 x^4+88 x^3-46 x^2+7 x-24\,,\nonumber\\
P_{627} &= 7 x^5+x^4+30 x^3-14 x^2+7 x+1\,,\nonumber\\
P_{628} &= 10 x^5-5 x^4-57 x^3+66 x^2-15 x-3\,,\nonumber\\
P_{629} &= 11 x^5-83 x^4+132 x^3-85 x^2+11 x-2\,,\nonumber\\
P_{630} &= 12 x^5-x^4+10 x^3-28 x^2-2 x-1\,,\nonumber\\
P_{631} &= 13 x^5+x^4+58 x^3-30 x^2+13 x+1\,,\nonumber\\
P_{632} &= 18 x^5-37 x^4-72 x^3+118 x^2-78 x+3\,,\nonumber\\
P_{633} &= 24 x^5-34 x^4-43 x^3+25 x^2-x-3\,,\nonumber\\
P_{634} &= 24 x^5-3 x^4+22 x^3-52 x^2-2 x-3\,,\nonumber\\
P_{635} &= 72 x^5+115 x^4-762 x^3+644 x^2-90 x-3\,,\nonumber\\
P_{636} &= 105 x^5-195 x^4+156 x^3-68 x^2+69 x-47\,,\nonumber\\
P_{637} &= 125 x^5-279 x^4+182 x^3-42 x^2-15 x-27\,,\nonumber\\
P_{638} &= -1367 x^6+2934 x^5-4861 x^4+8604 x^3-9253 x^2+5094 x-575\,,\nonumber\\
P_{639} &= -543 x^6+702 x^5-157 x^4-660 x^3+1035 x^2-450 x+137\,,\nonumber\\
P_{640} &= -89 x^6+402 x^5-499 x^4+354 x^3-37 x^2+24 x-11\,,\nonumber\\
P_{641} &= -43 x^6+6 x^5-59 x^4+104 x^3-x^2+6 x+15\,,\nonumber\\
P_{642} &= -29 x^6+240 x^5-287 x^4+170 x^3+19 x^2+26 x-11\,,\nonumber\\
P_{643} &= -27 x^6+2 x^5-27 x^4+56 x^3+7 x^2+2 x+7\,,\nonumber\\
P_{644} &= -13 x^6-59 x^5+14 x^4+33 x^3-30 x^2-20 x+11\,,\nonumber\\
P_{645} &= -6 x^6-x^5-8 x^4-28 x^3+16 x^2-x+18\,,\nonumber\\
P_{646} &= x^6-29 x^5+117 x^4-204 x^3+117 x^2-29 x+1\,,\nonumber\\
P_{647} &= x^6-27 x^5+174 x^4-276 x^3+166 x^2-27 x-7\,,\nonumber\\
P_{648} &= x^6-13 x^5+52 x^4-42 x^3+8 x^2-3 x+1\,,\nonumber\\
P_{649} &= x^6-8 x^5+33 x^4-52 x^3+29 x^2-8 x-3\,,\nonumber\\
P_{650} &= x^6-4 x^5+17 x^4-14 x^3+15 x^2-4 x-1\,,\nonumber\\
P_{651} &= x^6+2 x^5-8 x^4+11 x^3-8 x^2+2 x+1\,,\nonumber\\
P_{652} &= x^6+2 x^5-7 x^4+2 x^3-6 x^2+2 x+2\,,\nonumber\\
P_{653} &= x^6+2 x^5-7 x^4+7 x^3-9 x^2+2 x-1\,,\nonumber\\
P_{654} &= x^6+4 x^5-31 x^4+56 x^3-37 x^2+4 x-5\,,\nonumber\\
P_{655} &= x^6+5 x^5-35 x^4+36 x^3-27 x^2+5 x+9\,,\nonumber\\
P_{656} &= x^6+5 x^5-19 x^4+28 x^3-19 x^2+5 x+1\,,\nonumber\\
P_{657} &= x^6+9 x^5-76 x^4+148 x^3-96 x^2+9 x-19\,,\nonumber\\
P_{658} &= x^6+10 x^5-71 x^4+104 x^3-69 x^2+10 x+3\,,\nonumber\\
P_{659} &= x^6+36 x^5-208 x^4+664 x^3-330 x^2+36 x-121\,,\nonumber\\
P_{660} &= 2 x^6-19 x^5+78 x^4-138 x^3+78 x^2-19 x+2\,,\nonumber\\
P_{661} &= 2 x^6-11 x^5+44 x^4-78 x^3+44 x^2-11 x+2\,,\nonumber\\
P_{662} &= 2 x^6-2 x^5+13 x^4-26 x^3+12 x^2-2 x+1\,,\nonumber\\
P_{663} &= 2 x^6-2 x^5+17 x^4-32 x^3+17 x^2-2 x+2\,,\nonumber\\
P_{664} &= 2 x^6-x^5+6 x^4-16 x^3+5 x^2-x+1\,,\nonumber\\
P_{665} &= 2 x^6+2 x^5-7 x^4+40 x^3-19 x^2+2 x-10\,,\nonumber\\
P_{666} &= 2 x^6+3 x^5-16 x^4+12 x^3-14 x^2+3 x+4\,,\nonumber\\
P_{667} &= 2 x^6+5 x^5-33 x^4+48 x^3-33 x^2+5 x+2\,,\nonumber\\
P_{668} &= 2 x^6+18 x^5-147 x^4+168 x^3-127 x^2+18 x+22\,,\nonumber\\
P_{669} &= 2 x^6+35 x^5-235 x^4+356 x^3-225 x^2+35 x+12\,,\nonumber\\
P_{670} &= 3 x^6-22 x^5+169 x^4-264 x^3+167 x^2-22 x+1\,,\nonumber\\
P_{671} &= 3 x^6-8 x^5+35 x^4-58 x^3+35 x^2-8 x+3\,,\nonumber\\
P_{672} &= 3 x^6-7 x^5+31 x^4-60 x^3+32 x^2-7 x+4\,,\nonumber\\
P_{673} &= 3 x^6-7 x^5+54 x^4-84 x^3+50 x^2-7 x-1\,,\nonumber\\
P_{674} &= 3 x^6-6 x^5+50 x^4-80 x^3+48 x^2-6 x+1\,,\nonumber\\
P_{675} &= 3 x^6-5 x^5+41 x^4-20 x^3+17 x^2-5 x-21\,,\nonumber\\
P_{676} &= 3 x^6-4 x^5+25 x^4-52 x^3+25 x^2-4 x+3\,,\nonumber\\
P_{677} &= 3 x^6-x^5+6 x^4-20 x^3+6 x^2-x+3\,,\nonumber\\
P_{678} &= 3 x^6+4 x^5-13 x^4+4 x^3-13 x^2+4 x+3\,,\nonumber\\
P_{679} &= 3 x^6+7 x^5-48 x^4+84 x^3-56 x^2+7 x-5\,,\nonumber\\
P_{680} &= 3 x^6+35 x^5-207 x^4-118 x^3+355 x^2-x-19\,,\nonumber\\
P_{681} &= 3 x^6+68 x^5+139 x^4-76 x^3-81 x^2+152 x-37\,,\nonumber\\
P_{682} &= 3 x^6+70 x^5-519 x^4+776 x^3-513 x^2+70 x+9\,,\nonumber\\
P_{683} &= 4 x^6-26 x^5+173 x^4-296 x^3+173 x^2-26 x+4\,,\nonumber\\
P_{684} &= 4 x^6-7 x^5+28 x^4-72 x^3+36 x^2-7 x+12\,,\nonumber\\
P_{685} &= 4 x^6+3 x^5-12 x^4-18 x^3-8 x^2+3 x+8\,,\nonumber\\
P_{686} &= 4 x^6+37 x^5-261 x^4+372 x^3-247 x^2+37 x+18\,,\nonumber\\
P_{687} &= 5 x^6-28 x^5+115 x^4-190 x^3+115 x^2-28 x+5\,,\nonumber\\
P_{688} &= 5 x^6-15 x^5+82 x^4-180 x^3+90 x^2-15 x+13\,,\nonumber\\
P_{689} &= 5 x^6-12 x^5+47 x^4-106 x^3+51 x^2-12 x+9\,,\nonumber\\
P_{690} &= 5 x^6-10 x^5-60 x^4+59 x^3-16 x^2-7 x+5\,,\nonumber\\
P_{691} &= 5 x^6-8 x^5+49 x^4-104 x^3+51 x^2-8 x+7\,,\nonumber\\
P_{692} &= 5 x^6+3 x^5-15 x^4+12 x^3-17 x^2+3 x+3\,,\nonumber\\
P_{693} &= 5 x^6+4 x^5-11 x^4+4 x^3-15 x^2+4 x+1\,,\nonumber\\
P_{694} &= 5 x^6+10 x^5-58 x^4+32 x^3-40 x^2+10 x+23\,,\nonumber\\
P_{695} &= 6 x^6-7 x^5+30 x^4-56 x^3+26 x^2-7 x+2\,,\nonumber\\
P_{696} &= 7 x^6-38 x^5+287 x^4-456 x^3+285 x^2-38 x+5\,,\nonumber\\
P_{697} &= 7 x^6-21 x^5+95 x^4-122 x^3+63 x^2-21 x-25\,,\nonumber\\
P_{698} &= 7 x^6-8 x^5+71 x^4-112 x^3+65 x^2-8 x+1\,,\nonumber\\
P_{699} &= 7 x^6-2 x^5+16 x^4-40 x^3+10 x^2-2 x+1\,,\nonumber\\
P_{700} &= 7 x^6+6 x^5-33 x^4+24 x^3-31 x^2+6 x+9\,,\nonumber\\
P_{701} &= 7 x^6+6 x^5-29 x^4+24 x^3-31 x^2+6 x+5\,,\nonumber\\
P_{702} &= 7 x^6+76 x^5+231 x^4-952 x^3+961 x^2-292 x-15\,,\nonumber\\
P_{703} &= 8 x^6-59 x^5+410 x^4-644 x^3+392 x^2-59 x-10\,,\nonumber\\
P_{704} &= 10 x^6-55 x^5+230 x^4-422 x^3+230 x^2-55 x+10\,,\nonumber\\
P_{705} &= 10 x^6-33 x^5+143 x^4-127 x^3+106 x^2-33 x-27\,,\nonumber\\
P_{706} &= 10 x^6-27 x^5+116 x^4-210 x^3+120 x^2-27 x+14\,,\nonumber\\
P_{707} &= 10 x^6+9 x^5-52 x^4+52 x^3-54 x^2+9 x+8\,,\nonumber\\
P_{708} &= 10 x^6+9 x^5-46 x^4+52 x^3-48 x^2+9 x+8\,,\nonumber\\
P_{709} &= 11 x^6-96 x^5-53 x^4+946 x^3-911 x^2+166 x-47\,,\nonumber\\
P_{710} &= 11 x^6-51 x^5+349 x^4-564 x^3+333 x^2-51 x-5\,,\nonumber\\
P_{711} &= 11 x^6-36 x^5+165 x^4-246 x^3+133 x^2-36 x-21\,,\nonumber\\
P_{712} &= 11 x^6-23 x^5+163 x^4-292 x^3+163 x^2-23 x+11\,,\nonumber\\
P_{713} &= 11 x^6-18 x^5+76 x^4-159 x^3+75 x^2-18 x+10\,,\nonumber\\
P_{714} &= 11 x^6+32 x^5+107 x^4-728 x^3+597 x^2-72 x+37\,,\nonumber\\
P_{715} &= 11 x^6+42 x^5-285 x^4+424 x^3-287 x^2+42 x+9\,,\nonumber\\
P_{716} &= 12 x^6+x^5+12 x^4+28 x^3-22 x^2+x-22\,,\nonumber\\
P_{717} &= 12 x^6+29 x^5-242 x^4+300 x^3-236 x^2+29 x+18\,,\nonumber\\
P_{718} &= 13 x^6-57 x^5+233 x^4-408 x^3+237 x^2-57 x+17\,,\nonumber\\
P_{719} &= 13 x^6-17 x^5+117 x^4-228 x^3+113 x^2-17 x+9\,,\nonumber\\
P_{720} &= 13 x^6-3 x^5+17 x^4-56 x^3+15 x^2-3 x+11\,,\nonumber\\
P_{721} &= 13 x^6+2 x^5-37 x^4+236 x^3-337 x^2+282 x-63\,,\nonumber\\
P_{722} &= 13 x^6+7 x^5-36 x^4+28 x^3-36 x^2+7 x+13\,,\nonumber\\
P_{723} &= 13 x^6+36 x^5-141 x^4+246 x^3-157 x^2+36 x-3\,,\nonumber\\
P_{724} &= 13 x^6+39 x^5-133 x^4+196 x^3-165 x^2+39 x-19\,,\nonumber\\
P_{725} &= 13 x^6+118 x^5-869 x^4+1208 x^3-827 x^2+118 x+55\,,\nonumber\\
P_{726} &= 13 x^6+120 x^5-859 x^4+1264 x^3-837 x^2+120 x+35\,,\nonumber\\
P_{727} &= 14 x^6-9 x^5+91 x^4-148 x^3+81 x^2-9 x+4\,,\nonumber\\
P_{728} &= 14 x^6-8 x^5+57 x^4-144 x^3+57 x^2-8 x+14\,,\nonumber\\
P_{729} &= 14 x^6+2 x^5-3 x^4-16 x^3-9 x^2+2 x+8\,,\nonumber\\
P_{730} &= 15 x^6-28 x^5+127 x^4-240 x^3+125 x^2-28 x+13\,,\nonumber\\
P_{731} &= 15 x^6+20 x^5-122 x^4+152 x^3-124 x^2+20 x+13\,,\nonumber\\
P_{732} &= 15 x^6+426 x^5-543 x^4+320 x^3-71 x^2+230 x-41\,,\nonumber\\
P_{733} &= 16 x^6-43 x^5+186 x^4-346 x^3+190 x^2-43 x+20\,,\nonumber\\
P_{734} &= 16 x^6-27 x^5+114 x^4-220 x^3+110 x^2-27 x+12\,,\nonumber\\
P_{735} &= 17 x^6-166 x^5+489 x^4-716 x^3+475 x^2-158 x+11\,,\nonumber\\
P_{736} &= 17 x^6-18 x^5+121 x^4-220 x^3+105 x^2-18 x+1\,,\nonumber\\
P_{737} &= 17 x^6+21 x^5-71 x^4+122 x^3-87 x^2+21 x+1\,,\nonumber\\
P_{738} &= 17 x^6+28 x^5-183 x^4+224 x^3-177 x^2+28 x+23\,,\nonumber\\
P_{739} &= 17 x^6+34 x^5-230 x^4+304 x^3-228 x^2+34 x+19\,,\nonumber\\
P_{740} &= 18 x^6-16 x^5+73 x^4-158 x^3+67 x^2-16 x+12\,,\nonumber\\
P_{741} &= 18 x^6-9 x^5+94 x^4-172 x^3+88 x^2-9 x+12\,,\nonumber\\
P_{742} &= 18 x^6+x^5-68 x^3+2 x^2+x+20\,,\nonumber\\
P_{743} &= 19 x^6-73 x^5+303 x^4-544 x^3+307 x^2-73 x+23\,,\nonumber\\
P_{744} &= 19 x^6-8 x^5+43 x^4-128 x^3+37 x^2-8 x+13\,,\nonumber\\
P_{745} &= 19 x^6+24 x^5-73 x^4+100 x^3-105 x^2+24 x-13\,,\nonumber\\
P_{746} &= 19 x^6+42 x^5-147 x^4+196 x^3-137 x^2+42 x+29\,,\nonumber\\
P_{747} &= 19 x^6+62 x^5-433 x^4+536 x^3-395 x^2+62 x+57\,,\nonumber\\
P_{748} &= 20 x^6-303 x^5+1214 x^4-1692 x^3+1238 x^2-303 x+44\,,\nonumber\\
P_{749} &= 20 x^6-8 x^5+51 x^4-101 x^3+30 x^2-8 x-1\,,\nonumber\\
P_{750} &= 21 x^6-53 x^5+227 x^4-438 x^3+231 x^2-53 x+25\,,\nonumber\\
P_{751} &= 21 x^6-11 x^5+59 x^4-134 x^3+55 x^2-11 x+17\,,\nonumber\\
P_{752} &= 21 x^6+8 x^5-25 x^4+8 x^3-35 x^2+8 x+11\,,\nonumber\\
P_{753} &= 22 x^6-171 x^5+708 x^4-892 x^3+634 x^2-171 x-52\,,\nonumber\\
P_{754} &= 22 x^6-61 x^5+162 x^4-118 x^3-12 x^2+95 x-40\,,\nonumber\\
P_{755} &= 23 x^6+37 x^5-429 x^4+596 x^3-277 x^2-17 x+19\,,\nonumber\\
P_{756} &= 23 x^6+68 x^5-295 x^4+448 x^3-317 x^2+68 x+1\,,\nonumber\\
P_{757} &= 24 x^6-67 x^5+282 x^4-564 x^3+280 x^2-67 x+22\,,\nonumber\\
P_{758} &= 25 x^6-174 x^5+489 x^4-716 x^3+467 x^2-150 x+11\,,\nonumber\\
P_{759} &= 25 x^6-6 x^5+21 x^4-112 x^3+19 x^2-6 x+23\,,\nonumber\\
P_{760} &= 25 x^6-6 x^5+33 x^4-112 x^3+31 x^2-6 x+23\,,\nonumber\\
P_{761} &= 25 x^6+5 x^5-5 x^4+22 x^3-43 x^2+5 x-13\,,\nonumber\\
P_{762} &= 27 x^6-32 x^5+129 x^4-344 x^3+151 x^2-32 x+49\,,\nonumber\\
P_{763} &= 28 x^6+27 x^5-416 x^4+596 x^3-290 x^2-7 x+14\,,\nonumber\\
P_{764} &= 29 x^6-348 x^5+511 x^4-354 x^3+85 x^2-78 x+11\,,\nonumber\\
P_{765} &= 29 x^6+30 x^5-139 x^4+120 x^3-137 x^2+30 x+31\,,\nonumber\\
P_{766} &= 29 x^6+435 x^5-1475 x^4+1674 x^3-731 x^2-9 x+65\,,\nonumber\\
P_{767} &= 31 x^6+8 x^5-29 x^4-44 x^3-35 x^2+8 x+25\,,\nonumber\\
P_{768} &= 31 x^6+8 x^5-13 x^4-64 x^3-11 x^2+8 x+33\,,\nonumber\\
P_{769} &= 31 x^6+20 x^5-69 x^4+36 x^3-87 x^2+20 x+13\,,\nonumber\\
P_{770} &= 31 x^6+30 x^5-141 x^4+120 x^3-139 x^2+30 x+33\,,\nonumber\\
P_{771} &= 32 x^6-151 x^5+636 x^4-1240 x^3+650 x^2-151 x+46\,,\nonumber\\
P_{772} &= 34 x^6-91 x^5+382 x^4-746 x^3+384 x^2-91 x+36\,,\nonumber\\
P_{773} &= 34 x^6+x^5+8 x^4-68 x^3-14 x^2+x+12\,,\nonumber\\
P_{774} &= 35 x^6-90 x^5+388 x^4-718 x^3+378 x^2-90 x+25\,,\nonumber\\
P_{775} &= 37 x^6-118 x^5+343 x^4-716 x^3+639 x^2-206 x-27\,,\nonumber\\
P_{776} &= 37 x^6-59 x^5-233 x^4+714 x^3-541 x^2+73 x-31\,,\nonumber\\
P_{777} &= 37 x^6-57 x^5+371 x^4-668 x^3+339 x^2-57 x+5\,,\nonumber\\
P_{778} &= 38 x^6-107 x^5+466 x^4-932 x^3+456 x^2-107 x+28\,,\nonumber\\
P_{779} &= 39 x^6-112 x^5+413 x^4-900 x^3+769 x^2-228 x-21\,,\nonumber\\
P_{780} &= 40 x^6+69 x^5-463 x^4+628 x^3-465 x^2+69 x+38\,,\nonumber\\
P_{781} &= 42 x^6-41 x^5+182 x^4-380 x^3+166 x^2-41 x+26\,,\nonumber\\
P_{782} &= 42 x^6+2 x^5+3 x^4-40 x^3-41 x^2+2 x-2\,,\nonumber\\
P_{783} &= 43 x^6+20 x^5-27 x^4+16 x^3-73 x^2+20 x-3\,,\nonumber\\
P_{784} &= 44 x^6+67 x^5+304 x^4-1272 x^3+1264 x^2-371 x-20\,,\nonumber\\
P_{785} &= 46 x^6-39 x^5+271 x^4-420 x^3+197 x^2-39 x-28\,,\nonumber\\
P_{786} &= 47 x^6-154 x^5+53 x^4+580 x^3-701 x^2+110 x-23\,,\nonumber\\
P_{787} &= 50 x^6-55 x^5+170 x^4-732 x^3+814 x^2-333 x-2\,,\nonumber\\
P_{788} &= 50 x^6+37 x^5-310 x^4+364 x^3-348 x^2+37 x+12\,,\nonumber\\
P_{789} &= 51 x^6+32 x^5-117 x^4+32 x^3-107 x^2+32 x+61\,,\nonumber\\
P_{790} &= 53 x^6-33 x^5-279 x^4+596 x^3-427 x^2+53 x-11\,,\nonumber\\
P_{791} &= 55 x^6+105 x^5-1711 x^4+2400 x^3-889 x^2+135 x-71\,,\nonumber\\
P_{792} &= 56 x^6-141 x^5+930 x^4-1916 x^3+1004 x^2-141 x+130\,,\nonumber\\
P_{793} &= 59 x^6-446 x^5+203 x^4+872 x^3-1215 x^2+398 x+9\,,\nonumber\\
P_{794} &= 63 x^6-205 x^5+273 x^4+118 x^3-537 x^2+347 x-11\,,\nonumber\\
P_{795} &= 65 x^6+154 x^5-1187 x^4+1624 x^3-1209 x^2+154 x+43\,,\nonumber\\
P_{796} &= 67 x^6-105 x^5+485 x^4-938 x^3+465 x^2-105 x+47\,,\nonumber\\
P_{797} &= 71 x^6+132 x^5-461 x^4+508 x^3-535 x^2+132 x-3\,,\nonumber\\
P_{798} &= 77 x^6-144 x^5+1043 x^4-1760 x^3+969 x^2-144 x+3\,,\nonumber\\
P_{799} &= 81 x^6-571 x^5-129 x^4+1738 x^3-1887 x^2+461 x+43\,,\nonumber\\
P_{800} &= 91 x^6-48 x^5-99 x^4-638 x^3+595 x^2-50 x+29\,,\nonumber\\
P_{801} &= 96 x^6-163 x^5+758 x^4-1528 x^3+698 x^2-163 x+36\,,\nonumber\\
P_{802} &= 98 x^6-221 x^5+988 x^4-1976 x^3+944 x^2-221 x+54\,,\nonumber\\
P_{803} &= 111 x^6-139 x^5+111 x^4-1216 x^3+985 x^2-169 x+53\,,\nonumber\\
P_{804} &= 117 x^6+576 x^5-2249 x^4+3792 x^3-2523 x^2+576 x-157\,,\nonumber\\
P_{805} &= 119 x^6-132 x^5+603 x^4-1194 x^3+535 x^2-132 x+51\,,\nonumber\\
P_{806} &= 133 x^6-760 x^5+1463 x^4-1204 x^3+231 x^2+100 x-11\,,\nonumber\\
P_{807} &= 139 x^6+41 x^5-335 x^4+308 x^3-467 x^2+41 x+7\,,\nonumber\\
P_{808} &= 141 x^6+79 x^5-621 x^4+724 x^3-749 x^2+79 x+13\,,\nonumber\\
P_{809} &= 145 x^6-634 x^5+3441 x^4-7530 x^3+6725 x^2-2952 x+637\,,\nonumber\\
P_{810} &= 146 x^6+877 x^5-3462 x^4+3876 x^3-1458 x^2-263 x+200\,,\nonumber\\
P_{811} &= 181 x^6-72 x^5+599 x^4-1328 x^3+477 x^2-72 x+59\,,\nonumber\\
P_{812} &= 337 x^6-154 x^5-165 x^4-68 x^3+283 x^2-186 x+17\,,\nonumber\\
P_{813} &= 443 x^6+2898 x^5-5681 x^4+4104 x^3-2225 x^2+1314 x-565\,,\nonumber\\
P_{814} &= 649 x^6-210 x^5+2025 x^4-3960 x^3+1323 x^2-210 x-53\,.
\end{align}

\newpage

\begin{dmath*}
 F_{A,2}^{(1)} = C_F \Bigg[
%
%
  \Bigg\{
-8 x (1+x)^2 \eta ^2
-4 x (1+x)^2 \Big(
        3-2 x+3 x^2\Big) \eta ^3 H_0
\Bigg\}
+ \ep  \Bigg\{
-16 x (1+x)^2 \eta ^2
~~~~~~~~
\end{dmath*}
\vspace{-1.0cm}
\begin{dmath}
{\color{white}=}
+\eta ^3 \Big(
        -8 x (1+x)^2 \Big(
                3-2 x+3 x^2\Big) H_0
        +8 x (1+x)^2 \Big(
                3-2 x+3 x^2\Big) H_{-1} H_0
        -2 x (1+x)^2 \Big(
                3-2 x+3 x^2\Big) H_0^2
        -8 x (1+x)^2 \Big(
                3-2 x+3 x^2\Big) H_{0,-1}
        +4 x (1+x)^2 \Big(
                3-2 x+3 x^2\Big) \zeta_2
\Big)
\Bigg\}
+ \ep^2  \Bigg\{
-32 x (1+x)^2 \eta ^2
+\eta ^3 \Big(
        \Big(
                -16 x (1+x)^2 \Big(
                        3-2 x+3 x^2\Big)
                +16 x (1+x)^2 \Big(
                        3-2 x+3 x^2\Big) H_{-1}
                -8 x (1+x)^2 \Big(
                        3-2 x+3 x^2\Big) H_{-1}^2
                +2 x (1+x)^2 \Big(
                        3-2 x+3 x^2\Big) \zeta_2
        \Big) H_0
        +\Big(
                -4 x (1+x)^2 \Big(
                        3-2 x+3 x^2\Big)
                +4 x (1+x)^2 \Big(
                        3-2 x+3 x^2\Big) H_{-1}
        \Big) H_0^2
        -\frac{2}{3} x (1+x)^2 \Big(
                3-2 x+3 x^2\Big) H_0^3
        +\Big(
                -16 x (1+x)^2 \Big(
                        3-2 x+3 x^2\Big)
                +16 x (1+x)^2 \Big(
                        3-2 x+3 x^2\Big) H_{-1}
        \Big) H_{0,-1}
        -8 x (1+x)^2 \Big(
                3-2 x+3 x^2\Big) H_{0,0,-1}
        -16 x (1+x)^2 \Big(
                3-2 x+3 x^2\Big) H_{0,-1,-1}
        +4 x (1+x)^2 \Big(
                5-4 x+7 x^2\Big) \zeta_2
        -8 x (1+x)^2 \Big(
                3-2 x+3 x^2\Big) H_{-1} \zeta_2
        +8 x (1+x)^2 \Big(
                3-2 x+3 x^2\Big) \zeta_3
\Big)
\Bigg\}
+ \ep^3  \Bigg\{
-64 x (1+x)^2 \eta ^2
+\eta ^3 \Big(
        \Big(
                -32 x (1+x)^2 \Big(
                        3-2 x+3 x^2\Big)
                +\Big(
                        32 x (1+x)^2 \Big(
                                3-2 x+3 x^2\Big)
                        -4 x (1+x)^2 \Big(
                                3-2 x+3 x^2\Big) \zeta_2
                \Big) H_{-1}
                -16 x (1+x)^2 \Big(
                        3-2 x+3 x^2\Big) H_{-1}^2
                +
                \frac{16}{3} x (1+x)^2 \Big(
                        3-2 x+3 x^2\Big) H_{-1}^3
                +4 x (1+x)^2 \Big(
                        3-2 x+3 x^2\Big) \zeta_2
                +\frac{28}{3} x (1+x)^2 \Big(
                        3-2 x+3 x^2\Big) \zeta_3
        \Big) H_0
        +\Big(
                -8 x (1+x)^2 \Big(
                        3-2 x+3 x^2\Big)
                +8 x (1+x)^2 \Big(
                        3-2 x+3 x^2\Big) H_{-1}
                -4 x (1+x)^2 \Big(
                        3-2 x+3 x^2\Big) H_{-1}^2
                +x (1+x)^2 \Big(
                        3-2 x+3 x^2\Big) \zeta_2
        \Big) H_0^2
        +\Big(
                -\frac{4}{3} x (1+x)^2 \Big(
                        3-2 x+3 x^2\Big)
                +\frac{4}{3} x (1+x)^2 \Big(
                        3-2 x+3 x^2\Big) H_{-1}
        \Big) H_0^3
        -\frac{1}{6} x (1+x)^2 \Big(
                3-2 x+3 x^2\Big) H_0^4
        +\Big(
                -16 x (1+x)^2 \Big(
                        3-2 x+3 x^2\Big) \zeta_2
                -16 x (1+x)^2 \Big(
                        3-2 x+3 x^2\Big) \zeta_3
        \Big) H_{-1}
        +\Big(
                -32 x (1+x)^2 \Big(
                        3-2 x+3 x^2\Big)
                +32 x (1+x)^2 \Big(
                        3-2 x+3 x^2\Big) H_{-1}
                -16 x (1+x)^2 \Big(
                        3-2 x+3 x^2\Big) H_{-1}^2
                -4 x (1+x)^2 \Big(
                        3-2 x+3 x^2\Big) \zeta_2
        \Big) H_{0,-1}
        +\Big(
                -16 x (1+x)^2 \Big(
                        3-2 x+3 x^2\Big)
                +16 x (1+x)^2 \Big(
                        3-2 x+3 x^2\Big) H_{-1}
        \Big) H_{0,0,-1}
        +\Big(
                -32 x (1+x)^2 \Big(
                        3-2 x+3 x^2\Big)
                +32 x (1+x)^2 \Big(
                        3-2 x+3 x^2\Big) H_{-1}
        \Big) H_{0,-1,-1}
        -8 x (1+x)^2 \Big(
                3-2 x+3 x^2\Big) H_{0,0,0,-1}
        -16 x (1+x)^2 \Big(
                3-2 x+3 x^2\Big) H_{0,0,-1,-1}
        -32 x (1+x)^2 \Big(
                3-2 x+3 x^2\Big) H_{0,-1,-1,-1}
        +8 x (1+x)^2 \Big(
                5-4 x+7 x^2\Big) \zeta_2
        +8 x (1+x)^2 \Big(
                3-2 x+3 x^2\Big) H_{-1}^2 \zeta_2
        +\frac{28}{5} x (1+x)^2 \Big(
                3-2 x+3 x^2\Big) \zeta_2^2
        +\frac{8}{3} x (1+x)^2 \Big(
                19-12 x+17 x^2\Big) \zeta_3
\Big)
\Bigg\}
\Bigg] \,.
\end{dmath}

\begin{dmath*}
 F_{A,2}^{(2),\ns} =
%
C_F^2  \Bigg[
%
 \frac{1}{\ep}  \Bigg\{
        16 x (1+x)^2 \eta ^2
        +8 x (1+x)^2 \Big(
                5-2 x+5 x^2\Big) \eta ^3 H_0
        +8 x (1+x)^2 \Big(
                3-2 x+3 x^2\Big) \xi  \eta ^3 H_0^2
\Bigg\}
\end{dmath*}
\vspace{-1.0cm}
\begin{dmath*}
{\color{white}=}
+  \Bigg\{
                68 x (1+x)^2 \eta ^2
                +\eta ^3 \Big(
                        2 x (1+x)^2 \Big(
                                61-22 x+61 x^2\Big) H_0
                        +8 x (1+x)^2 \Big(
                                25-86 x+25 x^2\Big) H_{-1} H_0
                        +\Big(
                                128 x^2 (1+x)^2 H_0
                                -\frac{1088}{3} x^3 H_0^3
                        \Big) H_1
                        -128 x^2 (1+x)^2 H_{0,1}
                        -8 x (1+x)^2 \Big(
                                25-86 x+25 x^2\Big) H_{0,-1}
                        +\Big(
                                -2176 x^3 H_0 H_1
                                +2176 x^3 H_{0,1}
                        \Big) \zeta_2
                \Big)
                +\xi ^2 \Big(
                        -192 (-1+x) x^2 (1+x)^3 \eta ^4 \zeta_2 H_{0,-1}
                        -144 (-1+x) x (1+x)^3 \eta ^3 H_{-1}
                         \zeta_2
                \Big)
                +\xi  \Big(
                        \eta ^3 \Big(
                                \Big(
                                        128 (-1+x) x (1+x)^3 H_0
                                        -
                                        \frac{64}{3} (-1+x) x^2 (1+x) H_0^3
                                \Big) H_1
                                -128 (-1+x) x (1+x)^3 H_{0,1}
                                +\Big(
                                        -128 (-1+x) x^2 (1+x) H_0 H_1
                                        -576 x^2 (1+x)^2 H_{-1}
                                        +128 (-1+x) x^2 (1+x) H_{0,1}
                                \Big) \zeta_2
                        \Big)
                        -960 x^3 (1+x)^2 \eta ^4 H_{0,-1} \zeta_2
                \Big)
                +\eta ^5 \Big(
                        -64 x^2 (1+x)^2 H_0 H_{0,0,-1} P_{817}
                        -32 x^2 (1+x)^2 H_0^2 H_{0,1} P_{818}
                        +32 x^2 (1+x)^2 H_0^2 H_{0,-1} P_{820}
                        +\frac{4}{3} x^2 (1+x)^2 H_0^4 P_{835}
                        +64 x^2 (1+x)^2 H_{0,0,0,1} P_{843}
                        -\frac{8}{3} x (1+x)^2 H_0^3 P_{1018}
                        +\frac{32}{5} x^2 (1+x)^2 P_{816} \zeta_2^2
                        -1024 x^3 (1+x)^2 \Big(
                                5-7 x+5 x^2\Big) H_0 H_{0,0,1}
                        -384 x^3 (1+x)^2 \Big(
                                28-41 x+28 x^2\Big) H_{0,0,0,-1}
                        +\Big(
                                16 x (1+x)^2 H_0 P_{1037}
                                -96 x^3 (1+x)^2 \Big(
                                        2-9 x+2 x^2\Big) H_0^2
                        \Big) \zeta_2
                        -64 x^2 (1+x)^2 H_0 P_{821} \zeta_3
                \Big)
                +\eta ^4 \Big(
                        16 x (1+x)^2 H_0^2 H_1 P_{868}
                        -32 x (1+x)^2 H_0 H_{0,1} P_{900}
                        +32 x (1+x)^2 H_{0,0,1} P_{930}
                        -8 x (1+x)^2 H_{-1} H_0^2 P_{948}
                        +16 x (1+x)^2 H_0 H_{0,-1} P_{962}
                        -16 x (1+x)^2 H_{0,0,-1} P_{968}
                        +2 x (1+x)^2 H_0^2 P_{1002}
                        +4 x (1+x)^2 P_{985}
                         \zeta_2
                        +16 x (1+x)^2 P_{862} \zeta_3
                \Big)
                -192 x \log (2) x_+^2 \zeta_2
\Bigg\}
+ \ep  \Bigg\{
                \eta ^2  \Big( 32 c_1 x (x-1)^2 + 242 x (x+1)^2 \Big)
                +\xi  \Big(
                        \eta ^4 \Big(
                                \log (2) \zeta_2 \Big(
                                        -384 (-1+x) x^2 (1+x)^3 H_1
                                        -384 (-1+x) x^2 (1+x)^3 H_{-1}
                                \Big)
                                +64 (-1+x) x^2 (1+x)^3 H_{-1} H_0^2 H_1
                                +\Big(
                                        384 (-1+x) x^2 (1+x)^3 H_{-1} H_1
                                        -1920 x^3 (1+x)^2 H_{-1} H_{0,-1}
                                        -384 (-1+x) x^2 (1+x)^3 H_{-1,1}
                                \Big) \zeta_2
                        \Big)
                        +\eta ^3 \Big(
                                \Big(
                                        -768 (-1+x) x (1+x)^3 H_{-1} H_0
                                        -
                                        \frac{128}{3} (-1+x) x^2 (1+x) H_{-1} H_0^3
                                \Big) H_1
                                +\Big(
                                        128 (-1+x) x (1+x)^3 H_0
                                        +\frac{64}{3} (-1+x) x^2 (1+x) H_0^3
                                \Big) H_1^2
                                +\Big(
                                        -256 (-1+x) x (1+x)^3 H_1
                                        +768 (-1+x) x (1+x)^3 H_{-1}
                                \Big) H_{0,1}
                                +768 (-1+x) x (1+x)^3 H_1 H_{0,-1}
                                +256 (-1+x) x (1+x)^3 H_{0,1,1}
                                -768 (-1+x) x (1+x)^3 H_{0,1,-1}
                                -768 (-1+x) x (1+x)^3 H_{0,-1,1}
                                +\Big(
                                        \Big(
                                                -256 (-1+x) x^2 (1+x) H_{-1} H_1
                                                +128 (-1+x) x^2 (1+x) H_1^2
                                        \Big) H_0
                                        -128 (-1+x) x (1+x)^3 H_1
                                        +\Big(
                                                -256 (-1+x) x^2 (1+x) H_1
                                                +256 (-1+x) x^2 (1+x) H_{-1}
                                        \Big) H_{0,1}
                                        +256 (-1+x) x^2 (1+x) H_{0,1,1}
                                \Big) \zeta_2
                        \Big)
                        +384 (-1+x) x^2 (1+x)^3 \Big(
                                5-x+5 x^2\Big) \eta ^5 \zeta_2 H_{0,-1,-1}
                \Big)
                +\eta ^3 \Big(
                        5 x (1+x)^2 \Big(
                                67-18 x+67 x^2\Big) H_0
                        +16 x (1+x)^2 \Big(
                                75-166 x+75 x^2\Big) H_{-1} H_0
                        -8 x (1+x)^2 \Big(
                                127-298 x+127 x^2\Big) H_{-1}^2 H_0
                        +\Big(
                                -208 x (1+x)^2 \Big(
                                        3-2 x+3 x^2\Big) H_0
                                +\Big(
                                        -768 x^2 (1+x)^2 H_0
                                        -\frac{2176}{3} x^3 H_0^3
                                \Big) H_{-1}
                        \Big) H_1
                        +\Big(
                                128 x^2 (1+x)^2 H_0
                                +\frac{1088}{3} x^3 H_0^3
                        \Big) H_1^2
                        +\Big(
                                208 x (1+x)^2 \Big(
                                        3-2 x+3 x^2\Big)
                                -256 x^2 (1+x)^2 H_1
                                +768 x^2 (1+x)^2 H_{-1}
                        \Big) H_{0,1}
                        +\Big(
                                -16 x (1+x)^2 \Big(
                                        75-166 x+75 x^2\Big)
                                +768 x^2 (1+x)^2 H_1
\end{dmath*}
\begin{dmath*}
{\color{white}=}
                                +16 x (1+x)^2 \Big(
                                        127-298 x+127 x^2\Big) H_{-1}
                        \Big) H_{0,-1}
                        +256 x^2 (1+x)^2 H_{0,1,1}
                        -768 x^2 (1+x)^2 H_{0,1,-1}
                        -768 x^2 (1+x)^2 H_{0,-1,1}
                        -16 x (1+x)^2 \Big(
                                127-298 x+127 x^2\Big) H_{0,-1,-1}
                        +\Big(
                                \Big(
                                        -4352 x^3 H_{-1} H_1
                                        +2176 x^3 H_1^2
                                \Big) H_0
                                -128 x^2 (1+x)^2 H_1
                                +\Big(
                                        -4352 x^3 H_1
                                        +4352 x^3 H_{-1}
                                \Big) H_{0,1}
                                +4352 x^3 H_{0,1,1}
                        \Big) \zeta_2
                \Big)
                +\eta ^4 \Big(
                        384 x (1+x)^2 H_{0,0,1,-1} P_{837}
                        +384 x (1+x)^2 H_{0,0,-1,1} P_{837}
                        +192 x (1+x)^2 H_{0,0,1,1} P_{840}
                        +16 x (1+x)^2 H_0^2 H_1^2 P_{867}
                        +192 x (1+x)^2 H_{0,-1,0,1} P_{869}
                        +192 x (1+x)^2 H_0 H_{0,1,-1} P_{883}
                        +192 x (1+x)^2 H_0 H_{0,-1,1} P_{883}
                        +32 x (1+x)^2 H_{0,1}^2 P_{903}
                        -64 x (1+x)^2 H_0 H_{0,1,1} P_{904}
                        -96 x (1+x)^2 H_{0,0,-1,-1} P_{916}
                        -32 x (1+x)^2 H_0 H_{0,-1,-1} P_{960}
                        +8 x (1+x)^2 H_{-1}^2 H_0^2 P_{969}
                        +16 x (1+x)^2 H_{0,-1}^2 P_{983}
                        -4 x (1+x)^2 H_{-1} H_0^2 P_{1008}
                        +2 x (1+x)^2 H_0^2 P_{1012}
                        +\log (2) \zeta_2 \Big(
                                -384 x (1+x)^2 P_{865}
                                +1536 x^3 (1+x)^2 H_1
                                +1536 x^3 (1+x)^2 H_{-1}
                        \Big)
                        +\Big(
                                8 x (1+x)^2 H_0^2 P_{984}
                                -256 x^3 (1+x)^2 H_{-1} H_0^2
                        \Big) H_1
                        +\Big(
                                64 x (1+x)^2 H_{-1} H_0 P_{870}
                                -32 x (1+x)^2 H_0 P_{911}
                                -64 x (1+x)^2 H_{0,-1} P_{947}
                                +256 x^2 (1+x)^2 \Big(
                                        2+15 x+2 x^2\Big) H_0 H_1
                        \Big) H_{0,1}
                        +\Big(
                                -128 x (1+x)^2 H_{-1} H_0 P_{901}
                                -64 x (1+x)^2 H_0 H_1 P_{902}
                                -16 x (1+x)^2 H_0 P_{949}
                        \Big) H_{0,-1}
                        +\Big(
                                -384 x (1+x)^2 H_{-1} P_{837}
                                -192 x (1+x)^2 H_1 P_{840}
                                -16 x (1+x)^2 P_{957}
                        \Big) H_{0,0,1}
                        +\Big(
                                768 x (1+x)^2 H_1 P_{838}
                                +96 x (1+x)^2 H_{-1} P_{916}
                                +8 x (1+x)^2 P_{1009}
                        \Big) H_{0,0,-1}
                        +\Big(
                                -48 x (1+x)^2 H_{-1}^2 P_{929}
                                +8 x (1+x)^2 H_{-1} P_{997}
                                +4 x (1+x)^2 P_{1006}
                                -1536 x^3 (1+x)^2 H_{-1} H_1
                                +1536 x^3 (1+x)^2 H_{-1,1}
                        \Big) \zeta_2
                        +\Big(
                                -192 x (1+x)^2 H_1 P_{850}
                                +96 x (1+x)^2 H_{-1} P_{879}
                                -16 x (1+x)^2 P_{992}
                        \Big) \zeta_3
                \Big)
                +\eta ^5 \Big(
                        -768 x^2 (1+x)^2 H_0 H_{0,0,1,-1} P_{819}
                        -768 x^2 (1+x)^2 H_0 H_{0,0,-1,1} P_{819}
                        -1024 x^2 (1+x)^2 H_0 H_{0,0,1,1} P_{833}
                        +
                        \frac{4}{5} x^2 (1+x)^2 H_0^5 P_{834}
                        -512 x^2 (1+x)^2 H_{0,0,-1,0,-1} P_{845}
                        -256 x^2 (1+x)^2 H_0 H_{0,1}^2 P_{846}
                        -768 x^2 (1+x)^2 H_{0,0,0,-1,-1} P_{874}
                        +128 x^2 (1+x)^2 H_0 H_{0,0,-1,-1} P_{886}
                        +512 x^2 (1+x)^2 H_{0,0,0,1,-1} P_{887}
                        +512 x^2 (1+x)^2 H_{0,0,0,-1,1} P_{887}
                        -64 x^2 (1+x)^2 H_0^2 H_{0,-1,-1} P_{888}
                        +128 x^2 (1+x)^2 H_{0,0,1,0,-1} P_{905}
                        -128 x^2 (1+x)^2 H_0 H_{0,-1,0,1} P_{907}
                        -128 x^2 (1+x)^2 H_{0,0,1,0,1} P_{921}
                        +64 x^2 (1+x)^2 H_0^2 H_{0,1,1} P_{927}
                        -128 x^2 (1+x)^2 H_{0,0,0,1,1} P_{934}
                        +128 x^2 (1+x)^2 H_{0,0,-1,0,1} P_{943}
                        +64 x^2 (1+x)^2 H_{0,0,0,0,-1} P_{963}
                        -64 x^2 (1+x)^2 H_{0,0,0,0,1} P_{970}
                        -\frac{2}{3} x (1+x)^2 H_0^4 P_{1045}
                        -\frac{2}{3} x (1+x)^2 H_0^3 P_{1060}
                        +\log (2) \zeta_2 \Big(
                                1536 x^3 (1+x)^2 \Big(
                                        1+x+x^2\Big) H_{0,1}
                                +1536 x^3 (1+x)^2 \Big(
                                        1+x+x^2\Big) H_{0,-1}
                        \Big)
                        +\Big(
                                \frac{8}{3} x^2 (1+x)^2 H_0^4 P_{825}
                                -\frac{16}{3} x (1+x)^2 H_0^3 P_{1043}
                        \Big) H_1
                        +\Big(
                                \frac{8}{3} x^2 (1+x)^2 H_0^4 P_{835}
                                +\frac{8}{3} x (1+x)^2 H_0^3 P_{1050}
                        \Big) H_{-1}
                        +\Big(
                                -64 x^2 (1+x)^2 H_{-1} H_0^2 P_{818}
                                -128 x^2 (1+x)^2 H_0^2 H_1 P_{832}
                                +256 x^2 (1+x)^2 H_0 H_{0,-1} P_{875}
                                +16 x (1+x)^2 H_0^2 P_{1044}
                                +
                                \frac{64}{3} x^3 (1+x)^2 \Big(
                                        74-79 x+74 x^2\Big) H_0^3
                        \Big) H_{0,1}
                        +\Big(
                                64 x^2 (1+x)^2 H_{-1} H_0^2 P_{820}
                                -64 x^2 (1+x)^2 H_0^2 H_1 P_{823}
                                +\frac{32}{3} x^2 (1+x)^2 H_0^3 P_{906}
                                -8 x (1+x)^2 H_0^2 P_{1054}
                        \Big) H_{0,-1}
                        -384 x^3 (1+x)^2 \Big(
                                10-17 x+10 x^2\Big) H_0 H_{0,-1}^2
\end{dmath*}
\begin{dmath*}
{\color{white}=}
                        +\Big(
                                -128 x^2 (1+x)^2 H_{0,-1} P_{826}
                                +1024 x^2 (1+x)^2 H_0 H_1 P_{831}
                                -32 x^2 (1+x)^2 H_0^2 P_{873}
                                +128 x^2 (1+x)^2 H_{0,1} P_{922}
                                -64 x (1+x)^2 H_0 P_{1031}
                                -2048 x^3 (1+x)^2 \Big(
                                        5-7 x+5 x^2\Big) H_{-1} H_0
                        \Big) H_{0,0,1}
                        +\Big(
                                -128 x^2 (1+x)^2 H_{-1} H_0 P_{817}
                                -1024 x^2 (1+x)^2 H_0 H_1 P_{831}
                                +512 x^2 (1+x)^2 H_{0,-1} P_{839}
                                -96 x^2 (1+x)^2 H_0^2 P_{853}
                                -128 x^2 (1+x)^2 H_{0,1} P_{944}
                                +16 x (1+x)^2 H_0 P_{1049}
                        \Big) H_{0,0,-1}
                        -128 x^3 (1+x)^2 \Big(
                                26-25 x+26 x^2\Big) H_0^2 H_{0,1,-1}
                        -128 x^3 (1+x)^2 \Big(
                                26-25 x+26 x^2\Big) H_0^2 H_{0,-1,1}
                        +\Big(
                                128 x^2 (1+x)^2 H_{-1} P_{843}
                                -1024 x^2 (1+x)^2 H_1 P_{847}
                                +64 x^2 (1+x)^2 H_0 P_{937}
                                -32 x (1+x)^2 P_{1038}
                        \Big) H_{0,0,0,1}
                        +\Big(
                                64 x^2 (1+x)^2 H_0 P_{815}
                                +128 x^2 (1+x)^2 H_1 P_{965}
                                +16 x (1+x)^2 P_{1055}
                                -768 x^3 (1+x)^2 \Big(
                                        28-41 x+28 x^2\Big) H_{-1}
                        \Big) H_{0,0,0,-1}
                        +\Big(
                                64 x^2 (1+x)^2 H_{0,0,1} P_{841}
                                -512 x^2 (1+x)^2 H_{0,0,-1} P_{851}
                                -
                                \frac{32}{3} x^2 (1+x)^2 H_0^3 P_{852}
                                +128 x^2 (1+x)^2 H_{0,1,-1} P_{890}
                                +128 x^2 (1+x)^2 H_{0,-1,1} P_{891}
                                -8 x (1+x)^2 H_0^2 P_{1047}
                                +4 x (1+x)^2 H_0 P_{1059}
                                +\Big(
                                        -32 x^2 (1+x)^2 H_0^2 P_{889}
                                        -32 x (1+x)^2 H_0 P_{1035}
                                \Big) H_1
                                +\Big(
                                        -16 x (1+x)^2 H_0 P_{1048}
                                        -192 x^3 (1+x)^2 \Big(
                                                2-9 x+2 x^2\Big) H_0^2
                                \Big) H_{-1}
                                +\Big(
                                        128 x^2 (1+x)^2 H_0 P_{822}
                                        +32 x (1+x)^2 P_{1032}
                                \Big) H_{0,1}
                                +\Big(
                                        128 x^2 (1+x)^2 H_0 P_{877}
                                        -128 x^2 (1+x)^2 H_1 P_{891}
                                        +16 x (1+x)^2 P_{1026}
                                \Big) H_{0,-1}
                                +1536 x^3 (1+x)^2 \Big(
                                        5-x+5 x^2\Big) H_{0,-1,-1}
                                +64 x^2 (1+x)^2 P_{931} \zeta_3
                        \Big) \zeta_2
                        +\Big(
                                \frac{64}{5} x^2 (1+x)^2 H_{-1} P_{816}
                                +\frac{16}{5} x^2 (1+x)^2 H_0 P_{975}
                                -\frac{32}{5} x^2 (1+x)^2 H_1 P_{977}
                                -\frac{8}{5} x (1+x)^2 P_{1056}
                        \Big) \zeta_2^2
                        +\Big(
                                -128 x^2 (1+x)^2 H_{-1} H_0 P_{821}
                                -256 x^2 (1+x)^2 H_0 H_1 P_{831}
                                -64 x^2 (1+x)^2 H_{0,1} P_{844}
                                +128 x^2 (1+x)^2 H_{0,-1} P_{876}
                                -16 x^2 (1+x)^2 H_0^2 P_{909}
                                -16 x (1+x)^2 H_0 P_{1017}
                        \Big) \zeta_3
                        +96 x^2 (1+x)^2 P_{918} \zeta_5
                \Big)
                -384 (-1+x) x^2 (1+x)^3 \xi ^2 \eta ^4 \zeta_2 H_{-1}
                 H_{0,-1}
\Bigg\}
\Bigg]
\\
+ C_F C_A  \Bigg[
%
%
  \Bigg\{
                -
                \frac{968}{9} x (1+x)^2 \eta ^2
                +\eta  \Big(
                        16 x H_0 H_1
                        -16 x H_{0,1}
                \Big)
                +\eta ^3 \Big(
                        -\frac{2}{9} x (1+x)^2 \Big(
                                687+26 x+687 x^2\Big) H_0
                        +\frac{8}{3} x (1+x)^2 \Big(
                                3-88 x+3 x^2\Big) H_{-1} H_0
                        +\frac{32}{3} x^2 \Big(
                                3-8 x+3 x^2\Big) H_0^3 H_1
                        -\frac{8}{3} x (1+x)^2 \Big(
                                3-88 x+3 x^2\Big) H_{0,-1}
                        +\Big(
                                64 x^2 \Big(
                                        3-8 x+3 x^2\Big) H_0 H_1
                                -64 x^2 \Big(
                                        3-8 x+3 x^2\Big) H_{0,1}
                        \Big) \zeta_2
                \Big)
                +\xi ^2 \Big(
                        96 (-1+x) x^2 (1+x)^3 \eta ^4 \zeta_2 H_{0,-1}
                        +72 (-1+x) x (1+x)^3 \eta ^3 H_{-1} \zeta_2
                \Big)
                +\xi  \Big(
                        288 x^2 (1+x)^2 \eta ^3 H_{-1} \zeta_2
                        +480 x^3 (1+x)^2 \eta ^4 H_{0,-1} \zeta_2
                \Big)
                +\eta ^4 \Big(
                        -8 x (1+x)^2 H_{0,0,-1} P_{923}
                        +8 x (1+x)^2 H_0 H_{0,-1} P_{924}
                        -4 x (1+x)^2 H_{-1} H_0^2 P_{926}
                        -\frac{2}{3} x (1+x)^2 H_0^2 P_{978}
                        -48 x^2 (1+x)^2 \Big(
                                3-2 x+3 x^2\Big) H_0^2 H_1
                        +32 x^2 (1+x)^2 \Big(
                                7-2 x+7 x^2\Big) H_0 H_{0,1}
                        -32 x^2 (1+x)^2 \Big(
                                5+2 x+5 x^2\Big) H_{0,0,1}
                        -\frac{4}{3} x (1+x)^2 P_{1000} \zeta_2
                        +16 x (1+x)^2 P_{858} \zeta_3
                \Big)
                +\eta ^5 \Big(
                        -\frac{2}
                        {3} x^2 (1+x)^2 H_0^4 P_{835}
                        +128 x^2 (1+x)^2 H_0 H_{0,0,1} P_{881}
                        +16 x^2 (1+x)^2 H_0^2 H_{0,-1} P_{915}
                        -\frac{4}{3} x^3 (1+x)^2 H_0^3 P_{925}
                        -16 x^2 (1+x)^2 H_0^2 H_{0,1} P_{936}
                        +96 x^2 (1+x)^2 H_{0,0,0,-1} P_{941}
                        -32 x^2 (1+x)^2 H_0 H_{0,0,-1} P_{956}
                        -32 x^2 (1+x)^2 H_{0,0,0,1} P_{961}
                        +\frac{8}{5} x^2 (1+x)^2 P_{994} \zeta_2^2
                        +\Big(
                                8 x^2 (1+x)^2 H_0^2 P_{878}
                                -8 x^2 (1+x)^2 H_0 P_{1016}
                        \Big) \zeta_2
                        -32 x^2 (1+x)^2 H_0 P_{914} \zeta_3
                \Big)
                +96 x \log (2) x_+^2 \zeta_2
\Bigg\}
\end{dmath*}
\begin{dmath*}
{\color{white}=}
+ \ep  \Bigg\{
                \eta ^2 \Big( -16 c_1 x (x-1)^2 
                             - \frac{14872}{27} x (x+1)^2 \Big)
                +\xi  \Big(
                        \eta ^4 \Big(
                                \log (2) \zeta_2 \Big(
                                        192 (-1+x) x^2 (1+x)^3 H_1
                                        +192 (-1+x) x^2 (1+x)^3 H_{-1}
                                \Big)
                                -32 (-1+x) x^2 (1+x)^3 H_{-1} H_0^2 H_1
                                -64 (-1+x) x (1+x)^3 \Big(
                                        3-5 x+3 x^2\Big) H_0 H_1 H_{0,1}
                                +\Big(
                                        -192 (-1+x) x^2 (1+x)^3 H_{-1} H_1
                                        +960 x^3 (1+x)^2 H_{-1} H_{0,-1}
                                        +192 (-1+x) x^2 (1+x)^3 H_{-1,1}
                                \Big) \zeta_2
                        \Big)
                        +\eta ^3 \Big(
                                -8 (-1+x) x (1+x)^3 H_{-1}^2 H_0
                                +16 (-1+x) x (1+x)^3 H_{-1} H_{0,-1}
                                -16 (-1+x) x (1+x)^3 H_{0,-1,-1}
                                +624 x^2 (1+x)^2 H_1 \zeta_3
                        \Big)
                        -192 (-1+x) x^2 (1+x)^3 \Big(
                                5-x+5 x^2\Big) \eta ^5 \zeta_2 H_{0,-1,-1}
                \Big)
                +\eta ^3 \Big(
                        -\frac{1}{27} x (1+x)^2 \Big(
                                20559+1586 x+20559 x^2\Big) H_0
                        +\frac{4}{9} x (1+x)^2 \Big(
                                753+634 x+753 x^2\Big) H_{-1} H_0
                        +640 x^2 (1+x)^2 H_{-1}^2 H_0
                        +\Big(
                                32 x (1+x)^2 \Big(
                                        3-19 x+3 x^2\Big) H_0
                                +\frac{64}{3} x^2 \Big(
                                        3-8 x+3 x^2\Big) H_{-1} H_0^3
                        \Big) H_1
                        -\frac{32}{3} x^2 \Big(
                                3-8 x+3 x^2\Big) H_0^3 H_1^2
                        -32 x (1+x)^2 \Big(
                                3-19 x+3 x^2\Big) H_{0,1}
                        +\Big(
                                -
                                \frac{4}{9} x (1+x)^2 \Big(
                                        753+634 x+753 x^2\Big)
                                -1280 x^2 (1+x)^2 H_{-1}
                        \Big) H_{0,-1}
                        +1280 x^2 (1+x)^2 H_{0,-1,-1}
                        +\Big(
                                128 x^2 \Big(
                                        3-8 x+3 x^2\Big) H_{-1} H_0 H_1
                                -64 x^2 \Big(
                                        3-8 x+3 x^2\Big) H_0 H_1^2
                                +\Big(
                                        128 x^2 \Big(
                                                3-8 x+3 x^2\Big) H_1
                                        -128 x^2 \Big(
                                                3-8 x+3 x^2\Big) H_{-1}
                                \Big) H_{0,1}
                                -128 x^2 \Big(
                                        3-8 x+3 x^2\Big) H_{0,1,1}
                        \Big) \zeta_2
                \Big)
                +\eta  \Big(
                        -96 x H_{-1} H_0 H_1
                        +16 x H_0 H_1^2
                        +\Big(
                                -32 x H_1
                                +96 x H_{-1}
                        \Big) H_{0,1}
                        +96 x H_1 H_{0,-1}
                        +32 x H_{0,1,1}
                        -96 x H_{0,1,-1}
                        -96 x H_{0,-1,1}
                        -16 x H_1 \zeta_2
                \Big)
                +\eta ^4 \Big(
                        192 x (1+x)^2 H_{0,0,1,1} P_{830}
                        +64 x (1+x)^2 H_0 H_{0,1,-1} P_{856}
                        +64 x (1+x)^2 H_0 H_{0,-1,1} P_{856}
                        -64 x (1+x)^2 H_{0,0,1,-1} P_{857}
                        -64 x (1+x)^2 H_{0,0,-1,1} P_{857}
                        -64 x (1+x)^2 H_0 H_{0,1,1} P_{859}
                        -16 x (1+x)^2 H_0^2 H_1^2 P_{861}
                        -48 x (1+x)^2 H_0 H_{0,-1,-1} P_{908}
                        +12 x (1+x)^2 H_{-1}^2 H_0^2 P_{910}
                        -8 x (1+x)^2 H_{-1} H_0^2 P_{946}
                        +8 x (1+x)^2 H_{0,-1}^2 P_{959}
                        -\frac{2}{9} x (1+x)^2 H_0^2 P_{1011}
                        +\log (2) \zeta_2 \Big(
                                192 x (1+x)^2 P_{865}
                                -768 x^3 (1+x)^2 H_1
                                -768 x^3 (1+x)^2 H_{-1}
                        \Big)
                        +\Big(
                                -16 x (1+x)^2 H_0^2 P_{892}
                                +128 x^3 (1+x)^2 H_{-1} H_0^2
                        \Big) H_1
                        +\Big(
                                64 x (1+x)^2 H_0 P_{824}
                                -96 x (1+x)^2 H_{-1} H_0 P_{828}
                                -32 x (1+x)^2 H_{0,-1} P_{855}
                                -512 x^2 (1+x)^2 \Big(
                                        3-5 x+3 x^2\Big) H_0 H_1
                        \Big) H_{0,1}
                        -32 x^2 (1+x)^2 \Big(
                                11-10 x+11 x^2\Big) H_{0,1}^2
                        +\Big(
                                -32 x (1+x)^2 H_0 H_1 P_{854}
                                +8 x (1+x)^2 H_0 P_{991}
                                -64 x^2 (1+x)^2 \Big(
                                        5-4 x+5 x^2\Big) H_{-1} H_0
                        \Big) H_{0,-1}
                        +\Big(
                                -192 x (1+x)^2 H_1 P_{830}
                                +32 x (1+x)^2 P_{842}
                                +64 x (1+x)^2 H_{-1} P_{857}
                        \Big) H_{0,0,1}
                        +\Big(
                                192 x (1+x)^2 H_1 P_{827}
                                -16 x (1+x)^2 P_{980}
                                -16 x (1+x)^2 \Big(
                                        3-4 x+3 x^2
                                \Big)
\Big(7-68 x+7 x^2\Big) H_{-1}
                        \Big) H_{0,0,-1}
                        +16 x (1+x)^2 \Big(
                                3-4 x+3 x^2
                        \Big)
\Big(7-68 x+7 x^2\Big) H_{0,0,-1,-1}
                        -128 x^2 (1+x)^2 \Big(
                                5-x+5 x^2\Big) H_{0,-1,0,1}
                        +\Big(
                                -16 x (1+x)^2 H_{-1} P_{899}
                                +24 x (1+x)^2 H_{-1}^2 P_{929}
                                -
                                \frac{4}{9} x (1+x)^2 P_{1014}
                                +768 x^3 (1+x)^2 H_{-1} H_1
                                -768 x^3 (1+x)^2 H_{-1,1}
                        \Big) \zeta_2
                        +\Big(
                                -16 x (1+x)^2 H_{-1} P_{938}
                                +8 x (1+x)^2 P_{954}
                        \Big) \zeta_3
                \Big)
                +\xi ^2 \Big(
                        192 (-1+x) x^2 (1+x)^3 \eta ^4 \zeta_2 H_{-1} H_{0,-1}
                        -48 (-1+x) x (1+x)^3 \eta ^3 H_1 \zeta_3
                \Big)
\end{dmath*}
\begin{dmath*}
{\color{white}=}
                +\eta ^5 \Big(
                        512 x^2 (1+x)^2 H_0 H_{0,0,1,1} P_{833}
                        -\frac{2}{5} x^2 (1+x)^2 H_0^5 P_{834}
                        +32 x^2 (1+x)^2 H_0^2 H_{0,1,1} P_{863}
                        +\frac{1}{3} x^3 (1+x)^2 H_0^4 P_{872}
                        +96 x^2 (1+x)^2 H_0 H_{0,-1}^2 P_{880}
                        -320 x^2 (1+x)^2 H_0 H_{0,-1,0,1} P_{894}
                        -64 x^2 (1+x)^2 H_0 H_{0,1}^2 P_{919}
                        +32 x^2 (1+x)^2 H_0^2 H_{0,1,-1} P_{940}
                        +32 x^2 (1+x)^2 H_0^2 H_{0,-1,1} P_{940}
                        -448 x^2 (1+x)^2 H_{0,0,0,1,1} P_{945}
                        -128 x^2 (1+x)^2 H_{0,0,-1,0,-1} P_{951}
                        -32 x^2 (1+x)^2 H_0^2 H_{0,-1,-1} P_{967}
                        -128 x^2 (1+x)^2 H_0 H_{0,0,1,-1} P_{972}
                        -128 x^2 (1+x)^2 H_0 H_{0,0,-1,1} P_{972}
                        -64 x^2 (1+x)^2 H_{0,0,1,0,1} P_{981}
                        -192 x^2 (1+x)^2 H_{0,0,0,-1,-1} P_{982}
                        +64 x^2 (1+x)^2 H_{0,0,1,0,-1} P_{988}
                        +64 x^2 (1+x)^2 H_0 H_{0,0,-1,-1}
                         P_{989}
                        +64 x^2 (1+x)^2 H_{0,0,-1,0,1} P_{996}
                        +128 x^2 (1+x)^2 H_{0,0,0,1,-1} P_{998}
                        +128 x^2 (1+x)^2 H_{0,0,0,-1,1} P_{998}
                        +32 x^2 (1+x)^2 H_{0,0,0,0,1} P_{1004}
                        -32 x^2 (1+x)^2 H_{0,0,0,0,-1} P_{1007}
                        -
                        \frac{2}{3} x (1+x)^2 H_0^3 P_{1051}
                        +\log (2) \zeta_2 \Big(
                                -768 x^3 (1+x)^2 \Big(
                                        1+x+x^2\Big) H_{0,1}
                                -768 x^3 (1+x)^2 \Big(
                                        1+x+x^2\Big) H_{0,-1}
                        \Big)
                        +\Big(
                                \frac{4}{3} x^2 (1+x)^2 H_0^4 P_{912}
                                +\frac{16}{3} x (1+x)^2 H_0^3 P_{1030}
                        \Big) H_1
                        +\Big(
                                -\frac{4}{3} x^2 (1+x)^2 H_0^4 P_{835}
                                +\frac{4}{3} x (1+x)^2 H_0^3 P_{1052}
                        \Big) H_{-1}
                        +\Big(
                                64 x^2 (1+x)^2 H_0^2 H_1 P_{832}
                                -32 x^2 (1+x)^2 H_{-1} H_0^2 P_{936}
                                +64 x^2 (1+x)^2 H_0 H_{0,-1} P_{971}
                                -\frac{16}{3} x^2 (1+x)^2 H_0^3 P_{976}
                                +16 x (1+x)^2 H_0^2 P_{1022}
                        \Big) H_{0,1}
                        +\Big(
                                -32 x^2 (1+x)^2 H_0^2 H_1 P_{913}
                                +32 x^2 (1+x)^2 H_{-1} H_0^2 P_{915}
                                +\frac{16}{3} x^2 (1+x)^2 H_0^3 P_{979}
                                -4 x (1+x)^2 H_0^2 P_{1039}
                        \Big) H_{0,-1}
                        +\Big(
                                256 x^2 (1+x)^2 H_0 H_1 P_{849}
                                +256 x^2 (1+x)^2 H_{-1} H_0 P_{881}
                                -64 x^2 (1+x)^2 H_{0,-1} P_{935}
                                +64 x^2 (1+x)^2 H_{0,1} P_{966}
                                +16 x^2 (1+x)^2 H_0^2 P_{995}
                                -32 x (1+x)^2 H_0 P_{1033}
                        \Big) H_{0,0,1}
                        +\Big(
                                -256 x^2 (1+x)^2 H_{0,-1} P_{839}
                                -256 x^2 (1+x)^2 H_0 H_1 P_{849}
                                -64 x^2 (1+x)^2 H_{-1} H_0 P_{956}
                                -64 x^2 (1+x)^2 H_{0,1} P_{990}
                                -16 x^2 (1+x)^2 H_0^2 P_{999}
                                +8 x (1+x)^2 H_0 P_{1053}
                        \Big) H_{0,0,-1}
                        +\Big(
                                -128 x^2 (1+x)^2 H_1 P_{939}
                                -64 x^2 (1+x)^2 H_{-1} P_{961}
                                -32 x^2 (1+x)^2 H_0 P_{1001}
                                +32 x (1+x)^2 P_{1042}
                        \Big) H_{0,0,0,1}
                        +\Big(
                                192 x^2 (1+x)^2 H_{-1} P_{941}
                                +32 x^2 (1+x)^2 H_0 P_{1003}
                                -8 x (1+x)^2 P_{1057}
                                +64 x^2 (1+x)^2 \Big(
                                        3-11 x+3 x^2
                                \Big)
\Big(13-8 x+13 x^2\Big) H_1
                        \Big) H_{0,0,0,-1}
                        +\Big(
                                -32 x^2 (1+x)^2 H_{0,0,1} P_{841}
                                +256 x^2 (1+x)^2 H_{0,0,-1} P_{851}
                                +64 x^2 (1+x)^2 H_{0,-1,1} P_{860}
                                +64 x^2 (1+x)^2 H_{0,1,-1} P_{866}
                                +
                                \frac{8}{3} x^2 (1+x)^2 H_0^3 P_{950}
                                +4 x^2 (1+x)^2 H_0^2 P_{1023}
                                -2 x (1+x)^2 H_0 P_{1058}
                                +\Big(
                                        -16 x^2 (1+x)^2 H_0^2 P_{955}
                                        +32 x (1+x)^2 H_0 P_{1028}
                                \Big) H_1
                                +\Big(
                                        16 x^2 (1+x)^2 H_0^2 P_{878}
                                        +8 x (1+x)^2 H_0 P_{1036}
                                \Big) H_{-1}
                                +\Big(
                                        -32 x (1+x)^2 P_{1029}
                                        +32 x^2 (1+x)^2 \Big(
                                                10-37 x-37 x^3+10 x^4\Big) H_0
                                \Big) H_{0,1}
                                +\Big(
                                        -64 x^2 (1+x)^2 H_1 P_{860}
                                        -32 x^2 (1+x)^2 H_0 P_{882}
                                        +8 x (1+x)^2 P_{1046}
                                \Big) H_{0,-1}
                                -768 x^3 (1+x)^2 \Big(
                                        5-x+5 x^2\Big) H_{0,-1,-1}
                                -32 x^2 (1+x)^2 P_{953} \zeta_3
                        \Big) \zeta_2
                        +\Big(
                                \frac{16}{5} x^2 (1+x)^2 H_{-1} P_{994}
                                -\frac{16}{5} x^2 (1+x)^2 H_1 P_{1005}
                                -\frac{8}{5} x^2 (1+x)^2 H_0 P_{1010}
                                +\frac{4}{5} x (1+x)^2 P_{1041}
                        \Big) \zeta_2^2
                        +\Big(
                                -64 x^2 (1+x)^2 H_0 H_1 P_{849}
                                -8 x^2 (1+x)^2 H_0^2 P_{893}
                                -64 x^2 (1+x)^2 H_{-1} H_0 P_{914}
                                +96 x^2 (1+x)^2 H_{0,1} P_{917}
                                +64 x^2 (1+x)^2 H_{0,-1} P_{920}
                                -16 x (1+x)^2 H_0 P_{1027}
                        \Big) \zeta_3
                        +144 x^2 (1+x)^2 P_{952} \zeta_5
                \Big)
\Bigg\}
\Bigg]
\end{dmath*}
\begin{dmath}
{\color{white}=}
%
+ C_F n_l T_F  \Bigg[
%
%
  \Bigg\{
                \frac{304}{9} x (1+x)^2 \eta ^2
                +\eta ^3 \Big(
                        -\frac{16}{3} x (1+x)^2 \Big(
                                3-2 x+3 x^2\Big) \zeta_2
                        +\frac{8}{9} x (1+x)^2 \Big(
                                51-26 x+51 x^2\Big) H_0
                        -\frac{32}{3} x (1+x)^2 \Big(
                                3-2 x+3 x^2\Big) H_{-1} H_0
                        +\frac{8}{3} x (1+x)^2 \Big(
                                3-2 x+3 x^2\Big) H_0^2
                        +\frac{32}{3} x (1+x)^2 \Big(
                                3-2 x+3 x^2\Big) H_{0,-1}
                \Big)
\Bigg\}
+ \ep  \Bigg\{
                \frac{4664}{27} x (1+x)^2 \eta ^2
                +\eta ^3 \Big(
                        -32 x (1+x)^2 \Big(
                                3-2 x+3 x^2\Big) \zeta_3
                        +\frac{4}{27} x (1+x)^2 \Big(
                                1527-722 x+1527 x^2\Big) H_0
                        +32 x (1+x)^2 \Big(
                                3-2 x+3 x^2\Big) H_{-1}^2 H_0
                        +\frac{8}{9} x (1+x)^2 \Big(
                                69-38 x+69 x^2\Big) H_0^2
                        +\frac{8}{3} x (1+x)^2 \Big(
                                3-2 x+3 x^2\Big) H_0^3
                        +\Big(
                                -\frac{32}{9} x (1+x)^2 \Big(
                                        69-38 x+69 x^2\Big) H_0
                                -16 x (1+x)^2 \Big(
                                        3-2 x+3 x^2\Big) H_0^2
                        \Big) H_{-1}
                        +\Big(
                                \frac{32}{9} x (1+x)^2 \Big(
                                        69-38 x+69 x^2\Big)
                                -64 x (1+x)^2 \Big(
                                        3-2 x+3 x^2\Big) H_{-1}
                        \Big) H_{0,-1}
                        +32 x (1+x)^2 \Big(
                                3-2 x+3 x^2\Big) H_{0,0,-1}
                        +64 x (1+x)^2 \Big(
                                3-2 x+3 x^2\Big) H_{0,-1,-1}
                        +\Big(
                                -
                                \frac{32}{9} x (1+x)^2 \Big(
                                        21-19 x+48 x^2\Big)
                                +8 x (1+x)^2 \Big(
                                        3-2 x+3 x^2\Big) H_0
                                +32 x (1+x)^2 \Big(
                                        3-2 x+3 x^2\Big) H_{-1}
                        \Big) \zeta_2
                \Big)
\Bigg\}
\Bigg]
\\
+ C_F T_F  \Bigg[
%
%
%
 \Bigg\{
                \frac{128}{9} (-1+x) x (1+x) \xi  \eta ^2
                +\eta ^2 \Big(
                        -\frac{1024 x^2}{9}
                        +\Big(
                                \frac{8}{3} x H_0^2 P_{864}
                                +16 x P_{895} \zeta_2
                        \Big) x_+^2
                \Big)
                +\eta ^3 \Big(
                        \frac{8}{9} x H_0 P_{987}
                        -\frac{32}{3} x^3 H_0^3
                        -64 x^3 H_0 \zeta_2
                \Big)
\Bigg\}
+ \ep  \Bigg\{
                -128 x \Big(
                        3+2 x+3 x^2\Big) \log (2) x_+^4 \zeta_2
                +\eta ^2 \Big(
                        -\frac{16}{27} x \Big(
                                19-10 x+19 x^2\Big)
                        +\Big(
                                -\frac{16}{3} x H_{-1} H_0^2 P_{864}
                                +\frac{32}{3} x H_0^2 H_1 P_{864}
                                -\frac{64}{3} x H_0 H_{0,1} P_{864}
                                +\frac{64}{3} x H_{0,0,1} P_{864}
                                +\frac{32}{3} x H_{0,0,-1} P_{864}
                                -32 x H_{-1} P_{864} \zeta_2
                                +\frac{64}{3} x P_{932} \zeta_3
                        \Big) x_+^2
                \Big)
                +\eta ^3 \Big(
                        \frac{4}{27} x H_0 P_{1013}
                        -8 x^3 H_0^4
                        +\Big(
                                -\frac{16}{9} x H_0 P_{987}
                                -\frac{64}{3} x^3 H_0^3
                        \Big) H_{-1}
                        -128 x^3 H_0^2 H_{0,1}
                        +\Big(
                                \frac{16 x P_{987}}{9}
                                +128 x^3 H_0^2
                        \Big) H_{0,-1}
                        +512 x^3 H_0 H_{0,0,1}
                        -256 x^3 H_0 H_{0,0,-1}
                        -768 x^3 H_{0,0,0,1}
                        +128 x^3 H_{0,0,0,-1}
                        +\Big(
                                -\frac{8}{9} x H_0^3 P_{1020}
                                -\frac{8}{9} x H_0^2 P_{1021}
                                +\Big(
                                        -\frac{8}{3} x H_0 P_{1019}
                                        -\frac{16 x P_{1025}}{9}
                                \Big) \zeta_2
                        \Big) x_+
                        +\Big(
                                -128 x^3 H_{-1}
                                 H_0
                                -32 x^3 H_0^2
                                +512 x^3 H_{0,-1}
                        \Big) \zeta_2
                        +
                        \frac{32}{5} x^3 \zeta_2^2
                        +128 x^3 H_0 \zeta_3
                \Big)
\Bigg\}
\Bigg] \,.
\end{dmath}
%
%
\begin{dmath*}
 \hat F_{A,2}^{(2),\sing} =
%
C_F T_F  \Bigg[
%
%
  \Bigg\{
        \xi  \eta ^3 \Big(
                -256 (-1+x) x^2 (1+x) H_0 H_{0,0,1}
                -32 (-1+x) x^2 (1+x) H_0^2 \zeta_2
        \Big)
~~~~~~~~~~
\end{dmath*}
\vspace{-1.0cm}
\begin{dmath*}
{\color{white}=}
        +\eta ^4 \Big(
                32 x (1+x)^2 \zeta_2 \Big(
                        -1-6 x-6 x^2-14 x^3+3 x^4\Big)
                +8 x (1+x)^2 \Big(
                        -1+2 x-4 x^2-26 x^3+13 x^4\Big) H_0^2
        \Big)
        +\eta ^5 \Big(
                -128 x^2 (1+x)^2 \Big(
                        1+x+8 x^2+x^3+x^4\Big) \zeta_2 H_0
                -\frac{64}{3} x^3 (1+x)^2 \Big(
                        3+8 x-x^2+2 x^3\Big) H_0^3
        \Big)
        +\eta  \Big(
                128 x H_{-1} H_0
                -64 x H_0 H_1
                +32 x H_{0,1}
                -128 x H_{0,-1}
                +\Big(
                        -\frac{64}{3} x^2 H_0^3 H_1
                        +\Big(
                                -128 x^2 H_0 H_1
                                +128 x^2 H_{0,1}
                        \Big) \zeta_2
                \Big) x_+^2
        \Big)
        +\eta ^3 \Big(
                -16 x (1+x)^2 \Big(
                        3-2 x+3 x^2\Big) H_0
                +64 x^2 \Big(
                        3-2 x+3 x^2\Big) \zeta_3 H_0
                +\frac{4}{3} x^2 (1+x)^2 H_0^4
                +256 x^3 H_0^2 H_{0,1}
                -1536 x^3 H_0 H_{0,0,1}
                +1024 x^3 H_0 H_{0,0,-1}
\end{dmath*}
\begin{dmath*}
{\color{white}=}
                -64 (-7+x) x^2 (-1+7 x) H_{0,0,0,1}
                -3072 x^3 H_{0,0,0,-1}
                +192 x^3 H_0^2 \zeta_2
                +\frac{256}{5} x^2 \Big(
                        1+x+x^2\Big) \zeta_2^2
        \Big)
        +\eta ^2 \Big(
                -16 x \Big(
                        5+4 x+5 x^2\Big)
                -256 x^2 H_0^2 H_1
                +768 x^2 H_0 H_{0,1}
                -512 x^2 H_0 H_{0,-1}
                -1024 x^2 H_{0,0,1}
                +1024 x^2 H_{0,0,-1}
                +256 x^2 \zeta_3
        \Big)
\Bigg\}
+ \ep  \Bigg\{
        \xi  \eta ^3 \Big(
                -256 (-1+x) x^2 (1+x) H_0 H_{0,1}^2
                +\Big(
                        -512 (-1+x) x^2 (1+x) H_{-1} H_0
                        -128 (-1+x) x^2 (1+x) H_0^2
                \Big) H_{0,0,1}
                -128 (-1+x) x^2 (1+x) H_0^2 H_{0,0,-1}
                +128 (-1+x) x^2 (1+x) H_0^2 H_{0,1,1}
                +128 (-1+x) x^2 (1+x) H_0^2 H_{0,1,-1}
                +128 (-1+x) x^2 (1+x) H_0^2 H_{0,-1,1}
                +1024 (-1+x) x^2 (1+x) H_0 H_{0,0,1,-1}
                +1024 (-1+x) x^2 (1+x) H_0 H_{0,0,-1,1}
                +256 (-1+x) x^2 (1+x) H_0 H_{0,-1,0,1}
                -128 (-1+x) x^2 (1+x) H_{0,0,1,0,-1}
                +\Big(
                        -64 (-1+x) x^2 (1+x) H_{-1} H_0^2
                        -16 (-1+x) x^2 (1+x) H_0^3
                        +256 (-1+x) x^2 (1+x) H_0 H_{0,-1}
                        -1024 (-1+x) x^2 (1+x) H_{0,0,-1}
                        -64 (-1+x) x^2 (1+x) \zeta_3
                \Big) \zeta_2
        \Big)
        +\eta ^5 \Big(
                \frac{128}{3} x^2 (1+x)^2 H_{-1} H_0^3 P_{848}
                +256 x^2 (1+x)^2 H_0^2 H_{0,-1} P_{896}
                -128 x^2 (1+x)^2 H_0^2 H_{0,1} P_{898}
                +\frac{128}{3} x^2 (1+x)^2 H_0^3 H_1 P_{933}
                -\frac{8}{3} x^2 (1+x)^2 H_0^4 P_{942}
                -256 x^2 (1+x)^2 H_{0,0,0,-1} P_{958}
                +128 x^2 (1+x)^2 H_{0,0,0,1} P_{993}
                -\frac{16}{3} x (1+x)^2 H_0^3 P_{1040}
                -\frac{64}{5} x^2 (1+x)^2 P_{974} \zeta_2^2
                +\Big(
                        -256 x^2 (1+x)^2 H_{0,-1} P_{829}
                        +256 x^2 (1+x)^2 H_{-1} H_0 P_{848}
                        -256 x^2 (1+x)^2 H_{0,1} P_{884}
                        +256 x^2 (1+x)^2 H_0 H_1 P_{897}
                        +16 x (1+x)^2 H_0 P_{1034}
                \Big) \zeta_2
        \Big)
        +\eta  \Big(
                -512 x H_{-1}^2 H_0
                +384 x H_{-1} H_0 H_1
                -64 x H_0 H_1^2
                +\Big(
                        64 x H_1
                        -448 x H_{-1}
                \Big) H_{0,1}
                +\Big(
                        -384 x H_1
                        +1024 x H_{-1}
                \Big) H_{0,-1}
                -64 x H_{0,1,1}
                +448 x H_{0,1,-1}
                +448 x H_{0,-1,1}
                -1024 x H_{0,-1,-1}
                +\Big(
                        -\frac{128}{3} x^2 H_{-1} H_0^3 H_1
                        +\frac{64}{3} x^2 H_0^3 H_1^2
                        -256 x^2 H_0 H_{0,-1} H_{0,1}
                        +256 x^2 H_0^2 H_1 H_{0,-1}
                        -512 x^2 H_{0,0,-1,0,1}
                        +\Big(
                                -256 x^2 H_{-1} H_0 H_1
                                +128 x^2 H_0 H_1^2
                                +\Big(
                                        -256 x^2 H_1
                                        +256 x^2 H_{-1}
                                \Big) H_{0,1}
                                +1024 x^2 H_1 H_{0,-1}
                                +256 x^2 H_{0,1,1}
                                -1024 x^2 H_{0,-1,1}
                        \Big) \zeta_2
                \Big) x_+^2
                -64 x H_1 \zeta_2
        \Big)
        +\eta ^4 \Big(
                32 x (1+x)^2 H_0 H_{0,1} P_{871}
                +16 x (1+x)^2 H_0^2 H_1 P_{928}
                -16 x (1+x)^2 H_{-1} H_0^2 P_{964}
                +32 x (1+x)^2 H_{0,0,-1} P_{986}
                +8 x (1+x) H_0^2 P_{1024}
                +\Big(
                        64 x (1+x)^2 H_{-1} P_{836}
                        -32 x (1+x) P_{1015}
                \Big) \zeta_2
                -16 x (1+x)^2 P_{973} \zeta_3
        \Big)
        +\eta ^2 \Big(
                -120 x \Big(
                        3+4 x+3 x^2\Big)
                +\log (2) \zeta_2 \Big(
                        1536 x^2
                        -1536 x^2 H_1
                        -1536 x^2 H_{-1}
                \Big)
                +128 x^2 H_{-1}^2 H_0^2
                +256 x^2 H_{-1} H_0^2 H_1
                -384 x^2 H_0^2 H_1^2
                +\Big(
                        -1024 x^2 H_{-1} H_0
                        +1536 x^2 H_0 H_1
                        +2560 x^2 H_{0,-1}
                \Big) H_{0,1}
                -256 x^2 H_{0,1}^2
                +\Big(
                        \Big(
                                -32 x \Big(
                                        11+16 x+13 x^2\Big)
                                +512 x^2 H_{-1}
                        \Big) H_0
                        +512 x^2 H_0 H_1
                \Big) H_{0,-1}
                -1536 x^2 H_{0,-1}^2
                +\Big(
                        -64 x^2 (19+8 x)
                        -1536 x^2 H_1
                        +1536 x^2 H_{-1}
                \Big) H_{0,0,1}
                +\Big(
                        -1536 x^2 H_1
                        -1536 x^2 H_{-1}
                \Big) H_{0,0,-1}
                -1024 x^2 H_0 H_{0,1,1}
                -1536 x^2 H_0 H_{0,1,-1}
                -1536 x^2 H_0 H_{0,-1,1}
                +2560 x^2 H_0 H_{0,-1,-1}
                +1536 x^2 H_{0,0,1,1}
                -1536 x^2 H_{0,0,1,-1}
                -1536 x^2 H_{0,0,-1,1}
                +1536 x^2 H_{0,0,-1,-1}
                -1536 x^2 H_{0,-1,0,1}
                +\Big(
                        1536 x^2 H_{-1} H_1
                        +768 x^2 H_{-1}^2
                        -1536 x^2 H_{-1,1}
                \Big) \zeta_2
                -64 x^2 (3+5 x) H_0^2 x_+ \zeta_2
                +\Big(
                        2688 x^2 H_1
                        -384 x^2 H_{-1}
                \Big) \zeta_3
        \Big)
\end{dmath*}
\vspace{-1.0cm}
\begin{dmath}
{\color{white}=}
        +\eta ^3 \Big(
                -64 x^2 \Big(
                        11+31 x+11 x^2\Big) \zeta_5
                +\log (2) \zeta_2 \Big(
                        3072 x^3 H_{0,1}
                        +3072 x^3 H_{0,-1}
                \Big)
                -8 x (1+x) \Big(
                        33+15 x+27 x^2+29 x^3\Big) H_0
                +
                \frac{16}{15} x^2 (1+x)^2 H_0^5
                +\Big(
                        -128 (-2+x) x (1+x)^2 (-1+2 x) H_0
                        -\frac{8}{3} x^2 \Big(
                                5-14 x+5 x^2\Big) H_0^4
                \Big) H_1
                +\Big(
                        32 x (1+x)^2 \Big(
                                19-42 x+19 x^2\Big) H_0
                        +\frac{8}{3} x^2 (1+x)^2 H_0^4
                \Big) H_{-1}
                +\Big(
                        64 x P_{885}
                        +512 x^3 H_{-1} H_0^2
                        +\frac{64}{3} x^2 (3+x) (1+3 x) H_0^3
                        +512 x^3 H_0^2 H_1
                \Big) H_{0,1}
                -1536 x^3 H_0 H_{0,1}^2
                +\Big(
                        -32 x (1+x)^2 \Big(
                                19-42 x+19 x^2\Big)
                        -\frac{32}{3} x^2 \Big(
                                3-26 x+3 x^2\Big) H_0^3
                \Big) H_{0,-1}
                -1536 x^3 H_0 H_{0,-1}^2
                +\Big(
                        -256 x^2 \Big(
                                11-2 x+9 x^2\Big) H_0
                        -3072 x^3 H_{-1} H_0
                        -1536 x^3 H_0^2
                        -512 x^2 (1+x)^2 H_0 H_1
                        -256 x^2 \Big(
                                3-14 x+3 x^2\Big) H_{0,1}
                        +128 x^2 \Big(
                                3-50 x+3 x^2\Big) H_{0,-1}
                \Big) H_{0,0,1}
                +\Big(
                        -512 x^2 (2+x) (-3+2 x) H_0
                        +2048 x^3 H_{-1} H_0
                        +768 x^3 H_0^2
                        +512 x^2 (1+x)^2 H_0 H_1
                        +256 x^2 \Big(
                                3-2 x+3 x^2\Big) H_{0,1}
                        +7168 x^3 H_{0,-1}
                \Big) H_{0,0,-1}
                +768 x^3 H_0^2 H_{0,1,1}
                -768 x^3 H_0^2 H_{0,1,-1}
                -768 x^3 H_0^2 H_{0,-1,1}
                -512 x^3 H_0^2 H_{0,-1,-1}
                +\Big(
                        -128 x^2 \Big(
                                11-26 x+11 x^2\Big) H_0
                        +128 x^2 \Big(
                                13+10 x+13 x^2\Big) H_1
                        -128 (-7+x) x^2 (-1+7 x) H_{-1}
                \Big) H_{0,0,0,1}
                +\Big(
                        128 x^2 \Big(
                                5-38 x+5 x^2\Big) H_0
                        -256 x^2 \Big(
                                11+2 x+11 x^2\Big) H_1
                        -6144 x^3 H_{-1}
                \Big) H_{0,0,0,-1}
                +512 x^2 (1+x)^2 H_0 H_{0,0,1,1}
                +6144 x^3 H_0 H_{0,0,1,-1}
                +6144 x^3 H_0 H_{0,0,-1,1}
                -4096 x^3 H_0 H_{0,0,-1,-1}
                +3584 x^3 H_0 H_{0,-1,0,1}
                +64 x^2 \Big(
                        55+6 x+55 x^2\Big) H_{0,0,0,0,1}
                -64 x^2 \Big(
                        53-54 x+53 x^2\Big) H_{0,0,0,0,-1}
                +128 x^2 \Big(
                        5-94 x+5 x^2\Big) H_{0,0,0,1,1}
                +256 x^2 \Big(
                        5+14 x+5 x^2\Big) H_{0,0,0,1,-1}
                +256 x^2 \Big(
                        5+14 x+5 x^2\Big) H_{0,0,0,-1,1}
                -9216 x^3 H_{0,0,0,-1,-1}
                +256 x^2 \Big(
                        3-14 x+3 x^2\Big) H_{0,0,1,0,1}
                +3328 x^3 H_{0,0,1,0,-1}
                -5120 x^3 H_{0,0,-1,0,-1}
                +\Big(
                        384 x^3 H_{-1} H_0^2
                        +160 x^3 H_0^3
                        +128 x^2 (1+x)^2 H_0^2 H_1
                        +128 x^2 \Big(
                                3+2 x+3 x^2\Big) H_0 H_{0,1}
                        +1024 x^3 H_0 H_{0,-1}
                        -128 x^2 \Big(
                                7+2 x+7 x^2\Big) H_{0,0,1}
                        -4096 x^3 H_{0,0,-1}
                        -1024 x^2 \Big(
                                1+x+x^2\Big) H_{0,1,-1}
                        -3072 x^3 H_{0,-1,-1}
                        +768 x^3 \zeta_3
                \Big) \zeta_2
                +\Big(
                        -
                        \frac{32}{5} x^2 \Big(
                                5+66 x+5 x^2\Big) H_0
                        +\frac{64}{5} x^2 \Big(
                                5+14 x+5 x^2\Big) H_1
                        +\frac{512}{5} x^2 \Big(
                                1+x+x^2\Big) H_{-1}
                \Big) \zeta_2^2
                +\Big(
                        -128 x^2 \Big(
                                4+x+10 x^2\Big) H_0
                        +128 x^2 \Big(
                                3-2 x+3 x^2\Big) H_{-1} H_0
                        +16 x^2 \Big(
                                13+18 x+13 x^2\Big) H_0^2
                        +128 x^2 (1+x)^2 H_0 H_1
                        +128 x^2 \Big(
                                5-32 x+5 x^2\Big) H_{0,1}
                        -128 x^2 \Big(
                                3-8 x+3 x^2\Big) H_{0,-1}
                \Big) \zeta_3
        \Big)
        -32 x H_1 x_+^2
\Bigg\}
\Bigg] \,.
\end{dmath}
%
%
The polynomials are given by
\begin{align}
P_{815} &= x^4-661 x^3+786 x^2-661 x+1\,,\nonumber\\
P_{816} &= x^4-145 x^3+405 x^2-145 x+1\,,\nonumber\\
P_{817} &= x^4-93 x^3+136 x^2-93 x+1\,,\nonumber\\
P_{818} &= x^4-57 x^3+88 x^2-57 x+1\,,\nonumber\\
P_{819} &= x^4-53 x^3+78 x^2-53 x+1\,,\nonumber\\
P_{820} &= x^4-37 x^3+54 x^2-37 x+1\,,\nonumber\\
P_{821} &= x^4-33 x^3+52 x^2-33 x+1\,,\nonumber\\
P_{822} &= x^4-31 x^3+87 x^2-31 x+1\,,\nonumber\\
P_{823} &= x^4-29 x^3+38 x^2-29 x+1\,,\nonumber\\
P_{824} &= x^4-26 x^3+41 x^2-26 x+1\,,\nonumber\\
P_{825} &= x^4-25 x^3+18 x^2-25 x+1\,,\nonumber\\
P_{826} &= x^4-25 x^3+36 x^2-25 x+1\,,\nonumber\\
P_{827} &= x^4-23 x^3+26 x^2-23 x+1\,,\nonumber\\
P_{828} &= x^4-16 x^3+26 x^2-16 x+1\,,\nonumber\\
P_{829} &= x^4-14 x^3-32 x^2+6 x-9\,,\nonumber\\
P_{830} &= x^4-14 x^3+20 x^2-14 x+1\,,\nonumber\\
P_{831} &= x^4-13 x^3+30 x^2-13 x+1\,,\nonumber\\
P_{832} &= x^4-7 x^3+24 x^2-7 x+1\,,\nonumber\\
P_{833} &= x^4-5 x^3+20 x^2-5 x+1\,,\nonumber\\
P_{834} &= x^4-5 x^3+22 x^2-5 x+1\,,\nonumber\\
P_{835} &= x^4-5 x^3+38 x^2-5 x+1\,,\nonumber\\
P_{836} &= x^4-2 x^3+66 x^2+14 x-7\,,\nonumber\\
P_{837} &= x^4-x^3-22 x^2-x+1\,,\nonumber\\
P_{838} &= x^4-x^3+25 x^2-x+1\,,\nonumber\\
P_{839} &= x^4+2 x^3-9 x^2+2 x+1\,,\nonumber\\
P_{840} &= x^4+3 x^3+36 x^2+3 x+1\,,\nonumber\\
P_{841} &= x^4+7 x^3-94 x^2+7 x+1\,,\nonumber\\
P_{842} &= x^4+70 x^3-108 x^2+78 x-3\,,\nonumber\\
P_{843} &= x^4+107 x^3-144 x^2+107 x+1\,,\nonumber\\
P_{844} &= x^4+117 x^3-170 x^2+117 x+1\,,\nonumber\\
P_{845} &= 2 x^4-45 x^3+56 x^2-45 x+2\,,\nonumber\\
P_{846} &= 2 x^4-30 x^3+59 x^2-30 x+2\,,\nonumber\\
P_{847} &= 2 x^4-29 x^3+63 x^2-29 x+2\,,\nonumber\\
P_{848} &= 2 x^4-7 x^3-8 x^2+5 x-4\,,\nonumber\\
P_{849} &= 2 x^4-7 x^3-2 x^2-7 x+2\,,\nonumber\\
P_{850} &= 2 x^4-6 x^3+39 x^2-6 x+2\,,\nonumber\\
P_{851} &= 2 x^4-4 x^3+x^2-4 x+2\,,\nonumber\\
P_{852} &= 2 x^4+8 x^3+31 x^2+8 x+2\,,\nonumber\\
P_{853} &= 3 x^4-151 x^3+204 x^2-151 x+3\,,\nonumber\\
P_{854} &= 3 x^4-68 x^3+82 x^2-68 x+3\,,\nonumber\\
P_{855} &= 3 x^4-58 x^3+50 x^2-58 x+3\,,\nonumber\\
P_{856} &= 3 x^4-53 x^3+64 x^2-53 x+3\,,\nonumber\\
P_{857} &= 3 x^4-49 x^3+74 x^2-49 x+3\,,\nonumber\\
P_{858} &= 3 x^4-41 x^3+64 x^2-41 x+3\,,\nonumber\\
P_{859} &= 3 x^4-40 x^3+56 x^2-40 x+3\,,\nonumber\\
P_{860} &= 3 x^4-25 x^3+62 x^2-25 x+3\,,\nonumber\\
P_{861} &= 3 x^4-16 x^3+32 x^2-16 x+3\,,\nonumber\\
P_{862} &= 3 x^4-16 x^3+170 x^2-16 x+3\,,\nonumber\\
P_{863} &= 3 x^4-14 x^3-38 x^2-14 x+3\,,\nonumber\\
P_{864} &= 3 x^4-14 x^3-2 x^2-14 x+3\,,\nonumber\\
P_{865} &= 3 x^4-14 x^3+16 x^2-14 x+3\,,\nonumber\\
P_{866} &= 3 x^4-13 x^3+74 x^2-13 x+3\,,\nonumber\\
P_{867} &= 3 x^4-7 x^3-12 x^2-7 x+3\,,\nonumber\\
P_{868} &= 3 x^4-7 x^3+64 x^2-7 x+3\,,\nonumber\\
P_{869} &= 3 x^4-3 x^3+28 x^2-3 x+3\,,\nonumber\\
P_{870} &= 3 x^4-2 x^3-62 x^2-2 x+3\,,\nonumber\\
P_{871} &= 3 x^4+22 x^3-34 x^2+22 x+3\,,\nonumber\\
P_{872} &= 3 x^4+176 x^3-160 x^2+408 x-3\,,\nonumber\\
P_{873} &= 3 x^4+269 x^3-274 x^2+269 x+3\,,\nonumber\\
P_{874} &= 4 x^4-118 x^3+153 x^2-118 x+4\,,\nonumber\\
P_{875} &= 4 x^4-90 x^3+163 x^2-90 x+4\,,\nonumber\\
P_{876} &= 4 x^4-61 x^3+75 x^2-61 x+4\,,\nonumber\\
P_{877} &= 4 x^4-22 x^3+9 x^2-22 x+4\,,\nonumber\\
P_{878} &= 4 x^4-21 x^3+4 x^2-21 x+4\,,\nonumber\\
P_{879} &= 4 x^4-13 x^3-16 x^2-13 x+4\,,\nonumber\\
P_{880} &= 4 x^4-13 x^3+24 x^2-13 x+4\,,\nonumber\\
P_{881} &= 4 x^4-13 x^3+30 x^2-13 x+4\,,\nonumber\\
P_{882} &= 4 x^4-11 x^3-40 x^2-11 x+4\,,\nonumber\\
P_{883} &= 4 x^4-7 x^3+86 x^2-7 x+4\,,\nonumber\\
P_{884} &= 4 x^4-5 x^3+14 x^2-5 x+4\,,\nonumber\\
P_{885} &= 4 x^4-3 x^3-10 x^2-3 x+4\,,\nonumber\\
P_{886} &= 5 x^4-281 x^3+396 x^2-281 x+5\,,\nonumber\\
P_{887} &= 5 x^4-224 x^3+330 x^2-224 x+5\,,\nonumber\\
P_{888} &= 5 x^4-149 x^3+222 x^2-149 x+5\,,\nonumber\\
P_{889} &= 5 x^4-121 x^3+262 x^2-121 x+5\,,\nonumber\\
P_{890} &= 5 x^4-53 x^3+42 x^2-53 x+5\,,\nonumber\\
P_{891} &= 5 x^4-41 x^3+54 x^2-41 x+5\,,\nonumber\\
P_{892} &= 5 x^4-34 x^3+56 x^2-26 x+1\,,\nonumber\\
P_{893} &= 5 x^4-20 x^3+18 x^2-20 x+5\,,\nonumber\\
P_{894} &= 5 x^4-17 x^3+30 x^2-17 x+5\,,\nonumber\\
P_{895} &= 5 x^4-10 x^3+2 x^2-10 x+5\,,\nonumber\\
P_{896} &= 5 x^4-7 x^3+8 x^2+9 x-3\,,\nonumber\\
P_{897} &= 5 x^4-7 x^3+14 x^2-3 x+3\,,\nonumber\\
P_{898} &= 5 x^4-6 x^3+2 x^2+18 x-7\,,\nonumber\\
P_{899} &= 5 x^4+49 x^3+69 x^2-55 x+16\,,\nonumber\\
P_{900} &= 6 x^4-13 x^3+78 x^2-13 x+6\,,\nonumber\\
P_{901} &= 6 x^4-9 x^3+16 x^2-9 x+6\,,\nonumber\\
P_{902} &= 6 x^4-7 x^3+146 x^2-7 x+6\,,\nonumber\\
P_{903} &= 6 x^4-5 x^3+62 x^2-5 x+6\,,\nonumber\\
P_{904} &= 6 x^4+3 x^3+122 x^2+3 x+6\,,\nonumber\\
P_{905} &= 7 x^4-263 x^3+392 x^2-263 x+7\,,\nonumber\\
P_{906} &= 7 x^4-211 x^3+270 x^2-211 x+7\,,\nonumber\\
P_{907} &= 7 x^4-179 x^3+314 x^2-179 x+7\,,\nonumber\\
P_{908} &= 7 x^4-96 x^3+118 x^2-96 x+7\,,\nonumber\\
P_{909} &= 7 x^4-79 x^3+156 x^2-79 x+7\,,\nonumber\\
P_{910} &= 7 x^4-64 x^3+94 x^2-64 x+7\,,\nonumber\\
P_{911} &= 7 x^4-54 x^3+90 x^2-54 x+7\,,\nonumber\\
P_{912} &= 7 x^4-41 x^3+98 x^2-41 x+7\,,\nonumber\\
P_{913} &= 7 x^4-37 x^3+78 x^2-37 x+7\,,\nonumber\\
P_{914} &= 7 x^4-33 x^3+64 x^2-33 x+7\,,\nonumber\\
P_{915} &= 7 x^4-29 x^3+62 x^2-29 x+7\,,\nonumber\\
P_{916} &= 7 x^4-16 x^3-82 x^2-16 x+7\,,\nonumber\\
P_{917} &= 7 x^4-16 x^3+40 x^2-16 x+7\,,\nonumber\\
P_{918} &= 8 x^4-267 x^3+485 x^2-267 x+8\,,\nonumber\\
P_{919} &= 8 x^4-39 x^3+56 x^2-39 x+8\,,\nonumber\\
P_{920} &= 8 x^4-38 x^3+99 x^2-38 x+8\,,\nonumber\\
P_{921} &= 9 x^4-233 x^3+388 x^2-233 x+9\,,\nonumber\\
P_{922} &= 9 x^4-169 x^3+308 x^2-169 x+9\,,\nonumber\\
P_{923} &= 9 x^4-140 x^3+178 x^2-140 x+9\,,\nonumber\\
P_{924} &= 9 x^4-116 x^3+154 x^2-116 x+9\,,\nonumber\\
P_{925} &= 9 x^4-112 x^3+146 x^2-196 x+9\,,\nonumber\\
P_{926} &= 9 x^4-92 x^3+130 x^2-92 x+9\,,\nonumber\\
P_{927} &= 9 x^4-85 x^3+212 x^2-85 x+9\,,\nonumber\\
P_{928} &= 9 x^4-38 x^3+8 x^2-2 x-9\,,\nonumber\\
P_{929} &= 9 x^4-32 x^3+34 x^2-32 x+9\,,\nonumber\\
P_{930} &= 9 x^4-19 x^3+92 x^2-19 x+9\,,\nonumber\\
P_{931} &= 9 x^4-8 x^3-5 x^2-8 x+9\,,\nonumber\\
P_{932} &= 9 x^4-7 x^3+8 x^2-7 x+9\,,\nonumber\\
P_{933} &= 10 x^4-17 x^3+14 x^2+7 x-2\,,\nonumber\\
P_{934} &= 11 x^4-467 x^3+660 x^2-467 x+11\,,\nonumber\\
P_{935} &= 11 x^4-74 x^3+138 x^2-74 x+11\,,\nonumber\\
P_{936} &= 11 x^4-42 x^3+86 x^2-42 x+11\,,\nonumber\\
P_{937} &= 11 x^4+377 x^3-302 x^2+377 x+11\,,\nonumber\\
P_{938} &= 12 x^4-85 x^3+92 x^2-85 x+12\,,\nonumber\\
P_{939} &= 12 x^4-49 x^3+38 x^2-49 x+12\,,\nonumber\\
P_{940} &= 12 x^4-47 x^3+124 x^2-47 x+12\,,\nonumber\\
P_{941} &= 12 x^4-43 x^3+92 x^2-43 x+12\,,\nonumber\\
P_{942} &= 13 x^4-7 x^3+48 x^2+17 x+1\,,\nonumber\\
P_{943} &= 15 x^4-419 x^3+694 x^2-419 x+15\,,\nonumber\\
P_{944} &= 15 x^4-355 x^3+614 x^2-355 x+15\,,\nonumber\\
P_{945} &= 15 x^4-70 x^3+146 x^2-70 x+15\,,\nonumber\\
P_{946} &= 15 x^4-65 x^3+37 x^2+3 x+22\,,\nonumber\\
P_{947} &= 15 x^4-23 x^3+196 x^2-23 x+15\,,\nonumber\\
P_{948} &= 15 x^4-16 x^3+198 x^2-16 x+15\,,\nonumber\\
P_{949} &= 15 x^4+70 x^3-114 x^2+70 x+15\,,\nonumber\\
P_{950} &= 16 x^4-83 x^3+236 x^2-83 x+16\,,\nonumber\\
P_{951} &= 16 x^4-75 x^3+178 x^2-75 x+16\,,\nonumber\\
P_{952} &= 16 x^4-65 x^3+109 x^2-65 x+16\,,\nonumber\\
P_{953} &= 17 x^4-74 x^3+111 x^2-74 x+17\,,\nonumber\\
P_{954} &= 17 x^4+312 x^3-154 x^2+124 x+29\,,\nonumber\\
P_{955} &= 19 x^4-77 x^3+86 x^2-77 x+19\,,\nonumber\\
P_{956} &= 19 x^4-72 x^3+154 x^2-72 x+19\,,\nonumber\\
P_{957} &= 19 x^4+92 x^3-182 x^2+156 x-45\,,\nonumber\\
P_{958} &= 20 x^4-31 x^3+76 x^2-19 x+14\,,\nonumber\\
P_{959} &= 21 x^4-268 x^3+338 x^2-268 x+21\,,\nonumber\\
P_{960} &= 21 x^4-62 x^3+710 x^2-62 x+21\,,\nonumber\\
P_{961} &= 21 x^4-58 x^3+146 x^2-58 x+21\,,\nonumber\\
P_{962} &= 21 x^4-42 x^3+326 x^2-42 x+21\,,\nonumber\\
P_{963} &= 21 x^4+775 x^3-590 x^2+775 x+21\,,\nonumber\\
P_{964} &= 23 x^4-46 x^3-44 x^2+38 x-19\,,\nonumber\\
P_{965} &= 25 x^4-361 x^3+762 x^2-361 x+25\,,\nonumber\\
P_{966} &= 27 x^4-128 x^3+214 x^2-128 x+27\,,\nonumber\\
P_{967} &= 27 x^4-115 x^3+242 x^2-115 x+27\,,\nonumber\\
P_{968} &= 27 x^4-68 x^3+454 x^2-68 x+27\,,\nonumber\\
P_{969} &= 27 x^4-24 x^3+374 x^2-24 x+27\,,\nonumber\\
P_{970} &= 27 x^4+461 x^3-178 x^2+461 x+27\,,\nonumber\\
P_{971} &= 28 x^4-117 x^3+196 x^2-117 x+28\,,\nonumber\\
P_{972} &= 29 x^4-105 x^3+230 x^2-105 x+29\,,\nonumber\\
P_{973} &= 29 x^4-30 x^3+72 x^2+22 x+3\,,\nonumber\\
P_{974} &= 30 x^4-51 x^3+20 x^2+17 x-4\,,\nonumber\\
P_{975} &= 31 x^4+877 x^3-2470 x^2+877 x+31\,,\nonumber\\
P_{976} &= 36 x^4-149 x^3+364 x^2-149 x+36\,,\nonumber\\
P_{977} &= 37 x^4-1021 x^3+2202 x^2-1021 x+37\,,\nonumber\\
P_{978} &= 39 x^4+184 x^3-48 x^2+32 x+33\,,\nonumber\\
P_{979} &= 41 x^4-185 x^3+426 x^2-185 x+41\,,\nonumber\\
P_{980} &= 42 x^4+191 x^3-219 x^2+123 x+35\,,\nonumber\\
P_{981} &= 43 x^4-196 x^3+366 x^2-196 x+43\,,\nonumber\\
P_{982} &= 44 x^4-193 x^3+448 x^2-193 x+44\,,\nonumber\\
P_{983} &= 45 x^4-98 x^3+774 x^2-98 x+45\,,\nonumber\\
P_{984} &= 47 x^4-124 x^3+178 x^2-60 x-17\,,\nonumber\\
P_{985} &= 47 x^4+6 x^3-272 x^2-102 x+33\,,\nonumber\\
P_{986} &= 49 x^4-66 x^3-60 x^2+26 x+3\,,\nonumber\\
P_{987} &= 51 x^4-176 x^3-70 x^2-176 x+51\,,\nonumber\\
P_{988} &= 53 x^4-232 x^3+478 x^2-232 x+53\,,\nonumber\\
P_{989} &= 55 x^4-214 x^3+474 x^2-214 x+55\,,\nonumber\\
P_{990} &= 57 x^4-239 x^3+430 x^2-239 x+57\,,\nonumber\\
P_{991} &= 57 x^4+126 x^3-182 x^2+126 x+57\,,\nonumber\\
P_{992} &= 60 x^4-95 x^3+219 x^2+147 x-11\,,\nonumber\\
P_{993} &= 61 x^4-119 x^3+128 x^2-47 x+25\,,\nonumber\\
P_{994} &= 66 x^4-271 x^3+176 x^2-271 x+66\,,\nonumber\\
P_{995} &= 67 x^4-259 x^3+654 x^2-259 x+67\,,\nonumber\\
P_{996} &= 73 x^4-307 x^3+582 x^2-307 x+73\,,\nonumber\\
P_{997} &= 73 x^4-214 x^3+276 x^2+190 x+11\,,\nonumber\\
P_{998} &= 90 x^4-377 x^3+790 x^2-377 x+90\,,\nonumber\\
P_{999} &= 91 x^4-372 x^3+838 x^2-372 x+91\,,\nonumber\\
P_{1000} &= 93 x^4-700 x^3+588 x^2-476 x+63\,,\nonumber\\
P_{1001} &= 107 x^4-415 x^3+1090 x^2-415 x+107\,,\nonumber\\
P_{1002} &= 125 x^4-230 x^3+312 x^2-26 x+43\,,\nonumber\\
P_{1003} &= 151 x^4-593 x^3+1418 x^2-593 x+151\,,\nonumber\\
P_{1004} &= 163 x^4-661 x^3+1794 x^2-661 x+163\,,\nonumber\\
P_{1005} &= 171 x^4-695 x^3+814 x^2-695 x+171\,,\nonumber\\
P_{1006} &= 215 x^4-98 x^3-54 x^2-554 x+203\,,\nonumber\\
P_{1007} &= 221 x^4-875 x^3+2310 x^2-875 x+221\,,\nonumber\\
P_{1008} &= 233 x^4-510 x^3+356 x^2+182 x-117\,,\nonumber\\
P_{1009} &= 293 x^4-230 x^3-100 x^2+462 x-57\,,\nonumber\\
P_{1010} &= 303 x^4-1367 x^3+1474 x^2-1367 x+303\,,\nonumber\\
P_{1011} &= 348 x^4-1604 x^3+1989 x^2+398 x+885\,,\nonumber\\
P_{1012} &= 581 x^4-910 x^3+1242 x^2-142 x+125\,,\nonumber\\
P_{1013} &= 1527 x^4-440 x^3-2398 x^2-440 x+1527\,,\nonumber\\
P_{1014} &= 1866 x^4-7378 x^3+7947 x^2-5276 x+681\,,\nonumber\\
P_{1015} &= x^5+7 x^4+21 x^3-x^2+14 x-2\,,\nonumber\\
P_{1016} &= 9 x^5-94 x^4+110 x^3-196 x^2+45 x-18\,,\nonumber\\
P_{1017} &= 13 x^5+620 x^4-884 x^3+254 x^2-x+6\,,\nonumber\\
P_{1018} &= 14 x^5+175 x^4-56 x^3+14 x^2+6 x-9\,,\nonumber\\
P_{1019} &= 15 x^5-157 x^4+18 x^3-78 x^2-21 x-9\,,\nonumber\\
P_{1020} &= 15 x^5-109 x^4+36 x^3-60 x^2+27 x-9\,,\nonumber\\
P_{1021} &= 18 x^5+6 x^4+87 x^3+159 x^2+119 x-69\,,\nonumber\\
P_{1022} &= 25 x^5-251 x^4+518 x^3-318 x^2+105 x-3\,,\nonumber\\
P_{1023} &= 33 x^5-376 x^4+764 x^3-604 x^2+107 x+44\,,\nonumber\\
P_{1024} &= 45 x^5-81 x^4-110 x^3+26 x^2-15 x-9\,,\nonumber\\
P_{1025} &= 252 x^5-250 x^4-345 x^3+99 x^2+125 x-201\,,\nonumber\\
P_{1026} &= 3 x^6+86 x^5+575 x^4+416 x^3-131 x^2+146 x-39\,,\nonumber\\
P_{1027} &= 6 x^6-131 x^5+441 x^4-572 x^3+262 x^2-x-9\,,\nonumber\\
P_{1028} &= 6 x^6-77 x^5+251 x^4-402 x^3+212 x^2-65 x+3\,,\nonumber\\
P_{1029} &= 6 x^6-75 x^5+247 x^4-402 x^3+216 x^2-67 x+3\,,\nonumber\\
P_{1030} &= 6 x^6-38 x^5+215 x^4-402 x^3+248 x^2-104 x+3\,,\nonumber\\
P_{1031} &= 6 x^6+3 x^5-8 x^4+352 x^3-370 x^2+21 x-12\,,\nonumber\\
P_{1032} &= 6 x^6+9 x^5-410 x^4+588 x^3-320 x^2-29 x+12\,,\nonumber\\
P_{1033} &= 9 x^6-65 x^5-181 x^4+536 x^3-409 x^2+121 x-3\,,\nonumber\\
P_{1034} &= 9 x^6-14 x^5+259 x^4-156 x^3+27 x^2-14 x+1\,,\nonumber\\
P_{1035} &= 9 x^6+3 x^5-399 x^4+588 x^3-331 x^2-23 x+9\,,\nonumber\\
P_{1036} &= 9 x^6+28 x^5-89 x^4+392 x^3-221 x^2+196 x-27\,,\nonumber\\
P_{1037} &= 12 x^6-27 x^5-129 x^4+56 x^3-60 x^2+7 x-3\,,\nonumber\\
P_{1038} &= 15 x^6-86 x^5-307 x^4-678 x^3+815 x^2-44 x+21\,,\nonumber\\
P_{1039} &= 15 x^6-40 x^5-517 x^4+1832 x^3-1323 x^2+648 x-39\,,\nonumber\\
P_{1040} &= 19 x^6-35 x^5-150 x^4+130 x^3-19 x^2-3 x+2\,,\nonumber\\
P_{1041} &= 21 x^6-936 x^5+4040 x^4-7744 x^3+6563 x^2-2448 x+312\,,\nonumber\\
P_{1042} &= 21 x^6-232 x^5-5 x^4+456 x^3-521 x^2+152 x-3\,,\nonumber\\
P_{1043} &= 21 x^6-19 x^5-179 x^4+588 x^3-551 x^2-x-3\,,\nonumber\\
P_{1044} &= 24 x^6-29 x^5-178 x^4+674 x^3-652 x^2+27 x-18\,,\nonumber\\
P_{1045} &= 24 x^6+70 x^5+173 x^4+156 x^3+8 x^2+14 x-21\,,\nonumber\\
P_{1046} &= 27 x^6-336 x^5+219 x^4-920 x^3+25 x^2-80 x+9\,,\nonumber\\
P_{1047} &= 27 x^6+39 x^5-786 x^4+722 x^3-31 x^2-9 x+6\,,\nonumber\\
P_{1048} &= 33 x^6-80 x^5+279 x^4-112 x^3+99 x^2+120 x-51\,,\nonumber\\
P_{1049} &= 33 x^6-74 x^5+177 x^4+1940 x^3-2185 x^2+134 x-57\,,\nonumber\\
P_{1050} &= 33 x^6-20 x^5+627 x^4+112 x^3-1005 x^2-20 x-15\,,\nonumber\\
P_{1051} &= 35 x^6-448 x^5+653 x^4-400 x^3+235 x^2-12 x+33\,,\nonumber\\
P_{1052} &= 39 x^6-396 x^5+501 x^4+392 x^3-811 x^2+620 x-57\,,\nonumber\\
P_{1053} &= 45 x^6-284 x^5-1011 x^4+2752 x^3-2165 x^2+716 x-21\,,\nonumber\\
P_{1054} &= 69 x^6-166 x^5+205 x^4+952 x^3-1677 x^2+86 x-45\,,\nonumber\\
P_{1055} &= 75 x^6-256 x^5-543 x^4-3076 x^3+2529 x^2-124 x+51\,,\nonumber\\
P_{1056} &= 90 x^6+968 x^5-6009 x^4+8498 x^3-3484 x^2-498 x+243\,,\nonumber\\
P_{1057} &= 129 x^6-1128 x^5-981 x^4+3152 x^3-3337 x^2+824 x-3\,,\nonumber\\
P_{1058} &= 173 x^6-402 x^5+963 x^4-892 x^3+735 x^2-426 x+41\,,\nonumber\\
P_{1059} &= 245 x^6-222 x^5+531 x^4+92 x^3-41 x^2-22 x-7\,,\nonumber\\
P_{1060} &= 247 x^6-522 x^5+225 x^4+516 x^3-1091 x^2+158 x-109 \,. 
\end{align}

\newpage

\section{The scalar form factors up to two-loop} \label{app:sFF}

In this appendix, we present the scalar form factors $F_{S}^{(n)}$ up to two loops and ${\cal O}(\varepsilon)$. 

\begin{dmath*}
 F_{S}^{(1)} = C_F \Bigg[
\frac{1}{\ep}  \Bigg\{
-2
-2 \xi  H_0
\Bigg\}
+   \Bigg\{
-2
+\xi  \Big(
        4 H_{-1} H_0
        -H_0^2
        -4 H_{0,-1}
        +2 \zeta_2
\Big)
+8 x \eta  H_0
\Bigg\}
~~~~~~~~~~~~~~~
\end{dmath*}
\vspace{-1.0cm}
\begin{dmath}
{\color{white}=}
+ \ep  \Bigg\{
-4
+\eta  \Big(
        12 x H_0
        -16 x H_{-1} H_0
        +4 x H_0^2
        +16 x H_{0,-1}
        +\Big(
                -1-8 x+x^2\Big) \zeta_2
\Big)
+\xi  \Big(
        \Big(
                -2
                -4 H_{-1}^2
                +\zeta_2
        \Big) H_0
        +2 H_{-1} H_0^2
        -\frac{1}{3} H_0^3
        +8 H_{-1} H_{0,-1}
        -4 H_{0,0,-1}
        -8 H_{0,-1,-1}
        -4 H_{-1} \zeta_2
        +4 \zeta_3
\Big)
\Bigg\}
+ \ep^2  \Bigg\{
-8
+\eta  \Big(
        \Big(
                24 x
                -24 x H_{-1}
                +16 x H_{-1}^2
                -4 x \zeta_2
        \Big) H_0
        +\Big(
                -1
                +6 x
                -x^2
                -8 x H_{-1}
        \Big) H_0^2
        +\frac{4}{3} x H_0^3
        +\Big(
                24 x
                -32 x H_{-1}
        \Big) H_{0,-1}
        +16 x H_{0,0,-1}
        +32 x H_{0,-1,-1}
        +\Big(
                1-12 x+3 x^2\Big) \zeta_2
        +16 x H_{-1} \zeta_2
        -\frac{2}{3} \Big(
                -1+24 x+x^2\Big) \zeta_3
\Big)
+\xi  \Big(
        \Big(
                -4
                +\Big(
                        4
                        -2 \zeta_2
                \Big) H_{-1}
                +\frac{8}{3} H_{-1}^3
                +\frac{14}{3} \zeta_3
        \Big) H_0
        +\Big(
                -2 H_{-1}^2
                +\frac{\zeta_2}{2}
        \Big) H_0^2
        +\frac{2}{3} H_{-1} H_0^3
        -\frac{1}{12} H_0^4
        +\Big(
                -4
                -8 H_{-1}^2
                -2 \zeta_2
        \Big) H_{0,-1}
        +8 H_{-1} H_{0,0,-1}
        +16 H_{-1} H_{0,-1,-1}
        -4 H_{0,0,0,-1}
        -8 H_{0,0,-1,-1}
        -16 H_{0,-1,-1,-1}
        +4 H_{-1}^2 \zeta_2
        +\frac{14}{5} \zeta_2^2
        -8 H_{-1} \zeta_3
\Big)
\Bigg\}
+ \ep^3  \Bigg\{
-16
+\eta  \Big(
        \Big(
                48 x
                +\Big(
                        -48 x
                        +8 x \zeta_2
                \Big) H_{-1}
                +24 x H_{-1}^2
                -
                \frac{32}{3} x H_{-1}^3
                -6 x \zeta_2
                -\frac{56}{3} x \zeta_3
        \Big) H_0
        +\Big(
                12 x
                +\Big(
                        -12 x
                        +\Big(
                                -1-x^2\Big) \zeta_2
                \Big) H_{-1}
                +8 x H_{-1}^2
                -2 x \zeta_2
        \Big) H_0^2
        +\Big(
                \frac{1}{3} \Big(
                        -1+6 x-x^2\Big)
                -\frac{8}{3} x H_{-1}
        \Big) H_0^3
        +\frac{1}{3} x H_0^4
        +\Big(
                24 x \zeta_2
                +32 x \zeta_3
        \Big) H_{-1}
        +\Big(
                48 x
                -48 x H_{-1}
                +32 x H_{-1}^2
                +8 x \zeta_2
        \Big) H_{0,-1}
        +\Big(
                24 x
                -32 x H_{-1}
        \Big) H_{0,0,-1}
        +\Big(
                48 x
                -64 x H_{-1}
        \Big) H_{0,-1,-1}
        +16 x H_{0,0,0,-1}
        +32 x H_{0,0,-1,-1}
        +64 x H_{0,-1,-1,-1}
        +2 \Big(
                1-12 x+3 x^2\Big) \zeta_2
        -16 x H_{-1}^2 \zeta_2
        +\frac{1}{20} \Big(
                -9-224 x+9 x^2\Big) \zeta_2^2
        +\frac{2}{3} (-7+x) (-1+5 x) \zeta_3
\Big)
+\xi  \Big(
        \Big(
                -8
                +\Big(
                        8
                        -\frac{28 \zeta_3}{3}
                \Big) H_{-1}
                +\Big(
                        -4
                        +2 \zeta_2
                \Big) H_{-1}^2
                -\frac{4}{3} H_{-1}^4
                +\frac{47}{20} \zeta_2^2
                +\zeta_2
        \Big) H_0
        +\Big(
                -2
                +2 H_{-1}
                +\frac{4}{3} H_{-1}^3
                +\frac{7}{3} \zeta_3
        \Big) H_0^2
        +\Big(
                -\frac{2}{3} H_{-1}^2
                +\frac{1}{6} \zeta_2
        \Big) H_0^3
        +\frac{1}{6} H_{-1} H_0^4
        -\frac{1}{60} H_0^5
        +\Big(
                -4 \zeta_2
                -\frac{28 \zeta_2^2}
                {5}
        \Big) H_{-1}
        +\Big(
                -8
                +\Big(
                        8
                        +4 \zeta_2
                \Big) H_{-1}
                +\frac{16}{3} H_{-1}^3
                +\frac{4}{3} \zeta_3
        \Big) H_{0,-1}
        +\Big(
                -4
                -8 H_{-1}^2
                -2 \zeta_2
        \Big) H_{0,0,-1}
        +\Big(
                -8
                -16 H_{-1}^2
                -4 \zeta_2
        \Big) H_{0,-1,-1}
        +8 H_{-1} H_{0,0,0,-1}
        +16 H_{-1} H_{0,0,-1,-1}
        +32 H_{-1} H_{0,-1,-1,-1}
        -4 H_{0,0,0,0,-1}
        -8 H_{0,0,0,-1,-1}
        -16 H_{0,0,-1,-1,-1}
        -32 H_{0,-1,-1,-1,-1}
        -\frac{8}{3} H_{-1}^3 \zeta_2
        +8 H_{-1}^2 \zeta_3
        -\frac{8}{3} \zeta_2 \zeta_3
        +12 \zeta_5
\Big)
\Bigg\}
\Bigg] \,.
\end{dmath}

\begin{dmath*}
 F_{S}^{(2)} =
%
C_F^2  \Bigg[
\frac{1}{\ep^2}  \Bigg\{
        2
        +4 \xi  H_0
        +2 \xi ^2 H_0^2
\Bigg\}
+ \frac{1}{\ep}  \Bigg\{
        4
        +\xi  \Big(
                \eta  \Big(
                        -4 (-1+x) (1+x) H_0
                        -2 \Big(
                                -1+8 x+x^2\Big) H_0^2
                \Big)         
                -8 H_{-1} H_0
                +8 H_{0,-1}
                -4 \zeta_2
        \Big)
        +\xi ^2 \Big(
                -8 H_{-1} H_0^2
                +2 H_0^3
                +8 H_0 H_{0,-1}
                -4 H_0 \zeta_2
        \Big)
        -16 x \eta  H_0
\Bigg\}
\end{dmath*}
\begin{dmath*}
{\color{white}=}
+  \Bigg\{
                29
                +\eta ^2 \Big(
                        4 H_0 H_{0,-1} P_{1063}
                        +4 H_0^2 H_1 P_{1100}
                        -8 H_0 H_{0,1} P_{1115}
                        +8 H_{0,0,1} P_{1121}
                        +2 H_{-1} H_0^2 P_{1122}
                        -4 H_{0,0,-1} P_{1123}
                        +2 H_0^2 P_{1142}
                        +\Big(
                                \frac{16}{3} H_0^3 H_1 P_{1171}
                                -4 H_0^2 H_{0,-1} P_{1178}
                                -8 H_0 H_{0,0,-1} P_{1194}
                                +48 H_{0,0,0,-1} P_{1195}
                                -8 H_0^2 H_{0,1} P_{1196}
                                +16 H_0 H_{0,0,1} P_{1214}
                                -32 H_{0,0,0,1} P_{1215}
                                -\frac{2}{3} H_0^3 P_{1223}
                                +\frac{1}{6} H_0^4 P_{1281}
                                +\Big(
                                        16 H_0 H_1 P_{1170}
                                        -16 H_{0,1} P_{1170}
                                        -4 H_0 P_{1210}
                                        +4 H_0^2 P_{1216}
                                        -8 H_{0,-1} P_{1218}
                                \Big) \zeta_2
                                +\frac{2}{5} P_{1314} \zeta_2^2
                        \Big) x_+
                        -8 P_{1105} \zeta_3
                \Big)
                +\xi ^2 \Big(
                        -32 (-1+x)^2 (1+x)^2 \eta  x_+^2 \zeta_3 H_0
                        +16 H_{-1}^2 H_0^2
                        -8 H_{-1} H_0^3
                        +\Big(
                                32 H_0 H_1
                                -32 H_{0,-1}
                        \Big) H_{0,1}
                        +16 H_{0,1}^2
                        -32 H_{-1} H_0 H_{0,-1}
                        -32 H_0 H_1 H_{0,-1}
                        +8 H_{0,-1}^2
                        -64 H_1 H_{0,0,1}
                        +64 H_1 H_{0,0,-1}
                        -64 H_0 H_{0,1,1}
                        +32 H_0 H_{0,1,-1}
                        +32 H_0 H_{0,-1,1}
                        +16 H_0 H_{0,-1,-1}
                        +64 H_{0,0,1,1}
                        +32 H_{0,-1,0,1}
                        +16 H_{-1} H_0 \zeta_2
                        +16 H_1 \zeta_3
                \Big)
                +\eta  \Big(
                        x_+ \zeta_2 \Big(
                                -2 \Big(
                                        -3+75 x-53 x^2+5 x^3\Big)
                                -4 \Big(
                                        -5+49 x-53 x^2+x^3\Big) H_{-1}
                        \Big)
                        +4 (-3+x) (-1+3 x) H_0
                        +12 \Big(
                                5-14 x+5 x^2\Big) H_{-1} H_0
                        -12 \Big(
                                5-14 x+5 x^2\Big) H_{0,-1}
                        +160 x^2 H_0 x_+^2 \zeta_3
                \Big)
                +\xi  \Big(
                        8 H_{-1}^2 H_0
                        -16 H_{-1} H_{0,-1}
                        +16 H_{0,-1,-1}
                        +128 (-1+x) x (1+x) \eta  H_0 x_+^2 \zeta_3
                \Big)
                +\Big(
                        32 (-1+x) H_0 H_1
                        -32 (-1+x) H_{0,1}
                \Big) x_+
                +288 x \log (2) x_+^2 \zeta_2
\Bigg\}
+ \ep  \Bigg\{
                \log (2) x_+^2 \zeta_2 \Big(
                        24 \Big(
                                11+74 x+11 x^2\Big)
                        +48 \Big(
                                3+8 x+3 x^2\Big) H_1
                        +48 \Big(
                                3+8 x+3 x^2\Big) H_{-1}
                \Big)
                +\xi  \Big(
                        \eta  \Big(
                                16 (-1+x) (1+x) H_0 H_1
                                -16 (-1+x) (1+x) H_{0,1}
                                +256 (-1+x) x (1+x) H_{-1} H_0 x_+^2 \zeta_3
                        \Big)
                        -\frac{16}{3} H_{-1}^3 H_0
                        +16 H_{-1}^2 H_{0,-1}
                        -32 H_{-1} H_{0,-1,-1}
                        +32 H_{0,-1,-1,-1}
                \Big)
                +\eta  \Big(
                        \log (2) x_+^2 \zeta_2 \Big(
                                -192 x^2 H_{0,1}
                                -192 x^2 H_{0,-1}
                        \Big)
                        +12 (-3+x) (-1+3 x) H_0
                        +8 \Big(
                                3-86 x+3 x^2\Big) H_{-1} H_0
                        -4 \Big(
                                65-174 x+65 x^2\Big) H_{-1}^2 H_0
                        +256 x H_0 H_1
                        -256 x H_{0,1}
                        +\Big(
                                -8 \Big(
                                        3-86 x+3 x^2\Big)
                                +8 \Big(
                                        65-174 x+65 x^2\Big) H_{-1}
                        \Big) H_{0,-1}
                        -8 \Big(
                                65-174 x+65 x^2\Big) H_{0,-1,-1}
                        +\Big(
                                2 \Big(
                                        95-61 x-205 x^2+83 x^3\Big) H_{-1} H_0^2
                                -4 \Big(
                                        -21+93 x-101 x^2+13 x^3\Big) \zeta_3 H_{-1}
                                -8 \Big(
                                        -23+43 x-39 x^2+27 x^3\Big) H_{-1} H_{0,0,-1}
                                +8 \Big(
                                        -23+43 x-39 x^2+27 x^3\Big) H_{0,0,-1,-1}
                                +\Big(
                                        8 \Big(
                                                -5-90 x+67 x^2+4 x^3\Big)
                                        -4 \Big(
                                                29-115 x-7 x^2+5 x^3\Big) H_{-1}
                                        +4 \Big(
                                                -29+121 x-125 x^2+25 x^3\Big) H_{-1}^2
                                \Big) \zeta_2
                        \Big) x_+
                        +\Big(
                                -48 H_0 H_{0,-1}^2 P_{1081}
                                +128 H_0 H_1 H_{0,0,-1} P_{1085}
                                +32 H_0 H_{0,-1,0,1} P_{1125}
                                -32 H_{0,-1} H_{0,0,1} P_{1131}
                                +\Big(
                                        96 H_0^2 H_1 P_{1067}
                                        -32 H_0 H_{0,-1} P_{1095}
                                \Big) H_{0,1}
                                -32 x \Big(
                                        6-29 x+6 x^2\Big) H_0 H_{0,1}^2
                                -16 H_0 H_{0,1} P_{1134} \zeta_2
                                +320 x^2 H_{-1} H_0 \zeta_3
                        \Big) x_+^2
                \Big)
                +\eta ^2 \Big(
                        48 H_0 H_{0,1,-1} P_{1066}
                        +48 H_0 H_{0,-1,1} P_{1066}
                        +8 H_{0,1}^2 P_{1084}
                        +32 H_{0,-1,0,1} P_{1094}
                        +32 H_0 H_{0,1,1} P_{1096}
                        -8 H_0 H_{0,-1,-1} P_{1114}
                        +4 H_{0,-1}^2 P_{1120}
                        +4 H_0^2 H_1 P_{1135}
                        +2 H_0^2 P_{1136}
                        -2 H_{-1}^2 H_0^2 P_{1139}
                        +8 \Big(
                                2-3 x+2 x^2
                        \Big)
\Big(3+2 x+3 x^2\Big) H_0^2 H_1^2
                        +\Big(
                                -32 H_{0,-1} P_{1099}
                                -16 H_0 H_1 P_{1124}
                                +8 H_0 P_{1126}
                        \Big) H_{0,1}
                        +\Big(
                                -16 H_{-1} H_0 P_{1064}
                                -16 H_0 H_1 P_{1065}
                                -4 H_0 P_{1151}
                        \Big) H_{0,-1}
                        +\Big(
                                -8 P_{1144}
                                +32 \Big(
                                        1+x+x^2
                                \Big)
\Big(5-2 x+5 x^2\Big) H_1
                        \Big) H_{0,0,1}
                        +\Big(
                                16 H_1 P_{1098}
                                +4 P_{1157}
                        \Big) H_{0,0,-1}
                        -32 \Big(
                                1+x+x^2
                        \Big)
\Big(5-2 x+5 x^2\Big) H_{0,0,1,1}
\end{dmath*}
\begin{dmath*}
{\color{white}=}
                        +\Big(
                                -16 H_0^2 H_{0,1,-1} P_{1112}
                                -16 H_0^2 H_{0,-1,1} P_{1112}
                                +
                                \frac{32}{3} H_0^3 H_1^2 P_{1174}
                                +64 H_{0,0,1,0,1} P_{1180}
                                -48 H_0 H_{0,0,-1,-1} P_{1181}
                                -64 H_{0,0,1,0,-1} P_{1191}
                                +\frac{1}{6} H_0^4 P_{1193}
                                +16 H_0^2 H_{0,1,1} P_{1203}
                                -32 H_0 H_{0,0,1,-1} P_{1213}
                                -32 H_0 H_{0,0,-1,1} P_{1213}
                                +32 H_{0,0,-1,0,1} P_{1220}
                                +8 H_0^2 H_{0,-1,-1} P_{1230}
                                -32 H_{0,0,0,1,-1} P_{1240}
                                -32 H_{0,0,0,-1,1} P_{1240}
                                +32 H_{0,0,-1,0,-1} P_{1250}
                                +96 H_{0,0,0,-1,-1} P_{1252}
                                -32 H_0 H_{0,0,1,1} P_{1258}
                                +64 H_{0,0,0,1,1} P_{1265}
                                +\frac{1}{30} H_0^5 P_{1298}
                                +\frac{2}{3} H_0^3 P_{1306}
                                -16 H_{0,0,0,0,1} P_{1311}
                                +16 H_{0,0,0,0,-1} P_{1312}
                                +\Big(
                                        -\frac{64}{3} H_{-1} H_0^3 P_{1174}
                                        +\frac{2}{3} H_0^4 P_{1197}
                                        +\frac{8}{3} H_0^3 P_{1245}
                                \Big) H_1
                                +\Big(
                                        -\frac{2}{3} H_0^4 P_{1232}
                                        +\frac{2}{3} H_0^3 P_{1293}
                                \Big) H_{-1}
                                +\Big(
                                        \frac{8}{3} H_0^3 P_{1192}
                                        +16 H_{-1} H_0^2 P_{1200}
                                        -4 H_0^2 P_{1234}
                                \Big) H_{0,1}
                                +\Big(
                                        -16 x H_{-1} H_0^2 P_{1083}
                                        +16 H_0^2 H_1 P_{1176}
                                        -\frac{4}{3} H_0^3 P_{1261}
                                        +2 H_0^2 P_{1276}
                                \Big) H_{0,-1}                             
                                +\Big(
                                        32 H_{0,1} P_{1164}
                                        -64 H_{-1} H_0 P_{1177}
                                        +256 H_0 H_1 P_{1184}
                                        -8 H_0^2 P_{1233}
                                        -8 H_0 P_{1274}
                                \Big) H_{0,0,1}
                                +\Big(
                                        64 H_{-1} H_0 P_{1101}
                                        -32 H_{0,1} P_{1228}
                                        +4 H_0 P_{1231}
                                        +4 H_0^2 P_{1241}
                                        -16 H_{0,-1} P_{1285}
                                \Big) H_{0,0,-1}
                                +\Big(
                                        32 H_{-1} P_{1179}
                                        -64 H_1 P_{1247}
                                        +16 P_{1292}
                                        +16 H_0 P_{1300}
                                \Big) H_{0,0,0,1}
                                +\Big(
                                        -96 H_{-1} P_{1113}
                                        +32 H_1 P_{1290}
                                        -8 H_0 P_{1304}
                                        -4 P_{1313}
                                \Big) H_{0,0,0,-1}
                                +\Big(
                                        128 H_{0,1,1} P_{1174}
                                        +32 H_0 H_1^2 P_{1175}
                                        -32 H_{0,-1,1} P_{1224}
                                        -32 H_{0,1,-1} P_{1225}
                                        -16 H_{0,0,1} P_{1255}
                                        -4 H_0 P_{1271}
                                        +16 H_{0,-1,-1} P_{1282}
                                        -
                                        \frac{4}{3} H_0^3 P_{1296}
                                        +2 H_0^2 P_{1299}
                                        +8 H_{0,0,-1} P_{1301}
                                        +\Big(
                                                -64 H_{-1} H_0 P_{1175}
                                                -8 H_0^2 P_{1235}
                                                +8 H_0 P_{1260}
                                        \Big) H_1
                                        +\Big(
                                                -8 H_0^2 P_{1190}
                                                +4 H_0 P_{1227}
                                        \Big) H_{-1}
                                        +\Big(
                                                64 H_{-1} P_{1175}
                                                -32 H_1 P_{1199}
                                                -8 P_{1251}
                                        \Big) H_{0,1}
                                        +\Big(
                                                -32 H_{-1} P_{1185}
                                                -4 P_{1201}
                                                -16 H_0 P_{1209}
                                                +32 H_1 P_{1224}
                                        \Big) H_{0,-1}
                                        -4 P_{1297} \zeta_3
                                \Big) \zeta_2
                                +\Big(
                                        -\frac{8}{5} H_{-1} P_{1182}
                                        -\frac{8}{5} H_1 P_{1303}
                                        +\frac{2 P_{1315}}{5}
                                        -\frac{2}{5} H_0 P_{1316}
                                \Big) \zeta_2^2
                                +\Big(
                                        -16 H_0 H_1 P_{1204}
                                        -16 H_{0,-1} P_{1244}
                                        +16 H_{0,1} P_{1254}
                                        -\frac{8}{3} H_0^2 P_{1272}
                                        -\frac{4}{3} H_0 P_{1302}
                                \Big) \zeta_3
                                +6 P_{1307} \zeta_5
                        \Big) x_+
                        +\Big(
                                -4 H_1 P_{1148}
                                -\frac{2 P_{1159}}{3}
                        \Big) \zeta_3
                \Big)
                +\xi ^2 \Big(
                        -64 (-1+x)^2 (1+x)^2 \eta  x_+^2 \zeta_3 H_{-1} H_0
                        -\frac{64}{3} H_{-1}^3 H_0^2
                        +16 H_{-1}^2 H_0^3
                        +\Big(
                                128 H_{-1} H_0 H_{0,-1}
                                -96 H_{0,-1}^2
                        \Big) H_1
                        +\Big(
                                \Big(
                                        -128 H_{-1} H_0
                                        +64 H_{0,-1}
                                \Big) H_1
                                +64 H_0 H_1^2
                                +128 H_{-1} H_{0,-1}
                        \Big) H_{0,1}
                        +\Big(
                                32 H_1
                                -64 H_{-1}
                        \Big) H_{0,1}^2
                        +64 H_{-1}^2 H_0 H_{0,-1}
                        -64 H_0 H_1^2 H_{0,-1}
                        -32 H_{-1} H_{0,-1}^2
                        +\Big(
                                256 H_{-1} H_1
                                -128 H_1^2
                        \Big) H_{0,0,1}
                        +\Big(
                                -256 H_{-1} H_1
                                +128 H_1^2
                        \Big) H_{0,0,-1}
                        +\Big(
                                256 H_{-1} H_0
                                -192 H_0 H_1
                                -256 H_{0,-1}
                        \Big) H_{0,1,1}
                        +\Big(
                                -128 H_{-1} H_0
                                +64 H_0 H_1
                                -64 H_{0,1}
                                +192 H_{0,-1}
                        \Big) H_{0,1,-1}
                        +\Big(
                                -128 H_{-1} H_0
                                +64 H_0 H_1
                                +192 H_{0,1}
                                +192 H_{0,-1}
                        \Big) H_{0,-1,1}
                        +\Big(
                                -64 H_{-1} H_0
                                +64 H_0 H_1
                                -320 H_{0,1}
                                +32 H_{0,-1}
                        \Big) H_{0,-1,-1}
                        +\Big(
                                256 H_1
                                -256 H_{-1}
                        \Big) H_{0,0,1,1}
                        -256 H_1 H_{0,0,1,-1}
                        -256 H_1 H_{0,0,-1,1}
                        +256 H_1 H_{0,0,-1,-1}
                        +192 H_0 H_{0,1,1,1}
                        -64 H_0 H_{0,1,-1,-1}
                        -128 H_{-1} H_{0,-1,0,1}
                        -64 H_0 H_{0,-1,1,-1}
                        -64 H_0 H_{0,-1,-1,1}
                        +32 H_0 H_{0,-1,-1,-1}
                        -384 H_{0,0,1,1,1}
                        +768 H_{0,0,1,1,-1}
                        +256 H_{0,0,1,-1,1}
                        -256 H_{0,0,-1,1,1}
                        -64 H_{0,1,0,1,1}
                        +256 H_{0,1,0,1,-1}
                        -256 H_{0,-1,0,1,1}
                        +128 H_{0,-1,0,1,-1}
                        +128 H_{0,-1,0,-1,1}
                        -256 H_{0,-1,1,0,1}
                        +320 H_{0,-1,-1,0,1}
                        -32 H_{-1}^2 H_0 \zeta_2
                        +\Big(
                                -64 H_{-1} H_1
                                +32 H_1^2
                        \Big) \zeta_3
                \Big)
\end{dmath*}
\begin{dmath*}
{\color{white}=}
                +\Big(
                        -192 (-1+x) H_{-1} H_0 H_1
                        +32 (-1+x) H_0 H_1^2
                        +\Big(
                                -64 (-1+x) H_1
                                +192 (-1+x) H_{-1}
                        \Big) H_{0,1}
                        +192 (-1+x) H_1 H_{0,-1}
                        +64 (-1+x) H_{0,1,1}
                        -192 (-1+x) H_{0,1,-1}
                        -192 (-1+x) H_{0,-1,1}
                        -32 (-1+x) H_1 \zeta_2
                \Big) x_+
                -48 c_1 x x_+^2
                +\Big(
                        -
                        \frac{113}{2} (1+x)^2
                        -8 \Big(
                                3+8 x+3 x^2\Big) H_{-1} H_0^2 H_1
                        +16 \Big(
                                7+12 x+7 x^2\Big) H_{-1} H_0 H_{0,1}
                        -16 \Big(
                                11+16 x+11 x^2\Big) H_{-1} H_{0,0,1}
                        +16 \Big(
                                11+16 x+11 x^2\Big) H_{0,0,1,-1}
                        +16 \Big(
                                11+16 x+11 x^2\Big) H_{0,0,-1,1}
                        +\Big(
                                -48 \Big(
                                        3+8 x+3 x^2\Big) H_{-1} H_1
                                +48 \Big(
                                        3+8 x+3 x^2\Big) H_{-1,1}
                        \Big) \zeta_2
                \Big) x_+^2
\Bigg\}
\Bigg]
\\
+ C_F C_A  \Bigg[
\frac{1}{\ep^2}  \Bigg\{
        \frac{11}{3}
        +\frac{11 \xi  H_0}{3}
\Bigg\}
+ \frac{1}{\ep}  \Bigg\{
        -\frac{49}{9}
        +\xi  \Big(
                \eta  \Big(
                        -\frac{4}{3} x^2 H_0^3
                        -2 \Big(
                                -1+3 x^2\Big) H_0 \zeta_2
                \Big)
                -\frac{67}{9} H_0
                +4 H_{-1} H_0
                -4 H_0 H_1
                +4 H_{0,1}
                -4 H_{0,-1}
        \Big)
        +\eta  \Big(
                -4 x^2 H_0^2
                -2 \Big(
                        -1+3 x^2\Big) \zeta_2
        \Big)
        +\xi ^2 \Big(
                -4 H_0 H_{0,1}
                +4 H_0 H_{0,-1}
                +8 H_{0,0,1}
                -8 H_{0,0,-1}
                -2 \zeta_3
        \Big)
\Bigg\}
+  \Bigg\{
                -\frac{869}{27}
                +\eta ^2 x_+ \Big(
                        4 H_0^2 H_{0,-1} P_{1211}
                        -8 H_{0,0,0,-1} P_{1236}
                        +8 H_{0,0,0,1} P_{1248}
                        -\frac{1}{2} x^2 \Big(
                                1+7 x+4 x^2+4 x^3\Big) H_0^4
                        -32 x^2 \Big(
                                7-5 x+x^2+x^3\Big) H_0 H_{0,0,1}
                        +\Big(
                                4 H_{0,-1} P_{1217}
                                +2 \Big(
                                        1+x+3 x^2+x^3+x^4+x^5\Big) H_0^2
                        \Big) \zeta_2
                        +\frac{1}{5} P_{1162} \zeta_2^2
                        -2 H_0 P_{1243} \zeta_3
                \Big)
                +\xi ^2 \Big(
                        \Big(
                                -8 H_1
                                +8 H_{-1}
                        \Big) \zeta_3
                        +\Big(
                                16 H_{-1} H_0
                                -16 H_0 H_1
                                -8 H_{0,-1}
                        \Big) H_{0,1}
                        -4 H_{0,1}^2
                        -16 H_{-1} H_0 H_{0,-1}
                        +16 H_0 H_1 H_{0,-1}
                        +12 H_{0,-1}^2
                        +\Big(
                                32 H_1
                                -32 H_{-1}
                        \Big) H_{0,0,1}
                        +\Big(
                                -32 H_1
                                +32 H_{-1}
                        \Big) H_{0,0,-1}
                        +24 H_0 H_{0,1,1}
                        -8 H_0 H_{0,1,-1}
                        -8 H_0 H_{0,-1,1}
                        -8 H_0 H_{0,-1,-1}
                        -32 H_{0,0,1,1}
                        +32 H_{0,0,1,-1}
                        +32 H_{0,0,-1,1}
                        -32 H_{0,0,-1,-1}
                \Big)
                +\eta  \Big(
                        -\frac{22}{27} \Big(
                                11-114 x+11 x^2\Big) H_0
                        +\frac{2}{9} \Big(
                                53-102 x+53 x^2\Big) H_{-1} H_0
                        +8 \Big(
                                -3+x^2\Big) H_0^2 H_1
                        -\frac{2}
                        {9} \Big(
                                53-102 x+53 x^2\Big) H_{0,-1}
                        +\Big(
                                \frac{2}{3} \Big(
                                        49-17 x+79 x^2+13 x^3\Big) \zeta_3
                                +\frac{1}{9} \Big(
                                        -67+74 x+11 x^2+50 x^3\Big) H_0^2
                                +\frac{1}{3} \Big(
                                        61+175 x-35 x^2+79 x^3\Big) H_{-1} H_0^2
                                +16 \Big(
                                        -3-x+3 x^2+5 x^3\Big) H_{0,0,1}
                                -\frac{2}{3} \Big(
                                        -53-95 x+235 x^2+193 x^3\Big) H_{0,0,-1}
                                +\Big(
                                        \frac{2}{9} \Big(
                                                91+562 x-539 x^2+34 x^3\Big)
                                        -\frac{2}{3} \Big(
                                                25-137 x+169 x^2+7 x^3\Big) H_{-1}
                                \Big) \zeta_2
                        \Big) x_+
                        +\Big(
                                \frac{1}{9} H_0^3 P_{1061}
                                +8 H_0^2 H_{0,1} P_{1069}
                                +16 H_0 H_{0,0,-1} P_{1070}
                                -\frac{2}{3} H_0 P_{1079} \zeta_2
                        \Big) x_+^2
                \Big)
                +\xi  \Big(
                        8 H_0 H_1 \Big(
                                1+3 H_{-1}\Big)
                        +\eta  \Big(
                                \frac{4}{3} \Big(
                                        3+x^2\Big) H_{-1} H_0^3
                                +\frac{4}{3} \Big(
                                        -5+x^2\Big) H_0^3 H_1
                                +\Big(
                                        16 H_{-1} H_0
                                        +16 \Big(
                                                -2+x^2\Big) H_0 H_1
                                        -12 \Big(
                                                -3+x^2\Big) H_{0,1}
                                \Big) \zeta_2
                        \Big)
                        -\frac{104}{3} H_{-1}^2 H_0
                        -4 H_0 H_1^2
                        +\Big(
                                -8
                                +8 H_1
                                -24 H_{-1}
                        \Big) H_{0,1}
                        +\Big(
                                -24 H_1
                                +\frac{208 H_{-1}}{3}
                        \Big) H_{0,-1}
                        -8 H_{0,1,1}
                        +24 H_{0,1,-1}
                        +24 H_{0,-1,1}
                        -\frac{208}{3} H_{0,-1,-1}
                        +4 H_1 \zeta_2
                \Big)
                +\Big(
                        16 \Big(
                                3+5 x+3 x^2\Big) H_0 H_{0,1}
                        -2 \Big(
                                19+64 x+19 x^2\Big) H_0 H_{0,-1}
                \Big) x_+^2
                -144 x \log (2) x_+^2 \zeta_2
\Bigg\}
+ \ep  \Bigg\{
                \log (2) x_+^2 \zeta_2 \Big(
                        -12 \Big(
                                11+74 x+11 x^2\Big)
                        -24 \Big(
                                3+8 x+3 x^2\Big) H_1
                        -24 \Big(
                                3+8 x+3 x^2\Big) H_{-1}
                \Big)
                +\eta ^2 x_+ \Big(
                        H_{-1} H_0^4 P_{1187}
                        +128 H_0 H_{0,0,1,1} P_{1188}
                        -16 H_0^2 H_{0,1,1} P_{1189}
                        +8 H_0^2 H_{0,1,-1} P_{1208}
                        +8 H_0^2 H_{0,-1,1} P_{1208}
                        -32 H_{0,0,1,0,1} P_{1221}
                        -64 H_{0,0,-1,0,-1} P_{1222}
                        -8 H_0^2 H_{0,-1,-1} P_{1246}
                        +16 H_{0,0,-1,0,1} P_{1257}
\end{dmath*}
\begin{dmath*}
{\color{white}=}
                        +16 H_{0,0,1,0,-1} P_{1268}
                        -\frac{4}{3} H_0^3 H_{0,1} P_{1269}
                        +32 H_{0,0,0,1,-1} P_{1289}
                        +32 H_{0,0,0,-1,1} P_{1289}
                        -16 H_{0,0,0,1,1} P_{1305}
                        +8 H_{0,0,0,0,1} P_{1308}
                        -8 H_{0,0,0,0,-1} P_{1309}
                        -16 H_{0,0,0,-1,-1} P_{1310}
                        -\frac{1}{30} x^2 \Big(
                                5+43 x+24 x^2+24 x^3\Big) H_0^5
                        +\Big(
                                -8-8 x-11 x^2-5 x^3\Big) H_0^4 H_1
                        +\Big(
                                -8 H_0^2 H_1 P_{1202}
                                +
                                \frac{4}{3} H_0^3 P_{1295}
                                -8 \Big(
                                        8+8 x+15 x^2+x^3\Big) H_{-1} H_0^2
                        \Big) H_{0,-1}
                        +\Big(
                                32 H_{-1} H_0 P_{1173}
                                +32 H_{0,1} P_{1206}
                                -32 H_0 H_1 P_{1207}
                                -16 H_{0,-1} P_{1267}
                                +4 H_0^2 P_{1275}
                        \Big) H_{0,0,1}
                        +\Big(
                                -16 H_{0,1} P_{1238}
                                +32 H_{0,-1} P_{1242}
                                -8 H_0^2 P_{1249}
                        \Big) H_{0,0,-1}
                        +\Big(
                                -16 H_{-1} P_{1219}
                                +16 H_1 P_{1259}
                                -8 H_0 P_{1291}
                        \Big) H_{0,0,0,1}
                        +\Big(
                                16 H_{-1} P_{1229}
                                -16 H_1 P_{1239}
                                +8 H_0 P_{1287}
                        \Big) H_{0,0,0,-1}
                        +\Big(
                                -4 H_{-1} H_0^2 P_{1186}
                                -16 H_{0,0,-1} P_{1266}
                                +8 H_{0,-1,1} P_{1277}
                                +8 H_{0,1,-1} P_{1278}
                                -8 H_{0,-1,-1} P_{1283}
                                -4 H_0 H_{0,1} P_{1286}
                                +8 H_{0,0,1} P_{1288}
                                +\frac{2}{3} H_0^3 P_{1294}
                                +4 x^2 \Big(
                                        3+5 x+4 x^2+4 x^3\Big) H_0^2 H_1
                                +\Big(
                                        16 H_{-1} P_{1205}
                                        -16 H_1 P_{1237}
                                        +4 H_0 P_{1284}
                                \Big) H_{0,-1}
                                +2 P_{1273} \zeta_3
                        \Big) \zeta_2
                        +\Big(
                                \frac{4}{5} H_{-1} P_{1263}
                                -\frac{2}{5} H_0 P_{1264}
                                -\frac{4}{5} H_1 P_{1270}
                        \Big) \zeta_2^2
                        +\Big(
                                16 H_0 H_1 P_{1172}
                                -16 H_{-1} H_0 P_{1198}
                                -4 H_{0,1} P_{1256}
                                +2 H_0^2 P_{1279}
                                +4 H_{0,-1} P_{1280}
                        \Big) \zeta_3
                        +P_{1161} \zeta_5
                \Big)
                +\xi  \Big(
                        8 H_0 H_1^2 \Big(
                                1+3 H_{-1}\Big)
                        -24 H_0 H_1 H_{-1} \Big(
                                2+3 H_{-1}\Big)
                        +4 \Big(
                                -2+H_1\Big) H_1 \zeta_2
                        +\eta  \Big(
                                -
                                \frac{4}{3} \Big(
                                        3+5 x^2\Big) H_{-1}^2 H_0^3
                                +\Big(
                                        4 (-1+x) (1+x) H_0
                                        +\frac{8}{3} \Big(
                                                1+7 x^2\Big) H_{-1} H_0^3
                                \Big) H_1
                                -\frac{4}{3} \Big(
                                        1+7 x^2\Big) H_0^3 H_1^2
                                -4 (-1+x) (1+x) H_{0,1}
                                +\Big(
                                        -8 \Big(
                                                1+3 x^2\Big) H_{-1}^2 H_0
                                        +16 \Big(
                                                -1+5 x^2\Big) H_{-1} H_0 H_1
                                        -8 \Big(
                                                -1+5 x^2\Big) H_0 H_1^2
                                        +\Big(
                                                96 x^2 H_1
                                                -96 x^2 H_{-1}
                                        \Big) H_{0,1}
                                        -8 \Big(
                                                1+13 x^2\Big) H_{0,1,1}
                                \Big) \zeta_2
                        \Big)
                        +80 H_{-1}^3 H_0
                        -\frac{8}{3} H_0 H_1^3
                        +\Big(
                                -16 H_1 \Big(
                                        1+3 H_{-1}\Big)
                                +8 H_1^2
                                +48 H_{-1}
                                +72 H_{-1}^2
                        \Big) H_{0,1}
                        +\Big(
                                48 H_1 \Big(
                                        1+3 H_{-1}\Big)
                                -24 H_1^2
                                -240 H_{-1}^2
                        \Big) H_{0,-1}
                        +\Big(
                                16
                                -16 H_1
                                +48 H_{-1}
                        \Big) H_{0,1,1}
                        +\Big(
                                -48
                                +48 H_1
                                -144 H_{-1}
                        \Big) H_{0,1,-1}
                        +\Big(
                                -48
                                +48 H_1
                                -144 H_{-1}
                        \Big) H_{0,-1,1}
                        +\Big(
                                -144 H_1
                                +480 H_{-1}
                        \Big) H_{0,-1,-1}
                        +16 H_{0,1,1,1}
                        -48 H_{0,1,1,-1}
                        -48 H_{0,1,-1,1}
                        +144 H_{0,1,-1,-1}
                        -48 H_{0,-1,1,1}
                        +144 H_{0,-1,1,-1}
                        +144 H_{0,-1,-1,1}
                        -480 H_{0,-1,-1,-1}
                        +\Big(
                                -12 (-1+x) (1+x) H_0 H_{0,1}
                                -8 (-1+x) (1+x) H_{-1} H_0 H_{0,-1}
                                -80 (-1+x) (1+x) H_0 H_1 H_{0,-1}
                        \Big) x_+^2
                \Big)
                +\eta  \Big(
                        \log (2) x_+^2 \zeta_2 \Big(
                                96 x^2 H_{0,1}
                                +96 x^2 H_{0,-1}
                        \Big)
                        -
                        \frac{2}{81} \Big(
                                1061-16710 x+1061 x^2\Big) H_0
                        +\frac{4}{27} \Big(
                                341-1266 x+341 x^2\Big) H_{-1} H_0
                        +\frac{2}{9} \Big(
                                47+306 x+47 x^2\Big) H_{-1}^2 H_0
                        -24 x H_0 H_1
                        +24 x H_{0,1}
                        +\Big(
                                -\frac{4}{27} \Big(
                                        341-1266 x+341 x^2\Big)
                                -\frac{4}{9} \Big(
                                        47+306 x+47 x^2\Big) H_{-1}
                        \Big) H_{0,-1}
                        +\frac{4}{9} \Big(
                                47+306 x+47 x^2\Big) H_{0,-1,-1}
                        +\Big(
                                \frac{1}{27} \Big(
                                        -341+2599 x-1127 x^2-395 x^3\Big) H_0^2
                                +\frac{1}{9} \Big(
                                        -137-1037 x+115 x^2-173 x^3\Big) H_{-1} H_0^2
                                +\Big(
                                        -65-155 x-x^2-91 x^3\Big) H_{-1}^2 H_0^2
                                +\Big(
                                        2 \Big(
                                                -1+61 x-45 x^2+17 x^3\Big) H_0^2
                                        +4 \Big(
                                                15+17 x+7 x^2+9 x^3\Big) H_{-1} H_0^2
                                \Big) H_1
                                +2 \Big(
                                        -13+13 x-21 x^2+5 x^3\Big) H_0^2 H_1^2
                                +\Big(
                                        -4 \Big(
                                                5-65 x+81 x^2+11 x^3\Big)
                                        +8 \Big(
                                                3+45 x-37 x^2+5 x^3\Big) H_1
                                        -8 \Big(
                                                5+43 x-19 x^2+19 x^3\Big) H_{-1}
                                \Big) H_{0,0,1}
                                +\Big(
                                        \frac{2}{9} \Big(
                                                -277-1285 x+2207 x^2+587 x^3\Big)
                                        +8 \Big(
                                                -35-77 x+53 x^2+11 x^3\Big) H_1
                                        +4 \Big(
                                                61+191 x-35 x^2+95 x^3\Big) H_{-1}
                                \Big) H_{0,0,-1}
                                -8 \Big(
                                        3+45 x-37 x^2+5 x^3\Big) H_{0,0,1,1}
                                +8 \Big(
                                        5+43 x-19 x^2+19 x^3\Big) H_{0,0,1,-1}
                                +8 \Big(
                                        5+43 x-19 x^2+19 x^3\Big) H_{0,0,-1,1}
\end{dmath*}
\begin{dmath*}
{\color{white}=}
                                -4 \Big(
                                        61+191 x-35 x^2+95 x^3\Big) H_{0,0,-1,-1}
                                +\Big(
                                        -
                                        \frac{4}{27} \Big(
                                                -493-1480 x+2783 x^2+206 x^3\Big)
                                        -48 \Big(
                                                -1-2 x+3 x^2+2 x^3\Big) H_{-1} H_1
                                        -\frac{2}{9} \Big(
                                                61+349 x-623 x^2+277 x^3\Big) H_{-1}
                                        -6 \Big(
                                                -17+33 x-49 x^2+x^3\Big) H_{-1}^2
                                \Big) \zeta_2
                                +\Big(
                                        \frac{1}{9} \Big(
                                                1055+3683 x-4591 x^2-739 x^3\Big)
                                        -2 \Big(
                                                -95-53 x+9 x^2+51 x^3\Big) H_1
                                        -2 \Big(
                                                91-55 x+211 x^2+65 x^3\Big) H_{-1}
                                \Big) \zeta_3
                        \Big) x_+
                        +\Big(
                                \frac{1}{12} H_0^4 P_{1062}
                                +32 H_0 H_{0,0,1,-1} P_{1071}
                                +32 H_0 H_{0,0,-1,1} P_{1071}
                                +24 H_0 H_{0,-1}^2 P_{1082}
                                +16 H_0 H_{0,1}^2 P_{1086}
                                -16 H_0 H_{0,-1,0,1} P_{1102}
                                -32 H_0 H_{0,0,-1,-1} P_{1104}
                                +4 H_{0,0,0,1} P_{1137}
                                +\frac{2}{3} H_0^3 H_1 P_{1140}
                                +4 H_0 H_{0,0,1} P_{1141}
                                -2 H_{0,0,0,-1} P_{1143}
                                +\frac{1}{3} H_{-1} H_0^3 P_{1152}
                                +H_0^2 H_{0,-1} P_{1154}
                                +\frac{1}{27} H_0^3 P_{1158}
                                +\Big(
                                        -48 H_0^2 H_1 P_{1067}
                                        +16 H_{-1} H_0^2 P_{1069}
                                        -16 H_0 H_{0,-1} P_{1103}
                                        -2 H_0^2 P_{1145}
                                \Big) H_{0,1}
                                +\Big(
                                        -96 H_0 H_1 P_{1068}
                                        +32 H_{-1} H_0 P_{1070}
                                        -2 H_0 P_{1156}
                                \Big) H_{0,0,-1}
                                +\Big(
                                        4 H_{-1} H_0 P_{1080}
                                        -4 H_{0,1}
                                         P_{1097}
                                        +4 H_0 H_1 P_{1130}
                                        -4 H_{0,-1} P_{1138}
                                        +
                                        \frac{1}{2} H_0^2 P_{1150}
                                        +\frac{1}{9} H_0 P_{1153}
                                \Big) \zeta_2
                                +\frac{1}{5} P_{1155} \zeta_2^2
                                +\frac{4}{3} H_0 P_{1146} \zeta_3
                        \Big) x_+^2
                \Big)
                +\xi ^2 \Big(
                        \Big(
                                -64 H_{-1} H_0 H_{0,-1}
                                +48 H_{0,-1}^2
                        \Big) H_1
                        +\Big(
                                -32 H_{-1}^2 H_0
                                +\Big(
                                        64 H_{-1} H_0
                                        -32 H_{0,-1}
                                \Big) H_1
                                -32 H_0 H_1^2
                                +32 H_{-1} H_{0,-1}
                        \Big) H_{0,1}
                        +\Big(
                                -16 H_1
                                +16 H_{-1}
                        \Big) H_{0,1}^2
                        +32 H_{-1}^2 H_0 H_{0,-1}
                        +32 H_0 H_1^2 H_{0,-1}
                        -48 H_{-1} H_{0,-1}^2
                        +\Big(
                                -128 H_{-1} H_1
                                +64 H_1^2
                                +64 H_{-1}^2
                        \Big) H_{0,0,1}
                        +\Big(
                                128 H_{-1} H_1
                                -64 H_1^2
                                -64 H_{-1}^2
                        \Big) H_{0,0,-1}
                        +\Big(
                                -96 H_{-1} H_0
                                +96 H_0 H_1
                                +16 H_{0,1}
                                +80 H_{0,-1}
                        \Big) H_{0,1,1}
                        +\Big(
                                32 H_{-1} H_0
                                -32 H_0 H_1
                                +16 H_{0,1}
                                -48 H_{0,-1}
                        \Big) H_{0,1,-1}
                        +\Big(
                                32 H_{-1} H_0
                                -32 H_0 H_1
                                -48 H_{0,1}
                                -48 H_{0,-1}
                        \Big) H_{0,-1,1}
                        +\Big(
                                32 H_{-1} H_0
                                -32 H_0 H_1
                                +16 H_{0,1}
                                -48 H_{0,-1}
                        \Big) H_{0,-1,-1}
                        +\Big(
                                -128 H_1
                                +128 H_{-1}
                        \Big) H_{0,0,1,1}
                        +\Big(
                                128 H_1
                                -128 H_{-1}
                        \Big) H_{0,0,1,-1}
                        +\Big(
                                128 H_1
                                -128 H_{-1}
                        \Big) H_{0,0,-1,1}
                        +\Big(
                                -128 H_1
                                +128 H_{-1}
                        \Big) H_{0,0,-1,-1}
                        -112 H_0 H_{0,1,1,1}
                        +16 H_0 H_{0,1,1,-1}
                        +16 H_0 H_{0,1,-1,1}
                        +16 H_0 H_{0,1,-1,-1}
                        +16 H_0 H_{0,-1,1,1}
                        +16 H_0 H_{0,-1,1,-1}
                        +16 H_0 H_{0,-1,-1,1}
                        +16 H_0 H_{0,-1,-1,-1}
                        +128 H_{0,0,1,1,1}
                        -256 H_{0,0,1,1,-1}
                        -128 H_{0,0,1,-1,1}
                        +128 H_{0,0,1,-1,-1}
                        +128 H_{0,0,-1,1,-1}
                        +128 H_{0,0,-1,-1,1}
                        +256 H_{0,0,-1,-1,-1}
                        -64 H_{0,1,0,1,-1}
                        +64 H_{0,-1,0,1,1}
                        +192 H_{0,-1,0,-1,-1}
                        +64 H_{0,-1,1,0,1}
                        +\Big(
                                32 H_{-1} H_1
                                -16 H_1^2
                                -16 H_{-1}^2
                        \Big) \zeta_3
                \Big)
                +24 c_1 x x_+^2
                +\Big(
                        -
                        \frac{6437}{162} (1+x)^2
                        +\Big(
                                -240 x H_0
                                -8 \Big(
                                        5-8 x+5 x^2\Big) H_{-1} H_0
                                +8 \Big(
                                        5-24 x+5 x^2\Big) H_0 H_1
                                +40 \Big(
                                        5+12 x+5 x^2\Big) H_{0,-1}
                        \Big) H_{0,1}
                        -16 (2+x) (1+2 x) H_{0,1}^2
                        +2 \Big(
                                23+152 x+23 x^2\Big) H_0 H_{0,-1}
                        -64 x H_{-1} H_0 H_{0,-1}
                        +320 x H_0 H_1 H_{0,-1}
                        -2 \Big(
                                59+176 x+59 x^2\Big) H_{0,-1}^2
                        +8 \Big(
                                3+44 x+3 x^2\Big) H_0 H_{0,1,1}
                        -32 \Big(
                                5+17 x+5 x^2\Big) H_0 H_{0,1,-1}
                        -32 \Big(
                                5+17 x+5 x^2\Big) H_0 H_{0,-1,1}
                        +12 \Big(
                                19+64 x+19 x^2\Big) H_0 H_{0,-1,-1}
                        -8 \Big(
                                15+32 x+15 x^2\Big) H_{0,-1,0,1}
                        -24 \Big(
                                3+8 x+3 x^2\Big) H_{-1,1} \zeta_2
                \Big) x_+^2
\Bigg\}
\Bigg]
\\
+ C_F n_l T_F  \Bigg[
\frac{1}{\ep^2}  \Bigg\{
        -\frac{4}{3}
        -\frac{4 \xi  H_0}{3}
\Bigg\}
+ \frac{1}{\ep}  \Bigg\{
        \frac{20}{9}
        +\frac{20 \xi  H_0}{9}
\Bigg\}
+  \Bigg\{
                \frac{196}{27}
                +\eta  \Big(
                        \frac{16}{27} \Big(
                                7-48 x+7 x^2\Big) H_0
                        -\frac{16}{9} \Big(
                                5-12 x+5 x^2\Big) H_{-1} H_0
                        +\frac{4}{9} \Big(
                                5-12 x+5 x^2\Big) H_0^2
                        +\frac{16}{9} \Big(
                                5-12 x+5 x^2\Big) H_{0,-1}
                        -\frac{8}{9} \Big(
                                -1-12 x+11 x^2\Big) \zeta_2
                \Big)
\end{dmath*}
\begin{dmath*}
{\color{white}=}
                +\xi  \Big(
                        \frac{16}{3} H_{-1}^2 H_0
                        -\frac{8}{3} H_{-1} H_0^2
                        +\frac{4}{9} H_0^3
                        -\frac{32}{3} H_{-1} H_{0,-1}
                        +\frac{16}
                        {3} H_{0,0,-1}
                        +
                        \frac{32}{3} H_{0,-1,-1}
                        +\Big(
                                \frac{8}{3} H_0
                                +\frac{16}{3} H_{-1}
                        \Big) \zeta_2
                        -\frac{16 \zeta_3}{3}
                \Big)
\Bigg\}
+ \ep  \Bigg\{
                \frac{1706}{81}
                +\xi  \Big(
                        \Big(
                                -\frac{16 H_0}{3}
                                +32 H_{-1}
                        \Big) \zeta_3
                        -\frac{32}{3} H_{-1}^3 H_0
                        +8 H_{-1}^2 H_0^2
                        -\frac{8}{3} H_{-1} H_0^3
                        +32 H_{-1}^2 H_{0,-1}
                        -32 H_{-1} H_{0,0,-1}
                        -64 H_{-1} H_{0,-1,-1}
                        +16 H_{0,0,0,-1}
                        +32 H_{0,0,-1,-1}
                        +64 H_{0,-1,-1,-1}
                        +\Big(
                                -8 H_{-1} H_0
                                +2 H_0^2
                                -16 H_{-1}^2
                                +24 H_{0,-1}
                        \Big) \zeta_2
                        -\frac{96 \zeta_2^2}{5}
                \Big)
                +\eta  \Big(
                        \frac{16}{81} \Big(
                                77-636 x+77 x^2\Big) H_0
                        +\frac{32}{9} \Big(
                                5-18 x+5 x^2\Big) H_{-1}^2 H_0
                        +\frac{4}{27} \Big(
                                37-246 x+37 x^2\Big) H_0^2
                        +\frac{8}{27} \Big(
                                5-18 x+5 x^2\Big) H_0^3
                        +\frac{1}{3} \Big(
                                1+x^2\Big) H_0^4
                        +\Big(
                                -\frac{16}{27} \Big(
                                        37-246 x+37 x^2\Big) H_0
                                -\frac{16}{9} \Big(
                                        5-18 x+5 x^2\Big) H_0^2
                        \Big) H_{-1}
                        +\Big(
                                \frac{16}{27} \Big(
                                        37-246 x+37 x^2\Big)
                                -\frac{64}{9} \Big(
                                        5-18 x+5 x^2\Big) H_{-1}
                        \Big) H_{0,-1}
                        +\frac{32}{9} \Big(
                                5-18 x+5 x^2\Big) H_{0,0,-1}
                        +\frac{64}{9} \Big(
                                5-18 x+5 x^2\Big) H_{0,-1,-1}
                        +\Big(
                                -\frac{8}{27} \Big(
                                        -41-246 x+115 x^2\Big)
                                +\frac{4}{9} \Big(
                                        5-36 x+5 x^2\Big) H_0
                                +\frac{32}{9} \Big(
                                        5-18 x+5 x^2\Big) H_{-1}
                        \Big) \zeta_2
                        -
                        \frac{64}{9} \Big(
                                1-9 x+4 x^2\Big) \zeta_3
                \Big)
\Bigg\}
\Bigg]
\\
+ C_F T_F  \Bigg[
%
%
%
 \Bigg\{
                \frac{4}{27} \Big(
                        407+694 x+407 x^2\Big) x_+^2
                +\frac{64}{5} (-1+x) x (1+x) \xi ^2 \eta ^2 \zeta_2^2
                +\eta  \Big(
                        -128 x H_{-1} H_0
                        +64 x H_0 H_1
                        -32 x H_{0,1}
                        +128 x H_{0,-1}
                        +\Big(
                                \frac{32}{27} H_0 P_{1118}
                                +32 x^2 H_0^3 H_1
                                +\Big(
                                        192 x^2 H_0 H_1
                                        -192 x^2 H_{0,1}
                                \Big) \zeta_2
                        \Big) x_+^2
                        +\frac{4}{3} P_{1253} x_+^3 \zeta_2
                \Big)
                +\xi  \Big(
                        \eta  x_+^2 \Big(
                                -\frac{16}{3} (-1+x) x (1+x) H_0^3 H_1
                                +\Big(
                                        -32 (-1+x) x (1+x) H_0 H_1
                                        +32 (-1+x) x (1+x) H_{0,1}
                                \Big) \zeta_2
                        \Big)
                        +4 H_0 \zeta_2
                        -\frac{256}{5} x^2 \eta ^2 \zeta_2^2
                \Big)
                +\eta ^2 \Big(
                        -\frac{4}{9} \Big(
                                -1+48 x^2+x^4\Big) H_0^3
                        +256 x^2 H_0^2 H_1
                        -768 x^2 H_0 H_{0,1}
                        +512 x^2 H_0 H_{0,-1}
                        +1024 x^2 H_{0,0,1}
                        -1024 x^2 H_{0,0,-1}
                        +\frac{4}{9} H_0^2 P_{1319} x_+^2
                        -256 x^2 \zeta_3
                \Big)
                +\eta ^3 \Big(
                        -64 x H_0 H_{0,0,1} P_{1073}
                        +16 x H_{0,0,0,1} P_{1116}
                        -256 x^3 H_0^2 H_{0,1}
                        -1024 x^3 H_0 H_{0,0,-1}
                        +3072 x^3 H_{0,0,0,-1}
                        -
                        \frac{x H_0^4}{3 x_+^4}
                        -8 x H_0^2 P_{1075} \zeta_2
                        -16 x H_0 P_{1091} \zeta_3
                \Big)
\Bigg\}
+ \ep  \Bigg\{
                \xi  \Big(
                        \eta  x_+^2 \Big(
                                -\frac{32}{3} (-1+x) x (1+x) H_{-1} H_0^3 H_1
                                +\frac{16}{3} (-1+x) x (1+x) H_0^3 H_1^2
                                -64 (-1+x) x (1+x) H_0 H_{0,-1} H_{0,1}
                                +64 (-1+x) x (1+x) H_0^2 H_1 H_{0,-1}
                                -128 (-1+x) x (1+x) H_{0,0,-1,0,1}
                                +\Big(
                                        -64 (-1+x) x (1+x) H_{-1} H_0 H_1
                                        +32 (-1+x) x (1+x) H_0 H_1^2
                                        +\Big(
                                                -64 (-1+x) x (1+x) H_1
                                                +64 (-1+x) x (1+x) H_{-1}
                                        \Big) H_{0,1}
                                        +256 (-1+x) x (1+x) H_1 H_{0,-1}
                                        +64 (-1+x) x (1+x) H_{0,1,1}
                                        -256 (-1+x) x (1+x) H_{0,-1,1}
                                \Big) \zeta_2
                        \Big)
                        +\eta ^2 \Big(
                                1024 x^2 H_{0,1,-1} \zeta_2
                                -\frac{512}{5} x^2 H_{-1} \zeta_2^2
                        \Big)
                        -256 (-1+x) x^2 (1+x) \eta ^3 H_0 H_{0,-1,0,1}
                \Big)
                +\xi ^2 \Big(
                        \eta ^2 \Big(
                                -256 (-1+x) x (1+x) H_{0,1,-1} \zeta_2
                                +\frac{128}{5} (-1+x) x (1+x) H_{-1} \zeta_2^2
                        \Big)
                        +64 (-1+x)^2 x (1+x)^2 \eta ^3 H_0 H_{0,-1,0,1}
                \Big)
                +\eta  \Big(
                        \Big(
                                64 x H_1
                                +96 x H_{0,1}
                        \Big) \zeta_2
                        +\frac{32}{27} \Big(
                                6+203 x+257 x^2+6 x^3\Big) x_+ H_0
                        +512 x H_{-1}^2 H_0
                        +\Big(
                                512 x
                                -384 x H_{-1}
                        \Big) H_0 H_1
                        +64 x H_0 H_1^2
\end{dmath*}
\begin{dmath*}
{\color{white}=}
                        +\Big(
                                -64 x H_1
                                +448 x H_{-1}
                        \Big) H_{0,1}
                        +\Big(
                                384 x H_1
                                -1024 x H_{-1}
                        \Big) H_{0,-1}
                        +64 x H_{0,1,1}
                        -448 x H_{0,1,-1}
                        -448 x H_{0,-1,1}
                        +1024 x H_{0,-1,-1}
                        +\Big(
                                -
                                \frac{64}{27} H_{-1} H_0 P_{1119}
                                +64 x^2 H_{-1} H_0^3 H_1
                                -32 x^2 H_0^3 H_1^2
                                +\Big(
                                        -64 x \Big(
                                                7+12 x+7 x^2\Big)
                                        +384 x^2 H_0 H_{0,-1}
                                \Big) H_{0,1}
                                +\Big(
                                        \frac{64 P_{1119}}{27}
                                        -384 x^2 H_0^2 H_1
                                \Big) H_{0,-1}
                                +768 x^2 H_{0,0,-1,0,1}
                                +\Big(
                                        \frac{4 P_{1160}}{27}
                                        +384 x^2 H_{-1} H_0 H_1
                                        -192 x^2 H_0 H_1^2
                                        +\Big(
                                                384 x^2 H_1
                                                -384 x^2 H_{-1}
                                        \Big) H_{0,1}
                                        -1536 x^2 H_1 H_{0,-1}
                                        -384 x^2 H_{0,1,1}
                                        +1536 x^2 H_{0,-1,1}
                                \Big) \zeta_2
                        \Big) x_+^2
                \Big)
                +\eta ^2 \Big(
                        -128 x^2 H_{-1}^2 H_0^2
                        -\frac{8}{9} \Big(
                                -1-144 x^2+x^4\Big) H_{-1} H_0^3
                        +\Big(
                                -256 x^2 H_{-1} H_0^2
                                +16 x \Big(
                                        -1+16 x+x^2\Big) H_0^3
                        \Big) H_1
                        +384 x^2 H_0^2 H_1^2
                        +\Big(
                                1024 x^2 H_{-1} H_0
                                -1536 x^2 H_0 H_1
                                -2560 x^2 H_{0,-1}
                        \Big) H_{0,1}
                        +256 x^2 H_{0,1}^2
                        +\Big(
                                \Big(
                                        64 x \Big(
                                                1+27 x+2 x^2\Big)
                                        -512 x^2 H_{-1}
                                \Big) H_0
                                +\frac{16}{3} \Big(
                                        -1+192 x^2+x^4\Big) H_0^2
                                -512 x^2 H_0 H_1
                        \Big) H_{0,-1}
                        +1536 x^2 H_{0,-1}^2
                        +\Big(
                                1536 x^2 H_1
                                -1536 x^2 H_{-1}
                        \Big) H_{0,0,1}
                        +\Big(
                                1536 x^2 H_1
                                +1536 x^2 H_{-1}
                        \Big) H_{0,0,-1}
                        +1024 x^2 H_0 H_{0,1,1}
                        +1536 x^2 H_0 H_{0,1,-1}
                        +1536 x^2 H_0 H_{0,-1,1}
                        -2560 x^2 H_0 H_{0,-1,-1}
                        -1536 x^2 H_{0,0,1,1}
                        +1536 x^2 H_{0,0,1,-1}
                        +1536 x^2 H_{0,0,-1,1}
                        -1536 x^2 H_{0,0,-1,-1}
                        +1536 x^2 H_{0,-1,0,1}
                        +\Big(
                                \Big(
                                        1536 x^2 H_1
                                        +1536 x^2 H_{-1}
                                \Big) \zeta_2
                                +
                                \frac{64}{3} P_{1324} x_+^2 \zeta_2
                        \Big) \log (2)
                        -\frac{4}{27} H_0^2 P_{1262} x_+
                        +\Big(
                                \frac{16}{9} H_{0,0,-1} P_{1317}
                                -\frac{8}{9} H_{-1} H_0^2 P_{1318}
                                +\frac{16}{9} H_0^2 H_1 P_{1320}
                                -\frac{32}{9} H_0 H_{0,1} P_{1321}
                                +\frac{32}{9} H_{0,0,1} P_{1322}
                                +\frac{8}{27} H_0^3 P_{1325}
                                +\Big(
                                        \frac{16}{9} x H_0 P_{1226}
                                        -\frac{16}{3} H_{-1} P_{1323}
                                \Big) \zeta_2
                                -\frac{8}{9} P_{1326} \zeta_3
                        \Big) x_+^2
                        +\Big(
                                -\frac{8}{3} \Big(
                                        -1-288 x^2+x^4\Big) H_{-1} H_0
                                +\Big(
                                        32 x (3+x) (-1+3 x) H_0
                                        -1536 x^2 H_{-1}
                                \Big) H_1
                                -768 x^2 H_{-1}^2
                                +\frac{8}{3} \Big(
                                        -7-480 x^2+7 x^4\Big) H_{0,-1}
                                +1536 x^2 H_{-1,1}
                        \Big) \zeta_2
                        +\Big(
                                -2688 x^2 H_1
                                +384 x^2 H_{-1}
                        \Big) \zeta_3
                \Big)
                +\eta ^3 \Big(
                        -64 x H_0 H_{0,1}^2 P_{1073}
                        +32 x H_0^2 H_{0,1,1} P_{1073}
                        +256 x H_0 H_{0,0,1,-1} P_{1073}
                        +256 x H_0 H_{0,0,-1,1} P_{1073}
                        +32 x H_0^2 H_{0,1,-1} P_{1075}
                        +32 x H_0^2 H_{0,-1,1} P_{1075}
                        -32 x H_{0,0,1,0,-1} P_{1078}
                        +\frac{8}{3} x H_0^3 H_{0,-1} P_{1088}
                        -64 x H_{0,0,1,0,1} P_{1089}
                        -32 x H_{0,0,0,1,1} P_{1106}
                        +\frac{2}
                        {3} x H_0^4 H_1 P_{1109}
                        -64 x H_{0,0,0,1,-1} P_{1110}
                        -64 x H_{0,0,0,-1,1} P_{1110}
                        +16 x H_{0,0,0,0,-1} P_{1147}
                        -16 x H_{0,0,0,0,1} P_{1149}
                        +\log (2) \zeta_2 \Big(
                                -3072 x^3 H_{0,1}
                                -3072 x^3 H_{0,-1}
                        \Big)
                        +
                        \frac{1}{3} (1+x) \Big(
                                1-53 x^2+43 x^3+x^5\Big) H_0^4
                        +\Big(
                                -\frac{16}{3} x H_0^3 P_{1093}
                                +\frac{16}{3} (1+x) H_0^2 P_{1168}
                                -512 x^3 H_0^2 H_1
                        \Big) H_{0,1}
                        -512 x^3 H_{-1} H_0^2 H_{0,1}
                        +1536 x^3 H_0 H_{0,-1}^2
                        +\Big(
                                -32 x H_0^2 P_{1072}
                                -128 x H_{-1} H_0 P_{1073}
                                -32 x H_{0,-1} P_{1087}
                                +64 x H_{0,1} P_{1089}
                                -\frac{64}{3} (1+x) H_0 P_{1165}
                        \Big) H_{0,0,1}
                        +\Big(
                                -32 x H_0^2 P_{1075}
                                -64 x H_{0,1} P_{1091}
                                +\frac{32}{3} (1+x) H_0 P_{1167}
                                -2048 x^3 H_{-1} H_0
                                -7168 x^3 H_{0,-1}
                        \Big) H_{0,0,-1}
                        +512 x^3 H_0^2 H_{0,-1,-1}
                        +\Big(
                                32 x H_{-1} P_{1116}
                                +32 x H_0 P_{1127}
                                -32 x H_1 P_{1132}
                                +16 (1+x) P_{1183}
                        \Big) H_{0,0,0,1}
                        +\Big(
                                -32 x H_0 P_{1107}
                                +64 x H_1 P_{1128}
                                -\frac{16}{3} (1+x) P_{1163}
                                +6144 x^3 H_{-1}
                        \Big) H_{0,0,0,-1}
                        +4096 x^3 H_0 H_{0,0,-1,-1}
                        -3840 x^3 H_0 H_{0,-1,0,1}
                        +9216 x^3 H_{0,0,0,-1,-1}
                        +5120 x^3 H_{0,0,-1,0,-1}
                        + (1+x)^4 \Big( -\frac{2}{3} x H_{-1} H_0^4
                        -\frac{4}{15} x H_0^5
                        +128 x H_0 H_1 H_{0,0,1}
                        -128 x H_0 H_1 H_{0,0,-1}
                        -128 x H_0 H_{0,0,1,1}
                        -32 x H_0^2 H_1 \zeta_2
                        -32 x H_0 H_1 \zeta_3 \Big)
\end{dmath*}
\begin{dmath}
{\color{white}=}
                        +\Big(
                                64 x H_0 H_{0,-1} P_{1074}
                                -256 x H_{0,0,-1} P_{1074}
                                -16 x H_{-1} H_0^2 P_{1075}
                                -4 x H_0^3 P_{1076}
                                -32 x H_0 H_{0,1} P_{1092}
                                +32 x H_{0,0,1} P_{1117}                            
                                +2 (1+x) H_0^2 P_{1169}
                                +3072 x^3 H_{0,-1,-1}
                                -16 x P_{1077} \zeta_3
                        \Big) \zeta_2
                        +\Big(
                                -
                                \frac{16}{5} x H_1 P_{1110}
                                +\frac{8}{5} x H_0 P_{1111}
                                -\frac{8}{5} (1+x) P_{1166}
                        \Big) \zeta_2^2
                        +\Big(
                                32 x H_{0,-1} P_{1090}
                                -32 x H_{-1} H_0 P_{1091}
                                -32 x H_{0,1} P_{1108}
                                -4 x H_0^2 P_{1133}
                                -\frac{8}{9} (1+x) H_0 P_{1212}
                        \Big) \zeta_3
                        +16 x P_{1129} \zeta_5
                \Big)
                +\Big(
                        \frac{2}{27} \Big(
                                2107+4454 x+2107 x^2\Big)
                        -64 x H_1
                \Big) x_+^2
\Bigg\}
\Bigg] \,.
\end{dmath}
%
The polynomials are listed below
\begin{align}
P_{1061} &= -62 x^4+74 x^3+53 x^2+2 x-11\,,\nonumber\\
P_{1062} &= -32 x^4+56 x^3+49 x^2-11\,,\nonumber\\
P_{1063} &= x^4-102 x^3+42 x^2-102 x+1\,,\nonumber\\
P_{1064} &= x^4-42 x^3+18 x^2-42 x+1\,,\nonumber\\
P_{1065} &= x^4-28 x^3+6 x^2-28 x+1\,,\nonumber\\
P_{1066} &= x^4-20 x^3+6 x^2-20 x+1\,,\nonumber\\
P_{1067} &= x^4+2 x^3+4 x^2+2 x+1\,,\nonumber\\
P_{1068} &= x^4+2 x^3+6 x^2+2 x+1\,,\nonumber\\
P_{1069} &= x^4+2 x^3+8 x^2+2 x+1\,,\nonumber\\
P_{1070} &= x^4+2 x^3+14 x^2+2 x+1\,,\nonumber\\
P_{1071} &= x^4+2 x^3+35 x^2+2 x+1\,,\nonumber\\
P_{1072} &= x^4+4 x^3-50 x^2+4 x+1\,,\nonumber\\
P_{1073} &= x^4+4 x^3-26 x^2+4 x+1\,,\nonumber\\
P_{1074} &= x^4+4 x^3-18 x^2+4 x+1\,,\nonumber\\
P_{1075} &= x^4+4 x^3+22 x^2+4 x+1\,,\nonumber\\
P_{1076} &= x^4+4 x^3+38 x^2+4 x+1\,,\nonumber\\
P_{1077} &= x^4+4 x^3+46 x^2+4 x+1\,,\nonumber\\
P_{1078} &= x^4+4 x^3+102 x^2+4 x+1\,,\nonumber\\
P_{1079} &= 2 x^4-194 x^3-113 x^2-2 x+11\,,\nonumber\\
P_{1080} &= 2 x^4-19 x^3+26 x^2+61 x-18\,,\nonumber\\
P_{1081} &= 2 x^4+2 x^3+7 x^2+2 x+2\,,\nonumber\\
P_{1082} &= 2 x^4+4 x^3-x^2+4 x+2\,,\nonumber\\
P_{1083} &= 2 x^4+6 x^3-11 x^2+15 x-4\,,\nonumber\\
P_{1084} &= 3 x^4-20 x^3+2 x^2-20 x+3\,,\nonumber\\
P_{1085} &= 3 x^4+4 x^3+20 x^2+4 x+3\,,\nonumber\\
P_{1086} &= 3 x^4+6 x^3+x^2+6 x+3\,,\nonumber\\
P_{1087} &= 3 x^4+12 x^3-206 x^2+12 x+3\,,\nonumber\\
P_{1088} &= 3 x^4+12 x^3-110 x^2+12 x+3\,,\nonumber\\
P_{1089} &= 3 x^4+12 x^3-62 x^2+12 x+3\,,\nonumber\\
P_{1090} &= 3 x^4+12 x^3-38 x^2+12 x+3\,,\nonumber\\
P_{1091} &= 3 x^4+12 x^3-14 x^2+12 x+3\,,\nonumber\\
P_{1092} &= 3 x^4+12 x^3+2 x^2+12 x+3\,,\nonumber\\
P_{1093} &= 3 x^4+12 x^3+34 x^2+12 x+3\,,\nonumber\\
P_{1094} &= 4 x^4-15 x^3-2 x^2-15 x+4\,,\nonumber\\
P_{1095} &= 4 x^4-10 x^3+73 x^2-10 x+4\,,\nonumber\\
P_{1096} &= 4 x^4+9 x^3+6 x^2+9 x+4\,,\nonumber\\
P_{1097} &= 5 x^4-62 x^3+74 x^2-70 x+5\,,\nonumber\\
P_{1098} &= 5 x^4-54 x^3+2 x^2-54 x+5\,,\nonumber\\
P_{1099} &= 5 x^4-31 x^3+4 x^2-31 x+5\,,\nonumber\\
P_{1100} &= 5 x^4-16 x^3+6 x^2-16 x+5\,,\nonumber\\
P_{1101} &= 5 x^4-12 x^3+14 x^2-4 x+1\,,\nonumber\\
P_{1102} &= 5 x^4+10 x^3+x^2+10 x+5\,,\nonumber\\
P_{1103} &= 5 x^4+10 x^3+17 x^2+10 x+5\,,\nonumber\\
P_{1104} &= 5 x^4+10 x^3+43 x^2+10 x+5\,,\nonumber\\
P_{1105} &= 5 x^4+17 x^3-22 x^2+17 x+7\,,\nonumber\\
P_{1106} &= 5 x^4+20 x^3-386 x^2+20 x+5\,,\nonumber\\
P_{1107} &= 5 x^4+20 x^3-162 x^2+20 x+5\,,\nonumber\\
P_{1108} &= 5 x^4+20 x^3-138 x^2+20 x+5\,,\nonumber\\
P_{1109} &= 5 x^4+20 x^3-66 x^2+20 x+5\,,\nonumber\\
P_{1110} &= 5 x^4+20 x^3+46 x^2+20 x+5\,,\nonumber\\
P_{1111} &= 5 x^4+20 x^3+254 x^2+20 x+5\,,\nonumber\\
P_{1112} &= 6 x^4-17 x^3+x^2-14 x-8\,,\nonumber\\
P_{1113} &= 6 x^4-9 x^3+17 x^2-2 x+4\,,\nonumber\\
P_{1114} &= 7 x^4-146 x^3+54 x^2-146 x+7\,,\nonumber\\
P_{1115} &= 7 x^4-28 x^3+10 x^2-28 x+7\,,\nonumber\\
P_{1116} &= 7 x^4+28 x^3-214 x^2+28 x+7\,,\nonumber\\
P_{1117} &= 7 x^4+28 x^3-6 x^2+28 x+7\,,\nonumber\\
P_{1118} &= 7 x^4+50 x^3+146 x^2+50 x+7\,,\nonumber\\
P_{1119} &= 7 x^4+482 x^3+1010 x^2+482 x+7\,,\nonumber\\
P_{1120} &= 9 x^4-230 x^3+90 x^2-230 x+9\,,\nonumber\\
P_{1121} &= 9 x^4-40 x^3+14 x^2-40 x+9\,,\nonumber\\
P_{1122} &= 9 x^4+58 x^3-2 x^2+58 x+5\,,\nonumber\\
P_{1123} &= 11 x^4-146 x^3+82 x^2-146 x+7\,,\nonumber\\
P_{1124} &= 11 x^4-2 x^3+14 x^2-2 x+11\,,\nonumber\\
P_{1125} &= 11 x^4+6 x^3+77 x^2+6 x+11\,,\nonumber\\
P_{1126} &= 11 x^4+22 x^3-82 x^2+22 x+11\,,\nonumber\\
P_{1127} &= 11 x^4+44 x^3-126 x^2+44 x+11\,,\nonumber\\
P_{1128} &= 11 x^4+44 x^3-14 x^2+44 x+11\,,\nonumber\\
P_{1129} &= 11 x^4+44 x^3+102 x^2+44 x+11\,,\nonumber\\
P_{1130} &= 13 x^4-50 x^3+82 x^2-66 x+5\,,\nonumber\\
P_{1131} &= 13 x^4+32 x^3+12 x^2+32 x+13\,,\nonumber\\
P_{1132} &= 13 x^4+52 x^3+14 x^2+52 x+13\,,\nonumber\\
P_{1133} &= 13 x^4+52 x^3+46 x^2+52 x+13\,,\nonumber\\
P_{1134} &= 14 x^4+22 x^3+57 x^2+22 x+14\,,\nonumber\\
P_{1135} &= 15 x^4-60 x^3+66 x^2+4 x-17\,,\nonumber\\
P_{1136} &= 17 x^4-254 x^3+232 x^2-18 x+7\,,\nonumber\\
P_{1137} &= 17 x^4-64 x^3-188 x^2+256 x+15\,,\nonumber\\
P_{1138} &= 21 x^4+185 x^3+131 x^2+73 x-16\,,\nonumber\\
P_{1139} &= 23 x^4+102 x^3+10 x^2+102 x+19\,,\nonumber\\
P_{1140} &= 25 x^4-20 x^3+56 x^2-148 x-33\,,\nonumber\\
P_{1141} &= 25 x^4+24 x^3+148 x^2-200 x-21\,,\nonumber\\
P_{1142} &= 26 x^4-73 x^3+51 x^2-15 x+3\,,\nonumber\\
P_{1143} &= 33 x^4-344 x^3-1006 x^2+616 x+89\,,\nonumber\\
P_{1144} &= 37 x^4-16 x^3-98 x^2+48 x+5\,,\nonumber\\
P_{1145} &= 39 x^4-4 x^3+104 x^2-164 x-27\,,\nonumber\\
P_{1146} &= 47 x^4+268 x^3-263 x^2+40 x+38\,,\nonumber\\
P_{1147} &= 53 x^4+212 x^3-322 x^2+212 x+53\,,\nonumber\\
P_{1148} &= 55 x^4-138 x^3+70 x^2-138 x+55\,,\nonumber\\
P_{1149} &= 55 x^4+220 x^3-86 x^2+220 x+55\,,\nonumber\\
P_{1150} &= 67 x^4-98 x^3+88 x^2-2 x-11\,,\nonumber\\
P_{1151} &= 75 x^4-50 x^3-98 x^2-50 x+75\,,\nonumber\\
P_{1152} &= 87 x^4+4 x^3+206 x^2+388 x+35\,,\nonumber\\
P_{1153} &= 95 x^4-404 x^3+496 x^2+820 x-103\,,\nonumber\\
P_{1154} &= 115 x^4+100 x^3+116 x^2-364 x-43\,,\nonumber\\
P_{1155} &= 124 x^4+570 x^3+801 x^2-1158 x+325\,,\nonumber\\
P_{1156} &= 133 x^4+216 x^3+520 x^2-440 x-61\,,\nonumber\\
P_{1157} &= 233 x^4-388 x^3-52 x^2+56 x+55\,,\nonumber\\
P_{1158} &= 379 x^4+263 x^3+569 x^2+515 x-134\,,\nonumber\\
P_{1159} &= 683 x^4-672 x^3-1138 x^2+744 x+239\,,\nonumber\\
P_{1160} &= 895 x^4+3806 x^3+2288 x^2-2878 x-1007\,,\nonumber\\
P_{1161} &= -117 x^5-117 x^4+80 x^3-628 x^2-157 x-157\,,\nonumber\\
P_{1162} &= -71 x^5-71 x^4-32 x^3-116 x^2-3 x-3\,,\nonumber\\
P_{1163} &= x^5-73 x^4-216 x^3+72 x^2-73 x+1\,,\nonumber\\
P_{1164} &= x^5-17 x^4+96 x^3-80 x^2+25 x+7\,,\nonumber\\
P_{1165} &= x^5-16 x^4+15 x^3-9 x^2-16 x+1\,,\nonumber\\
P_{1166} &= x^5-13 x^4-100 x^3+172 x^2-13 x+1\,,\nonumber\\
P_{1167} &= x^5-13 x^4+180 x^3-204 x^2-13 x+1\,,\nonumber\\
P_{1168} &= x^5-7 x^4+138 x^3-150 x^2-7 x+1\,,\nonumber\\
P_{1169} &= x^5-5 x^4-156 x^3+100 x^2-5 x+1\,,\nonumber\\
P_{1170} &= x^5-3 x^4+22 x^3-18 x^2+5 x+1\,,\nonumber\\
P_{1171} &= x^5-x^4+12 x^3-8 x^2+3 x+1\,,\nonumber\\
P_{1172} &= x^5+x^4+5 x^3-7 x^2-2 x-2\,,\nonumber\\
P_{1173} &= x^5+x^4+16 x^3-8 x^2+3 x+3\,,\nonumber\\
P_{1174} &= x^5+2 x^4-3 x^3+7 x^2+1\,,\nonumber\\
P_{1175} &= x^5+3 x^4-8 x^3+12 x^2-x+1\,,\nonumber\\
P_{1176} &= x^5+5 x^4-21 x^3+5 x^2-13 x-9\,,\nonumber\\
P_{1177} &= x^5+5 x^4-2 x^3+14 x^2+x+5\,,\nonumber\\
P_{1178} &= x^5+9 x^4-28 x^3+24 x^2-11 x-3\,,\nonumber\\
P_{1179} &= x^5+11 x^4+6 x^3+34 x^2+9 x+19\,,\nonumber\\
P_{1180} &= x^5+14 x^4-57 x^3+47 x^2-19 x-6\,,\nonumber\\
P_{1181} &= x^5+21 x^4-58 x^3+54 x^2-23 x-3\,,\nonumber\\
P_{1182} &= x^5+35 x^4-27 x^3+271 x^2+87 x+121\,,\nonumber\\
P_{1183} &= 2 x^5-41 x^4-141 x^3+147 x^2-41 x+2\,,\nonumber\\
P_{1184} &= 2 x^5+x^4+9 x^3-7 x^2-1\,,\nonumber\\
P_{1185} &= 2 x^5+2 x^4-11 x^3+7 x^2-4 x-4\,,\nonumber\\
P_{1186} &= 2 x^5+2 x^4+5 x^3+3 x^2+2 x+2\,,\nonumber\\
P_{1187} &= 2 x^5+2 x^4+5 x^3+11 x^2+6 x+6\,,\nonumber\\
P_{1188} &= 2 x^5+2 x^4+7 x^3-5 x^2-x-1\,,\nonumber\\
P_{1189} &= 2 x^5+2 x^4+15 x^3-19 x^2-4 x-4\,,\nonumber\\
P_{1190} &= 2 x^5+4 x^4+x^3+23 x^2+8 x+10\,,\nonumber\\
P_{1191} &= 2 x^5+17 x^4-56 x^3+42 x^2-24 x-9\,,\nonumber\\
P_{1192} &= 2 x^5+20 x^4-51 x^3+11 x^2-40 x-22\,,\nonumber\\
P_{1193} &= 2 x^5+34 x^4-39 x^3-145 x^2-77 x+1\,,\nonumber\\
P_{1194} &= 3 x^5-17 x^4+54 x^3-50 x^2+19 x-1\,,\nonumber\\
P_{1195} &= 3 x^5-3 x^4+15 x^3-11 x^2+5 x-1\,,\nonumber\\
P_{1196} &= 3 x^5-3 x^4+22 x^3-14 x^2+7 x+1\,,\nonumber\\
P_{1197} &= 3 x^5-x^4+35 x^3+13 x^2+25 x+21\,,\nonumber\\
P_{1198} &= 3 x^5+3 x^4+5 x^3-7 x^2-4 x-4\,,\nonumber\\
P_{1199} &= 3 x^5+7 x^4-14 x^3+26 x^2-x+3\,,\nonumber\\
P_{1200} &= 3 x^5+9 x^4-10 x^3+26 x^2-x+5\,,\nonumber\\
P_{1201} &= 3 x^5+129 x^4-40 x^3-176 x^2-271 x+35\,,\nonumber\\
P_{1202} &= 4 x^5+4 x^4-x^3-15 x^2-12 x-12\,,\nonumber\\
P_{1203} &= 5 x^5-x^4+56 x^3-72 x^2-7 x-13\,,\nonumber\\
P_{1204} &= 5 x^5+x^4+30 x^3-34 x^2-3 x-7\,,\nonumber\\
P_{1205} &= 5 x^5+5 x^4-7 x^3+11 x^2-3 x-3\,,\nonumber\\
P_{1206} &= 5 x^5+5 x^4-x^3-3 x^2-7 x-7\,,\nonumber\\
P_{1207} &= 5 x^5+5 x^4+16 x^3-8 x^2-x-1\,,\nonumber\\
P_{1208} &= 5 x^5+5 x^4+17 x^3-25 x^2-9 x-9\,,\nonumber\\
P_{1209} &= 5 x^5+7 x^4-47 x^3+11 x^2-25 x-23\,,\nonumber\\
P_{1210} &= 6 x^5-44 x^4-27 x^3+47 x^2-11 x-3\,,\nonumber\\
P_{1211} &= 6 x^5+6 x^4+11 x^3-3 x^2-2 x-2\,,\nonumber\\
P_{1212} &= 7 x^5-133 x^4-234 x^3+630 x^2-133 x+7\,,\nonumber\\
P_{1213} &= 7 x^5-25 x^4+96 x^3-92 x^2+27 x-5\,,\nonumber\\
P_{1214} &= 7 x^5-x^4+22 x^3-10 x^2+7 x-1\,,\nonumber\\
P_{1215} &= 7 x^5+2 x^4+12 x^3-2 x^2+3 x-2\,,\nonumber\\
P_{1216} &= 7 x^5+5 x^4+17 x^3-5 x^2+x-1\,,\nonumber\\
P_{1217} &= 7 x^5+7 x^4-20 x^3+16 x^2-9 x-9\,,\nonumber\\
P_{1218} &= 7 x^5+7 x^4-16 x^3+20 x^2-5 x-5\,,\nonumber\\
P_{1219} &= 7 x^5+7 x^4+68 x^3-4 x^2+25 x+25\,,\nonumber\\
P_{1220} &= 8 x^5-36 x^4+191 x^3-155 x^2+54 x+10\,,\nonumber\\
P_{1221} &= 8 x^5+8 x^4+11 x^3-15 x^2-10 x-10\,,\nonumber\\
P_{1222} &= 9 x^5+9 x^4+10 x^3-4 x^2-6 x-6\,,\nonumber\\
P_{1223} &= 10 x^5-9 x^3+65 x^2+31 x-1\,,\nonumber\\
P_{1224} &= 10 x^5+14 x^4-27 x^3+31 x^2-12 x-8\,,\nonumber\\
P_{1225} &= 10 x^5+14 x^4-21 x^3+25 x^2-12 x-8\,,\nonumber\\
P_{1226} &= 10 x^5+197 x^4-200 x^3-442 x^2+54 x-51\,,\nonumber\\
P_{1227} &= 11 x^5-131 x^4+90 x^3-58 x^2-197 x+29\,,\nonumber\\
P_{1228} &= 11 x^5-25 x^4+173 x^3-141 x^2+41 x+5\,,\nonumber\\
P_{1229} &= 11 x^5+11 x^4+83 x^3-19 x^2+21 x+21\,,\nonumber\\
P_{1230} &= 11 x^5+43 x^4-112 x^3+116 x^2-41 x-9\,,\nonumber\\
P_{1231} &= 11 x^5+245 x^4+736 x^3-384 x^2+261 x-37\,,\nonumber\\
P_{1232} &= 12 x^5+12 x^4+25 x^3+43 x^2+22 x+22\,,\nonumber\\
P_{1233} &= 12 x^5+44 x^4-45 x^3+33 x^2-50 x-18\,,\nonumber\\
P_{1234} &= 13 x^5-81 x^4-280 x^3+136 x^2-57 x+13\,,\nonumber\\
P_{1235} &= 13 x^5+x^4+65 x^3-57 x^2+3 x-9\,,\nonumber\\
P_{1236} &= 13 x^5+13 x^4-35 x^3+67 x^2+3 x+3\,,\nonumber\\
P_{1237} &= 13 x^5+13 x^4-7 x^3+11 x^2-11 x-11\,,\nonumber\\
P_{1238} &= 13 x^5+13 x^4+19 x^3-27 x^2-17 x-17\,,\nonumber\\
P_{1239} &= 13 x^5+13 x^4+83 x^3-19 x^2+19 x+19\,,\nonumber\\
P_{1240} &= 13 x^5+113 x^4-330 x^3+286 x^2-135 x-35\,,\nonumber\\
P_{1241} &= 13 x^5+113 x^4-322 x^3+238 x^2-155 x-55\,,\nonumber\\
P_{1242} &= 14 x^5+14 x^4-x^3+13 x^2-8 x-8\,,\nonumber\\
P_{1243} &= 15 x^5+15 x^4+26 x^3-22 x^2-13 x-13\,,\nonumber\\
P_{1244} &= 15 x^5+27 x^4-53 x^3+57 x^2-25 x-13\,,\nonumber\\
P_{1245} &= 16 x^5-7 x^4-101 x^3+55 x^2-15 x+4\,,\nonumber\\
P_{1246} &= 16 x^5+16 x^4+27 x^3-19 x^2-12 x-12\,,\nonumber\\
P_{1247} &= 17 x^5+7 x^4+84 x^3-52 x^2+9 x-1\,,\nonumber\\
P_{1248} &= 17 x^5+17 x^4-20 x^3+52 x^2-x-1\,,\nonumber\\
P_{1249} &= 17 x^5+17 x^4+54 x^3-56 x^2-18 x-18\,,\nonumber\\
P_{1250} &= 17 x^5+37 x^4-78 x^3+66 x^2-43 x-23\,,\nonumber\\
P_{1251} &= 17 x^5+41 x^4-162 x^3+150 x^2+25 x-7\,,\nonumber\\
P_{1252} &= 17 x^5+43 x^4-87 x^3+83 x^2-45 x-19\,,\nonumber\\
P_{1253} &= 17 x^5+135 x^4+134 x^3+154 x^2-39 x-17\,,\nonumber\\
P_{1254} &= 18 x^5+8 x^4+21 x^3-9 x^2-2 x-12\,,\nonumber\\
P_{1255} &= 18 x^5+18 x^4+7 x^3-11 x^2-20 x-20\,,\nonumber\\
P_{1256} &= 19 x^5+19 x^4-80 x^3+60 x^2-29 x-29\,,\nonumber\\
P_{1257} &= 19 x^5+19 x^4+43 x^3-51 x^2-23 x-23\,,\nonumber\\
P_{1258} &= 19 x^5+19 x^4+54 x^3-42 x^2-13 x-13\,,\nonumber\\
P_{1259} &= 19 x^5+19 x^4+68 x^3-4 x^2+13 x+13\,,\nonumber\\
P_{1260} &= 19 x^5+31 x^4-170 x^3+142 x^2+15 x-5\,,\nonumber\\
P_{1261} &= 19 x^5+67 x^4-192 x^3+116 x^2-105 x-57\,,\nonumber\\
P_{1262} &= 19 x^5+4963 x^4+5538 x^3-1062 x^2-781 x-37\,,\nonumber\\
P_{1263} &= 20 x^5+20 x^4+95 x^3+53 x^2+54 x+54\,,\nonumber\\
P_{1264} &= 20 x^5+20 x^4+263 x^3-311 x^2-44 x-44\,,\nonumber\\
P_{1265} &= 20 x^5+49 x^4-87 x^3+89 x^2-48 x-19\,,\nonumber\\
P_{1266} &= 21 x^5+21 x^4-20 x^3+40 x^2-11 x-11\,,\nonumber\\
P_{1267} &= 21 x^5+21 x^4+14 x^3-6 x^2-17 x-17\,,\nonumber\\
P_{1268} &= 21 x^5+21 x^4+48 x^3-56 x^2-25 x-25\,,\nonumber\\
P_{1269} &= 22 x^5+22 x^4+61 x^3-57 x^2-20 x-20\,,\nonumber\\
P_{1270} &= 22 x^5+22 x^4+95 x^3+53 x^2+52 x+52\,,\nonumber\\
P_{1271} &= 23 x^5-52 x^4+50 x^3-74 x^2+43 x-6\,,\nonumber\\
P_{1272} &= 23 x^5+14 x^4+109 x^3-35 x^2+23 x+14\,,\nonumber\\
P_{1273} &= 23 x^5+23 x^4-152 x^3+196 x^2-x-1\,,\nonumber\\
P_{1274} &= 23 x^5+113 x^4+300 x^3-156 x^2+53 x-13\,,\nonumber\\
P_{1275} &= 24 x^5+24 x^4+123 x^3-119 x^2-22 x-22\,,\nonumber\\
P_{1276} &= 25 x^5-329 x^4-524 x^3+164 x^2-233 x+1\,,\nonumber\\
P_{1277} &= 25 x^5+25 x^4-16 x^3+20 x^2-23 x-23\,,\nonumber\\
P_{1278} &= 25 x^5+25 x^4-4 x^3+8 x^2-23 x-23\,,\nonumber\\
P_{1279} &= 25 x^5+25 x^4+56 x^3+20 x^2+13 x+13\,,\nonumber\\
P_{1280} &= 27 x^5+27 x^4-12 x^3-32 x^2-49 x-49\,,\nonumber\\
P_{1281} &= 27 x^5+27 x^4+52 x^3+16 x^2+7 x+7\,,\nonumber\\
P_{1282} &= 29 x^5+29 x^4-104 x^3+100 x^2-31 x-31\,,\nonumber\\
P_{1283} &= 31 x^5+31 x^4-104 x^3+100 x^2-33 x-33\,,\nonumber\\
P_{1284} &= 31 x^5+31 x^4-32 x^3+44 x^2-25 x-25\,,\nonumber\\
P_{1285} &= 31 x^5+31 x^4-30 x^3+26 x^2-33 x-33\,,\nonumber\\
P_{1286} &= 31 x^5+31 x^4+28 x^3+8 x^2-13 x-13\,,\nonumber\\
P_{1287} &= 32 x^5+32 x^4+223 x^3-251 x^2-46 x-46\,,\nonumber\\
P_{1288} &= 34 x^5+34 x^4+31 x^3+13 x^2-12 x-12\,,\nonumber\\
P_{1289} &= 35 x^5+35 x^4+106 x^3-86 x^2-25 x-25\,,\nonumber\\
P_{1290} &= 37 x^5+5 x^4+243 x^3-179 x^2+27 x-5\,,\nonumber\\
P_{1291} &= 38 x^5+38 x^4+229 x^3-209 x^2-28 x-28\,,\nonumber\\
P_{1292} &= 38 x^5+55 x^4+131 x^3-85 x^2+9 x-4\,,\nonumber\\
P_{1293} &= 39 x^5+189 x^4+258 x^3+110 x^2+119 x+53\,,\nonumber\\
P_{1294} &= 40 x^5+40 x^4+87 x^3-3 x^2+2 x+2\,,\nonumber\\
P_{1295} &= 42 x^5+42 x^4+59 x^3-27 x^2-26 x-26\,,\nonumber\\
P_{1296} &= 42 x^5+48 x^4+59 x^3+29 x^2-4 x+2\,,\nonumber\\
P_{1297} &= 43 x^5+27 x^4-48 x^3+140 x^2+19 x+3\,,\nonumber\\
P_{1298} &= 43 x^5+43 x^4+96 x^3+20 x^2+15 x+15\,,\nonumber\\
P_{1299} &= 45 x^5+101 x^4-211 x^3+107 x^2+16 x+6\,,\nonumber\\
P_{1300} &= 50 x^5+98 x^4+17 x^3+59 x^2-60 x-12\,,\nonumber\\
P_{1301} &= 51 x^5+51 x^4-130 x^3+110 x^2-61 x-61\,,\nonumber\\
P_{1302} &= 55 x^5+211 x^4-576 x^3+732 x^2-91 x+53\,,\nonumber\\
P_{1303} &= 56 x^5-48 x^4+425 x^3-573 x^2-26 x-130\,,\nonumber\\
P_{1304} &= 65 x^5+217 x^4-264 x^3+308 x^2-195 x-43\,,\nonumber\\
P_{1305} &= 67 x^5+67 x^4+134 x^3-94 x^2-47 x-47\,,\nonumber\\
P_{1306} &= 74 x^5-99 x^4-152 x^3+192 x^2-70 x+7\,,\nonumber\\
P_{1307} &= 77 x^5-35 x^4+516 x^3-368 x^2+109 x-3\,,\nonumber\\
P_{1308} &= 96 x^5+96 x^4+437 x^3-317 x^2-36 x-36\,,\nonumber\\
P_{1309} &= 98 x^5+98 x^4+495 x^3-407 x^2-54 x-54\,,\nonumber\\
P_{1310} &= 119 x^5+119 x^4+203 x^3-67 x^2-51 x-51\,,\nonumber\\
P_{1311} &= 139 x^5+207 x^4+169 x^3+95 x^2-75 x-7\,,\nonumber\\
P_{1312} &= 141 x^5+241 x^4+79 x^3+177 x^2-113 x-13\,,\nonumber\\
P_{1313} &= 147 x^5-63 x^4+894 x^3-550 x^2+203 x-55\,,\nonumber\\
P_{1314} &= 181 x^5+113 x^4+420 x^3-176 x^2+9 x-59\,,\nonumber\\
P_{1315} &= 202 x^5+1240 x^4-3283 x^3+679 x^2+583 x-157\,,\nonumber\\
P_{1316} &= 649 x^5+377 x^4+1488 x^3-84 x^2+325 x+53\,,\nonumber\\
P_{1317} &= 5 x^6-530 x^5-3241 x^4-4744 x^3-1585 x^2+298 x+5\,,\nonumber\\
P_{1318} &= 5 x^6-386 x^5-1009 x^4-640 x^3+503 x^2+370 x+5\,,\nonumber\\
P_{1319} &= 5 x^6-224 x^5-577 x^4-604 x^3-73 x^2+28 x+5\,,\nonumber\\
P_{1320} &= 5 x^6-170 x^5-19 x^4+314 x^3+305 x^2-8 x+5\,,\nonumber\\
P_{1321} &= 5 x^6-116 x^5+323 x^4+728 x^3+323 x^2-116 x+5\,,\nonumber\\
P_{1322} &= 5 x^6-44 x^5+755 x^4+1304 x^3+467 x^2-188 x+5\,,\nonumber\\
P_{1323} &= 5 x^6+40 x^5-157 x^4-640 x^3-349 x^2-56 x+5\,,\nonumber\\
P_{1324} &= 8 x^6+25 x^5-100 x^4-154 x^3-100 x^2+25 x+8\,,\nonumber\\
P_{1325} &= 15 x^6-183 x^5-1323 x^4-1604 x^3-121 x^2+43 x+5\,,\nonumber\\
P_{1326} &= 123 x^6-492 x^5-2159 x^4-1820 x^3-1223 x^2-24 x+123 \,.
\end{align}

\newpage

\section{The pseudo-scalar form factors up to two-loop} \label{app:pFF}

The non-singlet part of the pseudo-scalar form factor can be obtained using Eq.~(\ref{eq:cwiFF}).
The unrenormalized singlet contribution is given up to ${\cal O}(\varepsilon)$ by
\begin{dmath}
 \hat{F}_{P}^{(2),\sing} = C_F T_F \bigg[
\Bigg\{
        \eta ^2 \big(
                8 x (1+x)^2 H_0^2
                +32 x (1+x)^2 \zeta_2
        \big)
        +\eta ^3 \big(
                \frac{32}{3} x^2 (1+x)^2 H_0^3
                +64 x^2 (1+x)^2 H_0 \zeta_2
        \big)
        +\eta  \big(
                -\frac{1}{3} x H_0^4
                +\frac{16}{3} x H_0^3 H_1
                -64 x H_0 H_{0,0,1}
                +112 x H_{0,0,0,1}
                +\big(
                        -8 x H_0^2
                        +32 x H_0 H_1
                        -32 x H_{0,1}
                \big) \zeta_2
                -\frac{64}{5} x \zeta_2^2
                -48 x H_0 \zeta_3
        \big)
\Bigg\}
+ \ep  \Bigg\{
        \xi  \eta ^3 \big(
                -(-1+x) x (1+x)^3 H_0^4
                +32 (-1+x) x (1+x)^3 H_0^2 H_{0,1}
        \big)
        +\eta ^3 \big(
                \frac{32}{5} x (1+x)^2 \big(
                        7-13 x+7 x^2\big) \zeta_2^2
                -\frac{8}{3} x (1+x)^2 \big(
                        -3+11 x^2\big) H_0^3
                +\frac{64}{3} x^2 (1+x)^2 H_{-1} H_0^3
                +6 x^2 (1+x)^2 H_0^4
                -\frac{16}{3} x (1+x)^2 \big(
                        5-6 x+5 x^2\big) H_0^3 H_1
                +128 x^2 (1+x)^2 H_0^2 H_{0,1}
                -128 x^2 (1+x)^2 H_0^2 H_{0,-1}
                -16 x (1+x)^2 \big(
                        53-90 x+53 x^2\big) H_{0,0,0,1}
                +128 x (1+x)^2 \big(
                        3-x+3 x^2\big) H_{0,0,0,-1}
                +\big(
                        -16 x (1+x)^2 (-1+3 x) (1+3 x) H_0
                        +128 x^2 (1+x)^2 H_{-1} H_0
                        -32 x (1+x)^2 \big(
                                5-6 x+5 x^2\big) H_0 H_1
                        +32 x (1+x)^2 \big(
                                5-6 x+5 x^2\big) H_{0,1}
                        -512 x^2 (1+x)^2 H_{0,-1}
                \big) \zeta_2
        \big)
        +\eta ^2 \big(
                16 x (1+x)^2 H_0^2
                -32 x (1+x)^2 H_{-1} H_0^2
                +16 x (1+x)^2 H_0^2 H_1
                -32 x (1+x)^2 H_0 H_{0,-1}
                -32 x (1+x)^2 H_{0,0,1}
                +128 x (1+x)^2 H_{0,0,-1}
                -192 x (1+x)^2 H_{-1} \zeta_2
                +48 x (1+x)^2 \zeta_3
        \big)
        +\eta  \big(
                -\frac{2}{3} x H_{-1} H_0^4
                -\frac{4}{15} x H_0^5
                +\big(
                        \frac{32}{3} x H_{-1} H_0^3
                        +\frac{10}{3} x H_0^4
                \big) H_1
                -\frac{16}{3} x H_0^3 H_1^2
                +\big(
                        -16 x H_0^3
                        +64 x H_0 H_{0,-1}
                \big) H_{0,1}
                -64 x H_0 H_{0,1}^2
                +\big(
                        8 x H_0^3
                        -64 x H_0^2 H_1
                \big) H_{0,-1}
                +\big(
                        \big(
                                448 x
                                -128 x H_{-1}
                        \big) H_0
                        -32 x H_0^2
                        +128 x H_0 H_1
                        +192 x H_{0,1}
                        -96 x H_{0,-1}
                \big) H_{0,0,1}
                +\big(
                        -128 x H_0
                        -32 x H_0^2
                        -128 x H_0 H_1
                        -192 x H_{0,1}
                \big) H_{0,0,-1}
                +32 x H_0^2 H_{0,1,1}
                +32 x H_0^2 H_{0,1,-1}
                +32 x H_0^2 H_{0,-1,1}
                +\big(
                        352 x H_0
                        -416 x H_1
                        +224 x H_{-1}
                \big) H_{0,0,0,1}
                +\big(
                        -160 x H_0
                        +704 x H_1
                \big) H_{0,0,0,-1}
                -128 x H_0 H_{0,0,1,1}
                +256 x H_0 H_{0,0,1,-1}
                +256 x H_0 H_{0,0,-1,1}
                +64 x H_0 H_{0,-1,0,1}
                -880 x H_{0,0,0,0,1}
                +848 x H_{0,0,0,0,-1}
                -160 x H_{0,0,0,1,1}
                -320 x H_{0,0,0,1,-1}
                -320 x H_{0,0,0,-1,1}
                -192 x H_{0,0,1,0,1}
                -32 x H_{0,0,1,0,-1}
                +128 x H_{0,0,-1,0,1}
                +\big(
                        8 x H_0^2
                        -16 x H_{-1} H_0^2
                        -4 x H_0^3
                        +\big(
                                64 x H_{-1} H_0
                                -32 x H_0^2
                        \big) H_1
                        -32 x H_0 H_1^2
                        +\big(
                                -96 x H_0
                                +64 x H_1
                                -64 x H_{-1}
                        \big) H_{0,1}
                        +\big(
                                64 x H_0
                                -256 x H_1
                        \big) H_{0,-1}
                        +224 x H_{0,0,1}
                        -256 x H_{0,0,-1}
                        -64 x H_{0,1,1}
                        +256 x H_{0,1,-1}
                        +256 x H_{0,-1,1}
                        -16 x \zeta_3
                \big) \zeta_2
                +\big(
                        8 x H_0
                        -16 x H_1
                        -\frac{128}{5} x H_{-1}
                \big) \zeta_2^2
                +\big(
                        208 x H_0
                        -96 x H_{-1} H_0
                        -52 x H_0^2
                        -32 x H_0 H_1
                        -160 x H_{0,1}
                        +96 x H_{0,-1}
                \big) \zeta_3
                +176 x \zeta_5
        \big)
\Bigg\}
\Bigg] \,.
\end{dmath}

\providecommand{\href}[2]{#2}\begingroup\raggedright

\endgroup
\end{document}